\documentclass[12pt]{article}
\usepackage[a4paper]{geometry}
\usepackage{amsfonts,amssymb,amsmath,epsfig,enumerate,cases,lscape,booktabs}
\usepackage[all]{xy}
\DeclareMathAlphabet{\mathpzc}{OT1}{pzc}{m}{it}
\usepackage{ulem}
\usepackage{multirow}
\DeclareMathAlphabet{\mathpzc}{OT1}{pzc}{m}{it} 
\advance\oddsidemargin -1cm
\renewcommand{\thesection}{\Roman{section}}
\renewcommand{\theequation}{\thesection.\arabic{equation}}
\numberwithin{equation}{section}
\begin{document}
\newcommand{\HRule}{\rule{\linewidth}{0.1mm}}
\newcommand{\as}{\alpha_{\textrm{s}}\,}
\newcommand{\ds}{\displaystyle}
\newcommand{\scr}{\scriptstyle}
\newcommand{\aB}{a_{\textrm{B}}\,}
\newcommand{\Bstar}{ \overset{*\!\!}{B} }
\newcommand{\Gstar}{ \overset{*\!\!}{G} }
\newcommand{\lP}{l_{\wp}}
\newcommand{\nP}{n_{\wp}}
\newcommand{\SP}{S_\wp}
\newcommand{\GG}{\textnormal I \! \Gamma}
\newcommand{\A}{\mathcal{A}}
\newcommand{\B}{\mathcal{B}}
\newcommand{\D}{\mathcal{D}}
\newcommand{\F}{\mathcal{F}}
\newcommand{\M}{\mathcal{M}}
\newcommand{\vr}{\vec{r}}
\newcommand{\sdot}{\,{\scriptscriptstyle{}^{\bullet}}\,}
\newcommand{\R}{\mathcal{R}}
\newcommand{\J}{\mathcal{J}}
\renewcommand{\S}{\mathcal{S}}
\newcommand{\Z}{\mathcal{Z}}
\newcommand{\ZP}{\mathcal{Z}_{\wp}}

\newcommand{\boxes}{\rule{2mm}{2mm}\ \rule{2mm}{2mm}\ \rule{2mm}{2mm} }
\newcommand{\points}{$\bullet \bullet \bullet$}
\newcommand{\cross}{\textbf{xxx}}
\newcommand{\solid}{\rule[1mm]{4.0mm}{0.5mm}}
% [p]                                                
%    A   eckige Klammer links oben
\newcommand{\eklo}[2]{\, {}^{[\mathrm{#1}]}{\!#2}}

% (p)
%    A                                                                 
% ^(arg1) arg2  *r*unde *k*lammer *l*inks *o*ben  zB fürs Potential     
\newcommand{\rklo}[2]{\, {}^{(#1)}{ {#2} }}
%\newcommand{\rklo}[2]{\, {}^{({#1})}{\! {#2} }}

%	 [p]                                                
%      A_blah  eckige Klammer links oben rechts unten roman index
%
\newcommand{\ekloi}[3]{\, {}^{[\mathrm{#1}]}{#2}_\mathrm{#3}}

%	 (p)                                                
%      A_blah  runde Klammer links oben rechts unten roman index
%
\newcommand{\rkloi}[3]{\, {}^{(\mathrm{#1})}{#2}_\mathrm{#3}}

% X_[y]
%
% e*ckige *K*lammer *r*echts *u*nten       
\newcommand{\ekru}[2]{#1_\mathrm{[#2]}}

% ~X_[y]
%
% t*ilde e*ckige *K*lammer *r*echts *u*nten       
\newcommand{\tekru}[2]{\tilde{#1}_\mathrm{[#2]}}

%           (a)
%  mathbb Phi
%
% mathbb arg1 *o*ben *r*echts *r*unde *k*lammer   zB trial Funktion  
\newcommand{\orrk}[2]{ \mathbb{#1}^{(\textrm{#2})}  }

% vector E_p
%
% Vektorgrösse mit roman index rechts unten zB vektor 
\newcommand{\vri}[2]{ \vec{#1}_{\textrm{#2}}   }

% mathbb ~E_[T]
%
% *n*icht *r*elativ. Funktional:  tilde von mathbb arg1 index arg2 
\newcommand{\nrft}[2]{ \tilde{\mathbb{#1}}_{[\mathrm{#2}]} }

% mathbb E_T
%
% *n*icht *r*elativ. Funktional:  mathbb arg1_ mathrm arg2   
\newcommand{\nrf}[2]{ {\mathbb{{#1}}}_{\mathrm{ {#2}}} }

% varepsilon_x
%
% Energien  epsilon_roman text
\newcommand{\en}[1]{\varepsilon_{\textrm{#1}}}

%   (e)
%  E_R   
%     
% Energie oben rechts rund unten rechts roman  
\newcommand{\Ea}[3]{ #1^{(\mathrm{#2})}_{\mathrm{#3}}}

%   (e)
%  E_[R]   
%     
% Energie oben rechts mathrm rund unten rechts eckige klammer
\newcommand{\Eb}[3]{ #1^{(\mathrm{#2})}_{\mathrm{[#3]}}}

%   [e]
%  E_R   
%     
\newcommand{\Ec}[3]{ #1^{[{\mathrm #2}] }_{\mathrm{#3} } }

%   {j}
%  E_mathcal O   
%     
\newcommand{\Ed}[3]{ {\mathbb {#1}}^{{\{ #2 \}} }_{\mathcal{#3} } }

%   (j)
%  E_mathcal O   
%     
\newcommand{\Ee}[3]{ {\mathbb #1}^{{( #2 )} }_{\mathcal{#3} } }

% X_y
%
% *r*echts *u*nten arg1_mathrm arg2  
\newcommand{\ru}[2]{#1_\mathrm{#2}}

% ~X_y
%
% *t*ilde *r*echts *u*nten arg1_mathrm arg2  
\newcommand{\tru}[2]{\tilde{#1}_\mathrm{#2}}

%         [y]
% mathbb X
%
% rechts oben eckige klammer
\newcommand{\roek}[2]{ \mathbb{#1}^{[ \mathrm{#2} ] }  }

%  (y)
% X
%
% rechts oben runde klammer
\newcommand{\rork}[2]{ #1^{( \mathrm{#2} ) }  }

%   (y)
% ~X
%
% rechts oben runde klammer tilde
\newcommand{\rorkt}[2]{ \tilde{#1}^{( \mathrm{#2} ) }  }

%	 {p}                                                
%      A_blah  geschweifte Klammer links oben rechts unten roman index
%
\newcommand{\gkloi}[3]{\, {}^{\{\mathrm{#1}\}}{\!{#2}}_\mathrm{#3}}

%	 {p}                                                
%      ~A_blah  geschweifte Klammer links oben rechts unten roman index mit tilde
%
\newcommand{\gkloit}[3]{\, {}^{\{\mathrm{#1}\}}{ \tilde{#2} }_\mathrm{#3}}

% nü_stern^{n}
\newcommand{\nsn}{\nu_*^{\{n\}}}

%
%   X_{y}
%
% rechts unten geschw. klammer
\newcommand{\rugk}[2]{#1_{\{\rm #2\}} }

%     {y}
%   X
% rechts oben geschweifte klammer
\newcommand{\rogk}[2]{#1^{\{ {#2} \}} }

% {y} 
%    X
% links oben geschw. klammer
\newcommand{\logk}[2]{ {}^{\{#1\}}{\!#2}  }

\newenvironment{mysubsection}[1]
{  
\addcontentsline{toc}{subsection}{\emph{\textbf{#1}} } { \emph{\textbf{#1}} }\nopagebreak
}

\newenvironment{mysubsubsection}[1]
{  
\addcontentsline{toc}{subsubsection} {\emph{#1}}  { \emph{#1} }\nopagebreak
}

\newenvironment{mytable}[1]
{  
\addcontentsline{toc}{subsubsection}{#1} { }\nopagebreak
}

\newenvironment{myfigure}[1]
{  
\addcontentsline{toc}{subsubsection}{\emph{#1}} { }\nopagebreak
}

\newenvironment{myappendix}[1]
{\renewcommand{\thesection}{\Alph{section}}
\addcontentsline{toc}{subsection}{\textbf{\emph{\appendixname}}\
  \emph{\textbf{\thesection :}}\   \textbf{\emph{{#1}}}}  {}
}

%%% Local Variables: 
%%% mode: latex
%%% TeX-master: t
%%% End: 

%    'A   ein Strich (Apostroph) links oben
\newcommand{\slo}[1]{\,'{\!#1}}

%	 {p}                                                
%      A_blah  geschweifte Klammer links oben
%
\newcommand{\gklo}[2]{\, {}^{\{\mathrm{#1}\}}{{#2}}}

% (p)
%    A    runde Klammer mit roman links oben     
%
\newcommand{\rrklo}[2]{\, {}^{(\mathrm{#1})}{\! {#2} }}
% (p)
%    A    runde Klammer mit \sf (ohne Serifen) links oben     
%
\newcommand{\srklo}[2]{\, {}^{({\sf #1})}{\! {#2} }}

%%% Local Variables: 
%%% mode: latex
%%% TeX-master: t
%%% End: 

\newcommand{\e}{\operatorname{e}}
\newcommand{\crm}{\mathrm{c}}
\newcommand{\mustbe}{\stackrel{!}{=}}
\newcommand{\rt}{(r,\vartheta)}
\newcommand{\tablesize}{\scriptstyle}
\newcommand{\elp}{\ell_\mathcal{P}}
\newcommand{\elz}{\ell_z}
\newcommand{\spins}{s}
\newcommand{\ver}{\vec{\mathrm{e}}_r}
\newcommand{\vet}{\vec{\mathrm{e}}_\vartheta}
\newcommand{\vep}{\vec{\mathrm{e}}_\phi}

\newcommand{\dn}{\delta_{0}} %NEW
\newcommand{\dO}{\delta_{\Omega}}

\newcommand{\tanu}{\tilde{a}_\nu}
\newcommand{\tanuw}{\tilde{a}_{\nu w}}

\newcommand{\bjn}{{}^{(b)}\!j_0(\vec{r})} %NEW

\newcommand{\bko}{{}^{(b)}k_0}
\newcommand{\pko}{{}^{(p)}k_0}
\newcommand{\pgko}{{}^{\{p\}}k_0}
\newcommand{\pkr}{{}^{\{p\}}k_r}
\newcommand{\pkt}{{}^{\{p\}}k_\vartheta}
\newcommand{\pkp}{{}^{\{p\}}k_\phi}
\newcommand{\bkp}{{}^{\{b\}}k_\phi}
\newcommand{\gpko}{{}^{\{p\}}k_0}
\newcommand{\bbko}{{}^{[b]}k_0}
\newcommand{\bkn}{{}^{(b)}\!k_0\left(\vec{r}\right)}
\newcommand{\akn}{{}^{(a)}\!k_0\left(\vec{r}\right)}
\newcommand{\ak}{\vec{k}_a}
\newcommand{\akr}{{}^{(a)}\!k_r}
\newcommand{\akt}{{}^{(a)}\!k_\vartheta}
\newcommand{\akp}{{}^{(a)}\!k_\phi}
\newcommand{\bkv}{\vec{k}_b(\vec{r})} %NEW
\newcommand{\bkphi}{{}^{(b)\!}k_\phi}
\newcommand{\bk}{\vec{k}_b} %NEW
\newcommand{\bgkn}{{}^{\{b\}}\!k_0} %NEW

\newcommand{\bpp}{{}^{(b)}\varphi_{+}(\vec{r})}
\newcommand{\bpm}{{}^{(b)}\varphi_{-}(\vec{r})}
\newcommand{\bppm}{{}^{(b)}\varphi_{\pm}(\vec{r})}
\newcommand{\bppk}{{}^{(b)}\varphi_{+}^\dagger(\vec{r})}
\newcommand{\bpmk}{{}^{(b)}\varphi_{-}^\dagger(\vec{r})}

\newcommand{\tNGe}{{\tilde{N}_\textrm{G}^\textrm{(e)}}}
\newcommand{\NGe}{{N_\textrm{G}^\textrm{(e)}}}
\newcommand{\tNGee}{{\tilde{N}_\textrm{G}^\textrm{[e]}}}
\newcommand{\tNPhi}{{\tilde{N}_{\Phi}}}
\newcommand{\tNN}{\tilde{\mathbb{N}}}
\newcommand{\tNNO}{\tilde{\mathbb{N}}_{\Omega}}
\newcommand{\tNNPhi}{\tilde{\mathbb{N}}_{\Phi}}
\newcommand{\tNNGe}{{\tilde{\mathbb{N}}_\textrm{G}^\textrm{e}}} %NEW
\newcommand{\tNNGee}{{\tilde{\mathbb{N}}_\textrm{G}^\textrm{[e]}}}
\newcommand{\tNNGer}{{\tilde{\mathbb{N}}_\textrm{G}^\textrm{(e)}}} %NEW
\newcommand{\tNNGeg}{{\tilde{\mathbb{N}}_\textrm{G}^\textrm{\{e\}}}} %NEW
\newcommand{\tNNegan}{{\tilde{\mathbb{N}}^\textrm{\{e\}}_{\,\text{an}}}} %NEW
\newcommand{\tNNegiii}{{\tilde{\mathbb{N}}^\textrm{\{e\}}_{\mathsf{III}}}}
\newcommand{\tND}{\tilde{N}_\textrm{D}}

\newcommand{\Mee}{M^\textrm{(e)}}
\newcommand{\tMe}{\tilde{M}^\textrm{(e)}}
\newcommand{\tMee}{\tilde{M}^\textrm{[e]}}
\newcommand{\tMMee}{\tilde{\mathbb{M}}^\textrm{[e]}}
\newcommand{\tMMeg}{\tilde{\mathbb{M}}^\textrm{\{e\}}}
\newcommand{\tMMegiii}{\tMMeg_\mathsf{III}}
\newcommand{\tMMegv}{\tMMeg_\mathsf{V}}
\newcommand{\pptMMeg}{{}^{\prime\prime}\tMMeg}

\newcommand{\MRpm}{\mathcal{R}_\pm}
\newcommand{\MSpm}{\mathcal{S}_\pm}
\newcommand{\bMR}{{}^{(b)}\!\mathcal{R}}
\newcommand{\bMRpm}{{}^{(b)}\!\mathcal{R}_\pm}
\newcommand{\bMRp}{{}^{(b)}\!\mathcal{R}_+}
\newcommand{\bMRm}{{}^{(b)}\!\mathcal{R}_-}
\newcommand{\bMRpS}{{}^{(b)}\!{\overset{*}{\mathcal{R}}{} }_+}
\newcommand{\bMRmS}{{}^{(b)}\!{\overset{*}{\mathcal{R}}{} }_-}
\newcommand{\bMS}{{}^{(b)}\!\mathcal{S}}
\newcommand{\bMSpm}{{}^{(b)}\!\mathcal{S}_\pm}
\newcommand{\bMSp}{{}^{(b)}\!\mathcal{S}_+}
\newcommand{\bMSm}{{}^{(b)}\!\mathcal{S}_-}
\newcommand{\bMSpS}{{}^{(b)}\!{\overset{*}{\mathcal{S}}{} }_+}
\newcommand{\bMSmS}{{}^{(b)}\!{\overset{*}{\mathcal{S}}{} }_-}
\newcommand{\btRpm}{{}^{(b)}\!\tilde{R}_\pm}
\newcommand{\btRp}{{}^{(b)}\!\tilde{R}_+}
\newcommand{\btRm}{{}^{(b)}\!\tilde{R}_-}
\newcommand{\btSpm}{{}^{(b)}\!\tilde{S}_\pm}
\newcommand{\btSp}{{}^{(b)}\!\tilde{S}_+}
\newcommand{\btSm}{{}^{(b)}\!\tilde{S}_-}
\newcommand{\iiRpm}{{}^\mathsf{II}\!R_\pm}
\newcommand{\iiRp}{{}^\mathsf{II}\!R_+}
\newcommand{\iiRm}{{}^\mathsf{II}\!R_-}
\newcommand{\iiSpm}{{}^\mathsf{II}\!S_\pm}
\newcommand{\iiSp}{{}^\mathsf{II}\!S_+}
\newcommand{\iiSm}{{}^\mathsf{II}\!S_-}
\newcommand{\iitRpm}{{}^\mathsf{II}\!\tilde{R}_\pm}
\newcommand{\iitRp}{{}^\mathsf{II}\!\tilde{R}_+}
\newcommand{\iitRm}{{}^\mathsf{II}\!\tilde{R}_-}
\newcommand{\iitSpm}{{}^\mathsf{II}\!\tilde{S}_\pm}
\newcommand{\iitSp}{{}^\mathsf{II}\!\tilde{S}_+}
\newcommand{\iitSm}{{}^\mathsf{II}\!\tilde{S}_-}

\newcommand{\aMRpm}{{}^{(a)}\!\mathcal{R}_\pm}
\newcommand{\aMRp}{{}^{(a)}\!\mathcal{R}_+}
\newcommand{\aMRm}{{}^{(a)}\!\mathcal{R}_-}
\newcommand{\aMRpS}{{}^{(a)}\!{\overset{*}{\mathcal{R}}{} }_+}
\newcommand{\aMRmS}{{}^{(a)}\!{\overset{*}{\mathcal{R}}{} }_-}
\newcommand{\aMSpm}{{}^{(a)}\!\mathcal{S}_\pm}
\newcommand{\aMSp}{{}^{(a)}\!\mathcal{S}_+}
\newcommand{\aMSm}{{}^{(a)}\!\mathcal{S}_-}
\newcommand{\aMSpS}{{}^{(a)}\!{\overset{*}{\mathcal{S}}{} }_+}
\newcommand{\aMSmS}{{}^{(a)}\!{\overset{*}{\mathcal{S}}{} }_-}

\newcommand{\pMR}{{}^{(p)}\!\R}
\newcommand{\pMRpm}{{}^{(p)}\!\R_\pm}
\newcommand{\pMRp}{{}^{(p)}\!\R_+}
\newcommand{\pMRm}{{}^{(p)}\!\R_-}
\newcommand{\pMS}{{}^{(p)}\!\mathcal{S}}
\newcommand{\pMSpm}{{}^{(p)}\!\mathcal{S}_\pm}
\newcommand{\pMSp}{{}^{(p)}\!\mathcal{S}_+}
\newcommand{\pMSm}{{}^{(p)}\!\mathcal{S}_-}
\newcommand{\pSRpm}{{}^{(p)}\!\tilde{R}_\pm}
\newcommand{\pSRp}{{}^{(p)}\!\tilde{R}_+}
\newcommand{\pSRm}{{}^{(p)}\!\tilde{R}_-}
\newcommand{\pSSpm}{{}^{(p)}\!\tilde{S}_\pm}
\newcommand{\pSSp}{{}^{(p)}\!\tilde{S}_+}
\newcommand{\pSSm}{{}^{(p)}\!\tilde{S}_-}
\newcommand{\pMRpS}{{}^{(p)}\!{\overset{*}{\mathcal{R}}{} }_+}
\newcommand{\pMRmS}{{}^{(p)}\!{\overset{*}{\mathcal{R}}{} }_-}
\newcommand{\pMSpS}{{}^{(p)}\!{\overset{*}{\mathcal{S}}{} }_+}
\newcommand{\pMSmS}{{}^{(p)}\!{\overset{*}{\mathcal{S}}{} }_-}
\newcommand{\ptRpm}{{}^{(p)}\!\tilde{R}_\pm}
\newcommand{\ptRp}{{}^{(p)}\!\tilde{R}_+}
\newcommand{\ptRm}{{}^{(p)}\!\tilde{R}_-}
\newcommand{\ptSpm}{{}^{(p)}\!\tilde{S}_\pm}
\newcommand{\ptSp}{{}^{(p)}\!\tilde{S}_+}
\newcommand{\ptSm}{{}^{(p)}\!\tilde{S}_-}

\newcommand{\SPp}{\tilde{\Phi}_+}
\newcommand{\SPm}{\tilde{\Phi}_-}

\newcommand{\bpMR}{{}^{(b/p)}\!\R}
\newcommand{\bpMRpm}{{}^{(b/p)}\!\R_\pm}
\newcommand{\bpMS}{{}^{(b/p)}\!\mathcal{S}}
\newcommand{\bpMSpm}{{}^{(b/p)}\!\mathcal{S}_\pm}

\newcommand{\MCa}{\mathcal{C}_{(a)}}
\newcommand{\MCaS}{\overset{*}{\mathcal{C}}_{(a)}}

\newcommand{\Kb}{K_{\{b\}}} %NEW
\newcommand{\Aan}{A^{\text{an}}}
\newcommand{\peA}{{}^{[p]}\!A}
\newcommand{\peAo}{{}^{[p]}\!A_0} %NEW
\newcommand{\peAw}{{}^{[p]}\!A_w} %NEW
\newcommand{\bAe}{{}^{[b]}\!A_0}
\newcommand{\bAn}{{}^{(b)}\!A_0}
\newcommand{\pAn}{{}^{(p)}\!A_0}
\newcommand{\pgA}{{}^{\{p\}}\!A}
\newcommand{\beA}{{}^{[b]}\!A}
\newcommand{\pgAn}{{}^{\{p\}}\!A_0}
\newcommand{\pgAan}{{}^{\{p\}}\!A^{\text{an}}}
\newcommand{\pAvn}{{}^{(p)}\!\vec{A}_0}
\newcommand{\peAn}{{}^{[p]}\!A_0}
\newcommand{\bgAe}{{}^{\{b\}}\!A_0} %NEW
\newcommand{\bgAan}{{}^{\{b\}}\!A^{\text{an}}} %NEW
\newcommand{\bgA}{{}^{\{b\}}\!A} %NEW
\newcommand{\bgAiii}{\bgA^{\textsf{III}}} %NEW
\newcommand{\bgAv}{\bgA^{\textsf{V}}} %NEW
\newcommand{\bAmu}{{}^{(b)}\!A_\mu} %NEW
\newcommand{\Acnuiii}{\mathcal{A}_\nu^\mathsf{III}} %NEW
\newcommand{\Aciii}{\mathcal{A}^\mathsf{III}} %NEW
\newcommand{\bgAcnuiii}{{}^{\{b\}}\mathcal{A}_\nu^\mathsf{III}} %NEW
\newcommand{\bgAciii}{{}^{\{b\}}\mathcal{A}^\mathsf{III}} %NEW
\newcommand{\MMe}{\mathbb{M}^\textrm{(e)}}
\newcommand{\MMeg}{\mathbb{M}^\textrm{\{e\}}} %NEW
\newcommand{\tMMegan}{\tilde{\mathbb{M}}^\textrm{\{e\}}_{\ \text{an}}} %NEW
\newcommand{\mueg}{\mu^\textrm{\{e\}}}
\newcommand{\muegan}{\mu^\textrm{\{e\}}_{\ \text{an}}} %NEW
\newcommand{\muegiii}{\mu^\textrm{\{e\}}_{\mathsf{III}}} %NEW
\newcommand{\LLnu}{\mathbb{L}_\nu} %NEW
\newcommand{\KKnu}{\mathbb{K}_\nu} %NEW

\newcommand{\EE}{\mathbb{E}}
\newcommand{\tEE}{\tilde{\mathbb{E}}}
\newcommand{\eER}{E_\textrm{R}^\textrm{(e)}}
\newcommand{\ERe}{{E_\textrm{R}^\textrm{(e)}}}
\newcommand{\ERee}{{E_\textrm{R}^\textrm{[e]}}}
\newcommand{\eeER}{E_\textrm{R}^\textrm{[e]}} %NEW
\newcommand{\egER}{E_\textrm{R}^\textrm{\{e\}}} %NEW
\newcommand{\ET}{E_{\textrm{T}}}
\newcommand{\tETT}{\tilde{E}_{\textrm{[T]}}}
\newcommand{\tEePhi}{\tilde{E}_{\textrm{[$\Phi$]}}}
\newcommand{\tEEePhi}{\tilde{\mathbb{E}}_{\textrm{[$\Phi$]}}}
\newcommand{\EePhif}{E_{\textrm{[$\Phi$]}}^f}
\newcommand{\tEephif}{\tilde{E}_{\textrm{[$\phi$]}}^f}
\newcommand{\tEgT}{\tilde{E}_{\textrm{\{T\}}}}
\newcommand{\tEeT}{\tilde{E}_{\textrm{[T]}}}
\newcommand{\tEEeT}{\tilde{\mathbb{E}}_{\textrm{[T]}}}
\newcommand{\tEEgT}{\tilde{\mathbb{E}}_{\textrm{\{T\}}}}
\newcommand{\ETiv}{E^\textrm{(IV)}_{\textrm{[T]}}}
\newcommand{\EETiv}{\mathbb{E}^\textrm{(IV)}_{\textrm{[T]}}}
\newcommand{\EgTiv}{\mathbb{E}^\textrm{(IV)}_{\textrm{\{T\}}}}
\newcommand{\EET}{\mathbb{E}_\textrm{[T]}} %NEW
\newcommand{\EEt}{\mathbb{E}_\textrm{T}}
\newcommand{\EETT}{\mathbb{E}_\textrm{[T]}} %NEW
\newcommand{\EEivbn}{\mathbb{E}^\textrm{(IV)}(\beta,\nu)}
\newcommand{\EEeg}{\mathbb{E}^{\{\textrm{e}\}}} %NEW
\newcommand{\EEgiv}{\mathbb{E}^{\{\textsf{IV}\}}} %NEW
\newcommand{\tEEgiv}{\tilde{\mathbb{E}}^{\{\textsf{IV}\}}} %NEW
\newcommand{\EEgivbn}{\mathbb{E}^{\{\textsf{IV}\}}(\beta,\nu)} %NEW
\newcommand{\Egivbn}{E^{\{\textsf{IV}\}}(\beta,\nu)} %NEW
\newcommand{\EEeiv}{\mathbb{E}^{[\textsf{IV}]}} %NEW
\newcommand{\EEeivbn}{\mathbb{E}^{[\textsf{IV}]}(\beta,\nu)} %NEW
\newcommand{\EEO}{\mathbb{E}_\Omega} %NEW
\newcommand{\tEEO}{\tilde{\mathbb{E}}_{[\Omega]}}
\newcommand{\eEEO}{{}^{(e)}\!\mathbb{E}_{[\Omega]}}
\newcommand{\EEOe}{\mathbb{E}_{[\Omega]}} %NEW
\newcommand{\GtEEO}{{}^{(G)}\!\tEEO} %NEW
\newcommand{\GetEO}{{}^{[G]}\!\tEO} %NEW
\newcommand{\DtEEO}{{}^{(D)}\!\tEEO} %NEW
\newcommand{\etEEePhi}{{}^{(e)}\!\tilde{\mathbb{E}}_{[\Phi]}}
\newcommand{\etEEPhi}{{}^{(e)}\!\tilde{\mathbb{E}}_{\Phi}}
\newcommand{\DtEEPhi}{{}^{(D)}\!\tilde{\mathbb{E}}_{\Phi}}
\newcommand{\DtEEePhi}{{}^{(D)}\!\tilde{\mathbb{E}}_{[\Phi]}}
\newcommand{\GtEEePhi}{{}^{(G)}\!\tilde{\mathbb{E}}_{[\Phi]}}
\newcommand{\DEEO}{{}^{(D)}\!\EEO} %NEW
\newcommand{\DEEOe}{{}^{(D)}\!\EEOe} %NEW
\newcommand{\etEEO}{{}^{(e)}\!\tEEO} %NEW
\newcommand{\eetEEO}{{}^{[e]}\!\tEEO} %NEW
\newcommand{\antEEO}{{}^{\{\text{an}\}}\!\tEEO} %NEW
\newcommand{\anEEO}{{}^{\{\text{an}\}}\!\EE_{[\Omega]}}
\newcommand{\anrtEEO}{{}^{(\text{an})}\!\tEEO} %NEW
\newcommand{\eetEO}{{}^{[e]}\!\tEO} %NEW
\newcommand{\Egiv}{E_\textrm{\{T\}}^{\textrm{(IV)}}} %NEW
\newcommand{\EEP}{\mathbb{E}_\mathcal{P}} %NEW
\newcommand{\EEOn}{\EE_\mathcal{O}^{\{0\}}}
\newcommand{\EEOj}{\EE_\mathcal{O}^{\{j\}}}
\newcommand{\EEOon}{\EE_\mathcal{O}^{\{0,0\}}}
\newcommand{\ppE}{{}^{\prime\prime}\!E}
\newcommand{\ppEE}{{}^{\prime\prime}\EE}
\newcommand{\pptEEO}{{}^{\prime\prime}\tEEO}

\newcommand{\Ekin}{{E_\textrm{kin}}}
\newcommand{\EKIN}{{E_\textrm{KIN}}} %NEW
\newcommand{\ekin}{ {\varepsilon}_{\textrm{kin}} } %NEW
\newcommand{\eKIN}{ {\varepsilon}_{\textrm{KIN}} } %NEW
\newcommand{\pEkin}{{}^{(p)}\!\Ekin}
\newcommand{\rEkin}{{}^{(r)}\!\Ekin} %NEW
\newcommand{\rEKIN}{{}^{(r)}\!\EKIN} %NEW
\newcommand{\rekin}{{}^{(r)}\!\ekin} %NEW
\newcommand{\reKIN}{{}^{(r)}\!\eKIN} %NEW
\newcommand{\thEKIN}{{}^{(\vartheta)}\!\EKIN} %NEW
\newcommand{\thekin}{{}^{(\vartheta)}\!\ekin} %NEW
\newcommand{\theKIN}{{}^{(\vartheta)}\!\eKIN} %NEW

\newcommand{\epot}{ {\varepsilon}_{\textrm{pot}} }
\newcommand{\ePOT}{ {\varepsilon}_{\textrm{POT}} }

\newcommand{\etot}{ {\varepsilon}_{\textrm{tot}} }
\newcommand{\eeg}{ {\varepsilon}^{\{e\}}}
\newcommand{\eegtot}{ {\varepsilon}^{\{e\}}_{\textrm{tot}} }
\newcommand{\eagtot}{ {\varepsilon}^{\{a\}}_{\textrm{tot}} }

\newcommand{\tEEOo}{\tilde{\mathbb{E}}_\Omega} %NEW
\newcommand{\EReg}{{E_\textrm{R}^\textrm{\{e\}}}} %NEW
\newcommand{\Eegan}{{E^\textrm{\{e\}}_\text{an}}} %NEW
\newcommand{\Ergan}{{E^{\{r\}}_\text{an}}}
\newcommand{\Evgan}{{E^{\{\vartheta\}}_\text{an}}}
\newcommand{\Eean}{{E^\textrm{(e)}_\text{an}}}
\newcommand{\Eeg}{{E^\textrm{\{e\}}}}
\newcommand{\Erg}{E^{\{r\}}}
\newcommand{\Evg}{E^{\{\vartheta\}}}
\newcommand{\Eegiii}{{E_\mathsf{III}^\textrm{\{e\}}}}
\newcommand{\Ew}{{E_\textrm{w}}}
\newcommand{\Ewee}{{\Ew^\textrm{[e]}}} %NEW
\newcommand{\cEk}{{\cal{E}_\textrm{kin}}}

\newcommand{\tPhi}{\tilde{\Phi}}
\newcommand{\tPhipm}{\tilde{\Phi}_\pm}
\newcommand{\tPhip}{\tilde{\Phi}_+}
\newcommand{\tPhim}{\tilde{\Phi}_-}
\newcommand{\tPhib}{\tPhi_b} %NEW
\newcommand{\tPhinuw}{\tPhi_{\nu w}} %NEW

\newcommand{\tO}{\tilde{\Omega}}
\newcommand{\tOpm}{\tilde{\Omega}_\pm}
\newcommand{\tOp}{\tilde{\Omega}_+}
\newcommand{\tOm}{\tilde{\Omega}_-}

\newcommand{\lO}{\ell_\mathcal{O}}
\newcommand{\dlO}{\dot{\ell}_\mathcal{O}}
\newcommand{\ddlO}{\ddot{\ell}_\mathcal{O}}
\newcommand{\jO}{j_\mathcal{O}}
\newcommand{\nO}{n_\mathcal{O}}

\newcommand{\Sag}{S^{\{a\}}}
\newcommand{\Sjg}{S^{\{j\}}}
\newcommand{\SOag}{\Sag_\mathcal{O}}
\newcommand{\SOjg}{\Sjg_\mathcal{O}}

\newcommand{\lGe}{ {\lambda_\textrm{G}^{(\textrm{e})}}\!}

\newcommand{\Du}{\mathcal{D}_\mu}

\newcommand{\Tmunu}{T_{\mu\nu}}
\newcommand{\Too}{T_{00}}
\newcommand{\TTmunu}{{}^{(T)}\!\Tmunu}
\newcommand{\DTmunu}{{}^{(D)}\!\Tmunu}
\newcommand{\GTmunu}{{}^{(G)}\!\Tmunu}
\newcommand{\TToo}{{}^{(T)}\!\Too}
\newcommand{\DToo}{{}^{(D)}\!\Too}
\newcommand{\GToo}{{}^{(G)}\!\Too}

\newcommand{\ap}{{}^{(a)}\!\varphi_{\pm}(\vec{r})}
\newcommand{\app}{{}^{(a)}\!\varphi_{+}(\vec{r})}
\newcommand{\apm}{{}^{(a)}\!\varphi_{-}(\vec{r})}
\newcommand{\appm}{{}^{(a)}\!\varphi_{\pm}(\vec{r})}
\newcommand{\appmd}{{}^{(a)}\!\varphi_{\pm}^{\dagger}(\vec{r})}
\newcommand{\apmd}{{}^{(a)}\!\varphi_{-}^{\dagger}(\vec{r})}
\newcommand{\appd}{{}^{(a)}\!\varphi_{+}^{\dagger}(\vec{r})}
\newcommand{\bpx}{{}^{(b/p)}\!}
\newcommand{\bpxpp}{{}^{(b/p)}\!\varphi_{+}(\vec{r})}
\newcommand{\bpxpm}{{}^{(b/p)}\!\varphi_{-}(\vec{r})}
\newcommand{\pxppm}{{}^{(p)}\!\varphi_{\pm}(\vec{r})}
\newcommand{\pxpp}{{}^{(p)}\!\varphi_{+}(\vec{r})}
\newcommand{\pxpm}{{}^{(p)}\!\varphi_{-}(\vec{r})}
\newcommand{\pxppmd}{{}^{(p)}\!\varphi_{\pm}^{\dagger}(\vec{r})}
\newcommand{\pxpmd}{{}^{(p)}\!\varphi_{-}^{\dagger}(\vec{r})}
\newcommand{\pxppd}{{}^{(p)}\!\varphi_{+}^{\dagger}(\vec{r})}

\newcommand{\PsiPO}{\Psi_{\mathcal{P},\mathcal{O}}}
\newcommand{\PsiP}{\Psi_\mathcal{P}}
\newcommand{\PsiO}{\Psi_\mathcal{O}}

\newcommand{\zetaljm}{\zeta_{\ l}^{j,\;m}}
\newcommand{\zetaejm}{\zeta_{\ e}^{j,\;m}}
\newcommand{\zetapp}{\zeta^{\frac{1}{2},\frac{1}{2}}}
\newcommand{\zetapm}{\zeta^{\frac{1}{2},-\frac{1}{2}}}
\newcommand{\zetappm}{\zeta^{\frac{1}{2},\pm\frac{1}{2}}}
\newcommand{\xip}{\xi^{(+)}}
\newcommand{\xim}{\xi^{(-)}}
\newcommand{\etap}{\eta^{(+)}}
\newcommand{\etam}{\eta^{(-)}}
\newcommand{\omp}{\omega^{(+)}}
\newcommand{\omm}{\omega^{(-)}}

\newcommand{\Jhz}{\hat{\J}_z}
\newcommand{\Jhzpm}{\hat{\J}_z^{(\pm)}}
\newcommand{\Jhzp}{\hat{\J}_z^{(+)}}
\newcommand{\Jhzm}{\hat{\J}_z^{(-)}}
\newcommand{\Jlhz}{\hat{J}_z}
\newcommand{\Jlhzpm}{\hat{J}_z^{(\pm)}}
\newcommand{\Jlhzp}{\hat{J}_z^{(+)}}
\newcommand{\Jlhzm}{\hat{J}_z^{(-)}}
\newcommand{\Lhz}{\hat{\mathcal{L}}_z}
\newcommand{\Lhzp}{\hat{\mathcal{L}}_z^{(+)}}
\newcommand{\Lhzm}{\hat{\mathcal{L}}_z^{(-)}}
\newcommand{\Llhz}{\hat{L}_z}
\newcommand{\Llhzp}{\hat{L}_z^{(+)}}
\newcommand{\Llhzm}{\hat{L}_z^{(-)}}
\newcommand{\Shz}{\hat{\mathcal{S}}_z}

\newcommand{\ajz}{{}^{(a)}\!j_z}
\newcommand{\pjz}{{}^{(p)}\!j_z}
\newcommand{\bjz}{{}^{(b)}\!j_z}

\newcommand{\ptTkin}{{}^{(p)}\!\tilde{T}_\text{kin}}
\newcommand{\pTkin}{{}^{(p)}\!T_\text{kin}}
\newcommand{\ptTkinePhi}{{}^{(p)}\!\tilde{T}_{\text{kin}\,[\Phi]}}
\newcommand{\ptTr}{{}^{(p)}\!\tilde{T}_r}
\newcommand{\pTr}{{}^{(p)}\!T_r}
\newcommand{\ptTth}{{}^{(p)}\!\tilde{T}_\vartheta}
\newcommand{\pTth}{{}^{(p)}\!T_\vartheta}
\newcommand{\ptTphi}{{}^{(p)}\!\tilde{T}_\phi}
\newcommand{\pTphi}{{}^{(p)}\!T_\phi}

\newcommand{\Zp}{\mathcal{Z}_\mathcal{P}}
\newcommand{\tZp}{\tilde{\mathcal{Z}}_\mathcal{P}}

\newcommand{\ugR}{{}^{(\text{u})}\!g_R}
\newcommand{\ugS}{{}^{(\text{u})}\!g_S}
\newcommand{\lgR}{{}^{(\text{l})}\!g_R}
\newcommand{\lgS}{{}^{(\text{l})}\!g_S}

\newcommand{\tNf}{\tilde{N}_f}

\newcommand{\pr}{{}^{(p)}\!}
\newcommand{\br}{{}^{(b)}\!}
\newcommand{\bpr}{{}^{(b,p)}\!}
\newcommand{\bg}{{}^{\{b\}}\!}

\newcommand{\Pnuiii}{\mathcal{P}_\nu^\mathsf{III}}
\newcommand{\bgP}{\bg \mathcal{P}}

\newcommand{\bbar}{{\mathchoice
{{\vcenter{\offinterlineskip\vskip.1ex\hbox{$\,\tilde{}$}\vskip-1.85ex\hbox{$b$}\vskip.4ex}}}
{{\vcenter{\offinterlineskip\vskip.1ex\hbox{$\,\tilde{}$}\vskip-1.75ex\hbox{$b$}\vskip.4ex}}}
{{\vcenter{\offinterlineskip\vskip.1ex\hbox{$\scriptstyle\,\tilde{}$}\vskip-1.2ex\hbox{$\scriptstyle b$}\vskip.2ex}}}
{{\vcenter{\offinterlineskip\vskip.1ex\hbox{$\scriptscriptstyle\,\tilde{}$}\vskip-.9ex\hbox{$\scriptscriptstyle b$}\vskip.4ex}}}
}}

\enlargethispage{\baselineskip} 
\title{\bf Dimorphism of Ortho-Positronium\\ in\\ Relativistic Schr\"odinger Theory}
\author{M.\ Mattes and M.\ Sorg} %\\[1cm] II.\ Institut f\"ur Theoretische Physik der Universit\"at Stuttgart\\ Pfaffenwaldring 57 \\ D 70550 Stuttgart, Germany
\date{ }
\maketitle
\begin{abstract}
  The non-relativistic energy levels of ortho-positronium are calculated in the quadrupole
  and octupole approximations for the interaction potential. For this purpose, the RST
  eigenvalue problem of angular momentum is illustratively solved for the quantum numbers
  $\jO=0,1,2,3,4$ and $\bjz=\pm 1$. This eigenvalue problem admits \emph{ambiguous}
  solutions for $0<|\bjz|<\jO$ whereas the solutions for $\bjz=0$ and $\bjz=\pm \jO$ are
  \emph{unique}. In order to attain some (at least approximative) solutions of the energy
  eigenvalue problem one tries a factorized ansatz for the wave function and thus splits
  off the angular problem (with its ambiguous solutions) from the residual radial
  problem. The latter does, as usual, finally fix the energy eigenvalues. But it is just
  by this procedure that the ambiguity of the angular problem is transferred to most of
  the energy levels which thereby become doubled. The corresponding doubling energy
  amounts to (roughly) one percent of the total binding energy and is, however, of purely
  \emph{electric} origin, since magnetism is completely neglected. Indeed, the charge
  distributions of both positronium constituents (i.e.\ electron and positron) do inherit
  their ambiguity from the ambiguous solution of the angular eigenvalue problem
  ($\leadsto$ charge ``dimorphism''); and naturally the dimorphic configurations must then
  possess slightly different interaction energies of the electrostatic type.
  \vspace{2.5cm}
 \noindent


\end{abstract}

%%% Local Variables: 
%%% mode: latex
%%% TeX-master: "main"
%%% End: 

\begin{center}
  {\Large\textbf{Contents}}\\[3em]
\end{center}
\begin{itemize}
\item[\textbf{I.}] {\large \textbf{Introduction and Survey of Results \dotfill 8 }}
  \begin{itemize}
  \item[]
    \begin{itemize}
    \item[] {\large \emph{Particle-Wave Duality\dotfill 10}}
    \item[] {\large \emph{Probabilistic Point-Particle Picture \dotfill 12  }}
    \item[] {\large \emph{Complementarity and Environment\dotfill 17 }}
    \item[] {\large \emph{Fluid-Dynamic Character of RST\dotfill  20}}
    \item[] {\large \emph{Ortho-Dimorphism\dotfill 21 }}
    \item[] {\large \emph{Induced Energy Difference\dotfill 22 }}
    \end{itemize}
  \end{itemize}
\item[\textbf{II.}] {\large \textbf{Positronium Eigenvalue Problem \dotfill 25 }}
  \begin{itemize}
  \item[\textbf{1.}] \emph{\textbf{Relativistic Schr\"odinger Equation} \dotfill 25  }
  \item[\textbf{2.}] \emph{\textbf{Dirac Equation} \dotfill 26 }
  \item[\textbf{3.}] \emph{\textbf{Maxwell Equations} \dotfill 28 }
  \item[\textbf{4.}] \emph{\textbf{Conservation Laws} \dotfill 29 }
  \item[\textbf{5.}] \emph{\textbf{Stationary Field Configurations} \dotfill 32  }
    \begin{itemize}
    \item[]{\large \emph{Gauge Field Subsystem \dotfill 33 }} 
    \item[]{\large \emph{Matter Subsystem \dotfill 34 }} 
    \end{itemize}
  \end{itemize}
\item[\textbf{III.}] {\large \textbf{Ortho/Para Dichotomy \dotfill 37 }}
  \begin{itemize}
  \item[\textbf{1.}] \emph{\textbf{Mass Eigenvalue Equations} \dotfill 38 }
  \item[\textbf{2.}] \emph{\textbf{Poisson Equations} \dotfill 42 }
  \item[\textbf{3.}] \emph{\textbf{Non-Unique Spinor Fields} \dotfill 43 }
    \begin{itemize}
    \item[]{\large \emph{Fermionic States in RST \dotfill 45 }}
    \item[]{\large \emph{Bosonic States in RST \dotfill 47 }}  
    \item[]{\large \emph{Uniqueness of the Physical Densities \dotfill 51 }} 
    \end{itemize}
  \end{itemize}
\item[\textbf{IV.}] {\large \textbf{Ortho-Positronium \dotfill 54 }}
  \begin{itemize}
  \item[\textbf{1.}] \emph{\textbf{Mass Eigenvalue Equations for the Amplitude Fields} \dotfill 54 }
    \begin{itemize}
    \item[]{\large\emph{Product Ansatz \dotfill 57}}
    \end{itemize}
  \item[\textbf{2.}] \emph{\textbf{Angular Momentum Quantization in RST\dotfill 61}}
    \begin{itemize}
    \item[]{\large\emph{Anisotropy of the Gauge Potential\dotfill 61}}
    \item[]{\large\emph{Solving the Eigenvalue Problem of Angular Momentum\dotfill 63}}
    \item[]{\large\emph{Ortho-Dimorphism for $\jO=2$\dotfill 69}}
    \item[]{\large\emph{Fig.IV.A: Rolling out of the Rotating Charge Clouds\dotfill 73}}
    \item[]{\large\emph{Fig.IV.B: Dimorphism of Angular Density  $\gklo{b}{k}_0(r)$ \\ \phantom{Fig.IV.B:} for
          $\jO=3,\bjz=\pm 2$\dotfill 75}}
    \end{itemize}
  \end{itemize}
\item[\textbf{V.}] {\large \textbf{Anisotropy of the Gauge Potential
      $\boldsymbol{\rklo{b}\!{A}_0(\vec{r})}$ \dotfill 76 }}
  \begin{itemize}
  \item[\textbf{1.}]{\emph{\textbf{Spherical Symmetry as a First Approximation \dotfill 76}}}
  \item[\textbf{2.}]{\emph{\textbf{Anisotropy Corrections \dotfill 79}}}
  \end{itemize}
\item[\textbf{VI.}] {\large \textbf{Energy of Matter and Gauge Fields \dotfill 85 }}
  \begin{itemize}
  \item[\textbf{1.}]{\emph{\textbf{Energy of the Matter Fields\dotfill 87}}}
  \item[\textbf{2.}]{\emph{\textbf{Energy of the Gauge Fields\dotfill 88}}}
  \item[\textbf{3.}]{\emph{\textbf{Mass Equivalent\dotfill 90}}}
  \item[\textbf{4.}]{\emph{\textbf{Principle of Minimal Energy\dotfill 92}}}
    \begin{itemize}
    \item[]{\large\emph{Isotropic Energy Function $\roek{E}{\sf IV}(\beta,\nu)$\dotfill 93}}
    \item[]{\large\emph{Anisotropic Energy Function $\rogk{\mathbb{E}}{\sf
            IV}(\beta,\nu)$\dotfill 96}}
    \end{itemize}
  \end{itemize}
\item[\textbf{VII.}] {\large \textbf{Energy Difference of Dimorphic Partners \dotfill 99  }}
  \begin{itemize}
  \item[\textbf{1.}]{\emph{\textbf{Quadrupole Corrections\dotfill 100}}}
    \begin{itemize}
    \item[]{\large\emph{Quadrupole Equation\dotfill 101}}
    \end{itemize}
  \item[\textbf{2.}]{\emph{\textbf{Energy Function of the Anisotropic
          Configurations\dotfill 104}}}
  \item[\textbf{3.}]{\emph{\textbf{Comparison of Ortho- and Para Levels\dotfill 107}}}
    \begin{itemize}
    \item[]{\large\emph{Fig.VII.A: Comparison of Ortho- and Para-Levels for \\
          Principal Quantum Numbers $n=4$ and $n=5$\dotfill 108}}
    \end{itemize}
  \end{itemize}
\item[\textbf{VIII.}] {\large \textbf{Multipole Solutions \dotfill 112 }}
  \begin{itemize}
  \item[\textbf{1.}]{\emph{\textbf{Alternative Multipole Expansion\dotfill 112}}}
    \begin{itemize}
    \item[]{\large\emph{Properties of the New Basis Functions\dotfill 113}}
    \item[]{\large\emph{Comparison of the Expansion Coefficients\dotfill 117}}
    \end{itemize}
  \item[\textbf{2.}]{\emph{\textbf{Multipole Equations\dotfill 118}}}
    \begin{itemize}
    \item[]{\large\emph{Anisotropy Energy\dotfill 119}}
    \item[]{\large\emph{General Multipole Equation\dotfill 120}}
    \item[]{\large\emph{Quadrupole Approximation \dotfill 121}}
    \end{itemize}
  \item[\textbf{3.}]{\emph{\textbf{Octupole Approximation\dotfill 124}}}
    \begin{itemize}
    \item[]{\large\emph{Octupole Energy\dotfill 125}}
    \item[]{\large\emph{Conjugate Potentials\dotfill 126}}
    \item[]{\large\emph{Hybrid Method\dotfill 129}}
    \item[]{\large\emph{Separative Method\dotfill 131}}
    \item[]{\large\emph{Separative Energy Functional\dotfill 133}}
    \end{itemize}
  \item[\textbf{4.}]{\emph{\textbf{Magnitude of Octupole Splitting\dotfill 135}}}
    \begin{itemize}
    \item[]{\large\emph{Quadrupole Part\dotfill 136}}
    \item[]{\large\emph{Octupole Energy\dotfill 137}}
    \item[]{\large\emph{Schr\"odinger Equation with Octupole Interaction\dotfill 138}}
    \item[]{\large\emph{Variational Procedure\dotfill 139}}
    \item[]{\large\emph{Numerical Results\dotfill 143}}
    \end{itemize}
    \vspace{10mm}
    {\centerline *}
  \end{itemize}
\item[] {\large \textbf{Appendix A:}}
  \item[]{\large\textbf{General Properties of the Quadrupole Approximation\dotfill 146}}
    \begin{itemize}
    \item[]{\large\emph{General Validity of the Ratios
          (\ref{eq:VI.17a})-(\ref{eq:VI.17b})\dotfill 146}}
    \item[]{\large\emph{No Dimorphism for $\bjz=0$\dotfill 147}}
    \item[]{\large\emph{No Dimorphism for $\bjz=\pm\jO$\dotfill 149}}
    \end{itemize}\vspace{\baselineskip}
\item[] {\large \textbf{Appendix B:}}
\item[]{\large\textbf{Groundstate ($\boldsymbol{\nO=1\Rightarrow\jO=0,\bjz=0}$)\dotfill 153}}\vspace{\baselineskip}
\item[] {\large \textbf{Appendix C:}}
\item[]{\large\textbf{No Dimorphism for first Excited State ($\boldsymbol{\nO=2}$)\dotfill
    157}}\vspace{\baselineskip}
\item[] {\large \textbf{Appendix D:}}
\item[]{\large\textbf{Dimorphism vs.\ Elimination of $\boldsymbol j_z$-Degeneracy\dotfill 161}}
  \begin{itemize}
  \item[]
    \begin{itemize}
    \item[]{\large\emph{Lowering of Groundstate Energy\dotfill 165}}
    \item[]{\large\emph{No Dimorphism for $\jO=2$\dotfill 165}}
    \item[]{\large\emph{Equality of Ortho- and Para-Levels\dotfill 167}}
    \end{itemize}
  \end{itemize}\vspace{5mm}
\item[]{\large\textbf{Appendix E:}}
\item[]{\large\textbf{Octupole Solution\dotfill 169}}
  \begin{itemize}
  \item []
    \begin{itemize}
    \item[]{\large\emph{Boundary Conditions\dotfill 170}}
    \item[]{\large\emph{Exact Octupole Solution\dotfill 172}}
    \item[]{\large\emph{Mass-Equivalent Function $\rogk{\mu}{e}_4(\nu)$\dotfill 174}}
    \item[]{\large\emph{Relative Magnitude of Quadrupole and Octupole
          \\Corrections\dotfill 175}}
    \end{itemize}
  \end{itemize}
\item[]{\large\textbf{Fig.E.I: Multipole Solutions $\boldsymbol{\gklo{b}{\cal P}_2(y)}$
      (\ref{eq:E.37}) and \\ \phantom{Fig.E.I}$\boldsymbol{\gklo{b}{\cal P}_4(y)}$ (\ref{eq:E.26}) for
      $\boldsymbol \nu=3$\dotfill 179}}
\item[]{\large\textbf{Fig.E.II: Quadrupole Function $\boldsymbol{\rogk{\mu}{e}_2(\nu)=\frac{9}{4}\rogk{\mu}{e}_\mathsf{III}(\nu)}$
      (\ref{eq:VII.15}) \\ \phantom{Fig.E.II} and  Octupole Function
      $\boldsymbol{\rogk{\mu}{e}_4(\nu)}$ (\ref{eq:E.27}) \dotfill 181 }} \vspace{\baselineskip}
\item[]{\large\textbf{References\dotfill 182}}
\end{itemize}

%%% Local Variables: 
%%% mode: latex
%%% TeX-master: "main"
%%% End: 

\section{Introduction and Survey of Results}
\indent

The present paper aims at the further elaboration of \emph{Relativistic Schr\"{o}dinger
  Theory} (RST), i.\,e.\ a \emph{fluid-dynamic} version of relativistic few-particle
quantum theory [1,2]. This fluid-dynamic approach is applied here in order to study the
non-relativistic energy spectrum of ortho-positronium; and it is found that there does
occur a somewhat strange phenomenon, i.\,e. the doubling of the number of the energy
levels. But since such a fluid-dynamic view on the physics of elementary particles is
rather unusual, it may perhaps appear desirable to first premise some conceptual and
epistemic remarks about its relation to the conventional quantum theory which is strongly
dominated by the \emph{probabilistic point-particle picture} of quantum matter.

As desirable and instructive as such a comparison of our fluid-dynamic concepts to the
conventional point-particle logic may be, there unfortunately arises a certain difficulty
here because the point-particle proponents seem still to disagree about their fundamental
notions.

The birth of quantum mechanics, some nine decades ago, has first been accompanied by
considerable confusion and many controversies about the epistemic status of the new
theory, see ref. [3] for a historic account of the notorious Bohr-Einstein debate. But now
that almost a century has passed away, and a considerable technical and cultural progress
could be made thanks to the overwhelming applications of the new theory (think, e.\,g., of
nuclear and laser technology, semiconductors, materials science, supernovae astrophysics,
big bang cosmology etc.) one should nowadays think that the teething troubles of the
theory's childhood have been successfully overcome already long time ago. Indeed, it is
hardly conceivable how such an immense civilizing progress could have been made without
possessing a correspondingly profound theoretical basis. In this sense, one expects that
present-day quantum theory has been anchored in the mean time on solid fundaments,
comparable to the situation with classical electromagnetism or thermodynamics, and thus
will be able to provide us with hopeful future developments.

However, this impression is grossly fallacious. Indeed, a superficial inspection of some
modern presentations of the fundamentals of conventional quantum theory is sufficient in
order to ascertain that the controversies about the true meaning of this theory and its
conceptional framework are becoming now rather more violent than dying out. This regrettable
fact is brought to light mostly on the occasions of celebrating the very birth of quantum
mechanics, e.g.\  Heisenberg's \emph{matrix mechanics} of 1925. Usually, the inventor (or
discoverer) of a new fruitful theory (such as, e.g., Galilei [4], Newton [5] or Maxwell
[6]) enjoys later on much reputation by the scientific descendants; and the corresponding
first texts about the original new idea are frequently celebrated by historic
reprints. However, concerning Heisenberg's \emph{matrix mechanics} of 1925, the situation
is rather ambiguous: whereas some people do hold it adequate to celebrate Heisenberg's
invention (discovery) by dedicating a book to his original proposal of matrix mechanics
[7], others argue that this proposal ``\emph{consisted essentially in the introduction of
  novel algorithms (i.\,e. his matrix mechanics) and even renowned and formalistic
  physicists do nowadays confirm that this paper is actually unintelligible}'' (ref.[8],
 p.\ 39; authors' translation).

A further subject of never-ending controversies concerns even the central concept of
quantum theory: i.e.\ the \emph{wave function} (or more generally: \emph{quantum state}). Is the
wave function to be associated to some real property of the observed system and thus does
exist also when the system is not observed? Or does the wave function merely reflect the
state of information of the observer? In the latter case, the wave function is thought to
``collapse'' whenever the observer's state of information undergoes a sudden change
(e.\,g. by reading off the pointer of some measurement device). A competent opinion looks
as follows: ``\emph{In the first place, it is often argued that the wave function itself
  should not be regarded as giving an objective description of the world (or part of it)
  but as providing information merely of 'one's state of knowledge' about the world. This
  view I really cannot accept.}'' (R. Penrose in ref. [9], p. 121). Another opinion is the
following: ``\emph{The claim, that the wave function be not real but describe
  'information', reminds me of arguments which are widespread in homeopathy}'' (ref. [8],
p.\ 79; authors' translation).

Some co-authors do also frankly concede that they disagree even about the fundamental
quantum concepts: ``\emph{Although we have tried to write a 'coherent' book, our reader
  will soon notice that our conceptions vary on some basic notions. Characteristic are our
  different opinions on the relevance of mathematical concepts for the interpretation of
  quantum mechanics and hence different inclinations to make active use of these concepts
  in physical arguments.}'' [10]

It should be evident that such a loose bedrock is not suited to erect on it a continual
scientific and technological progress; and it seems that already the fathers of the new
theory (especially Bohr and Heisenberg) were aware of such a grave drawback of their new
theory and therefore they tried to cobble together some minimalistic interpretation of
their quantum formalism (i.\,e. the \emph{kopenhagen interpretation}, see, e.\,g.,
ref. [9]) which should suffice to handle with the quantum formalism ``for all practical
purposes''. However, the Bohr-Heisenberg interpretative proposal of their formalism evokes
nowadays some harsh critique: ``\emph{If a physicist declares that he consistently applies
  the kopenhagen interpretation, then this does mean nothing else than that he
  consistently always refers to the same kind of conceptual inconsistency -- but this does
  not yield a consistent theory}'' (ref.[8], p.\ 51; authors' translation).

\begin{center}
  \large{\textit{Wave-Particle Duality}}
\end{center}

One of the key concepts of the kopenhagen interpretation does refer to the
\emph{complementarity principle}, with the \emph{wave-particle duality} representing the
most conspicuous exemplification thereof. Indeed, the progress of the past decades was
made by extensive use of \emph{both} the particle \emph{and} the wave concept in like
manner. Both concepts have been thought to be unrenouncable for the description of
elementary matter and its interactions. This fact becomes immediately evident by merely
glimpsing at the titles of some of the competent text books, see ref.s [11]--[25]. On the
other hand, there seem to arise now serious arguments which point to the possibility that
'particles' (in the classical and quantum sense) do not really exist but merely are some
kind of \emph{fiction} (if not illusion): ``\emph{There are no quantum jumps, nor are
  there particles}'' [8,26]. That might be true. But even if the long-lived concept of a
'particle' (classical and quantum) ultimately should turn out as a mere fiction, then it
is surely a very useful fiction since it advanced the progress of the past decades
considerably. If now a new time should dawn with the particle concept being superseded by
some novel, more powerful concept (perhaps \emph{decoherence} [8,10]?, or \emph{emergence}
[27]?) then that novel concept will also possess the epistemic status of being not more
than a \emph{fiction}! Indeed, it is hold quite generally that the limited epistemic
abilities of a human being admits to describe 'reality' at most in terms of
\emph{fictions} [28] which, however, must then of course be required to be free of logical
and observational contradictions.  \vskip 1cm
\begin{center}
*
\end{center}

Thus it seems to us that the notorious quantum controversies are nothing else than some
kind of useless struggles for domination among different fictions. In place of opposing
these different fictions against each other in an irreconcilable way, one should rather
consider them as helpful complementations of each other, so that they together can mediate
to us a more integral picture of what is really going on in the quantum world.

Bearing this situation in mind, one may doubt whether the ardently desired and unique
`\emph{theory of everything}' [29] will ever be discovered (or invented, resp.) in order
to supersede Bohr's complementarity principle. So long as we must be content with theories
having a loose end, we should rather recall Bohr's viewpoint on this question to our
mind. Indeed, it seems that Bohr was the first who noticed clearly that the behaviour of
elementary particles could not be satisfactorily described in terms of one single logical
framework. More concretely, he thought that the (self-suggesting) \emph{probabilistic
  point-particle} description of an elementary particle should be complemented by a
\emph{fluid-dynamic wave} description, namely in order to better understand and manage the
notorious \emph{wave-particle duality}. This original idea of Bohr is now known as the
`\emph{complementarity principle}' [9] (see also the corresponding remarks in the
precedent paper [1]). According to this principle (if it is understood in the right sense)
one needs two logical systems (or ``pictures''), excluding one another to a certain
degree, in order to become able to predict and `understand' the outcomes of the considered
experiments.

For instance, in the well-known two-slit interference experiment for electrons one becomes
forced to find some plausible explanation why the electrons do form the well-known
interference pattern on the detection screen after having passed the two-slit
arrangement. The solution of this problem is based upon the postulate that the electrons
do pass the two-slit region in form of a wave (with appropriate wave length); but their
arrival on the detection screen occurs in form of point-like particles [30].

In order to manage intelectually such a highly ambiguous phenomenon it seems wise to us to
resort to Bohr's complementarity principle and develop two complementary logical systems
(or `pictures'), i.\,e. the probabilistic point-particle picture and the fluid-dynamic
wave picture. This means that we have to specify two mathematical formalisms together with
their associated physical interpretations so that we can `understand' and predict the
outcomes of physical experiments and/or observations. Whether or not a future \emph{theory
of everything} will be able to incorporate simultaneously both pictures (or create a
completely new picture) must be left unclarified for the time being.

\begin{center} \large{\textit{Probabilistic Point-Particle Picture}}
\end{center}

The necessity for developing additionally also a fluid-dynamic picture of the quantum
objects becomes perhaps most clear when one reconsiders the somewhat artificial manner in
which the point-particle theory tries to explain the undeniable wave-like aspects which
become evident in the notorious double-slit experiment. 

A closer inspection of the interference pattern and the associated double-slit geometry in
terms of the particle picture may suggest the following conclusion: The confining of the
particle position to a spatial volume measured by a certain linear dimension (slit distance
$\Delta x$, say) must induce in some (mysterious) way a momentum uncertainty ($\Delta
p_x$, say) so that both \emph{point-particle quantities} will obey the Heisenberg
uncertainty relation
\begin{equation}
\label{eq:I.1} \Delta x \cdot \Delta p_x \gtrsim \frac{\hbar}{2} \;,
\end{equation} being needed for explaining the appearance of the interference pattern (see
any textbook about elementary quantum mechanics which treats the double-slit experiment in
the point-particle picture, e.\,g. [31]). Furthermore, since the interference pattern on
the detection screen is built up by a huge number of particle-like impacts, one thought
that one must describe such a phenomenon in terms of a \emph{statistical} theory. For such
a purpose, the appropriate mathematical formalism appeared to be the Hilbert space of
quantum states $|\Psi\!>$ with the classical observables (such as position $x$ and
momentum $p$) being transcribed to the corresponding operators (here $\hat{x}$ and
$\hat{p}$) acting over that Hilbert space. The particles passing through the double slit
and striking at the detection screen do then build up a statistical ensemble which is to
be characterized by the statistical operator (density matrix) $\hat{\rho}$, with
$\text{tr}\ \hat{\rho} = 1$. Thus, let the quantum state of a point particle, located
point-like at $\xi$ ($0 \lesssim \xi \lesssim \Delta x$), be denoted by the Hilbert space
vector $|\xi\!>$
\begin{equation}
\label{eq:I.2} \hat{x}\;|\xi\!>\; = \xi\,|\xi\!> \;,
\end{equation} then the statistical operator $\hat{\rho}$ for such an ensemble of
particles with \emph{unknown} (but well determined) locations in the double-slit region
($0 \lesssim \xi \lesssim \Delta x$) is adopted as
\begin{equation}
\label{eq:I.3} \hat{\rho} = \int\limits_0^{\Delta x} d\xi\;w(\xi)\;|\xi\!>\,<\!\xi|\;.
\end{equation} Here, the probability $w(\xi)$ for the space point $\xi$ ($0 \lesssim \xi
\lesssim \Delta x$) being occupied by some particle is to be normalized to unity as usual
\begin{equation}
\label{eq:I.4} \int\limits_0^{\Delta x} d\xi\;w(\xi) = 1 \;.
\end{equation}

Finally, it remains to demonstrate how the desired Heisenberg uncertainty relation
(\ref{eq:I.1}) can arise from such a mathematical construction. For this purpose, one
merely has to interprete the position and momentum uncertainties $\Delta x$ and $\Delta p$
as the corresponding mean square deviations (``variances'')
\begin{subequations}
\begin{align}
\label{eq:I.5a} \left( \Delta x \right)^2 &\doteqdot \text{tr}\left[ \hat{\rho}\,\left(
\hat{x} - <x> \right)^2 \right] \\
\label{eq:I.5b} \left( \Delta p \right)^2 &\doteqdot \text{tr}\left[ \hat{\rho}\,\left(
\hat{p}_x - <p_x> \right)^2 \right] \;,
\end{align}
\end{subequations} and additionally one postulates the Heisenberg commutation relations
for the operators $\hat{x}$ and $\hat{p}$ as follows:
\begin{equation}
\label{eq:I.6} \left[ \hat{x}, \hat{p}_x \right] = i\hbar \cdot \mathbf{1} \;.
\end{equation}

But the crucial point is now that this probability construction
(\ref{eq:I.2})-(\ref{eq:I.6}) cannot bring forth that desired uncertainty relation
(\ref{eq:I.1}) which one would like to see being validated in order to become able to
explain the appearance of the interference pattern. Here, the problematic constituent of
the whole probability construction is the assumption (\ref{eq:I.3}) which means that
anyone of the point particles does occupy a unique (albeit unknown) point of
space. Namely, such a collection of pure point particles does not develop the right
lateral pressure for generating a transverse momentum~$p_x$ across their direction of
flight ($y$-axis, say) which is necessary in order to produce just the observed
interference pattern on the detection screen. Indeed, if this assumption (\ref{eq:I.3}) is
rejected and replaced by the assumption that any ``point particle'' is in a (exotic) state
$|\Psi\!>$ which is some ``\emph{superposition}'' of the states $|\xi\!>$ (\ref{eq:I.2}),
being themselves due to the classical localization of a real point particle, then it is no
problem to deduce the wanted uncertainty relation (\ref{eq:I.1}). Namely, the statistical
operator $\hat{\rho}$ becomes now
\begin{equation}
\label{eq:I.7} \hat{\rho} = |\Psi\!>\,<\!\Psi|
\end{equation} with the superposition $|\Psi\!>$ of the classical configurations $|\xi\!>$
being defined through
\begin{equation}
\label{eq:I.8} |\Psi\!> = \int\limits_0^{\Delta x} d\xi\;\psi(\xi)\,|\xi\!> \;.
\end{equation} 
Obviously, this is just the place where the point-particle picture has to introduce
fluid-dynamic concepts, since the particle's spatial presence is (in some mysterious way)
now smeared out over some finite region. And consequently, the variances
(\ref{eq:I.5a})--(\ref{eq:I.5b}) emerge now in the position representation as
\begin{subequations}
\begin{align}
\label{eq:I.9a} \left( \Delta x \right)^2 &= \int\limits_0^{\Delta x}
dx\;\psi^*(x)\,\left( x - <\!x\!> \right)^2 \psi(x) \\
\label{eq:I.9b} \left( \Delta p_x \right)^2 &= \int\limits_0^{\Delta x}
dx\;\psi^*(x)\,\left( \hat{p}_x - <\!p_x\!> \right)^2 \psi(x) \;,
\end{align}
\end{subequations} where the mean values $<\!x\!>$, $<\!p_x\!>$ are defined as usual
through
\begin{subequations}
\begin{align}
\label{eq:I.10a} <\!x\!> &\doteqdot \int\limits_0^{\Delta x} dx\;\psi^*(x)\,x\,\psi(x) \\
\label{eq:I.10b} <\!p_x\!> &\doteqdot \int\limits_0^{\Delta x}
dx\;\psi^*(x)\,\hat{p}_x\,\psi(x) \\ &\left( \hat{p}_x =
\frac{\hbar}{i}\,\frac{\partial}{\partial x} \right) \;. \nonumber
\end{align}
\end{subequations}

Now, whenever the considered system (here: a point particle confined to the
one-di\-men\-sio\-nal interval $\left[ 0, \Delta x \right]$) is in such a superposition
$|\Psi\!>$ of classical configurations $|\xi\!>$ and therefore can be described by a
``\emph{wave function}'' $\psi(x)$ $\left( \doteqdot <\!x|\Psi\!> \right)$, then a quite
general theorem ensures that an inequality of the following type must hold (Schwarz'
inequality, e.\,g. [32])
\begin{equation}
\label{eq:I.11} \left( \Delta x \right)^2 \cdot \left( \Delta p \right)^2 \geq \left(
\frac{1}{2}\,\left|<\!C\!>\right| \right)^2 \;,
\end{equation} where $<\!C\!>$ is the expectation value of the commutator $\hat{C}$
\begin{equation}
\label{eq:I.12} \hat{C} \doteqdot \left[ \hat{x}, \hat{p}_x \right] = i\hbar\,\mathbf{1}
\;.
\end{equation} Obviously, the result (\ref{eq:I.11})--(\ref{eq:I.12}) for the variances
$\Delta x$, $\Delta p_x$ is now just the wanted uncertainty relation (\ref{eq:I.1}).

It appears natural that such a successful invention of a mathematical formalism (together
with the physical interpretation of the inherent mathematical quantities) will be
celebrated as a great progress in '\emph{understanding}' the working in the
micro-world. So much the more the formalism can be extended in order to deal with all
possible physical situations. It is merely necessary to set up a Hilbert space whose basis
vectors may be selected in such a way that they describe the classically realizable
configurations ($|\xi\!>$, say); but other basis systems are also possible. The Hilbert
space for the many-particle systems is the tensor product of the one-particle Hilbert
spaces. Next, one writes down the general state vector $|\Psi\!>$, if desired as a
superposition of the classical states $|\xi\!>$, cf. (\ref{eq:I.8}); and then one lets
this state vector $|\Psi\!>$ evolve in time according to the time-dependent
Schr\"{o}dinger equation
\begin{equation}
\label{eq:I.13} i\hbar\,\frac{\partial}{\partial t}\,|\Psi\!>_t = \hat{H}\,|\Psi\!>_t \;.
\end{equation}

For the stationary states one puts
\begin{equation}
\label{eq:I.14} |\Psi\!>_t = \e^{-i\,Et/\hbar}\,|\Psi\!>_0
\end{equation} which then yields the time-independent Schr\"{o}dinger equation
\begin{equation}
\label{eq:I.15} \hat{H}\,|\Psi\!>_0 = E\,|\Psi\!>_0
\end{equation} which, e.\,g., admits to determine the energy spectrum $\{E\}$ of bound
systems. This quantization formalism may not only be applied to mechanical systems but
also to field systems and then entails the well-known \emph{quantum field theory}. But
despite its overwhelming success the latter quantum-formalism continues to own the
epistemic status of a \emph{probabilistic point-particle fiction}; and therefore the
question must be asked whether there do perhaps exist other `fictions' which, after
thorough elaboration, do also yield a comprehensive and consistent picture of the
micro-world?
\newpage
\begin{center} \large{\textit{Complementarity and Environment}}
\end{center}

As already mentioned at various occasions, the precedent point-particle picture seems to
us to be plagued by a certain deficiency: i.e.\ the very concept of a ``point''
particle. In the original classical sense, a point particle does occupy some point
$\vec{x}$ of 3-space and possesses a definite momentum $\vec{p}$. Concerning the precise
position, the corresponding state vector $|\xi\!>$ in the one-particle Hilbert-space obeys
the eigenvalue equation (\ref{eq:I.2}). However, by means of the set (\ref{eq:I.3}) of
such state vectors one cannot explain the interference pattern on the detection screen!
Therefore one brings into play the conception of a coherent superposition $|\Psi\!>$
(\ref{eq:I.8}) of all classically possible positions $\xi$ ($0 \leq \xi \leq \Delta x$) so
that this range $\Delta x$ of all those possible locations together with the range of all
possible momenta $\Delta p$ can obey the required uncertainty relation (\ref{eq:I.1}). It
seems that such a strange superposition of classical locations has been considered because
one felt being forced to take account of the particle's finite ``\emph{extension}'' within
the logical framework of \emph{point} particles! Indeed, this auxiliary construction may
be understood in the sense that the point particle becomes somehow ``smeared'' over the
whole interval $\Delta x$ and thus appears as a kind of droplet of some esoteric fluid;
whereas on other occasions it may be treated as a true point particle! It surely does not
come as a surprise that such a self-contradictory and mysterious idea of a point particle
has provoked the question: ``\emph{Is this now profound thoughtfulness or the ultimate degree
  of craziness?}''
(ref.[8], p. 52; authors' translation).

Most textbook authors prefer to evade the notion of a point particle with unknowable
position and momentum and they rather try to explain the double-slit experiment completely
in terms of the wave picture [30]. But then the problem becomes urgent anew when one
resolves microscopically the interference pattern, being thought to arise in the wave
picture by the superposition of different wave trains. However, one finds the pattern
being composed of point-like impacts on the detection screen. So one feels oneself being
thrown back again to the notorious phenomenon of wave-particle duality.

In such a confused situation it seems reasonable to suppose that the dominance of either
the particle aspect or the wave aspect is brought forth cooperatively by the object itself
\emph{plus} its surroundings. Such a holistic theory (if feasible) could then interpolate
between the pure particle-like and the pure wave-like aspects of the considered physical
situation. But when one cuts the system off its environment ($\Rightarrow$ closed system)
the balance of the particle-like and wave-like aspect goes lost and one has to decide
whether one wants to neglect the particle aspect or the wave aspect. In many situations,
one of both aspects will be the dominant one and therefore suggests itself to become
preferred over the other one for the description of the isolated system. In other
instances, both aspects may be mutually counterbalanced; and if one nevertheless cuts the
system off its surroundings and favours only one of the complementary aspects, then there
will arise those notorious ``quantum paradoxes'', e.\,g. the mentioned double-slit
phenomenon. In recent time, there were put forth some proposals of treating the system in
combination with its surroundings (see, e.\,g., ref.s [8,10,27]) but presently a final
judgement seems premature.

Historically, the method of neglecting the system's environment, together with the
preference of the probabilistic point-particle picture, has been the prevailing world view
and the corresponding success is impressive, see ref.s [11]--[25]. But, if concentrating
on the system alone by neglecting its surroundings, one should not despise the possibility
of describing the isolated system also in fluid-dynamic terms according to the
complementary wave picture. One cannot exclude that the more comprehensive view on
``system plus environment'' can also be built up by starting from the side of the
fluid-dynamic approach.

A proposal of the latter kind is the ``\emph{Relativistic Schr\"{o}dinger Theory}'' [1,2],
i.\,e. a fluid-dynamic version of relativistic few-particle quantum mechanics. The
subsequent investigation is an application of this theory to \textbf{ortho-positronium} to
be conceived as a closed system (\emph{para-positronium} has been the subject of the
preceding paper [2]). More concretely, the present paper is a study of the question to
what extent the non-relativistic RST spectrum of ortho-positronium does coincide with that
of the conventional quantum theory (which is a probabilistic point-particle theory). The
main result of this study is that most of the conventional energy levels become
\emph{duplicated} as an indirect consequence of the specific way in which the angular
momenta of the constituents (i.\,e. electron and positron) are to be composed in RST to
the total angular momentum. Unfortunately, the corresponding energy eigenvalue problem is
in RST much more complicated than in the conventional quantum theory so that we have to
resort to appropriate perturbation techniques which we develop up to the ``octupole''
approximation [33]. Thus our obtained result of level duplication emerges here in the
quadrupole and octupole approximations and therefore might eventually be an artefact of
those perturbation orders, i.e.\ possibly not being present in the \emph{exact} RST
solution of the eigenvalue problem for ortho-positronium. However, the main effect of
level doubling occurs in the quadrupole approximation (\textbf{Fig. VII.A} on
p.\pageref{fig7a}) and the relative doubling energy amounts to \emph{more} than 1\%; on
the other hand, the octupole correction of the levels in the quadrupole approximation
amounts to \emph{less} than 1 percent (table on p.~\pageref{table8}). This hints at level
doubling being a true effect in RST which receives merely some minor corrections from the
higher-multipole approximations.  \vskip 1cm
\begin{center} *
\end{center}

It is true, the origin of this level duplication can be traced back to the fact that
ortho-positronium has unity of spin ($\leadsto s_\mathcal{O} = 1$); whereas
para-positronium has zero spin ($\leadsto s_\mathcal{P} = 0$) and therefore does not
undergo that phenomenon of level duplication. But it is important to note that the energy
difference of both levels (emerging by duplication) is not of magnetic but rather of
\emph{electric} origin! Namely, for the ortho-spin $s_\mathcal{O} = 1$, the eigenvalue
problem for angular momentum admits (mostly) two different solutions belonging to the same
values of total angular momentum $j_\mathcal{O}$ and its $z$-component $\bjz$. This then
yields two different angular distributions of the corresponding charge densities
($\leadsto$ \emph{dimorphism}, see \textbf{Fig.IV.B} on p.~\pageref{fig4b}) which thereby
acquire different interaction energies of the \emph{electrostatic} type.

Subsequently, these results will be elaborated in detail along the following
\mbox{arrangement.}

\begin{center} \large{\textit{Fluid-Dynamic Character of RST}}
\end{center}

As proponents of a fluid-dynamic description of the elementary matter we tend to the
hypothesis that `matter' (in its most general sense) is always spread out over some finite
region of three-space and can never be concentrated in a truly point-like manner. The
fiction of a point-like particle appears to us as a (more or less realistic) idealization;
and consequently the finite-size effects of matter must be somehow simulated in the
point-particle picture, preferably by saying that the state $|\Psi\!>$ of the
point-particle refers to a ``superposition'' of states being due to really point-like
positions $|\xi\!>$, see equation (\ref{eq:I.8}). By this construction it may seem now
that the concept of \emph{wavefunction} $\psi(x)$ ($\doteqdot <\!x|\Psi\!>$) is the
crucial point in the probabilistic approach and thus must be made responsible for its
overwhelming success.

But actually RST is also based essentially on the use of wavefunctions $\Psi(x)$, namely
in order to generate the physical densities of the considered system such as, e.\,g., the
total current density $j_\mu$ or the energy-momentum density $\Tmunu$ of the considered
system, see equations (\ref{eq:II.29}) and (\ref{eq:II.35}) below. Thus, in contrast to
the situation with the point-particle case, the RST wave function $\Psi$ owns a
well-defined meaning, namely to generate all the physical densities of the considered
system (Sect.s II and III are taken over from the precedent paper [2] in order to
elucidate the common basis of both para- and orthopositronium $\leadsto$ \emph{ortho/para
  dichotomy}). The RST handling with the wave functions $\Psi(x)$ is quite different from
the probabilistic point-particle case when many-particle systems are considered: here, the
conventional theory relies on the tensor product of the one-particle Hilbert spaces and
therefore sets up the dynamics (\ref{eq:I.13}) in the \emph{configuration space}. However,
RST adopts the Whitney sum of the one-particle fibre bundles
\begin{equation}
\label{eq:I.16} \Psi(x) = \psi_1(x) \oplus \psi_2(x) \oplus \psi_3(x) \oplus \ldots
\end{equation} which then admits to set up the dynamics in the \emph{real four-dimensional space-time},
namely in form of the \emph{Relativistic Schr\"odinger Equation} for the wave function
$\Psi(x)$ in combination with the (non-abelian) \emph{Maxwell equations} for the bundle
connection $\A_\mu(x)$, see equations (\ref{eq:II.1}) and (\ref{eq:II.15}) below.

This formal difference of the dynamical equations of both approaches entails an important
physical consequence: whereas (in the non-relativistic approximation) the probabilistic
conventional approach takes the predetermined Coulomp potential as the basis for the
interaction mechanism, the corresponding RST mechanism retains the interaction field as a
truly dynamical object being equipped with its own field equation. For instance, for the
positronium system (to be considered subsequently) the non-relativistic Hamiltonian
$\hat{H}$ (\ref{eq:I.15}) contains the non-dynamical Coulomb potential in a manifest way
\begin{equation}
\label{eq:I.17} \hat{H} = \frac{\hat{p}_1^2}{2M} + \frac{\hat{p}_2^2}{2M} -
\frac{\e^2}{||\vec{r}_1 - \vec{r}_2||} \;.
\end{equation} On the other hand, the corresponding RST interaction potential $\left(
\bgAe(\vr) \right)$, say, must obey the Poisson equation (\ref{eq:IV.19}) below, where the
source is given by the electric charge density $\left( \bgkn(\vr) \right)$ generated by
the RST wave function $\Psi$. Through this arrangement the RST eigenvalue problem becomes a
system of coupled differential equations, which (in the spherically symmetric
approximation) consists of the Schr\"odinger equation (\ref{eq:IV.17}) and the Poisson
equation (\ref{eq:IV.18}). Naturally, such a coupled eigenvalue system is much more
difficult to solve than the simple conventional problem (\ref{eq:I.15}) which for the
positronium Hamiltonian (\ref{eq:I.17}) admits to determine exactly the energy spectrum
$E_C^{(n)}$ in terms of the principal quantum number $n$:
\begin{equation}
\label{eq:I.18} E_C^{(n)} = -\frac{\e^2}{4a_B} \cdot \frac{1}{n^2} = -
\frac{6{,}8029\ldots}{n^2}\,\text{[eV]} \;.
\end{equation}
\\[-10mm]
\begin{center} \large{\textit{Ortho-Dimorphism}}
\end{center}

In contrast to this simple situation in the conventional probabilistic theory, the
solution of the corresponding RST eigenvalue problem can be worked out only
approximately. The difficulty refers here to the fact that the wave function becomes
considerably anisotropic, according to the value of angular momentum $\bjz$; as an example
see \textbf{Fig.IV.A} below. But since the wave function generates the charge density
which then acts as the source of the interaction potential, the anisotropy of the wave
function is transferred to the interaction potential via the Poisson equation so that we
ultimately have to solve the (RST form of the) Schr\"odinger equation for an
\emph{anisotropic} potential. Clearly, this can be attained only in an approximative way;
and most of the paper is concerned with setting up an adequate approximation procedure
(\textbf{Sect.VIII}) for managing this anisotropy effect. The key point is here a suitable
factorization of the wave function in a radial part and an angular-dependent part which
then also becomes transferred to the charge density, cf. its product form
(\ref{eq:IV.19})--(\ref{eq:IV.21}) below.

As usual, the (approximate) product form of the wave function splits up the energy
eigenvalue problem in two subproblems, namely (\textbf{i}) the radial problem which
depends upon the specific physical system to be considered and which ultimately yields the
energy eigenvalues, and (\textbf{ii}) the eigenvalue problem of angular momentum which is
of quite general nature and thus is independent of the details of the considered
system. Now it is just this latter problem, cf. (\ref{eq:IV.6a})--(\ref{eq:IV.6b}), which
provides us with the origin of the ortho-dimorphism. Namely, for the same values of the
angular-momentum quantum numbers $\left\{ \jO, \bjz \right\}$ with $(-\jO) \leq \bjz \leq
\jO$ there do exist \emph{two} solutions of the eigenvalue problem. This ambiguity then is
transferred to the angular part $\bgkn(\vartheta)$ of the charge density, see
\textbf{Fig.IV.B} below.

\begin{center} \large{\textit{Induced Energy Difference of Dimorphic Partners}}
\end{center}

Since the charge density is the source of the electrostatic interaction potential, cf. the
Poisson equation (\ref{eq:II.43}), the angular ambiguity of the charge density is
immediately transferred to the interaction potential which thereby inherits the anisotropy
of the charge density. More precisely, for any pair of angular-momentum quantum numbers
$\left\{ \jO, \bjz \right\}$ the interaction potential $\bgAe(\vr)$ acquires a
corresponding angular dependence; and the elaboration of this interconnection necessitates
an extensive search for the most rational representation of the angular dependency of the
interaction potential (\textbf{Sect. VII--VIII}). There are various possibilities, but
ultimately it turns out that the \emph{separative method} provides the most pleasant
representation of the wanted multipole expansion because here the energy contributions of
the various multipole modes can be clearly separated. Consequently, one can subdivide the
total anisotropy energy into the set of contributions of any mode ($\leadsto$ quadrupole
energy, octupole energy, \ldots).

But if once the anisotropy energy is determined, one can substitute this (together with
the isotropy energy and the kinetic energy) in the RST energy functional and try to
extremalize this by use of an appropriate trial ansatz. For our present purpose, we are
satisfied with a very simple ansatz for the radial part $\tO(r)$ of the wave amplitude,
cf. (\ref{eq:VI.1a})--(\ref{eq:VI.1b}), with only two trial parameters $(\beta, \nu)$. The
value of the RST energy functional $\tEEO$ (\ref{eq:VI.25}) on the chosen trial
configuration yields the corresponding energy function $\tEEgiv(\beta,\nu)$ as a function
of both trial parameters $\beta$ and $\nu$, cf.~(\ref{eq:VI.36}). After minimalization of
this function with respect to the first trial parameter $\beta$ we are left with the
problem of determining the minimal value of the reduced energy function $\EEOj(\nu)$
(\ref{eq:VIII.110}) or, resp., the maximal value of the associated spectral function
$\SOjg(\nu)$ (\ref{eq:VIII.112}). This function contains both quantum numbers $\jO$ and
$\bjz$ ($\Leftrightarrow \bg m_1, \bg m_2$) of angular momentum while the principal
quantum number $\nO$ is restricted to $\nO = \jO + 1$ on account of our too simple ansatz
(\ref{eq:VI.1a})--(\ref{eq:VI.1b}). Thus, one finally obtains the energy spectrum of
ortho-positronium simply by looking for the maximal value of the spectral function
$\SOjg(\nu)$ (\ref{eq:VIII.112}), namely by admitting all possible quantum numbers $\jO =
0,1,2,3,\ldots$ and $\bjz$, with $-\jO \leq \bjz \leq \jO$. This may be performed by means
of some appropriate numerical program, and the most important results are the following
(\textbf{Fig.VII.A} and table on p.~\pageref{table8}):

The energy spectrum resembles the conventional one (\ref{eq:I.18}) but its ($n^2$)-fold 
degeneracy becomes eliminated. Such a result has already been found for the spectrum of
para-positronium [2]; but for ortho-positronium there additionally occurs now a doubling
of most of the energy levels. The origin of this strange effect traces back to the
specific spin composition for ortho-positronium ($s_\mathcal{O}=1$), where two different
solutions of the eigenvalue problem for angular momentum are mostly possible for fixed
values of quantum numbers~$\jO$ and~$\bjz$. The corresponding energy difference (due to
level splitting) amounts to (roughly) 1\% of the binding energy, see the table
on p~\pageref{table8}; it is not caused by
magnetism but is a purely electric effect due to the ambiguity of the electric charge
distribution, see \textbf{Fig.IV.B},~p.~\pageref{fig4b}.

\section{Positronium Eigenvalue Problem}
\indent

In order that the paper be sufficiently self-contained, it may appear useful to mention
briefly some fundamental facts about RST. As its very notation says, the central idea is
the Relativistic Schr\"odinger Equation (\ref{eq:II.1}) which leads one in a rather
straight-forward way to the Dirac equation (\ref{eq:II.13}) for few-particle (or
many-particle) systems. In order to ultimately end up with a closed dynamical system for
the fluid-dynamic quantum matter, one adds the (generally non-Abelian) Maxwell equations
(\ref{eq:II.15}) where this coupled system of matter and gauge field dynamics
automatically entails certain conservation laws, such as those for charge (\ref{eq:II.25})
or energy-momentum (\ref{eq:II.32}). This fundamental structure of RST is then
subsequently specialized down to the non-relativistic positronium system, especially to its
ortho-form, with the main interest aiming at its energy spectrum.

\begin{center}
  \emph{\textbf{1.\ Relativistic Schr\"odinger Equation}}
\end{center}

A subset of problems within the general framework of RST concerns the (stationary) bound
systems. The simplest of those systems is positronium which consists of two oppositely
charged particles of the same rest mass ($M$). The physical behaviour of its matter
subsystem is assumed here to obey the \emph{Relativistic Schr\"odinger Equation}
\begin{equation}
\label{eq:II.1}
i\hbar\crm\;\Du\,\Psi = \mathcal{H}_\mu\,\Psi
\end{equation}
where the two-particle wave function $\Psi(x)$ is the direct sum of the two one-particle wave functions $\psi_a(x)\ (a=1,2)$
\begin{equation}
\label{eq:II.2}
\Psi(x) = \psi_1(x) \oplus \psi_2(x) \;.
\end{equation}
The gauge-covariant derivative $\D$ on the left-hand side of the basic wave equation (\ref{eq:II.1}) is defined in terms of the $\mathfrak{u}(2)$-valued gauge potential $\A_\mu$ as usual
\begin{equation}
\label{eq:II.3}
\Du\,\Psi = \partial_\mu\,\Psi + \A_\mu\,\Psi \;,
\end{equation}
or rewritten in component form
\begin{subequations}
\begin{align}
\label{eq:II.4a}
D_\mu\,\psi_1 &= \partial_\mu\,\psi_1 - i\,A^2_\mu\,\psi_1 \\
\label{eq:II.4b}
D_\mu\,\psi_2 &= \partial_\mu\,\psi_2 - i\,A^1_\mu\,\psi_2 \;.
\end{align}
\end{subequations}
Here, the electromagnetic four-potentials $A^a_\mu\ (a=1,2)$ are the components of the
original gauge potential $\A_\mu$ with respect to some suitable basis $\tau_\alpha\
(\alpha=1, \ldots 4)$ of the $\mathfrak{u}(2)$-algebra
\begin{equation}
\label{eq:II.5}
\A_\mu (x) = A^\alpha_\mu (x)\,\tau_\alpha = A^a_\mu (x)\,\tau_a + B_\mu(x)\,\chi - B^*_\mu(x)\,\bar{\chi} \;.
\end{equation}
Here, the electromagnetic generators $\tau_a\ (a=1,2)$ do commute
\begin{equation}
\label{eq:II.6}
\left[ \tau_1, \tau_2 \right] = 0
\end{equation}
and the exchange potential $B_\mu$ is put to zero ($\leadsto B_\mu(x) \equiv 0$) because the two positronium constituents (i.\,e. electron and positron) do count as \emph{non-identical} particles. Recall that the exchange effects, being mediated by the exchange potential $B_\mu(x)$, do occur exclusively for \emph{identical} particles so that $B_\mu(x)$ is inactive for the positronium constituents ($\leadsto B_\mu(x) \equiv 0$). Thus the bundle connection $\A_\mu(x)$ (\ref{eq:II.5}) becomes reduced to its $\mathfrak{u}(1) \oplus \mathfrak{u}(1)$ projection
\begin{equation}
\label{eq:II.7}
\A_\mu(x)\ \Rightarrow\ A^a_\mu(x)\,\tau_a \;.
\end{equation}

\begin{center}
  \emph{\textbf{2.\ Dirac Equation}}
\end{center}

For Dirac particles, which are to be described by four-spinors $\psi_a(x)$, the Hamiltonian $\mathcal{H}_\mu$ in the Relativistic Schr\"odinger Equation (\ref{eq:II.1}) obeys the relation
\begin{equation}
\label{eq:II.8}
\GG^\mu\,\mathcal{H}_\mu = \M\crm^2 \;,
\end{equation}
where $\GG^\mu$ is the total velocity operator and thus is the direct sum of the Dirac matrices $\gamma^\mu$
\begin{equation}
\label{eq:II.9}
\GG^\mu = \left( -\gamma^\mu \right) \oplus \gamma^\mu \;.
\end{equation}
The mass operator $\M$ specifies the two particle masses $M^a\ (a=1,2)$
\begin{equation}
\label{eq:II.10}
\M = i\,M^a\,\tau_a
\end{equation}
and is required to be Hermitian ($\bar{\M} = \M$) and covariantly constant
\begin{equation}
\label{eq:II.11}
\Du\,\M \equiv 0 \;.
\end{equation}
This requirement is trivially satisfied for particles of identical rest masses ($M^1 = M^2 \doteqdot M$) since for such a situation the mass operator becomes proportional to the identity operator
\begin{equation}
\label{eq:II.12}
\M = M\ \mathbf{1} \;.
\end{equation}
Thus the result is that, by virtue of the relation (\ref{eq:II.8}), the Relativistic Schr\"odinger Equation (\ref{eq:II.1}) becomes the two-particle Dirac equation
\begin{equation}
\label{eq:II.13}
i\hbar\mathrm{c}\,\GG^\mu\,\D_\mu\,\Psi = \M\crm^2\,\Psi \;,
\end{equation}
or in component form
\begin{subequations}
\begin{align}
\label{eq:II.14a}
i\hbar\mathrm{c}\,\gamma^\mu\,D_\mu\,\psi_1 &= -M\crm^2\,\psi_1 \\
\label{eq:II.14b}
i\hbar\mathrm{c}\,\gamma^\mu\,D_\mu\,\psi_2 &= M\crm^2\,\psi_2 \;,
\end{align}
\end{subequations}
where the gauge-covariant derivatives ($D$) of the single-particle wave functions $\psi_a(x)$ are given by equations (\ref{eq:II.4a})--(\ref{eq:II.4b}).

For \emph{identical} particles, the Dirac equations (\ref{eq:II.14a})--(\ref{eq:II.14b}) would couple both particles much more directly since the exchange potential $B_\mu$ is generated cooperatively by both particles and simultaneously does act back on any individual particle which then entails the phenomenon of self-coupling. However, for the present situation of \emph{non-identical} particles the coupling is more indirect: any particle does generate a Dirac four-current $k_{a\,\mu}(x)\ (a=1,2)$ which is the source of the four-potential $A^a_\mu(x)$ (see below). And then this four-potential $A^a_\mu$ of the $a$-th particle acts on the wave-function $\psi_b(x)$ of the other particle ($b \neq a$) as shown by equations (\ref{eq:II.14a})--(\ref{eq:II.14b}) in connection with the gauge-covariant derivatives $D$ (\ref{eq:II.4a})--(\ref{eq:II.4b}).

\begin{center}
  \emph{\textbf{3.\ Maxwell Equations}}
\end{center}

\begin{sloppypar}
The bundle connection $\A_\mu(x)$ (\ref{eq:II.5}) is itself a dynamical object of the theory (just as is the wave function $\Psi(x)$) and therefore must be required to obey some field equation. This is the (generally non-Abelian) Maxwell equation
\begin{gather}
\label{eq:II.15}
\D^\mu\F_{\mu\nu} = -4\pi i\as\J_\nu \\
\left(\as \doteqdot \frac{e^2}{\hbar \crm} \right) \;. \nonumber
\end{gather}
Here, the bundle curvature $\F_{\mu\nu}$ is defined in terms of the bundle connection $\A_\mu$ as usual, i.\,e.
\begin{equation}
\label{eq:II.16}
\F_{\mu\nu} = \nabla_\mu\A_\nu - \nabla_\nu\A_\mu + \left[\A_\mu,\A_\nu \right] \;.
\end{equation}
For the present situation of non-identical particles, the connection $\A_\mu$ becomes reduced to its (Abelian) $\mathfrak{u}(1) \oplus \mathfrak{u}(1)$ projection, cf. (\ref{eq:II.7}), which then also holds for its curvature $\F_{\mu\nu}$
\begin{equation}
\label{eq:II.17}
\F_{\mu\nu}\ \Rightarrow\ \nabla_\mu\A_\nu - \nabla_\nu\A_\mu \;.
\end{equation}
Decomposing here both the curvature $\F_{\mu\nu}$ and current operator $\J_\mu$ with respect to the chosen basis of commuting generators $\tau_a\ (a=1,2)$
\begin{subequations}
\begin{align}
\label{eq:II.18a}
\F_{\mu\nu}\ &\Rightarrow\ F^a_{\mu\nu} \tau_a \\
\label{eq:II.18b}
\J_\nu\ &\Rightarrow\ i {j^a}_\nu \tau_a \;,
\end{align}
\end{subequations}
one obtains the Maxwell equations (\ref{eq:II.15}) in component form as
\begin{equation}
\label{eq:II.19}
\nabla^\mu F^a_{\mu\nu} = 4\pi \as j^a_\nu \;.
\end{equation}
Since for the present Abelian situation the curvature components $F^a_{\mu\nu}$ (\ref{eq:II.18a}) are linked to the connection components $A^a_\mu$ (\ref{eq:II.7}) as usual in Maxwellian electrodynamics (in its Abelian form)
\begin{equation}
\label{eq:II.20}
F^a_{\mu\nu} = \nabla_\mu A^a_\nu - \nabla_\nu A^a_\mu \;,
\end{equation}
the Maxwell equations (\ref{eq:II.19}) for the field strengths $F^a_{\mu\nu}$ become converted to the d'Alembert equations for the four-potentials $A^a_\mu$
\begin{equation}
\label{eq:II.21}
\square\,A^a_\mu = 4\pi \as j^a_\mu \;,
\end{equation}
provided the gauge potentials $A^a_\mu$ do obey the Lorentz gauge condition
\begin{equation}
\label{eq:II.22}
\nabla^\mu A^a_\mu \equiv 0 \;.
\end{equation}
\end{sloppypar}
\begin{center}
  \emph{\textbf{4.\ Conservation Laws}}
\end{center}

One of the most striking features in the description of physical systems is that both classical and quantum matter do obey certain conservation laws. For the presently considered Relativistic Schr\"odinger Theory, as a \emph{fluid-dynamic} theory, this means that there should exist certain \emph{local} conservation laws, preferably concerning charge and energy-momentum. Moreover, these local laws should turn out as an immediate consequence of the basic dynamical equations, i.\,e. the Relativistic Schr\"odinger Equation (\ref{eq:II.1}) and the Maxwell equations (\ref{eq:II.15}).

In this regard, a very satisfying feature of the Relativistic Schr\"odinger Theory is now that such conservation laws are automatically implied by the dynamical equations themselves. In order to elaborate this briefly, consider first the conservation of total charge which as a local law reads
\begin{equation}
\label{eq:II.23}
\nabla^\mu j_\mu \equiv 0 \;.
\end{equation}
But such a continuity equation for the total four-current $j_\mu$ can easily be deduced from \emph{both} the matter equation (\ref{eq:II.1}) and the gauge field equations (\ref{eq:II.15}); and this fact signals the internal consistency of the RST dynamics. First, consider the gauge field dynamics (\ref{eq:II.15}) and observe here the identity
\begin{equation}
\label{eq:II.24}
\D^\mu\D^\nu\F_{\mu\nu} \equiv 0
\end{equation}
which holds in any flat space-time. Obviously, the combination of this identity with the Maxwell equations (\ref{eq:II.15}) yields the following continuity equation in operator form
\begin{equation}
\label{eq:II.25}
\D^\mu\J_\mu \equiv 0 \;.
\end{equation}
Decomposing here the current operator in component form yields
\begin{equation}
\label{eq:II.26}
D^\mu j^\alpha_\mu \equiv 0
\end{equation}
which furthermore simplifies to
\begin{gather}
\label{eq:II.27}
\nabla^\mu\;j^a_\mu \equiv 0 \\
(a = 1,2) \nonumber
\end{gather}
under the Abelian reduction (\ref{eq:II.18a})--(\ref{eq:II.18b}). But when the individual Maxwell currents $j^a_\mu$ do obey such a continuity equation (\ref{eq:II.27}), then the total current $j_\mu$
\begin{equation}
\label{eq:II.28}
j_\mu \doteqdot \sum_{a=1}^{2}\, j^a_\mu
\end{equation}
must also obey a continuity equation which is just the requirement (\ref{eq:II.23}).

On the other hand, we can start also from the matter dynamics (\ref{eq:II.1}) and can define the total current $j_\mu$ by
\begin{equation}
\label{eq:II.29}
j_\mu \doteqdot \bar{\Psi} \GG_\mu \Psi \;.
\end{equation}
The divergence of this current is
\begin{equation}
\label{eq:II.30}
\nabla^\mu j_\mu = \left( \D^\mu \bar{\Psi} \right) \GG_\mu + \bar{\Psi} \GG_\mu \left( \D^\mu \Psi \right) + \bar{\Psi} \left( \D^\mu \GG_\mu \right) \Psi \;.
\end{equation}
Here, one requires that the gauge-covariant derivative of the total velocity operator $\GG_\mu$ is covariantly constant
\begin{equation}
\label{eq:II.31}
\D^\mu \GG_\mu \equiv 0 \;,
\end{equation}
and furthermore one evokes the Relativistic Schr\"odinger equation together with the Hamiltonian condition (\ref{eq:II.8}) which then ultimately yields again the desired continuity equation (\ref{eq:II.23}). Thus, the conservation of total charge is actually deducible from both subdynamics of the whole RST system; and this fact supports the mutual compatibility of both subdynamics (i.\,e. the matter dynamics (\ref{eq:II.1}) and the gauge field dynamics (\ref{eq:II.15})).

A further important conservation law does refer to the energy-momentum content of the considered physical system. Aiming again at a \emph{local} law, one may think of a continuity equation of the following form
\begin{equation}
\label{eq:II.32}
\nabla^\mu\, \TTmunu \equiv 0 \;,
\end{equation}
where $\TTmunu$ is the total energy-momentum density, i.\,e. the sum of the Dirac matter part $\left( \DTmunu \right)$ and the gauge field part $\left( \GTmunu \right)$:
\begin{equation}
\label{eq:II.33}
\TTmunu = \DTmunu + \GTmunu \;.
\end{equation}
Clearly, the validity of the total law (\ref{eq:II.32}) does not require an analogous law for the subdensities but merely requires the right balance of the energy-momentum exchange between the subsystems, i.\,e.
\begin{equation}
\label{eq:II.34}
\nabla^\mu \, \DTmunu = - \nabla^\mu \, \GTmunu \;.
\end{equation}

Indeed, the matter part has been identified as
\begin{equation}
\label{eq:II.35}
\DTmunu = \frac{i\hbar\crm}{4} \left[ \bar{\Psi} \GG_\mu (\D_\nu \Psi) - (\D_\nu \bar{\Psi}) \GG_\mu \Psi + \bar{\Psi}\GG_\nu (\D_\mu \Psi ) - (\D_\mu \bar{\Psi}) \GG_\nu \Psi \right]
\end{equation}
and the gauge field part by
\begin{equation}
\label{eq:II.36}
\GTmunu = \frac{\hbar\crm}{4\pi\as} K_{\alpha\beta} \left( F^\alpha_{\ \mu\lambda} F^{\beta\ \lambda}_{\ \nu} - \frac{1}{4} g_{\mu\nu} F^{\alpha}_{\ \sigma\lambda} F^{\beta\sigma\lambda} \right) \;,
\end{equation}
where $K_{\alpha\beta}$ is the fibre metric in the associated Lie algebra bundle. The (local) conservation law (\ref{eq:II.32}) comes now actually about through the mutual annihilation (\ref{eq:II.34}) of the sources of both energy-momentum densities, i.\,e.
\begin{equation}
\label{eq:II.37}
\nabla^\mu\,\DTmunu = -\nabla^\mu\,\GTmunu = \hbar\crm\, F^{\alpha}_{\ \mu\nu}\, j^{\ \mu}_{\alpha} \;.
\end{equation}
Obviously the sources of the partial densities $\DTmunu$ and $\GTmunu$ are just the well-known \emph{Lorentz forces} in non-Abelian form.

It should now appear self-suggesting that the definition of the total energy ($\ET$) of an RST field configuration is to be based upon the time component $\TToo$ of the energy-momentum density $\TTmunu$, i.\,e.
\begin{equation}
\label{eq:II.38}
E_T \doteqdot \int d^3\vr\;\TToo(\vr) \;.
\end{equation}
But since the total density $\TTmunu$ is the sum of a matter part and a gauge field part, cf. (\ref{eq:II.33}), the total energy $\ET$ (\ref{eq:II.38}) naturally breaks up in an analogous way
\begin{equation}
\label{eq:II.39}
\ET = E_\mathrm{D} + E_\mathrm{G} \;,
\end{equation}
with the self-evident definitions
\begin{subequations}
\begin{align}
\label{eq:II.40a}
E_\mathrm{D} &\doteqdot \int d^3\vr\;\DToo(\vr) \\
\label{eq:II.40b}
E_\mathrm{G} &\doteqdot \int d^3\vr\;\GToo(\vr) \;.
\end{align}
\end{subequations}
But clearly, such a preference of the time component $\Too$ among all the other components $\Tmunu$ entails the selection of a special time axis for the space-time manifold. This then induces a similar space-time splitting of all the other objects in the theory, i.\,e. we have to consider now \emph{stationary} field configurations which are generally thought to represent the basis of the energy spectra of the bound systems.

\begin{center}
  \emph{\textbf{5.\ Stationary Field Configurations}}
\end{center}

In the present context, the notion of stationarity is coined with regard to the time-independence of the physical observables of the theory, i.\,e. the physical densities and the electromagnetic fields generated by them. In contrast to this, the wave functions do not count as observables and therefore are not required to be time-independent. But their time-dependence must be in such a way that the associated densities become truly time-independent.

\begin{center}
  \large{\textit{Gauge-Field Subsystem}}
\end{center}

The simplest space-time splitting refers to the four-potentials $A^a_{\ \mu}$, which for the stationary states become time-independent and thus appear in the following form:
\begin{gather}
\label{eq:II.41}
A^a_{\ \mu}(x)\ \Rightarrow\ \left\{ {}^{(a)}\!A_0(\vr)\;;\ -\vec{A}_a(\vr) \right\} \\
(a = 1,2) \;. \nonumber
\end{gather}
A similar arrangement does apply also to the Maxwell four-currents $j^a_{\ \mu}$
\begin{equation}
\label{eq:II.42}
j^a_{\ \mu}\ \Rightarrow\ \left\{ {}^{(a)}\!j_0(\vr)\;;\ -\vec{j}_a(\vr) \right\} \;,
\end{equation}
so that the d'Alambert equations (\ref{eq:II.21}) become split up into the Poisson equations, for both the scalar potentials ${}^{(a)}\!A_0(\vr)$
\begin{equation}
\label{eq:II.43}
\Delta {}^{(a)}\!A_0(\vr) = -4\pi\as {}^{(a)}\!j_0(\vr)
\end{equation}
and the three-vector potentials $\vec{A}_a(\vr)$
\begin{equation}
\label{eq:II.44}
\Delta \vec{A}_a(\vr) = -4\pi\as \vec{j}_a(\vr) \;.
\end{equation}
Recall here that the standard solutions of these equations are formally given by
\begin{subequations}
\begin{align}
\label{eq:II.45a}
{}^{(a)}\!A_0(\vr) &= \as \int d^3\vr\,'\;\frac{{}^{(a)}\!j_0(\vr\,')}{||\vr-\vr\,'||} \\
\label{eq:II.45b}
\vec{A}_a(\vr) &= \as \int d^3\vr\,'\;\frac{\vec{j}_a(\vr\,')}{||\vr-\vr\,'||} \;.
\end{align}
\end{subequations}
Similar arguments would apply also to the exchange potential $B_\mu = \left\{ B_0, -\vec{B} \right\}$ but since we are dealing with non-identical particles the exchange potential must be put to zero $(B_\mu(x) \equiv 0)$.

It is true, the particle interactions are organized here via the (electromagnetic and exchange) \emph{potentials} which, according to the \emph{principle of minimal coupling}, are entering the covariant derivatives $D_\mu \psi_a$ of the wave functions $\psi_a$ as shown by equations (\ref{eq:II.4a})--(\ref{eq:II.4b}). But nevertheless it is very instructive to glimpse also at the \emph{field strengths} $F^a_{\ \mu\nu}$. Their space-time splitting is given by
\begin{subequations}
\begin{align}
\label{eq:II.46a}
\vec{E}_a = \left\{ {}^{(a)}\!E^j \right\} &\doteqdot \left\{ F^a_{\ 0j} \right\} \\
\label{eq:II.46b}
\vec{H}_a = \left\{ {}^{(a)}\!H^j \right\} &\doteqdot \left\{ \frac{1}{2}\, \varepsilon^{jk}_{\ \ l}\, F^{a\ l}_{\ k} \right\}
\end{align}
\end{subequations}
and thus the \emph{linear} Maxwell equations (\ref{eq:II.19}) do split up in three-vector form $(a=1,2)$ to the scalar equations for the electric fields
\begin{equation}
\label{eq:II.47}
\vec{\nabla} \sdot \vec{E}_a = 4\pi\as {}^{(a)}\!j_0
\end{equation}
and to the curl equations for the magnetic fields
\begin{equation}
\label{eq:II.48}
\vec{\nabla} \times \vec{H}_a = 4\pi\as \vec{j}_a \;.
\end{equation}
There is a pleasant consistency check for these linearized (but still relativistic) field equations in three-vector form; namely one may first link the field-strengths to the potentials in three-vector notation (cf. (\ref{eq:II.20}) for the corresponding relativistic link):
\begin{subequations}
\begin{align}
\label{eq:II.49a}
\vec{E}_a(\vr) &= - \vec{\nabla} {}^{(a)}\!A_0(\vr) \\
\label{eq:II.49b}
\vec{H}_a(\vr) &= \vec{\nabla} \times \vec{A}_a(\vr)
\end{align}
\end{subequations}
and then one substitutes these three-vector field strengths into their source and curl equations (\ref{eq:II.47})--(\ref{eq:II.48}). In this way one actually recovers the Poisson equations (\ref{eq:II.43})--(\ref{eq:II.44}) for the electromagnetic potentials ${}^{(a)}\!A_0, \vec{A}_a$.

\begin{center}
  \large{\textit{Matter Subsystem}}
\end{center}

Concerning now the stationary form of the matter dynamics, one resorts of course to the generally used factorization of the wave functions $\psi_a(\vr, t)$ into a time and a space factor
\begin{equation}
\label{eq:II.50}
\psi_a(\vr, t) = \exp\left[ -i\, \frac{M_a\crm^2}{\hbar}t \right] \cdot \psi_a(\vr) \;.
\end{equation}
Here, the mass eigenvalues $M_a\ (a=1,2)$ are the proper objects to be determined from the mass eigenvalue equations which we readily put forward now. For this purpose, observe first that the Dirac four-spinors $\psi_a$ may be conceived as the direct sum of Pauli two-spinors ${}^{(a)}\!\varphi_\pm$:
\begin{equation}
\label{eq:II.51}
\psi_a(\vr) = {}^{(a)}\!\varphi_+(\vr) \oplus {}^{(a)}\!\varphi_-(\vr) \;.
\end{equation}
Consequently, the Dirac mass eigenvalue equations (to be deduced from the general Dirac equations (\ref{eq:II.14a})--(\ref{eq:II.14b}) by means of the factorization ansatz (\ref{eq:II.50})) are recast to their equivalent Pauli form for the two-spinors ${}^{(a)}\varphi_\pm(\vr)$
\begin{subequations}
\begin{align}
\label{eq:II.52a}
i\, \vec{\sigma} \sdot \vec{\nabla}\;{}^{(1)}\!\varphi_\pm(\vr) + {}^{(2)}\!A_0(\vr) \cdot
{}^{(1)}\!\varphi_\mp(\vr) &= \frac{\pm M_p - M_1}{\hbar}\,\crm \cdot {}^{(1)}\!\varphi_\mp(\vr) \\
\label{eq:II.52b}
i\, \vec{\sigma} \sdot \vec{\nabla}\;{}^{(2)}\!\varphi_\pm(\vr) + {}^{(1)}\!A_0(\vr) \cdot
{}^{(2)}\!\varphi_\mp(\vr) &= -\frac{M_2 \pm M_e}{\hbar}\,\crm \cdot {}^{(2)}\!\varphi_\mp(\vr) \;.
\end{align}
\end{subequations}
(Observe here that we do neglect for the moment the magnetic effects by putting the three-vector potentials $\vec{A}_a(\vr)$ (\ref{eq:II.41}) to zero: $\vec{A}_a(\vr) \Rightarrow 0$). The \emph{mass eigenvalue} for the positron (with rest mass $M_p$) is denoted by $M_1$ and for the electron (with rest mass $M_e$) by $M_2$.

Summarizing, the RST eigenvalue system consists of the \emph{mass eigenvalue equations}
(\ref{eq:II.52a})--(\ref{eq:II.52b}) for the Pauli spinors ${}^{(a)}\!\varphi_\pm(\vr)$ in
combination with the Poisson equations (\ref{eq:II.43}). Since the magnetic effects are
neglected, the Poisson equations (\ref{eq:II.44}) need not be considered here. But what is
necessary in order to close the whole eigenvalue problem is the prescription for the link
of the Pauli spinors ${}^{(a)}\!\varphi_\pm(\vr)$ to the Maxwell charge densities
${}^{(a)}\!j_0(\vr)$, or more generally to the Maxwellian four-currents $j^a_{\ \mu}$
(\ref{eq:II.42}) as the sources of the four-potentials $A^a_{\ \mu}$, cf. the d'Alembert
equations (\ref{eq:II.21}). Surely, such a link between the wave functions $\psi_a$ and
the currents $j^a_{\ \mu}$ will have something to do with the Dirac four-currents
$k_{a\mu}$ which are usually defined by $(a=1,2)$
\begin{equation}
\label{eq:II.53}
k_{a\mu} \doteqdot \bar{\psi_a}\, \gamma_u\, \psi_a \;.
\end{equation}
Indeed, a more profound scrutiny reveals the following link~\cite{34}
\begin{subequations}
\begin{align}
\label{eq:II.54a}
j^1_{\ \mu} &\equiv k_{1\mu} = \bar{\psi_1}\, \gamma_u\, \psi_1 \\
\label{eq:II.54b}
j^2_{\ \mu} &\equiv -k_{2\mu} = -\bar{\psi_2}\, \gamma_u\, \psi_2 \;.
\end{align}
\end{subequations}
The change in sign of both Dirac currents reflects the positive and negative charge of both particles. In terms of the Pauli spinors $\appm\ (a=1,2)$ the space and time components of the Dirac currents read
\begin{subequations}
\begin{align}
\label{eq:II.55a}
\akn &= \appd \, \app + \apmd \, \apm \\
\label{eq:II.55b}
\vec{k}_a(\vr) &= \appd \, \vec{\sigma} \, \apm + \apmd \, \vec{\sigma} \, \apm \;.
\end{align}
\end{subequations}
Although the magnetic effects, which originate from the three-currents $\vec{k}_a(\vr)$ via the magnetic Poisson equations (\ref{eq:II.44}), are neglected in the present paper these current densities nevertheless play now an important part for identifying two essentially different kinds of positronium.

\section{Ortho/Para Dichotomy}
\indent

In the conventional theory, the manifestation of two principally different kinds of
positronium is traced back to the two possibilities of combining the spins of the electron
(e) and positron (p); if both spins $\spins_e$ and $\spins_p$ add up to the
total Spin $S = 1$
\begin{gather}
\label{eq:III.1}
s_O = \spins_p + \spins_e = 1 \\
\left( \spins_p = \spins_e = \frac{1}{2} \right) \;, \nonumber
\end{gather}
one has \emph{ortho-positronium}; and zero spin
\begin{equation}
\label{eq:III.2}
s_\wp = \spins_p - \spins_e = 0
\end{equation}
yields \emph{para-positronium}. This is the well known ortho/para dichotomy mentioned in
any textbook on relativistic quantum mechanics. In contrast to this (generally valid)
composition rule for angular momenta, the ortho/para dichotomy \emph{in RST} is based upon
the (anti) parallelity of the Maxwellian three-currents $\vec{j}_a(\vr)$. Here, it is
assumed that both particles do occupy physically equivalent one-particle states
$\psi_a(\vr)\ (a=1,2)$ in the sense that the Dirac currents (and therefore also the
Maxwell currents) are either parallel or antiparallel. Thus we propose the following
characterization of \textbf{ortho-positronium}~\cite{34}:
\begin{subequations}
\begin{align}
\label{eq:III.3a}
\vec{k}_1(\vr) &\equiv -\vec{k}_2(\vr) \doteqdot \vec{k}_b(\vr) \\
\label{eq:III.3b}
\vec{j}_1(\vr) &\equiv \vec{j}_2(\vr) \doteqdot \vec{j}_b(\vr) \equiv \vec{k}_b(\vr) \\
\label{eq:III.3c}
\vec{A}_1(\vr) &\equiv \vec{A}_2(\vr) \doteqdot \vec{A}_b(\vr) \\
\label{eq:III.3d}
\vec{H}_1(\vr) &\equiv \vec{H}_2(\vr) \doteqdot \vec{H}_b(\vr) \;,
\end{align}
\end{subequations}
and analogously for \textbf{para-positronium}:
\begin{subequations}
\begin{align}
\label{eq:III.4a}
\vec{k}_1(\vr) &\equiv \vec{k}_2(\vr) \doteqdot \vec{k}_p(\vr) \\
\label{eq:III.4b}
\vec{j}_1(\vr) &\equiv -\vec{j}_2(\vr) \doteqdot \vec{j}_p(\vr) \equiv \vec{k}_p(\vr) \\
\label{eq:III.4c}
\vec{A}_1(\vr) &\equiv -\vec{A}_2(\vr) \doteqdot \vec{A}_p(\vr) \\
\label{eq:III.4d}
\vec{H}_1(\vr) &\equiv -\vec{H}_2(\vr) \doteqdot \vec{H}_p(\vr) \;.
\end{align}
\end{subequations}

The interesting point with such a subdivision of positronium into two classes is the fact
that this subdivision is based upon the \emph{magnetic} effects which however are
\emph{neglected} for the present paper; but despite this neglection the subdivision is of
great relevance also for the presently considered \emph{electrostatic} approximation!
Namely, even within the framework of the latter approximation scheme, there do emerge
different distributions of electrostatic charge $\left( {}^{(a)}\!k_0(\vr) \right)$ for
ortho- and para-positronium with \emph{corresponding} quantum numbers ($\leadsto$
\emph{ortho/para dichotomy}); and moreover there does arise also a certain ambiguity of
the electric charge distribution even within the subclass of the ortho-configurations due
to the same quantum number ($\leadsto$ \emph{ortho-dimorphism}). In order to elaborate
these effects it is necessary to first specify the general eigenvalue problem down to the
subcases and then to look for the corresponding solutions.

\begin{center}
  \emph{\textbf{1.\ Mass Eigenvalue Equations}}
\end{center}

The hypothesis of physically equivalent states for both positronium constituents entails that both mass eigenvalues $M_a$ are actually identical, i.\,e.
\begin{equation}
\label{eq:III.5}
M_1 = -M_2 \doteqdot -M_* \;.
\end{equation}
Next, one considers the Maxwellian charge densities ${}^{(a)}\!j_0(\vr)$ which must of course differ in sign for oppositely charged particles
\begin{equation}
\label{eq:III.6}
{}^{(1)}\!j_0(\vr) = -{}^{(2)}\!j_0(\vr)
\end{equation}
and this must be true for both ortho- and para-positronium. On the other hand, the link (\ref{eq:II.54a})--(\ref{eq:II.54b}) of the Maxwellian currents $j^a_{\ \mu}$ to the Dirac currents $k_{a\mu}$ shows that the requirement (\ref{eq:III.6}) demands the identity of the Dirac densities ${}^{(a)}\!k_0(\vr)$, i.\,e.
\begin{equation}
\label{eq:III.7}
{}^{(1)}\!k_0(\vr) \equiv {}^{(2)}\!k_0(\vr) \;,
\end{equation}
or rewritten in terms of the Pauli spinors (\ref{eq:II.55a})
\begin{equation}
\label{eq:III.8}
{}^{(1)}\!\varphi_+^\dagger(\vr) \, {}^{(1)}\!\varphi_+(\vr) + {}^{(1)}\!\varphi_-^\dagger(\vr) \, {}^{(1)}\!\varphi_-(\vr) = {}^{(2)}\!\varphi_+^\dagger(\vr) \, {}^{(2)}\!\varphi_+(\vr) + {}^{(2)}\!\varphi_-^\dagger(\vr) \, {}^{(2)}\!\varphi_-(\vr) \;.
\end{equation}

However, the situation is different for the Dirac three-currents $\vec{k}_a(\vr)$, since
they may differ in sign, cf. (\ref{eq:III.3a}) vs. (\ref{eq:III.4a}). Expressing the
(anti) parallelity of both Dirac currents in terms of the Pauli spinors,
cf. (\ref{eq:II.55b}), one requires
\begin{align}
\label{eq:III.9}
&{}^{(1)}\!\varphi_+^\dagger(\vr)\,\vec{\sigma}\,{}^{(1)}\!\varphi_-(\vr) + {}^{(1)}\!\varphi_-^\dagger(\vr)\,\vec{\sigma}\,{}^{(1)}\!\varphi_+(\vr)  \\
&= \mp \left\{ {}^{(2)}\!\varphi_+^\dagger(\vr)\,\vec{\sigma}\,{}^{(2)}\!\varphi_-(\vr) + {}^{(2)}\!\varphi_-^\dagger(\vr)\,\vec{\sigma}\,{}^{(2)}\!\varphi_+(\vr) \right\} \;, \nonumber
\end{align}
where the upper/lower sign refers to the ortho/para case, resp. Both conditions (\ref{eq:III.8}) and (\ref{eq:III.9}) can be satisfied by putting
\begin{subequations}
\begin{align}
\label{eq:III.10a}
{}^{(1)}\!\varphi_+(\vr) &= \mp i\,{}^{(2)}\!\varphi_+(\vr) \doteqdot {}^{(b/p)}\!\varphi_+(\vr) \\
\label{eq:III.10b}
{}^{(1)}\!\varphi_-(\vr) &= i\,{}^{(2)}\!\varphi_-(\vr) \doteqdot {}^{(b/p)}\!\varphi_-(\vr)
\end{align}
\end{subequations}
where the upper/lower sign refers again to ortho/para-positronium, resp.

It is true, the disposal (\ref{eq:III.10a})--(\ref{eq:III.10b}) satisfies both algebraic
requirements (\ref{eq:III.8}) and (\ref{eq:III.9}), but additionally there must be
satisfied also a differential requirement: actually, the spinor identifications
(\ref{eq:III.10a})--(\ref{eq:III.10b}) leave us with just one spinor field
(i.\,e. ${}^{(b)}\!\varphi_\pm(\vr)$ for ortho-positronium and
${}^{(p)}\!\varphi_\pm(\vr)$ for para-positronium); and therefore both spinor equations
(\ref{eq:II.52a})--(\ref{eq:II.52b}) must collapse without contradiction to only one
spinor equation (either for ${}^{(b)}\!\varphi_\pm(\vr)$ or
${}^{(p)}\!\varphi_\pm(\vr)$). In order to validate this requirement, we have to make a
disposal also for the electrostatic gauge potentials ${}^{(a)}\!A_0(\vr)$. But this can
easily be done by observing the link between the Dirac densities ${}^{(a)}\!k_0(\vr)$ and
${}^{(a)}\!A_0(\vr)$ as it is implemented by the Poisson equations
(\ref{eq:II.43}). Indeed, this link entails that the potentials ${}^{(a)}\!A_0(\vr)$ must
differ (or not) in sign when this is (or is not) the case also for the Maxwell densities
${}^{(a)}\!j_0(\vr)$. Therefore one concludes that for the positronium situation both
electrostatic potentials must \emph{always} differ in sign
\begin{equation}
\label{eq:III.11}
{}^{(1)}\!A_0(\vr) = -{}^{(2)}\!A_0(\vr) \doteqdot {}^{(b/p)}\!A_0(\vr) \;.
\end{equation}
But when this circumstance is duly respected, both mass eigenvalue equations (\ref{eq:II.52a})--(\ref{eq:II.52b}) actually do collapse to a single one for \textbf{ortho-positronium}:
\begin{equation}
\label{eq:III.12}
i\vec{\sigma}\sdot  \vec{\nabla} \, {}^{(b)}\!\varphi_\pm(\vr) - {}^{(b)}\!A_0(\vr) \cdot {}^{(b)}\!\varphi_\mp(\vr) = \frac{M_* \pm M}{\hbar}\crm \cdot {}^{(b)}\!\varphi_\mp(\vr) \;.
\end{equation}
The existence of one and the same eigenvalue equation (\ref{eq:III.12}) for both ortho-positronium constituents thus validates our original hypothesis that both the electron and the positron should occupy physically equivalent states.

The relativistic pair (\ref{eq:III.12}) of Pauli equations has a single Schr\"odinger-like
equation as its non-relativistic limit. Indeed, assuming that the ``negative''
Pauli-spinor ${}^{(b)}\!\varphi_-(\vr)$ is always considerably smaller than its
``positive'' companion ${}^{(b)}\!\varphi_+(\vr)$ one can solve the upper one of the
equations (\ref{eq:III.12}) for ${}^{(b)}\!\varphi_-(\vr)$ approximately in the following
form:
\begin{equation}
\label{eq:III.13}
{}^{(b)}\!\varphi_-(\vr) \simeq \frac{i\hbar}{2M\crm}\,\vec{\sigma} \sdot \vec{\nabla}\,{}^{(b)}\!\varphi_+(\vr) \;,
\end{equation}
and if this is substituted into the lower equation (\ref{eq:III.12}) one finally ends up with the following non-relativistic eigenvalue equation of the Pauli form
\begin{equation}
\label{eq:III.14}
-\frac{\hbar^2}{2M} \, \Delta \, {}^{(b)}\!\varphi_+(\vr) - {}^{(b)}\!A_0(\vr) \cdot {}^{(b)}\!\varphi_+(\vr) = E_* \cdot {}^{(b)}\!\varphi_+(\vr) \;.
\end{equation}
Here, the non-relativistic eigenvalue $E_*$ emerges as the difference of the rest mass $M$ and the relativistic mass eigenvalue $M_*$, i.\,e.
\begin{equation}
\label{eq:III.15}
E_* \doteqdot \left( M_* - M \right) \crm^2 \;.
\end{equation}
Subsequently, we will be satisfied with clarifying the phenomenon of the ortho-dimorphism in the non-relativistic version (\ref{eq:III.14}) of the original relativistic eigenvalue equation (\ref{eq:III.12}).

But the case of para-positronium is a little bit more complicated. To begin with the
\emph{positron} equation (\ref{eq:II.52a}), this becomes transcribed by the
identifications (\ref{eq:III.10a})--(\ref{eq:III.10b}), lower case, to the following form
(\textbf{para-positronium}):
\begin{equation}
\label{eq:III.16}
i\,\vec{\sigma} \sdot \vec{\nabla}\,{}^{(p)}\!\varphi_\pm(\vr) - {}^{(p)}\!A_0(\vr) \cdot {}^{(p)}\!\varphi_\mp(\vr) = \frac{M_* \pm M}{\hbar}\crm \cdot {}^{(p)}\!\varphi_\mp(\vr) \;.
\end{equation}
Obviously, this positron equation is just of the same form as the joint positron/electron
equation (\ref{eq:III.12}) for ortho-positronium. However, the \emph{electron} equation
(\ref{eq:II.52b}) of para-positronium becomes transcribed by the identifications
(\ref{eq:III.10a})--(\ref{eq:III.10b}) to a somewhat different form:
\begin{equation}
\label{eq:III.17}
i\,\vec{\sigma} \sdot \vec{\nabla} \, {}^{(p)}\!\varphi_\pm(\vr) + {}^{(p)}\!A_0(\vr) \cdot {}^{(p)}\!\varphi_\mp(\vr) = -\frac{M_* \pm M}{\hbar}\crm \cdot {}^{(p)}\!\varphi_\mp(\vr) \;.
\end{equation}
Since the sign of the potential term $\left( \sim {}^{(p)}\!A_0(\vr) \right)$ is reversed
here in comparison to the positron equation (\ref{eq:III.16}), the latter electron
equation (\ref{eq:III.17}) is the ''\emph{charge conjugated}'' form of the first equation
(\ref{eq:III.16}).

The \emph{charge conjugation} is defined here by the following replacements:
\begin{subequations}
\begin{align}
\label{eq:III.18a}
{}^{(p)}\!\varphi_+(\vr)\ &\Rightarrow\ {}^{(p)}\!\varphi_-(\vr)\quad,\quad {}^{(p)}\!\varphi_-(\vr)\ \Rightarrow\ {}^{(p)}\!\varphi_+(\vr) \\
\label{eq:III.18b}
{}^{(p)}\!A_0(\vr)\ &\Rightarrow\ -{}^{(p)}\!A_0(\vr) \\
\label{eq:III.18c}
M_*\ &\Rightarrow\ -M_* \;.
\end{align}
\end{subequations}
Indeed, one is easily convinced that the two forms of eigenvalue equations
(\ref{eq:III.16}) and (\ref{eq:III.17}) for para-positronium are transcribed to one
another by these replacements (\ref{eq:III.18a})--(\ref{eq:III.18c}). This means that any
solution of the positron equation (\ref{eq:III.16}) can be interpreted also to be a
solution of the electron equation (\ref{eq:III.17}); however, the charge conjugation does
not leave invariant the Poisson equations, cf. (\ref{eq:III.21})-(\ref{eq:III.22})
below. Therefore, we prefer here the use of solutions with the same non-relativistic
limit! Indeed, it follows from both equations (\ref{eq:III.16}) and (\ref{eq:III.17}) that
the ``negative'' Pauli spinor ${}^{(p)}\!\varphi_-(\vr)$ can be approximately expressed in
terms of the ``positive'' spinor ${}^{(p)}\!\varphi_+(\vr)$ through
\begin{equation}
\label{eq:III.19}
{}^{(p)}\!\varphi_-(\vr) \simeq \pm\frac{i\hbar}{2M\crm}\,\vec{\sigma} \sdot \vec{\nabla}\,{}^{(p)}\!\varphi_+(\vr) \;,
\end{equation}
where the upper case refers to (\ref{eq:III.16}) and the lower case to (\ref{eq:III.17}). This result may then be substituted in either residual equation (\ref{eq:III.16}) and (\ref{eq:III.17}) which in both cases yields the \emph{same} Schr\"odinger-like equation for the ``positive'' spinor ${}^{(p)}\!\varphi_+(\vr)$:
\begin{equation}
\label{eq:III.20}
-\frac{\hbar^2}{2M}\,\Delta\,{}^{(p)}\!\varphi_+(\vr) - \hbar\crm {}^{(p)}\!A_0(\vr) \cdot {}^{(p)}\!\varphi_+(\vr) = E_* \cdot {}^{(p)}\!\varphi_+(\vr) \;.
\end{equation}

Thus it becomes again evident that the corresponding solutions of (\ref{eq:III.16}) and
(\ref{eq:III.17}) do actually describe physically equivalent states. The notion of
``physical equivalence'' is meant here to refer in first line to the numerical identity of
the energy being carried by anyone of the constituents of para-positronium, see below.

\begin{center}
  \emph{\textbf{2.\ Poisson Equations}}
\end{center}

\begin{sloppypar}
  The mass eigenvalue equations do not yet represent a closed system and therefore cannot
  be solved before an equation for the interaction potential ${}^{(b/p)}\!A_0(\vr)$ has
  been specified. On principle, this has already been done in form of equation
  (\ref{eq:II.43}) so that we merely have to further specify that equation in agreement
  with the ortho/para dichotomy. Observing here the circumstance that the Maxwellian
  current of the first particle ($a=1$, positron) agrees with the Dirac current,
  cf. (\ref{eq:II.54a}) and (\ref{eq:II.55a}), the Poisson equation (\ref{eq:II.43}) reads
  in terms of the Pauli spinors
\begin{align}
\label{eq:III.21}
\Delta\,{}^{(b/p)}\!A_0(\vr) &= -4\pi\as {}^{(b/p)}\!k_0(\vr) \\
&= - 4\pi\as \left\{ {}^{(b/p)}\!\varphi_+^\dagger(\vr)\,{}^{(b/p)}\!\varphi_+(\vr) + {}^{(b/p)}\!\varphi_-^\dagger(\vr)\,{}^{(b/p)}\!\varphi_-(\vr) \right\} \;. \nonumber
\end{align}
This Poisson equation closes the relativistic eigenvalue systems, both for
ortho-positronium (\ref{eq:III.12}) and for para-positronium
(\ref{eq:III.16})--(\ref{eq:III.17}). For the non-relativistic limit, one merely
suppresses the ``negative'' Pauli spinors ${}^{(b/p)}\!\varphi_-(\vr)$ so that the
relativistic Poisson equations (\ref{eq:III.21}) simplify to
\begin{equation}
\label{eq:III.22}
\Delta\,{}^{(b/p)}\!A_0(\vr) = -4\pi\as\,{}^{(b/p)}\!\varphi_+^\dagger(\vr)\,{}^{(b/p)}\!\varphi_+(\vr)
\end{equation}
which then closes both the non-relativistic eigenvalue equations (\ref{eq:III.14}) for ortho-positronium and (\ref{eq:III.20}) for para-positronium.
\end{sloppypar}

Clearly, these coupled systems of eigenvalue and Poisson equations cannot be solved exactly (though exact solutions do surely exist), and consequently we have to resort to some adequate approximation procedure. But this suggests itself when we subsequently will establish the variational \emph{principle of minimal energy}.
\pagebreak

\begin{center}
  \emph{\textbf{3.\ Non-Unique Spinor Fields}}
\end{center}

Despite the fact that we originally subdivided the whole set of positronium configurations
into two subclasses, i.\,e. ortho- and para-positronium
(\ref{eq:III.3a})--(\ref{eq:III.4d}), it may seem now that by the neglection of the
magnetic forces we ended up with an eigenvalue problem, which does no longer offer any
handle for sticking to that original subdivision into two peculiar subsets. Indeed, the
adopted \emph{electrostatic approximation} does admit exclusively an interaction force of
the purely electric type (being described by the electric potential
${}^{(b/p)}\!A_0(\vr)$), whereas the original ortho/para dichotomy was based upon the
three-currents $\ak(\vr)$ as the curls of the magnetic fields, cf. (\ref{eq:II.48}). As a
result of this neglection of magnetism, the Poisson equation (\ref{eq:III.21}), or
(\ref{eq:III.22}), resp., holds equally well for both the ortho-configurations (b) and the
para-configurations (p). But also the mass eigenvalue equations, especially in their
non-relativistic forms (\ref{eq:III.14}) and (\ref{eq:III.20}) are formally the same for
ortho- and para-positronium. Does this mean that, through passing over to the
electrostatic approximation, the difference between ortho- and para-positronium has gone
lost? This is actually not the case because the difference of total angular momentum (in
combination with the hypothesis of the physical equivalence of both constituent states)
leaves its footprint also on the electrostatic approximation.

The crucial point here refers to the fluid-dynamic character of RST, as opposed to the
probabilistic character of the conventional quantum theory. This entails that in RST the
angular momenta of the subsystems cannot be combined (to the total angular momentum of the
whole system) in such a way as it is the case in the conventional theory ($\leadsto$
addition theorem for angular momenta). More concretely: if we wish to insist on the
viewpoint that the (observable) angular momentum of the considered two-particle system
should emerge as the eigenvalue $j_z$ of the angular momentum operator $\Jhz = \Lhz +
\Shz$, i.\,e.
\begin{equation}
\label{eq:III.23}
\Jhz\,\Psi_{b/p}(\vr) = {}^{(b/p)}\!j_z\,\hbar \cdot \Psi_{b/p}(\vr) \;,
\end{equation}
then the quantum number ${}^{(b/p)}\!j_z$ due to the whole two-particle system must be
carried already by any individual particle! Namely, the wave function $\Psi_{b/p}(\vr)$
refers here to the two-particle system as a whole and thus, according to the RST
philosophy, is to be conceived as the direct (Whitney) sum of the one-particle constituent
wave functions $\psi_a(\vr)\ (a=1,2)$
\begin{equation}
\label{eq:III.24}
\Psi_{b/p}(\vr) = {}^{(b/p)}\!\psi_1(\vr) \oplus {}^{(b/p)}\!\psi_2(\vr) \;.
\end{equation}

According to this sum structure, the total angular momentum $\Jhz$ is also the sum of the individual angular momenta
\begin{equation}
\label{eq:III.25}
\Jhz = {}^{(1)}\!\hat{J}_z \oplus {}^{(2)}\!\hat{J}_z \;.
\end{equation}
Therefore the eigenvalue equations (\ref{eq:III.23}) for the total angular momentum $\hat{\J}_z$ can be decomposed as follows
\begin{equation}
\label{eq:III.26}
\Jhz\,\Psi_{b/p}(\vr) = \left( {}^{(1)}\!\hat{J}_z\,\psi_1(\vr) \right) \oplus \left( {}^{(2)}\!\hat{J}_z\,\psi_2(\vr) \right) \;.
\end{equation}
Consequently both one-particle spinors $\psi_1(\vr)$ and $\psi_2(\vr)$ must obey the same eigenvalue equation as the total wave function (\ref{eq:III.24}), i.\,e.
\begin{subequations}
\begin{align}
\label{eq:III.27a}
{}^{(1)}\!\hat{J}_z\,{}^{(b/p)}\!\psi_1(\vr) &= {}^{(b/p)}\!j_z\,\hbar \cdot {}^{(b/p)}\!\psi_1(\vr) \\
\label{eq:III.27b}
{}^{(2)}\!\hat{J}_z\,{}^{(b/p)}\!\psi_2(\vr) &= {}^{(b/p)}\!j_z\,\hbar \cdot {}^{(b/p)}\!\psi_2(\vr) \;.
\end{align}
\end{subequations}

Furthermore, the Dirac four-spinors $\psi_a(\vr)$ can also be conceived as the direct sum of Pauli two-spinors $\app$ and $\apm$, i.\,e.
\begin{equation}
\label{eq:III.28}
\psi_a(\vr) = \app \oplus \apm \;.
\end{equation}
Therefore the same eigenvalue equation must also hold for the individual Pauli spinors, especially after those identifications (\ref{eq:III.10a})--(\ref{eq:III.10b}):
\begin{subequations}
\begin{align}
\label{eq:III.29a}
\Jlhzp\;\bpxpp &= \bpx j_z\, \hbar\; \bpxpp \\
\label{eq:III.29b}
\Jlhzm\;\bpxpm &= \bpx j_z\, \hbar\; \bpxpm \;.
\end{align}
\end{subequations}
This means mathematically that the eigenvalue $\bpx j_z$ of the two-particle state
$\Psi_{b/p}$ (\ref{eq:III.23}) becomes transferred to any \emph{individual} Pauli
component of this two-particle state! In physical terms, the bosonic or fermionic character
of the two-particle state $\Psi$ becomes thus incorporated in any individual constituent
of the two-particle system.

But now it is clear that positronium as a whole carries bosonic properties ($\leadsto$
$j_z$ is integer-valued); and this must therefore hold also for any Pauli constituent
$\appm$ of both particles $(a = 1,2)$, cf. (\ref{eq:III.29a})--(\ref{eq:III.29b}). On the
other hand, it is well known that the Pauli spinors do form a half-integer representation
of the rotation group SO(3). This means that one can select in any two-dimensional spinor
space a certain spinor basis $\left\{ \zetaejm \right\}$ with the following eigenvalue
properties:
\begin{subequations}
\begin{align}
\label{eq:III.30a}
\hat{\vec{J}}\,^2\;\zetaljm &= j(j+1) \hbar^2\,\zetaljm \\
\label{eq:III.30b}
\hat{J}_z\;\zetaljm &= m \hbar\,\zetaljm \\
\label{eq:III.30c}
\hat{\vec{L}}\,^2\;\zetaljm &= \ell(\ell+1) \hbar^2\,\zetaljm \\
\label{eq:III.30d}
\hat{\vec{S}}\,^2\;\zetaljm &= s(s+1) \hbar^2\,\zetaljm \;.
\end{align}
\end{subequations}
Here the electron/positron spin is $s=\frac{1}{2}$; the orbital angular momentum is $\ell
= 0,1,2,3,\ldots$ and thus the lowest possible value of $j(=\ell \pm s)$ is
$j=\frac{1}{2}$ with $\ell=0$ or $\ell=1$. Therefore we have two basis systems for
$j=\frac{1}{2}$, namely $\left\{ \zetapp_0; \zetapm_0 \right\}$ and $\left\{ \zetapp_1;
  \zetapm_1 \right\}$.

\begin{center}
  \large{\textit{Fermionic States}}
\end{center}

The existence of these two basis systems admits us to decompose now a \emph{fermionic} state in the following way
\begin{subequations}
\begin{align}
\label{eq:III.31a}
\app &= \aMRp(\vr) \cdot \zetapp_0 + \aMSp(\vr) \cdot \zetapm_0 \\
\label{eq:III.31b}
\apm &= -i \left\{ \aMRm(\vr) \cdot \zetapp_1 + \aMSm(\vr) \cdot \zetapm_1 \right\} \;.
\end{align}
\end{subequations}
The action of the angular momentum operator $\Jlhz$ on such a fermionic state is obviously
\begin{subequations}
\begin{align}
\label{eq:III.32a}
\Jlhzp\,\app &= \left( \Llhz\,\aMRp(\vr) \right) \cdot \zetapp_0 + \left( \Llhz\,\aMSp(\vr) \right) \cdot \zetapm_0 \\
&\quad + \aMRp(\vr) \cdot \left( \Jlhzp\,\zetapp_0 \right) + \aMSp(\vr) \cdot \left( \Jlhzp\,\zetapm_0 \right) \nonumber \\
&= \left[ \left( \Llhz + \frac{\hbar}{2} \right) \aMRp(\vr) \right] \cdot \zetapp_0 + \left[ \left( \Llhz - \frac{\hbar}{2} \right) \aMSp(\vr) \right] \cdot \zetapm_0 \nonumber \quad\quad \\[1em]
\label{eq:III.32b}
\Jlhzm\,\apm &= -i \left( \Llhz\,\aMRm(\vr) \right) \cdot \zetapp_1 - i \left( \Llhz\,\aMSm(\vr) \right) \cdot \zeta_1^{\frac{1}{2},-\frac{1}{2}} \\
&\quad - i \, \aMRm(\vr) \cdot \left( \Jlhzm\,\zetapp_1 \right) - i \, \aMSm(\vr) \cdot \left( \Jlhzm\,\zetapm_1 \right) \nonumber \\
&= -i \left[ \left( \Llhz + \frac{\hbar}{2} \right) \aMRm(\vr) \right] \cdot \zetapp_1 - i \left[ \left( \Llhz - \frac{\hbar}{2} \right) \aMSm(\vr) \right] \cdot \zetapm_1 \;. \nonumber \quad\quad
\end{align}
\end{subequations}
The required results (\ref{eq:III.29a})--(\ref{eq:III.29b}) of the action of the angular momentum operator $\Jlhz$ on the Pauli spinors $\appm$ (with $a=1,2$ or $a=b/p$) are now deducible from the present equations (\ref{eq:III.32a})--(\ref{eq:III.32b}) by making the following arrangements:
\begin{subequations}
\begin{align}
\label{eq:III.33a}
\Llhz\,\aMRpm(\vr) &= \elz \hbar \cdot \aMRpm(\vr) \\
\label{eq:III.33b}
\Llhz\,\aMSpm(\vr) &= (\elz + 1) \hbar \cdot \aMSpm(\vr) \;.
\end{align}
\end{subequations}
Indeed, with these disposals the equations (\ref{eq:III.32a})--(\ref{eq:III.32b}) adopt the required form of the eigenvalue equations (\ref{eq:III.29a})--(\ref{eq:III.29b}) with the eigenvalue of angular momentum being found as
\begin{equation}
\label{eq:III.34}
\ajz = \elz + \frac{1}{2} \;.
\end{equation}
Since the quantum number of orbital angular momentum is adopted as integer $(\elz = 0, \pm 1, \pm 2, \pm 3, \ldots)$, we actually end up with half-integer quantum numbers $\ajz$ (\ref{eq:III.34}) for \emph{fermionic} states!

These fermionic states can obviously be realized by use of \emph{unique} amplitude fields $\aMRpm(\vr), \aMSpm(\vr)$ and also \emph{unique} spinor basis fields $\zetappm_0, \zetappm_1$. Evidently, the latter fields work as the carriers of the spin, whereas the amplitude fields do contribute the orbital angular momentum. If this philosophy is tried also for the bosonic states we are forced to give up the uniqueness of the spinor basis!

\begin{center}
  \large{\textit{Bosonic States}}
\end{center}

\begin{sloppypar}
  Joining here the general conviction that bosonic states should have integer quantum
  numbers $\ajz$ (\ref{eq:III.29a})--(\ref{eq:III.29b}), i.\,e. $\ajz = 0, \pm 1, \pm 2,
  \pm 3, \ldots$, it suggests itself to think that the amplitude fields should furthermore
  carry integer quantum numbers $(\elz)$ of orbital angular momentum; i.\,e. such
  equations as (\ref{eq:III.33a})--(\ref{eq:III.33b}) should persist also for the bosonic
  states. Thus the necessary modification must refer to the basis spinor fields $\zetaljm$
  (\ref{eq:III.30a})--(\ref{eq:III.30d}). More concretely, we think of four basis spinor
  fields $\xip_0, \xim_0, \xip_1, \xim_1$ which for \textbf{para-positronium} obey the
  following eigenvalue equations
\begin{equation}
\label{eq:III.35}
\Jlhzp\,\xip_0 = \Jlhzp\,\xim_0 = \Jlhzm\,\xip_1 = \Jlhzm\,\xim_1 = 0 \;;
\end{equation}
and similarly for \textbf{ortho-positronium} one wishes to work with a spinor basis $\etap_0, \etam_0, \etap_1, \etam_1$ of the following kind:
\begin{subequations}
\begin{align}
\label{eq:III.36a}
\Jlhzp\,\etap_0 &= \hbar\,\etap_0 \\
\label{eq:III.36b}
\Jlhzp\,\etam_0 &= - \hbar\,\etam_0 \\
\label{eq:III.36c}
\Jlhzm\,\etap_1 &= \hbar\,\etap_1 \\
\label{eq:III.36d}
\Jlhzm\,\etam_1 &= - \hbar\,\etam_1 \;.
\end{align}
\end{subequations}
This says that for the para-case (\ref{eq:III.35}) the spins $\spins_a\
\left(=\frac{1}{2}\right)$ of both constituent particles ($a=1,2$) do combine to zero spin
quantum number $\spins_\mathcal{P}$ of the para-type $(\spins_\mathcal{P} \doteqdot
\spins_1 - \spins_2 = 0,$ $\spins_1 = \spins_2 = \frac{1}{2})$; whereas for the ortho-case
(\ref{eq:III.36a})--(\ref{eq:III.36d}) the individual spins combine to unity
$(\spins_\mathcal{O} \doteqdot \spins_1 + \spins_2 = 1)$. Thus both basis systems
(\ref{eq:III.35}) and (\ref{eq:III.36a})--(\ref{eq:III.36d}) carry integer spin and
therefore may be used for the corresponding decomposition of the Pauli spinors $\appm$ due
to a bound two-particle system.
\end{sloppypar}

For para-positronium $(\spins_\mathcal{P} = 0)$ one has now in place of the fermionic
situation (\ref{eq:III.31a})--(\ref{eq:III.31b}) the following decomposition
\begin{subequations}
\begin{align}
\label{eq:III.37a}
\pxpp &= \pMRp(\vr) \cdot \xip_0 + \pMSp(\vr) \cdot \xim_0 \\
\label{eq:III.37b}
\pxpm &= -i \left\{ \pMRm(\vr) \cdot \xip_1 + \pMSm(\vr) \cdot \xim_1 \right\} \;.
\end{align}
\end{subequations}
Here, the action of the angular momentum operator $\Jlhz$ looks now as follows
\begin{subequations}
\begin{align}
\label{eq:III.38a}
\Jlhzp\,\pxpp &= \left( \Llhz\,\pMRp(\vr) \right) \cdot \xip_0 + \left( \Llhz\,\pMSp(\vr) \right) \cdot \xim_0 \\
\label{eq:III.38b}
\Jlhzm\,\pxpm &= -i \left( \Llhz\,\pMRm(\vr) \right) \cdot \xip_1 - i \left( \Llhz\,\pMSm(\vr) \right) \cdot \xim_1 \;.
\end{align}
\end{subequations}
Consequently in order to satisfy again the para-form ($p$) of the eigenvalue equations (\ref{eq:III.29a})--(\ref{eq:III.29b}) one puts here
\begin{subequations}
\begin{align}
\label{eq:III.39a}
\Llhz\,\pMRpm(\vr) &= \elz\,\hbar \cdot \pMRpm(\vr) \\
\label{eq:III.39b}
\Llhz\,\pMSpm(\vr) &= \elz\,\hbar \cdot \pMSpm(\vr) \;,
\end{align}
\end{subequations}
so that the quantum number $\pjz$ (\ref{eq:III.29a})--(\ref{eq:III.29b}) is solely due to \emph{orbital} angular momentum:
\begin{gather}
\label{eq:III.40}
\pjz \equiv \elz \\
(\elz = 0, \pm 1, \pm 2, \pm 3, \ldots) \nonumber
\end{gather}

For ortho-positronium $(\spins_\mathcal{O} = 1)$, the situation is somewhat different. First, the decomposition of the ortho-spinors $\bppm$ looks quite similar to the para-case (\ref{eq:III.37a})--(\ref{eq:III.37b}):
\begin{subequations}
\begin{align}
\label{eq:III.41a}
\bpp &= \bMRp(\vr) \cdot \etap_0 + \bMSp(\vr) \cdot \etam_0 \\
\label{eq:III.41b}
\bpm &= -i \left\{ \bMRm(\vr) \cdot \etap_1 + \bMSm(\vr) \cdot \etam_1 \right\} \;.
\end{align}
\end{subequations}
But the action of the angular momentum operator $\Jlhzpm$ on these ortho-states looks now as follows
\begin{subequations}
\begin{align}
\label{eq:III.42a}
\Jlhzp\,\bpp &= \left[ \left( \Llhz + \hbar \right) \bMRp(\vr) \right] \cdot \etap_0 + \left[ \left( \Llhz - \hbar \right) \bMSp(\vr) \right] \cdot \etam_0 \\
\label{eq:III.42b}
\Jlhzm\,\bpm &= -i \left\{ \left[ \left( \Llhz + \hbar \right) \bMRm(\vr) \right] \cdot \etap_1 + \left[ \left( \Llhz - \hbar \right) \bMSm(\vr) \right] \cdot \etam_1 \right\} \;.
\end{align}
\end{subequations}
For satisfying here again the eigenvalue requirement (\ref{eq:III.29a})--(\ref{eq:III.29b}) in its ortho-form ($b$) one puts in a self-evident way
\begin{subequations}
\begin{align}
\label{eq:III.43a}
\Llhz\,\bMRpm &= \elz\,\hbar \cdot \bMRpm \\
\label{eq:III.43b}
\Llhz\,\bMSpm &= (\elz+2)\,\hbar \cdot \bMSpm \;.
\end{align}
\end{subequations}
This arrangement yields then the following eigenvalue equations for the ortho-spinors
\begin{equation}
\label{eq:III.44}
\hat{J}_z^{\pm} \,\bppm = (\elz + 1)\,\hbar \cdot \bppm \;.
\end{equation}
Thus one finds the quantum numbers of the ortho-system:
\begin{gather}
\label{eq:III.45}
\bjz = \elz + 1 \\
(\bjz = 0, \pm 1, \pm 2, \pm 3, \ldots) \;. \nonumber
\end{gather}

\begin{sloppypar}
For a realization of the required basis spinors one takes the fermionic basis $\left\{ \zetappm_0, \zetappm_1 \right\}$ as the point of departure and introduces a general spinor basis $\left\{ \omp_0, \omm_0, \omp_1, \omm_1 \right\}$ through
\begin{subequations}
\begin{align}
\label{eq:III.46a}
\omp_0 &= \e^{-i \bbar \phi} \cdot \zetapp_0 \\
\label{eq:III.46b}
\omm_0 &= \e^{i \bbar \phi} \cdot \zetapm_0 \\
\label{eq:III.46c}
\omp_1 &= \e^{-i \bbar \phi} \cdot \zetapp_1 \\
\label{eq:III.46d}
\omm_1 &= \e^{i \bbar \phi} \cdot \zetapm_1 \;.
\end{align}
\end{subequations}
Here it is easy to see emerging the following relations concerning angular momentum
\begin{subequations}
\begin{align}
\label{eq:III.47a}
\Jlhzp\,\omp_0 &= - \left( \bbar - \frac{1}{2} \right)\,\hbar \cdot \omp_0 \\
\label{eq:III.47b}
\Jlhzp\,\omm_0 &= \left( \bbar - \frac{1}{2} \right)\,\hbar \cdot \omm_0 \\
\label{eq:III.47c}
\Jlhzm\,\omp_1 &= - \left( \bbar - \frac{1}{2} \right)\,\hbar \cdot \omp_1 \\
\label{eq:III.47d}
\Jlhzm\,\omm_1 &= \left( \bbar - \frac{1}{2} \right)\,\hbar \cdot \omm_1 \;.
\end{align}
\end{subequations}
Obviously, this $\omega$-basis contains some free parameter $\bbar$ (i.\,e. the
\emph{boson number}) and if this is chosen as $\bbar = \frac{1}{2}$ we obtain the
desired $\xi$-basis (\ref{eq:III.35}) for para-positronium; and if $\bbar$ is chosen as
$\bbar = -\frac{1}{2}$ we obtain the $\eta$-basis (\ref{eq:III.36a})--(\ref{eq:III.36d})
for ortho-positronium. For $\bbar = 0$ we get back the purely fermionic $\zeta$-basis
(\ref{eq:III.30a})--(\ref{eq:III.30d}).
\end{sloppypar}

With the choice of the $\omega$-basis (\ref{eq:III.46a})--(\ref{eq:III.46d}) the loss of
uniqueness becomes now evident: since the original $\zeta$-basis
(\ref{eq:III.30a})--(\ref{eq:III.30d}) is \emph{unique} (i.\,e. the $\zetaljm(\vartheta,
\phi)$ constitute a ``unique'' spinor field on the 2-sphere $S^2$), the other two basis
systems $\xi$ (\ref{eq:III.35}) and $\eta$ (\ref{eq:III.36a})--(\ref{eq:III.36d}) are
double-valued. More concretely, for both the $\xi$- and the $\eta$-basis $\left( \leadsto
  \bbar = \pm \frac{1}{2} \right)$ one finds by performing one revolution around the
$z$-axis $\left( 0 \leq \phi \leq 2\pi \right)$:
\begin{subequations}
\begin{align}
\label{eq:III.48a}
\xi^{(\pm)}_{0,1}(\phi + 2\pi) &= \e^{\pm i \pi} \cdot \xi^{(\pm)}_{0,1}(\phi) = -\xi^{(\pm)}_{0,1} \\
\label{eq:III.48b}
\eta^{(\pm)}_{0,1}(\phi + 2\pi) &= \e^{\pm i \pi} \cdot \eta^{(\pm)}_{0,1}(\phi) = -\eta^{(\pm)}_{0,1} \;,
\end{align}
\end{subequations}
and this says that we need two revolutions around the $z$-axis $\left( 0 \leq \phi \leq
  4\pi \right)$ in order to return to the original basis configurations. Since we adopt
all the amplitude fields $\bpMRpm(\vr), \bpMSpm(\vr)$ to be unique scalar fields, the
double-valuedness of the para- and ortho-basis becomes transferred to the para-spinors
$\pxppm$ (\ref{eq:III.37a})--(\ref{eq:III.37b}) and ortho-spinors $\bppm$
(\ref{eq:III.41a})--(\ref{eq:III.41b}) and from here ultimately to the Dirac spinors
$\PsiPO$
\begin{subequations}
\begin{align}
\label{eq:III.49a}
\PsiP(\vr) &= \pxpp \oplus \pxpm \\
\label{eq:III.49b}
\PsiO(\vr) &= \bpp \oplus \bpm \;.
\end{align}
\end{subequations}
Summarizing, the Dirac wave functions $\PsiP(\vr)$ and $\PsiO(\vr)$ for ortho- and
para-positronium must in RST be double-valued in the following sense:
\begin{subequations}
\begin{align}
\label{eq:III.50a}
\PsiP(r, \vartheta, \phi+2\pi) &= - \PsiP(r, \vartheta, \phi) \\
\label{eq:III.50b}
\PsiO(r, \vartheta, \phi+2\pi) &= - \PsiO(r, \vartheta, \phi) \;.
\end{align}
\end{subequations}
\pagebreak
\begin{center}
  \large{\textit{Uniqueness of the Physical Densities}}
\end{center}

Naturally, in a fluid-dynamic theory (such as the present RST) the proper observable
objects are the physical densities, such as those of charge, current, energy, linear and
angular momentum, etc. A plausible condition on these densities is surely given by the
demand that these physical densities should be \emph{single-valued} tensor fields. But, as
we will readily demonstrate, this condition can be satisfied also by \emph{non-unique}
wave functions; and this fact allows us to actually deal with such non-unique wave
functions, as given for example by the double-valued positronium states
(\ref{eq:III.50a})--(\ref{eq:III.50b}). Therefore one wishes to have some condition on the
wave functions which on the one hand admits their non-uniqueness but on the other hand
ensures the uniqueness of the associated physical densities!

Now according to the present RST philosophy, the non-uniqueness of the (Dirac) wave
functions is to be traced back to the spinor basis, whereas the amplitude fields are
furthermore required to be unique. Therefore the non-uniqueness of the wave functions is
measured by the boson number $\bbar$, cf. (\ref{eq:III.46a})--(\ref{eq:III.46d}); and thus
the condition of uniqueness of the densities is to be retraced to some condition for the
fixation of the boson number $\bbar$. Such a fixation may be attained now by considering
specifically the Dirac density $\akn$ and the three-current $\ak(\vr)$ which read in terms
of the Pauli spinors $\appm$ as shown by equations
(\ref{eq:II.55a})--(\ref{eq:II.55b}). Decomposing these Pauli spinors with respect to the
$\omega$-basis (\ref{eq:III.46a})--(\ref{eq:III.46d}) lets then appear the (Dirac) charge
densities (\ref{eq:II.55a}) in the following form:
\vspace{0.5em}
\begin{equation}
\label{eq:III.51}
\akn = \frac{\aMRpS \cdot \aMRp + \aMSpS \cdot \aMSp + \aMRmS \cdot \aMRm + \aMSmS \cdot \aMSm}{4\pi} \;.
\vspace{0.5em}
\end{equation}
Evidently, these charge densities are unique in any case and therefore do not yet provide
an immediate handle for fixing the parameter $\bbar$.

This situation changes now when one considers also the Dirac currents $\ak$
(\ref{eq:II.55b}), which by their very definitions are always real-valued objects:
\vspace{0.5em}
\begin{subequations}
\begin{align}
\label{eq:III.52a}
\akr &= \frac{i}{4\pi} \left\{ \aMRpS \cdot \aMRm + \aMSpS \cdot \aMSm - \aMRmS \cdot \aMRp - \aMSmS \cdot \aMSp \right\} \\[0.5em]
\label{eq:III.52b}
\akt &= - \frac{i}{4\pi} \left\{ \e^{2i(\bbar - \frac{1}{2})\phi} \cdot \;\MCa - \e^{-2i(\bbar - \frac{1}{2})\phi} \cdot \;\MCaS \right\} \\
&( \MCa \doteqdot \aMRpS \cdot \aMSm + \aMRmS \cdot \aMSp ) \nonumber \\[0.5em]
\label{eq:III.52c}
\akp &= \frac{\sin \vartheta}{4\pi} \left\{ \aMRpS \cdot \aMRm + \aMRmS \cdot \aMRp - \aMSpS \cdot \aMSm - \aMSmS \cdot \aMSp \right\} \\
&\quad - \frac{\cos \vartheta}{4\pi} \left\{ \e^{2i(\bbar - \frac{1}{2})\phi} \cdot \;\MCa + \e^{-2i(\bbar - \frac{1}{2})\phi} \cdot \;\MCaS \right\} \;. \nonumber
\end{align}
\end{subequations}
But here a nearby restriction upon the parameter $\bbar$ suggests itself, namely through the plausible demand that the Dirac currents $\ak$ (\ref{eq:II.55b}), with their components being specified by (\ref{eq:III.52a})--(\ref{eq:III.52c}), must be \emph{unique} (!) vector fields over three-space (albeit only apart from the origin $r = 0$ and the $z$ axis $\vartheta = 0,\pi$). Evidently this demand of uniqueness reads in terms of the spherical polar coordinates $\{ r, \vartheta, \phi \}$
\begin{equation}
\label{eq:III.53}
\ak(r, \vartheta, \phi + 2\pi) = \ak(r, \vartheta, \phi) \;,
\end{equation}
and thus the values of $\bbar$ become restricted to the range
\begin{gather}
\label{eq:III.54}
\bbar = \frac{1}{2} \left( n + 1 \right) \\
(n = 0, \pm 1, \pm 2, \pm 3, \ldots) \nonumber
\end{gather}
which then entails also \emph{(half-)integer} quantum numbers for the $z$ component of the angular momentum (\ref{eq:III.47a})--(\ref{eq:III.47d}) of the spinor basis:
\begin{equation}
\label{eq:III.55}
s_z = \pm \left( \bbar - \frac{1}{2} \right) = \pm \frac{n}{2} \;.
\end{equation}
Here it will suffice to admit for the spinor basis (of the two-particle systems)
exclusively the values $\bbar=\pm\frac{1}{2}$; other values of~$\bbar$ come into play for
bound systems of more than two fermions.

Notice here that this \emph{(half-)integrity} arises as a consequence of the demand of
uniqueness with respect to certain physical \emph{densities} (i.\,e. Dirac current),
whereas the corresponding \emph{integral} quantum numbers of conventional non-relativistic
quantum mechanics are mostly traced back in the textbooks to the uniqueness requirement
for the \emph{wave functions} themselves (not the \emph{densities}). The lowest values of
$s_z$ (\ref{eq:III.55}) are $s_z = \pm\frac{1}{2}$ for $\bbar = 0$ and $s_z = 0, \pm 1$
for $\bbar = \pm \frac{1}{2}$. Thus for the first case $(\bbar = 0)$ we have a
\emph{fermionic basis} and for the second case $\left( \bbar = \pm\frac{1}{2} \right)$ one
deals with a \emph{bosonic basis}. In this sense, a Dirac particle is said to occupy a
fermionic quantum state $\psi$ if the ``\emph{boson number}'' $\bbar$ of its spinor basis
is zero ($\bbar = 0$), and a bosonic quantum state if the boson number $\bbar$ equals
$\pm\frac{1}{2}$. Observe that through this arrangement the fermionic or bosonic character
of the quantum state of a Dirac particle is defined by reference to the corresponding
spinor \emph{basis}.

\section{Ortho-Positronium ($\bbar = -\frac{1}{2}$)}
\indent

After the general RST logic for the occurence of the ortho/para dichotomy is sufficiently
displayed, we will now turn (for the remainder of the paper) to ortho-positronium. For
this specific two-particle system there occurs a further ambiguity, i.\,e. the
``\emph{ortho-dimorphism}'', which is not present in the para-configuration. The effect of
dimorphism consists in the circumstance that the angular momentum quantization admits the
emergence of two different charge distributions $\bgkn(\vr)$ which, however, are both due
the \emph{same} configuration of quantum numbers! Since these two electrostatic charge
distributions are differing slightly, they carry a slightly different electrostatic
interaction energy, and this causes a slightly different binding energy. In this way,
ortho-positronium does appear in RST in form of \emph{electrostatic} doublets. Observe
here that, for our subsequent discussion of this ortho-dimorphism, we will be satisfied
with the \emph{electrostatic approximation}, where magnetic effects are neglected
completely!

\begin{center}
  \emph{\textbf{1.\ Mass Eigenvalue Equations for the Amplitude Fields}}
\end{center}

The point of departure for our quantization of angular momentum is the mass eigenvalue
equation (\ref{eq:III.12}) in Pauli form. Here, we will adopt for the moment the
spherically symmetric approximation where the interaction potential $\bAn(\vr)$ appears to
be spherically symmetric (i.\,e. $\bAn(\vr) \Rightarrow \bAe(r)$, $r \doteqdot
||\vr||$). For this situation, the amplitude fields $\bMRpm$, $\bMSpm$
(\ref{eq:III.41a})--(\ref{eq:III.41b}) can be assumed to be the product of a purely
angular-dependent factor and a purely radial factor which then entails the splitting of
the original eigenvalue system into a purely angular and a purely radial problem. As
usual, the solution of the radial problem yields then the energy spectrum
(i.\,e. quantization of energy).

To begin with, one inserts the decomposition of the Pauli spinors $\bppm$ (\ref{eq:III.41a})--(\ref{eq:III.41b}) into the Pauli equations (\ref{eq:III.12}) and thereby obtains the following system of eigenvalue equations for the amplitude fields $\bMRpm(r,\vartheta,\phi)$, $\bMSpm(r,\vartheta,\phi)$~\cite{34}
\begin{subequations}
\begin{align}
\label{eq:IV.1a}
\frac{\partial\,\bMRp}{\partial r} + \frac{i}{r}\,\frac{\partial\,\bMRp}{\partial \phi} - \frac{1}{2r}\,\bMRp - \bAn \cdot \bMRm & \\
+ \e^{-2i\phi} \left\{ \frac{1}{r}\,\frac{\partial\,\bMSp}{\partial \vartheta} - \frac{\cot \vartheta}{r} \cdot \left[ \frac{1}{2}\,\bMSp + i\,\frac{\partial\,\bMSp}{\partial \phi} \right] \right\} &= \frac{M + M_*}{\hbar}\,\crm \cdot \bMRm \nonumber \\[3em]
\label{eq:IV.1b}
\frac{\partial\,\bMSp}{\partial r} - \frac{i}{r}\,\frac{\partial\,\bMSp}{\partial \phi} - \frac{1}{2r}\,\bMSp - \bAn \cdot \bMSm & \\
- \e^{2i\phi} \left\{ \frac{1}{r}\,\frac{\partial\,\bMRp}{\partial \vartheta} - \frac{\cot \vartheta}{r} \cdot \left[ \frac{1}{2}\,\bMRp - i\,\frac{\partial\,\bMRp}{\partial \phi} \right] \right\} &= \frac{M + M_*}{\hbar}\,\crm \cdot \bMSm \nonumber \\[3em]
\label{eq:IV.1c}
\frac{\partial\,\bMRm}{\partial r} - \frac{i}{r}\,\frac{\partial\,\bMRm}{\partial \phi} + \frac{5}{2r}\,\bMRm + \bAn \cdot \bMRp & \\
- \e^{-2i\phi} \left\{ \frac{1}{r}\,\frac{\partial\,\bMSm}{\partial \vartheta} - \frac{\cot \vartheta}{r} \cdot \left[ \frac{1}{2}\,\bMSm + i\,\frac{\partial\,\bMSm}{\partial \phi} \right] \right\} &= \frac{M - M_*}{\hbar}\,\crm \cdot \bMRp \nonumber \\[3em]
\label{eq:IV.1d}
\frac{\partial\,\bMSm}{\partial r} + \frac{i}{r}\,\frac{\partial\,\bMSm}{\partial \phi} + \frac{5}{2r}\,\bMSm - \bAn \cdot \bMSp & \\
+ \e^{2i\phi} \left\{ \frac{1}{r}\,\frac{\partial\,\bMRm}{\partial \vartheta} - \frac{\cot \vartheta}{r} \cdot \left[ \frac{1}{2}\,\bMRm - i\,\frac{\partial\,\bMRm}{\partial \phi} \right] \right\} &= \frac{M - M_*}{\hbar}\,\crm \cdot \bMSp  \;. \nonumber
\end{align}
\end{subequations}
This may appear as a relatively complicated system but it can be simplified by imposing
some plausible requirements on the desired solutions. Naturally, these requirements will
refer to angular momentum, which usually serves to classify the solutions of the energy
eigenvalue problems for the few-particle systems. Thus it is surely reasonable to demand
that the eigensolutions of the present ortho-system (\ref{eq:IV.1a})--(\ref{eq:IV.1d}) be
classifyable by means of the associated eigenvalues $\bjz$ (\ref{eq:III.45}) of $\Jhz
(\doteqdot \hat{J}_z^{(+)} \oplus \hat{J}_z^{(-)})$. Recall here our hypothesis that the
total spin of a bound system agrees with the individual spins of both Pauli two-spinors
which build up the common basis for each of the Dirac four-spinors (in the present case
$s_\mathcal{O} = s_1 + s_2 = 1$, see the discussion below (\ref{eq:III.53})).

In this sense, we try now to obtain solutions for the present ortho-system
(\ref{eq:IV.1a})--(\ref{eq:IV.1d}) which are of the following form:
\begin{subequations}
\begin{align}
\label{eq:IV.2a}
\bMRpm(r,\vartheta,\phi) &= \frac{\e^{i\elz\phi}}{\sin \vartheta \sqrt{r\,\sin \vartheta}} \cdot \btRpm\rt \\[1.5em]
\label{eq:IV.2b}
\bMSpm(r,\vartheta,\phi) &= \sqrt{\frac{\sin \vartheta}{r}}\,\e^{i(\elz + 2)\phi} \cdot \btSpm\rt \;.
\end{align}
\end{subequations}
Indeed, by this ansatz the eigenvalue equations (\ref{eq:III.43a})--(\ref{eq:III.45}) for angular momentum are actually satisfied; and furthermore the original mass eigenvalue system (\ref{eq:IV.1a})--(\ref{eq:IV.1d}) becomes transcribed to the new amplitude fields $\btRpm\rt, \btSpm\rt$ and thus reappears now as follows:
\begin{subequations}
\begin{align}
\label{eq:IV.3a}
\frac{\partial\,\btRp\rt}{\partial r} - \frac{\bjz}{r} \cdot \btRp\rt - \bAn\rt \cdot \btRm\rt & \\
+ \frac{\sin^2 \vartheta}{r} \cdot \frac{\partial\,\btSp\rt}{\partial\vartheta} + \frac{\bjz + 1}{r}\,\sin \vartheta \cos \vartheta \cdot \btSp\rt &= \frac{M + M_*}{\hbar}\,\crm \cdot \btRm\rt \nonumber \\[3em]
\label{eq:IV.3b}
\frac{\partial\,\btSp\rt}{\partial r} + \frac{\bjz}{r} \cdot \btSp\rt - \bAn\rt \cdot \btSm\rt & \\
- \frac{1}{r\,\sin^2\vartheta} \cdot \frac{\partial\,\btRp\rt}{\partial\vartheta} + \frac{\bjz + 1}{r\,\sin^2\vartheta}\,\cot \vartheta \cdot \btRp\rt &= \frac{M + M_*}{\hbar}\,\crm \cdot \btSm\rt \nonumber \\[3em]
\label{eq:IV.3c}
\frac{\partial\,\btRm\rt}{\partial r} + \frac{\bjz+1}{r} \cdot \btRm\rt + \bAn\rt \cdot \btRp\rt & \\
- \frac{\sin^2 \vartheta}{r} \cdot \frac{\partial\,\btSm\rt}{\partial\vartheta} - \frac{\sin\vartheta\,\cos\vartheta}{r}\,\left( \bjz + 1 \right) \cdot \btSm\rt &= \frac{M - M_*}{\hbar}\,\crm \cdot \btRp\rt \nonumber \\[3em]
\label{eq:IV.3d}
\frac{\partial\,\btSm\rt}{\partial r} - \frac{\bjz-1}{r} \cdot \btSm\rt + \bAn\rt \cdot \btSp\rt & \\
+ \frac{1}{r\,\sin^2\vartheta} \cdot \frac{\partial\,\btRm\rt}{\partial\vartheta} - \frac{\bjz + 1}{r\,\sin^2\vartheta}\,\cot \vartheta \cdot \btRm\rt &= \frac{M - M_*}{\hbar}\,\crm \cdot \btSp\rt \;. \nonumber
\end{align}
\end{subequations}

This is still a system too complicated in order that there could be hope to find exact solutions, even if one could consider the ortho-potential $\bAn\rt$ being prescribed from the outside. However, this potential does actually couple back to the ortho-wave function $\psi_b(\vr)$, via the Poisson equation (\ref{eq:III.21}), i.\,e. in terms of the new amplitude fields $\btRpm\rt$ and $\btSpm\rt$ (\ref{eq:IV.2a})--(\ref{eq:IV.2b})
\begin{align}
\label{eq:IV.4}
\Delta\,\bAn\rt = &-\as\, \Bigg\{ \frac{\left[ \btRp\rt \right]^2 + \left[ \btRm\rt \right]^2}{r\,\sin^3\vartheta} \\
&+ \frac{\sin\vartheta}{r}\,\left( \left[ \btSp\rt \right]^2 + \left[ \btSm\rt \right]^2 \right) \Bigg\} \,. \nonumber
\end{align}
\vskip 2em

\begin{center}
  \large{\textit{Product Ansatz}}
\end{center}

In view of these complications it should be obvious that it is necessary to resort to some approximative procedure in order to extract the physically relevant solutions of that intricately coupled system (\ref{eq:IV.3a})--(\ref{eq:IV.4}). For this purpose, the \emph{spherically symmetric approximation} suggests itself where the angular dependence of the gauge potential $\bAn\rt$ is neglected, i.\,e. we replace $\bAn\rt$ by its spherically symmetric approximation $\bAe(r)$. But observe here that this approximation assumption does \emph{not} entail the spherical symmetry of the new wave amplitudes $\btRpm\rt$ and $\btSpm\rt$ (\ref{eq:IV.2a})--(\ref{eq:IV.2b})! However, the angular dependence of the wave amplitudes becomes now manageable, namely by applying the following product ansatz:
\begin{subequations}
\begin{align}
\label{eq:IV.5a}
\btRpm\rt &= g_R(\vartheta) \cdot \iiRpm(r) \\
\label{eq:IV.5b}
\btSpm\rt &= g_S(\vartheta) \cdot \iiSpm(r) \;.
\end{align}
\end{subequations}

Indeed, substituting this ansatz into the eigenvalue system (\ref{eq:IV.3a})--(\ref{eq:IV.3d}) enables us to separate this system into two first-order subsystems, namely the angular system
\begin{subequations}
\begin{align}
\label{eq:IV.6a}
\frac{d\,g_R(\vartheta)}{d\vartheta} - \left( \bjz + 1 \right) \,\cot\vartheta \cdot g_R(\vartheta) &= \ddot{\lO}\,\sin^2\vartheta \cdot g_S(\vartheta) \\
\label{eq:IV.6b}
\frac{d\,g_S(\vartheta)}{d\vartheta} + \left( \bjz + 1 \right)\,\cot\vartheta \cdot g_S(\vartheta) &= \frac{\dot{\lO}}{\sin^2\vartheta} \cdot g_R(\vartheta)
\end{align}
\end{subequations}
and the radial system
\begin{subequations}
\begin{align}
\label{eq:IV.7a}
\frac{d\,\iiRp(r)}{dr} - \frac{\bjz}{r} \cdot \iiRp(r) - \bAe(r) \cdot \iiRm(r) &+ \frac{\dot{\lO}}{r} \cdot \iiSp(r) \\
&= \frac{M + M_*}{\hbar}\,\crm \cdot \iiRm(r) \nonumber \\[2em]
\label{eq:IV.7b}
\frac{d\,\iiSp(r)}{dr} + \frac{\bjz}{r} \cdot \iiSp(r) - \bAe(r) \cdot \iiSm(r) &- \frac{\ddot{\lO}}{r} \cdot \iiRp(r) \\
&= \frac{M + M_*}{\hbar}\,\crm \cdot \iiSm(r) \nonumber \\[2em]
\label{eq:IV.7c}
\frac{d\,\iiRm(r)}{dr} + \frac{\bjz+1}{r} \cdot \iiRm(r) + \bAe(r) \cdot \iiRp(r) &- \frac{\dot{\lO}}{r} \cdot \iiSm(r) \\
&= \frac{M - M_*}{\hbar}\,\crm \cdot \iiRp(r) \nonumber \\[2em]
\label{eq:IV.7d}
\frac{d\,\iiSm(r)}{dr} - \frac{\bjz-1}{r} \cdot \iiSm(r) + \bAe(r) \cdot \iiSp(r) &+ \frac{\ddot{\lO}}{r} \cdot \iiRm(r) \\
&= \frac{M - M_*}{\hbar}\,\crm \cdot \iiSp(r) \nonumber \;.
\end{align}
\end{subequations}
The meaning of the newly introduced constants $\dot{\lO}$ and $\ddot{\lO}$ will readily be
clarified, but first we will perform a further simplification of the present ortho-system
(\ref{eq:IV.7a})--(\ref{eq:IV.7d}).

Indeed, it may appear somewhat strange that all four equations
(\ref{eq:IV.7a})--(\ref{eq:IV.7b}) for the four variables $\iiRpm$, $\iiSpm$ are necessary
in order to determine the one single mass eigenvalue $M_*$. Such a system surely appears
to be overdetermined; and therefore the overdetermination must be eliminated, namely by
\emph{identifying} the amplitude fields in such a way that there remain two independent
amplitude fields ($\tOpm(r)$, say) which satisfy a coupled first-order system of only
two equations. The wanted identification of the amplitude fields is the following:
\begin{subequations}
\begin{align}
\label{eq:IV.8a}
\iiRp(r) \equiv \iiSp(r) &\doteqdot \tOp(r) \\
\label{eq:IV.8b}
\iiRm(r) \equiv \iiSm(r) &\doteqdot \tOm(r) \;.
\end{align}
\end{subequations}
By this identification requirement, the four equations (\ref{eq:IV.7a})--(\ref{eq:IV.7d})
become concentrated to only two equations, namely
\begin{subequations}
\begin{align}
\label{eq:IV.9a}
\frac{d\,\tOp(r)}{dr} + \frac{\dot{\lO} - \ddot{\lO}}{2r} \cdot \tOp(r) - \bAe(r) \cdot \tOm(r) &= \frac{M + M_*}{\hbar}\,\crm \cdot \tOm(r) \\
\label{eq:IV.9b}
\frac{d\,\tOm(r)}{dr} + \frac{2-\left(\dot{\lO} - \ddot{\lO}\right)}{2r} \cdot \tOm(r) + \bAe(r) \cdot \tOp(r) &= \frac{M - M_*}{\hbar}\,\crm \cdot \tOp(r) \;.
\end{align}
\end{subequations}
Moreover, the identification requirement (\ref{eq:IV.8a})--(\ref{eq:IV.8b}) yields also the constraint
\begin{equation}
\label{eq:IV.10}
\bjz = \frac{1}{2}\,\left( \dot{\lO} + \ddot{\lO} \right) \;.
\end{equation}

Of course, it is self-suggesting that those newly introduced constants $\dot{\lO}$ and $\ddot{\lO}$ (\ref{eq:IV.6a})--(\ref{eq:IV.6b}) must have to do something with the quantum number ($j_\mathcal{O}$, say) of the total angular momentum of the ortho-states. In order to preliminarily reveal that supposed interrelationship we pass over to the non-relativistic approximation of the latter eigenvalue system (\ref{eq:IV.9a})--(\ref{eq:IV.9b}) for the amplitude fields $\tOpm(r)$. This non-relativistic limit may be attained by approximately solving the first-equation (\ref{eq:IV.9a}) for $\tOm(r)$, yielding
\begin{equation}
\label{eq:IV.11}
\tOm(r) \simeq \frac{\hbar}{2M\crm}\,\left\{ \frac{d\,\tOp(r)}{dr} + \frac{\dot{\lO} - \ddot{\lO}}{2r} \cdot \tOp(r) \right\} \;,
\end{equation}
and substituting this into the second equation (\ref{eq:IV.9b}). The result is a decoupled second-order equation for $\tOp(r)$ (being simply denoted by $\tO(r)$) which then looks as follows:
\begin{align}
\label{eq:IV.12}
- \frac{\hbar^2}{2\,M}\,\left( \frac{d^2\,\tO(r)}{dr^2} + \frac{1}{r}\,\frac{d\,\tO(r)}{dr} \right) &+ \frac{\hbar^2}{2\,Mr^2}\,\left( \frac{\dot{\lO} - \ddot{\lO}}{2} \right)^2 \cdot \tO(r) \\
{}- \hbar\crm\,\bAe(r) \cdot \tO(r) &= \left( M_* - M \right)\,\crm^2 \cdot \tO(r) \;. \nonumber
\end{align}
Obviously this looks like a Schr\"{o}dinger-like eigenvalue equation for the determination of the (non-relativistic) energy eigenvalue $E_*$
\begin{equation}
\label{eq:IV.13}
E_* \doteqdot \left( M_* - M \right)\,\crm^2 \;,
\end{equation}
where the effect of angular momentum is expressed by the term containing $\dot{\lO}$ and
$\ddot{\lO}$. In view of this strong analogy it seems reasonable to define the quantum
number ($\jO$) of the \emph{total} (i.\,e. spin plus orbital) angular momentum for the
solution of the Schr\"{o}dinger equation (\ref{eq:IV.12}) by
\begin{equation}
\label{eq:IV.14}
\jO \doteqdot \left| \frac{\dot{\lO} - \ddot{\lO}}{2} \right| \;.
\end{equation}

Thus we have now two equations, namely (\ref{eq:IV.10}) and (\ref{eq:IV.14}), which relate
the quantum numbers of angular momentum $\bjz$ and $\jO$ to the constants $\dot{\lO}$ and
$\ddot{\lO}$; and this then admits to express the latter in terms of the first ones
(although in an ambiguous way) as
\begin{subequations}
\begin{align}
\label{eq:IV.15a}
\dot{\lO} &= \bjz \pm \jO \\
\label{eq:IV.15b}
\ddot{\lO} &= \bjz \mp \jO \;.
\end{align}
\end{subequations}
But obviously, the product of both constants is unambiguous:
\begin{equation}
\label{eq:IV.16}
\dot{\lO} \cdot \ddot{\lO} = \bjz^2 - \jO^2 \;.
\end{equation}
Anticipating here the fact that the magnitude $\jO$ of angular momentum is never smaller than its $z$-component $\bjz$ (i.\,e. $\jO \geq \left| \bjz \right|$), we infer that both constants $\dot{\lO}$ and $\ddot{\lO}$ must always be of opposite sign ($\dot{\lO} \cdot \ddot{\lO} \leq 0$).

Thus, one formally obtains from (\ref{eq:IV.12}) a true Schr\"{o}dinger equation
\begin{align}
\label{eq:IV.17}
-\frac{\hbar^2}{2\,M}\,\left( \frac{d^2\,\tO(r)}{dr^2} + \frac{1}{r}\,\frac{d\,\tO(r)}{dr} \right) + \frac{\hbar^2}{2\,Mr^2}\, \jO^2 \cdot \tO(r) 
&- \hbar\crm\,\bAe(r) \cdot \tO(r) \\ &= E_* \cdot \tO(r) \;. \nonumber
\end{align}
However, one should keep in mind that the potential $\bAe(r)$ is not fixed from the
outside (as in the conventional Schr\"{o}dinger theory) but must be considered the
solution of the non-relativistic Poisson equation (\ref{eq:IV.4}), i.\,e.
\begin{equation}
\label{eq:IV.18}
\Delta\,\bAe(r) = -\as\,\frac{\tO^2(r)}{r} \;.
\end{equation}
One guesses from this fact that the interaction potential $\bAe(r)$ will depend on the
ortho-state $\tO(r)$ itself so that the eigenvalue problem appears highly
intricate. Different states $\tO(r)$ will generate different potentials $\bAe(r)$
according to (\ref{eq:IV.18}) so that the energy eigenvalues $E_*$ due to different states
will be found to be different, too ($\Rightarrow$ elimination of the conventional
degeneracy (\ref{eq:I.18})); and thus the RST energy spectrum of positronium must be
expected to be much more intricate than predicted by the conventional theory
(\ref{eq:I.18}); already in the non-relativistic domain and in the electrostatic
approximation.

\begin{center}
  \emph{\textbf{2.\ Angular Momentum Quantization in RST}}
\end{center}

It is rather evident that the off-separated angular system (\ref{eq:IV.6a})--(\ref{eq:IV.6b}) represents an extra problem which may be tackled independently of the proper energy eigenvalue problem (\ref{eq:IV.17})--(\ref{eq:IV.18}). On the other hand, the eigenvalues $E_*$ of the latter problem will clearly be influenced by the solutions of the angular problem (\ref{eq:IV.6a})--(\ref{eq:IV.6b}), namely via the quantum number $\jO$ of angular momentum (\ref{eq:IV.14}) and in general also via the interaction potential $\bAn(\vr)$ since this potential ``feels'' the angular behaviour of the amplitude fields via the Poisson equation (\ref{eq:IV.4}).

\begin{center}
  \large{\textit{Anisotropy of the Gauge Potential}}
\end{center}

This anisotropic influence can be expressed even more distinctly by observing our product ansatz (\ref{eq:IV.5a})--(\ref{eq:IV.5b}) together with the identifications (\ref{eq:IV.8a})--(\ref{eq:IV.8b}) which then recasts that former Poisson equation (\ref{eq:IV.4}) to the more concise form
\begin{equation}
\label{eq:IV.19}
\Delta \bAn\rt = -4\pi\as \, \bko\rt = -4\pi\as\,\bgkn(\vartheta) \cdot \bgkn(r) \;,
\end{equation}
i.\,e. the charge density $\bkn$ (\ref{eq:III.51}) becomes now also factorized ($\bko\rt =
\bgkn(\vartheta) \cdot \bgkn(r)$) with the angular factor $\bgkn(\vartheta)$ being
specified by
\begin{equation}
\label{eq:IV.20}
\bgkn(\vartheta) = \frac{1}{4\pi}\,\left\{ \frac{g_R^2(\vartheta)}{\sin^3\vartheta} +
  \sin\vartheta\cdot g_S^2(\vartheta) \right\}
\end{equation}
and the radial factor $\bgkn(r)$ by
\begin{equation}
\label{eq:IV.21}
\bgkn(r) = \frac{\tOp^2(r) + \tOm^2(r)}{r} \;.
\end{equation}

This factorization effect will subsequently allow us to develop a suitable approximation
procedure in order to manage the anisotropy of the interaction potential
$\bAn\rt$. Namely, the formal solution of the Poisson equation (\ref{eq:IV.19}) reads
\begin{align}
\label{eq:IV.22}
\bAn\rt &= \as \int d^3 \vr\,' \, \frac{\bko(r', \vartheta')}{||\vr - \vr\,'||} \\
&= \as \int \frac{d\,\Omega'}{4\pi} \, \left\{ \frac{g_R^2(\vartheta')}{\sin^3\vartheta'} + \sin\vartheta' \cdot g_S^2(\vartheta') \right\} \; \int dr' \, r' \, \frac{\tOp^2(r') + \tOm^2(r')}{||\vr - \vr\,'||} \;, \nonumber
\end{align}
where the charge normalization condition
\begin{equation}
\label{eq:IV.23}
\int d^3\vr \, \bko\rt = 1
\end{equation}
allows now also a factorized form:
\begin{subequations}
\begin{align}
\label{eq:IV.24a}
\int d\Omega \; \bgkn(\vartheta) &= \int \frac{d\,\Omega}{4\pi} \, \left\{ \frac{g_R^2(\vartheta)}{\sin^3\vartheta} + \sin\vartheta \cdot g_S^2(\vartheta) \right\} = 1 \\
\label{eq:IV.24b}
\int dr \, r^2 \; \bgkn(r) &= \int dr \, r \, \left\{ \tOp^2(r) + \tOm^2(r) \right\} = 1 \;.
\end{align}
\end{subequations}
This, however, helps us now to get approximate results for the gauge potential $\bAn\rt$ from its integral representation (\ref{eq:IV.22}). For instance, the roughest approximation consists in neglecting completely the angular dependence of the \emph{radial} integral in (\ref{eq:IV.22}) (by replacing it through its angular average) so that the angular normalization condition (\ref{eq:IV.24a}) can be immediately applied which then yields
\begin{equation}
\label{eq:IV.25}
\bAn\rt \Rightarrow \bAe(r) = \frac{\as}{4\pi} \int\limits_0^\infty \frac{d^3\,\vr\,'}{r'}\,\frac{\tOp^2(r') + \tOm^2(r')}{||\vr - \vr\,'||}
\end{equation}
whose non-relativistic approximation reads
\begin{equation}
\label{eq:IV.26}
\bAe(r) = \frac{\as}{4\pi} \int \frac{d^3\,\vr\,'}{r'} \, \frac{\tO^2(r')}{||\vr - \vr\,'||}
\end{equation}
and thus is the solution of the ``spherically symmetric'' Poisson equation (\ref{eq:IV.18}). Consequently, if we wish to go beyond this spherically symmetric approximation for the interaction potential $\bAn\rt$ (\ref{eq:IV.22}) we are forced to first work out the solutions $g_R(\vartheta)$ and $g_S(\vartheta)$ of the original angular eigenvalue problem (\ref{eq:IV.6a})--(\ref{eq:IV.6b}).

\begin{center}
  \large{\textit{Solving the Eigenvalue Problem of Angular Momentum}}
\end{center}

The close interrelationship between angular momentum of both particles and anisotropy of their interaction potential $\bgAe\rt$ (\ref{eq:IV.22}) necessitates now to explicitly solve the coupled angular system (\ref{eq:IV.6a})--(\ref{eq:IV.6b}). For this purpose, one first decouples that system by differentiating once more which yields
\begin{subequations}
\begin{align}
\label{eq:IV.27a}
\frac{d^2\,g_R(\vartheta)}{d\vartheta^2} - 2 \cot \vartheta \cdot \frac{d\,g_R(\vartheta)}{d\vartheta} + \left\{ \left( \jO^2 - 1 \right) + \frac{\left( \frac{3}{2} \right)^2 - \left( \bjz - \frac{1}{2} \right)^2}{\sin^2\vartheta} \right\} \cdot g_R(\vartheta) &= 0 \\
\label{eq:IV.27b}
\frac{d^2\,g_S(\vartheta)}{d\vartheta^2} + 2 \cot \vartheta \cdot \frac{d\,g_S(\vartheta)}{d\vartheta} + \left\{ \left( \jO^2 - 1 \right) + \frac{\left( \frac{1}{2} \right)^2 - \left( \bjz + \frac{1}{2} \right)^2}{\sin^2\vartheta} \right\} \cdot g_S(\vartheta) &= 0
\end{align}
\end{subequations}
where $\bjz$ is defined by equation (\ref{eq:III.45}) and $\jO$ by (\ref{eq:IV.14}).

Once we have arrived at two decoupled equations one can try to separately solve each of both by some transformation of variables which renders the problem more manageable. Thus, putting
\begin{subequations}
\begin{align}
\label{eq:IV.28a}
x &\doteqdot \sin\vartheta \\
\label{eq:IV.28b}
g_R(\vartheta) &\Rightarrow G_R(x) \\
\label{eq:IV.28c}
g_S(\vartheta) &\Rightarrow G_S(x) \;,
\end{align}
\end{subequations}
the decoupled second-order equations (\ref{eq:IV.27a})--(\ref{eq:IV.27b}) adopt the following form
\begin{subequations}
\begin{align}
\label{eq:IV.29a}
\left( 1 - x^2 \right) \frac{d^2\,G_R(x)}{dx^2} &+ \left( x - \frac{2}{x} \right) \frac{d\,G_R(x)}{dx} + \\
&+ \left\{ \left( \jO^2 - 1 \right) + \frac{\left( \frac{3}{2} \right)^2 - \left( \bjz - \frac{1}{2} \right)^2}{x^2} \right\} \cdot G_R(x) = 0 \nonumber \\[2em]
\label{eq:IV.29b}
\left( 1 - x^2 \right) \frac{d^2\,G_S(x)}{dx^2} &+ \left( \frac{2}{x} - 3x \right) \frac{d\,G_S(x)}{dx} + \\
&+ \left\{ \left( \jO^2 - 1 \right) + \frac{\left( \frac{1}{2} \right)^2 - \left( \bjz + \frac{1}{2} \right)^2}{x^2} \right\} \cdot G_S(x) = 0 \;. \nonumber
\end{align}
\end{subequations}
It is true, the coupled first-order system (\ref{eq:IV.6a})--(\ref{eq:IV.6b}) does not allow that \emph{both} angular functions $g_R(\vartheta)$ \emph{and} $g_S(\vartheta)$ can simultaneously be written in the form claimed by equations (\ref{eq:IV.28a})--(\ref{eq:IV.28c}); but the fact, that each one of these two equations (\ref{eq:IV.6a}) and (\ref{eq:IV.6b}) can emerge in the form (\ref{eq:IV.29a})--(\ref{eq:IV.29b}), means that always \emph{one} of both equations (\ref{eq:IV.29a}) or (\ref{eq:IV.29b}) may be realized. Thus, our strategy of elaborating the desired solutions works as follows: first, select \emph{one} of both equations (\ref{eq:IV.29a})--(\ref{eq:IV.29b}) and determine its solution as a function of $x$ (\ref{eq:IV.28a}). Then determine the associated angular function $g_R(\vartheta)$, or $g_S(\vartheta)$, resp., according to the relations (\ref{eq:IV.28b})--(\ref{eq:IV.28c}); and finally determine the residual angular function $g_S(\vartheta)$, or $g_R(\vartheta)$, resp., from the corresponding other angular equation (\ref{eq:IV.6a}), or (\ref{eq:IV.6b}), resp.

The very form of the second-order equations (\ref{eq:IV.29a})--(\ref{eq:IV.29b}) suggests to try a power series
\begin{subequations}
\begin{align}
\label{eq:IV.30a}
G_R(x) &= \sum_n \rho_n \cdot x^n \\
\label{eq:IV.30b}
G_S(x) &= \sum_n \sigma_n \cdot x^n \;.
\end{align}
\end{subequations}
If this is substituted in the second-order equations (\ref{eq:IV.29a})--(\ref{eq:IV.29b}), one gets by the standard methods the following recurrence formulae
\begin{subequations}
\begin{align}
\label{eq:IV.31a}
\rho_{n+2} &= \frac{(n-1)^2 - \jO^2}{(n+2)(n-1) - (\bjz + 1)(\bjz - 2)} \cdot \rho_n \\[1em]
\label{eq:IV.31b}
\sigma_{n+2} &= \frac{n(n+2) - (\jO + 1)(\jO - 1)}{(n+2)(n+3) - \bjz (\bjz + 1)} \cdot \sigma_n \;.
\end{align}
\end{subequations}
In order that these solutions be unique, we next have to fix the lowest order coefficients of both series. For the first series (\ref{eq:IV.30a}) we obtain for the lowest power $n_\text{min}$
\begin{equation}
\label{eq:IV.32}
n_\text{min} = \begin{cases}
\ \bjz + 1\;,\quad &\bjz \geq 0 \\
\ -(\bjz - 2)\;,\quad &\bjz < 0
\end{cases}
\end{equation}
and for the second series (\ref{eq:IV.30b})
\begin{equation}
\label{eq:IV.33}
n_\text{min} = \begin{cases}
\ \bjz\;,\quad &\bjz \geq 0 \\
\ -(\bjz + 1)\;,\quad &\bjz < 0 \;.
\end{cases}
\end{equation}

On the other hand, for $n \rightarrow \infty$ the coefficients $\rho_{n+2}$ and $\sigma_{n+2}$ would tend to $\rho_n$ and $\sigma_n$, resp.; and therefore both sums (\ref{eq:IV.30a})--(\ref{eq:IV.30b}) cannot adopt definite values if they do not stop at some finite $n_\text{max}$. However, such a maximal value $n_\text{max}$ of the power $n$ can immediately be read off from the recurrence formulae (\ref{eq:IV.31a})--(\ref{eq:IV.31b})
\begin{equation}
\label{eq:IV.34}
n_\text{max} = \begin{cases}
\ \jO + 1\;,\quad &\text{(\ref{eq:IV.31a})} \\
\ \jO - 1\;,\quad &\text{(\ref{eq:IV.31b})} \;.
\end{cases}
\end{equation}
These results enable us to closer specify now our original ans\"{a}tze (\ref{eq:IV.30a})--(\ref{eq:IV.30b}); namely
\begin{equation}
\label{eq:IV.35}
G_R(x) = \begin{cases}
\ \sum\limits_{n = \bjz + 1}^{\jO + 1} \rho_n \cdot x^n\;,\quad &\bjz \geq 0 \\[2em]
\ \sum\limits_{n = -(\bjz - 2)}^{\jO + 1} \rho_n \cdot x^n\;,\quad &\bjz < 0
\end{cases}
\end{equation}
and
\begin{equation}
\label{eq:IV.36}
G_S(x) = \begin{cases}
\ \sum\limits_{n = \bjz}^{\jO - 1} \sigma_n \cdot x^n\;,\quad &\bjz \geq 0 \\[2em]
\ \sum\limits_{n = -(\bjz + 1)}^{\jO - 1} \sigma_n \cdot x^n\;,\quad &\bjz < 0 \;.
\end{cases}
\end{equation}
From these results one concludes that the range of quantum numbers $\jO$ and $\bjz$ is the following:
\begin{align*}
-(\jO - 1) &\leq \bjz \leq \jO &\text{for } G_R(x) \text{ (\ref{eq:IV.35})}\\
-\jO &\leq \bjz \leq (\jO - 1) &\text{for } G_S(x) \text{ (\ref{eq:IV.36})}
\end{align*}
with $\jO$ being an integer $(\jO = 0,1,2,3,\ldots)$.

As a brief concrete demonstration of the general method, we consider the situation with $\jO = 2 \Rightarrow -2 \leq \bjz \leq +2$. Starting here with $\bjz = 0$, we first conclude from the arrangements (\ref{eq:IV.35})--(\ref{eq:IV.36}) that the first case (\ref{eq:IV.35}) does apply where the function $G_R(x)$ must then look as follows
\begin{equation}
\label{eq:IV.37}
G_R(x) = \sum_{n=1}^3 \rho_n \cdot x^n = \rho_1 x + \rho_3 x^3 \;,
\end{equation}
or by means of the recurrence formula (\ref{eq:IV.31a})
\begin{equation}
\label{eq:IV.38}
G_R(x) = \rho_1 (x - 2x^3)
\end{equation}
with the constraint $\rho_1$ to be fixed (below) by the normalization condition. Next, since the counterpart $G_S(x)$ of $G_R(x)$ cannot exist for $\jO = 2, \bjz = 0$ we have to determine the second angular function $g_S(\vartheta)$ directly from the first-order equation (\ref{eq:IV.6a}) with $g_R(\vartheta)$ being deduced from $G_R(x)$ (\ref{eq:IV.38}) as
\begin{equation}
\label{eq:IV.39}
g_R(\vartheta) = \rho_1 \sin\vartheta \, \left( 1 - 2 \sin^2\vartheta \right) \;.
\end{equation}
This yields
\begin{equation}
\label{eq:IV.40}
g_S(\vartheta) = - \frac{4 \rho_1}{\ddlO} \cos\vartheta \;.
\end{equation}
Now that both angular functions are fixed up to the normalization constant $\rho_1$, we can pin the latter down by the normalization condition (\ref{eq:IV.24a}) which adopts the following form for the present situation
\begin{align}
\label{eq:IV.41}
1 &= \int \frac{d\,\Omega}{4\pi}\,\left\{ \frac{g_R^2(\vartheta)}{\sin^3\vartheta} + \sin\vartheta \cdot g_S^2(\vartheta) \right\} \\
&= \frac{1}{2} \, \rho_1^2 \int\limits_0^\pi d\vartheta\,\left\{ \left[ 1 -
    2\sin^2\vartheta \right]^2 + \left( \frac{4}{\ddlO} \right)^2
  \sin^2\vartheta\cos^2\vartheta \right\}\ . \nonumber
\end{align}
This ascribes to $\rho_1$ the value
\begin{equation}
\label{eq:IV.42}
\rho_1^2 = \frac{4}{\pi\,\left( 1 + \frac{4}{\ddlO^2} \right)} \;.
\end{equation}
Here, one wishes to express the constant $\ddlO$ through $\jO$ and $\bjz$, see equation (\ref{eq:IV.15b}), which, however, offers two possibilities, namely either
\begin{subequations}
\begin{align}
\label{eq:IV.43a}
\dlO &= \bjz + \jO\ \Rightarrow\ +2 \\
\label{eq:IV.43b}
\ddlO &= \bjz - \jO\ \Rightarrow\ -2
\end{align}
\end{subequations}
or
\begin{subequations}
\begin{align}
\label{eq:IV.44a}
\dlO &= \bjz - \jO\ \Rightarrow\ -2 \\
\label{eq:IV.44b}
\ddlO &= \bjz + \jO\ \Rightarrow\ +2 \;.
\end{align}
\end{subequations}

But these two possibilities differ merely in sign and therefore do not generate
essentially different solutions of the eigenvalue problem under consideration (in contrast
to the other cases with $j_z \neq 0$, see the table below). Indeed, both situations
(\ref{eq:IV.43a})--(\ref{eq:IV.43b}) \emph{and} (\ref{eq:IV.44a})--(\ref{eq:IV.44b}) yield
the same normalization constant $\rho_1$ (\ref{eq:IV.42}), i.\,e.
\begin{equation}
\label{eq:IV.45}
\rho_1 = \pm \sqrt{\frac{2}{\pi}} \;,
\end{equation}
where the ambiguity in sign is immaterial. Thus the solutions $g_R(\vartheta)$ (\ref{eq:IV.39}) and $g_S(\vartheta)$ (\ref{eq:IV.40}) adopt their final form as
\begin{subequations}
\begin{align}
\label{eq:IV.46a}
g_R(\vartheta) &= \sqrt{\frac{2}{\pi}} \, \sin\vartheta\,\left( 1 - 2\sin^2\vartheta \right) \\
\label{eq:IV.46b}
g_S(\vartheta) &= - 2 \, \sqrt{\frac{2}{\pi}} \, \cos\vartheta \\
&\boldsymbol{\left( \bjz = 0, \jO = 2 \right)} \;. \nonumber
\end{align}
\end{subequations}
The peculiar point with this result is that the \emph{essentially} ambiguous situation
(\ref{eq:IV.15a})--(\ref{eq:IV.15b}) does not entail the existence of two essentially
different solutions for \emph{all} allowed values of $\bjz$ but only for \emph{some} $\bjz
\neq 0$ (see the table below). Indeed, that essential ambiguity of some solutions of the
angular eigenvalue problem (\ref{eq:IV.6a})--(\ref{eq:IV.6b}) is just the core of the
\emph{ortho-dimorphism}, i.\,e. the dimorphism of the ortho-positronium
configurations. Naturally, the most urgent question arising in this context must refer to
the energy difference of such a dimorphic pair, which will readily be worked out.

\vskip 0.8cm
\begin{center}
$\boldsymbol{\jO = 2}$
\vskip 0.5cm
\label{table4a}
\begin{tabular}{|c|c|c||c|c||c|c|}
\hline
$\bjz$ & $\dlO$ & $\ddlO$ & $G_R(x)$ & $G_S(x)$ & $g_R(\vartheta)$ & $g_S(\vartheta)$ \\
\hline\hline
\multirow{2}{*}{$2$} & \multirow{2}{*}{$0$} & \multirow{2}{*}{$4$} & \multirow{2}{*}{$\rho_3 \cdot x^3$} & \multirow{2}{*}{---} & \multirow{2}{*}{$\frac{4}{\sqrt{3\pi}}\,\sin^3\vartheta$} & \multirow{2}{*}{$0$} \\
& & & & & & \\
\specialrule{1.pt}{0pt}{0pt}
\multirow{2}{*}{$1$} & $3$ & $-1$ & \multirow{2}{*}{---} & \multirow{2}{*}{$\sigma_1 \cdot x$} & $\frac{2}{\sqrt{\pi}}\,\sin^2\vartheta\cos\vartheta$ & $\frac{2}{\sqrt{\pi}}\,\sin\vartheta$ \\
\cline{2-3}\cline{6-7}
& $-1$ & $3$ & & & $-\frac{6}{\sqrt{3\pi}}\,\sin^2\vartheta\cos\vartheta$ & $\frac{2}{\sqrt{3\pi}}\,\sin\vartheta$ \\
\specialrule{1.2pt}{0pt}{0pt}
\multirow{2}{*}{$0$} & $2$ & $-2$ & \multirow{2}{*}{$\rho_1 \, \left( x - 2x^3 \right)$} & \multirow{2}{*}{---} & \multirow{2}{*}{$\sqrt{\frac{2}{\pi}}\,\sin\vartheta\,\left( 1 - 2\sin^2\vartheta \right)$} & \multirow{2}{*}{$-2\,\sqrt{\frac{2}{\pi}}\,\cos\vartheta$} \\
\cline{2-3}
& $-2$ & $2$ & & & & \\
\specialrule{1.2pt}{0pt}{0pt}
\multirow{2}{*}{$-1$} & $1$ & $-3$ & \multirow{2}{*}{$\rho_3 \cdot x^3$} & \multirow{2}{*}{---} & $\frac{2}{\sqrt{\pi}}\,\sin^3\vartheta$ & $-\frac{2}{\sqrt{\pi}}\,\cos\vartheta$ \\
\cline{2-3}\cline{6-7}
& $-3$ & $1$ & & & $\frac{2}{\sqrt{3\pi}}\,\sin^3\vartheta$ & $\frac{6}{\sqrt{3\pi}}\,\cos\vartheta$ \\
\specialrule{1.2pt}{0pt}{0pt}
\multirow{2}{*}{$-2$} & \multirow{2}{*}{$-4$} & \multirow{2}{*}{$0$} & \multirow{2}{*}{---} & \multirow{2}{*}{$\sigma_1 \cdot x$} & \multirow{2}{*}{$0$} & \multirow{2}{*}{$\frac{4}{\sqrt{3\pi}}\,\sin\vartheta$} \\
& & & & & & \\
\hline
\end{tabular}
\end{center}
\vskip 1cm

\newpage
\begin{center}
  \large{\textit{\textbf{Ortho-Dimorphism for}}} $\boldsymbol{\jO = 2}$
\end{center}

The table (precedent page) displays the possible configurations of the pair $\dlO, \ddlO$
corresponding to the quantum numbers $\bjz = -2,-1,0,1,2$ due to $\boldsymbol{\jO =
  2}$. The associated angular functions $g_R(\vartheta), g_S(\vartheta)$ as solutions of
the first-order system (\ref{eq:IV.6a})--(\ref{eq:IV.6b}) emerge \emph{ambiguously} for
$\bjz = \pm 1$ but are \emph{unique} for $\bjz = 0, \pm 2$. The case with $\bjz = -1$ is
the simplest one and has already been treated in ref.~\cite{34} (in that paper the present
quantum number $\jO$ has there been denoted by $\lO$).

The table of angular functions $g_R(\vartheta), g_S(\vartheta)$ (for $\jO = 2$) obviously
displays some regularities (or symmetries, resp.) as far as the permutation of both
functions $g_R(\vartheta)$ and $g_S(\vartheta)$ is concerned. Surely, this must have to do
something with the isotropy of three-space, especially with respect to the replacement
$\bjz \Leftrightarrow -\bjz$. Indeed, if this replacement is associated with the following
replacements (irrespective of an irrelevant change of sign)
\begin{subequations}
  \begin{align}
    \label{eq:IV.47a}
    g_R(\vartheta) &\Rightarrow g_S(\vartheta)\cdot\sin^2\vartheta\\*
    \label{eq:IV.47b}
    g_S(\vartheta) &\Rightarrow -\frac{g_R(\vartheta)}{\sin^2(\vartheta)}\\*
    \label{eq:IV.47c}
    \dlO & \Rightarrow -\ddlO\ ,\ \ddlO  \Rightarrow -\dlO
  \end{align}
\end{subequations}
then both eigenvalue equations (\ref{eq:IV.6a})-(\ref{eq:IV.6b}) become merely interchanged and
the angular density~$\bgkn(\vartheta)$ is left invariant.

In order to get a broader basis for our inductive reasoning, we extend the calculations to
$\jO = 3$, see the table below. Combining both tables for $\jO = 2$ and $\jO = 3$ we
conclude that the angular functions $g_R(\vartheta)$ and $g_S(\vartheta)$ are always
\emph{unique} for $\bjz = 0$ and for $\bjz = \pm\jO$ for arbitrary $\jO$
$(=1,2,3,4\ldots)$. This is proven quite generally in \textbf{App.A}. For the other values
of $\bjz$ (i.\,e. $0 < |\bjz| < \jO$) the angular functions $g_R(\vartheta)$,
$g_S(\vartheta)$ are generally expected to be ambiguous. However, the extent of this
ambiguity deserves an extra discussion where perhaps the energy difference of the
ambiguous configurations may be taken as a quantitative measure of just that ambiguity.

%\vskip 1cm
\begin{landscape}
\begin{center}
\label{table4b}
\begin{tabular}{|c|c|c||c|c||c|c|}
\hline
$\bjz$ & $\dlO$ & $\ddlO$ & $G_R(x)$ & $G_S(x)$ & $g_R(\vartheta)$ & $g_S(\vartheta)$ \\
\hline\hline
\multirow{2}{*}{$3$} & \multirow{2}{*}{$0$} & \multirow{2}{*}{$6$} & \multirow{2}{*}{$\rho_4 \, x^4$} & \multirow{2}{*}{---} & \multirow{2}{*}{$\sqrt{\frac{32}{5\pi}}\,\sin^4\vartheta$} & \multirow{2}{*}{$0$} \\
& & & & & & \\
\specialrule{1.2pt}{0pt}{0pt}
\multirow{2}{*}{$2$} & $5$ & $-1$ & \multirow{2}{*}{---} & \multirow{2}{*}{$\sigma_2 \, x^2$} & $\sqrt{\frac{16}{3\pi}}\,\sin^3\vartheta\cos\vartheta$ & $\sqrt{\frac{16}{3\pi}}\,\sin^2\vartheta$ \\
\cline{2-3}\cline{6-7}
& $-1$ & $5$ & & & $-5 \sqrt{\frac{16}{15\pi}}\,\sin^3\vartheta\cos\vartheta$ & $\sqrt{\frac{16}{15\pi}}\,\sin^2\vartheta$ \\
\specialrule{1.2pt}{0pt}{0pt}
\multirow{2}{*}{$1$} & $4$ & $-2$ & \multirow{2}{*}{$\rho_2\left( x^2 - \frac{4}{3} x^4 \right)$} & \multirow{2}{*}{---} & $\sqrt{\frac{12}{\pi}}\,\sin^2\vartheta\,\left( 1 - \frac{4}{3}\sin^2\vartheta \right)$ & $\frac{4}{3}\sqrt{\frac{12}{\pi}}\,\sin\vartheta\cos\vartheta$ \\
\cline{2-3}\cline{6-7}
& $-2$ & $4$ & & & $\sqrt{\frac{24}{\pi}}\,\sin^2\vartheta\,\left( 1 - \frac{4}{3}\sin^2\vartheta \right)$ & $-\frac{2}{3}\sqrt{\frac{24}{\pi}}\,\sin\vartheta\cos\vartheta$ \\
\specialrule{1.2pt}{0pt}{0pt}
\multirow{2}{*}{$0$} & $\multirow{2}{*}{3}$ & $\multirow{2}{*}{-3}$ & \multirow{2}{*}{---} & \multirow{2}{*}{$\sigma_0\left( 1 - \frac{4}{3} x^2 \right)$} & \multirow{2}{*}{$\sqrt{\frac{2}{\pi}}\,\sin\vartheta\cos\vartheta\,\left( 1 - 4\sin^2\vartheta \right)$} & \multirow{2}{*}{$\sqrt{\frac{18}{\pi}}\,\left( 1 - \frac{4}{3}\sin^2\vartheta \right)$} \\
&  &  & & & & \\
\specialrule{1.2pt}{0pt}{0pt}
\multirow{2}{*}{$-1$} & $2$ & $-4$ & \multirow{2}{*}{---} & \multirow{2}{*}{$\sigma_0 \left( 1 - \frac{4}{3} x^2 \right)$} & $-\frac{4}{3}\sqrt{\frac{12}{\pi}}\,\sin^3\vartheta\cos\vartheta$ & $\sqrt{\frac{12}{\pi}}\,\left( 1 - \frac{4}{3}\sin^2\vartheta \right)$ \\
\cline{2-3}\cline{6-7}
& $-4$ & $2$ & & & $\frac{2}{3}\sqrt{\frac{24}{\pi}}\,\sin^3\vartheta\cos\vartheta$ & $\sqrt{\frac{24}{\pi}}\,\left( 1 - \frac{4}{3}\sin^2\vartheta \right)$ \\
\specialrule{1.2pt}{0pt}{0pt}
\multirow{2}{*}{$-2$} & $1$ & $-5$ & \multirow{2}{*}{$\rho_4 \, x^4$} & \multirow{2}{*}{---} & $\sqrt{\frac{16}{3\pi}}\,\sin^4\vartheta$ & $-\sqrt{\frac{16}{3\pi}}\,\sin\vartheta\cos\vartheta$ \\
\cline{2-3}\cline{6-7}
& $-5$ & $1$ & & & $\sqrt{\frac{16}{15\pi}}\,\sin^4\vartheta$ & $5\sqrt{\frac{16}{15\pi}}\,\sin\vartheta\cos\vartheta$ \\
\specialrule{1.2pt}{0pt}{0pt}
\multirow{2}{*}{$-3$} & \multirow{2}{*}{$-6$} & \multirow{2}{*}{$0$} & \multirow{2}{*}{---} & \multirow{2}{*}{$\sigma_2 \, x^2$} & \multirow{2}{*}{$0$} & \multirow{2}{*}{$\sqrt{\frac{32}{5\pi}}\,\sin^2\vartheta$} \\
& & & & & & \\
\hline
\end{tabular}
\end{center}
\vskip 0.5cm

\begin{center}
  \large{\textit{\textbf{Ortho-Dimorphism for}}} $\boldsymbol{\jO = 3}$
\end{center}

The possible configurations due to $\boldsymbol{\jO = 3}$ are $\bjz = 0, \pm 1,\pm 2, \pm
3$ and display again the corresponding ambiguities (for $\bjz = \pm 1, \pm 2$) and
uniqueness (for $\bjz = 0$ and $\bjz = \pm\jO$). Thus, one concludes that for general
$\jO$ ($= 1,2,3,4,\ldots$) one always has $2\jO + 1 - 3 = 2(\jO - 1)$ \emph{dimorphic}
pairs and 3 \emph{solitary} angular distributions.
\end{landscape}

The somewhat hidden symmetries of the pair of functions $g_R(\vartheta)$ and
$g_S(\vartheta)$ due to the interchange of the positive and negative $z$-direction
($\vartheta = 0 \Leftrightarrow \vartheta = \pi$) becomes now more obvious by looking at
the angular densities $\bgkn(\vartheta)$ (\ref{eq:IV.20}). The subsequent table presents a
collection of all these angular objects due to $\jO = 1,2,3$. From here it becomes
immediately obvious that the ambiguous pair $g_R(\vartheta), g_S(\vartheta)$ due to a
given $\bjz$ (with $0 < |\bjz| < \jO$) does generate the same couple of angular densities
$\bgkn(\vartheta)$ as does the pair $g_R(\vartheta), g_S(\vartheta)$ due to
$-\bjz$. Consequently, the \emph{electric} charge density $\bgkn(\vr)$ does not single out
a preferential direction on the $z$-axis (contrary to the \emph{current} density $\bkv$).

\vskip 1cm
\begin{center}
\label{table4c}
\begin{tabular}{|c|c||c|}
\hline
$\jO$ & $\bjz$ & $\bgkn(\vartheta)$ (\ref{eq:IV.20}) \\
\hline\hline
$1$ & $0$ & $\frac{1}{2\pi^2} \cdot \frac{1}{\sin\vartheta}$ \\
\hline
$1$ & $\pm 1$ & $\frac{\sin\vartheta}{\pi^2}$ \\
\specialrule{1.2pt}{0pt}{0pt}
$2$ & $0$ & $\frac{1}{2\pi^2} \cdot \frac{1}{\sin\vartheta}$ \\
\hline
\multirow{2}{*}{$2$} & \multirow{2}{*}{$\pm 1$} & $\frac{1}{\pi^2} \sin\vartheta$ \\
\cline{3-3}
& & $\frac{1}{3\pi^2}\,\left\{ 9\sin\vartheta - 8\sin^3\vartheta \right\}$ \\
\hline
$2$ & $\pm 2$ & $\frac{4}{3\pi^2} \sin^3\vartheta$ \\
\specialrule{1.2pt}{0pt}{0pt}
$3$ & $0$ & $\frac{1}{2\pi^2} \cdot \frac{1}{\sin\vartheta}$ \\
\hline
\multirow{2}{*}{$3$} & \multirow{2}{*}{$\pm 1$} & $\frac{3}{\pi^2}\,\left\{ \sin\vartheta - \frac{8}{9} \sin^3\vartheta \right\}$ \\
\cline{3-3}
& & $\frac{6}{\pi^2}\,\left\{ \sin\vartheta - \frac{20}{9} \sin^3\vartheta + \frac{4}{3}\sin^5\vartheta \right\}$ \\
\hline
\multirow{2}{*}{$3$} & \multirow{2}{*}{$\pm 2$} & $\frac{4}{3\pi^2}\,\sin^3\vartheta$ \\
\cline{3-3}
& & $\frac{4}{\pi^2}\,\left\{ \frac{5}{3} \sin^3\vartheta - \frac{8}{5}\sin^5\vartheta \right\}$ \\
\hline
$3$ & $\pm 3$ & $\frac{8}{5\pi^2}\,\sin^5\vartheta$ \\
\hline
\end{tabular}
\end{center}
\vskip 1cm

\begin{center}
  \mbox{\large{\textit{\textbf{Angular Density $\boldsymbol{\bgkn(}\vartheta)$ (\ref{eq:IV.20})
        for the States Due to $\boldsymbol{\jO = 1,2,3}$}}}}
\end{center}
\pagebreak 
Quite intuitively, one would suppose that the shape of the charge density
$\bgkn(\vartheta)$ is influenced by both the azimuthal quantum number $\bjz$ and the total
number $\jO$. However, the above table says that the angular densities $\bgkn(\vartheta)$
are the same for $\jO = 2, \bjz = 0$ and $\jO = 3, \bjz = 0$ and also
for~$\jO=1,\bjz=0$. Indeed, one can show that the angular charge density
$\bgkn(\vartheta)$ for $\bjz = 0$ is the same for all values of $\jO$, see equation
(\ref{eq:A.5}) of \textbf{App.A}. Thus one becomes tempted to attribute the deformations
of the angular charge distributions $\bgkn(\vartheta)$ to the action of ``centrifugal
forces'', which are linked to the azimuthal quantum number $\bjz$ rather than to the total
number $\jO$! In this sense, one expects that for $\bjz = 0$ the charge distribution is
concentrated in the vicinity of the rotational axis ($\vartheta = 0,\pi$); whereas for
maximal $\bjz$ ($\Rightarrow \bjz = \pm \jO$) the charge distribution $\bgkn$ becomes
rolled out to a disc-like shape. Indeed, the occurrence of this effect is demonstrated by
the \textbf{Fig.IV.A} below. Moreover it should be considered a matter of course that such
a deformation of the charge clouds will entail some change of the electrostatic
interaction which then implies that the \emph{$j_z$-degeneracy becomes eliminated} (see
below). This elimination of degeneracy, however, has nothing to do with the
ortho-dimorphism which occurs for dimorphic partners belonging to the \emph{same} pair of
quantum numbers~$\jO,\bjz$; see \textbf{Fig.IV.B} below for a simple demonstration.

\begin{center}
\epsfig{file=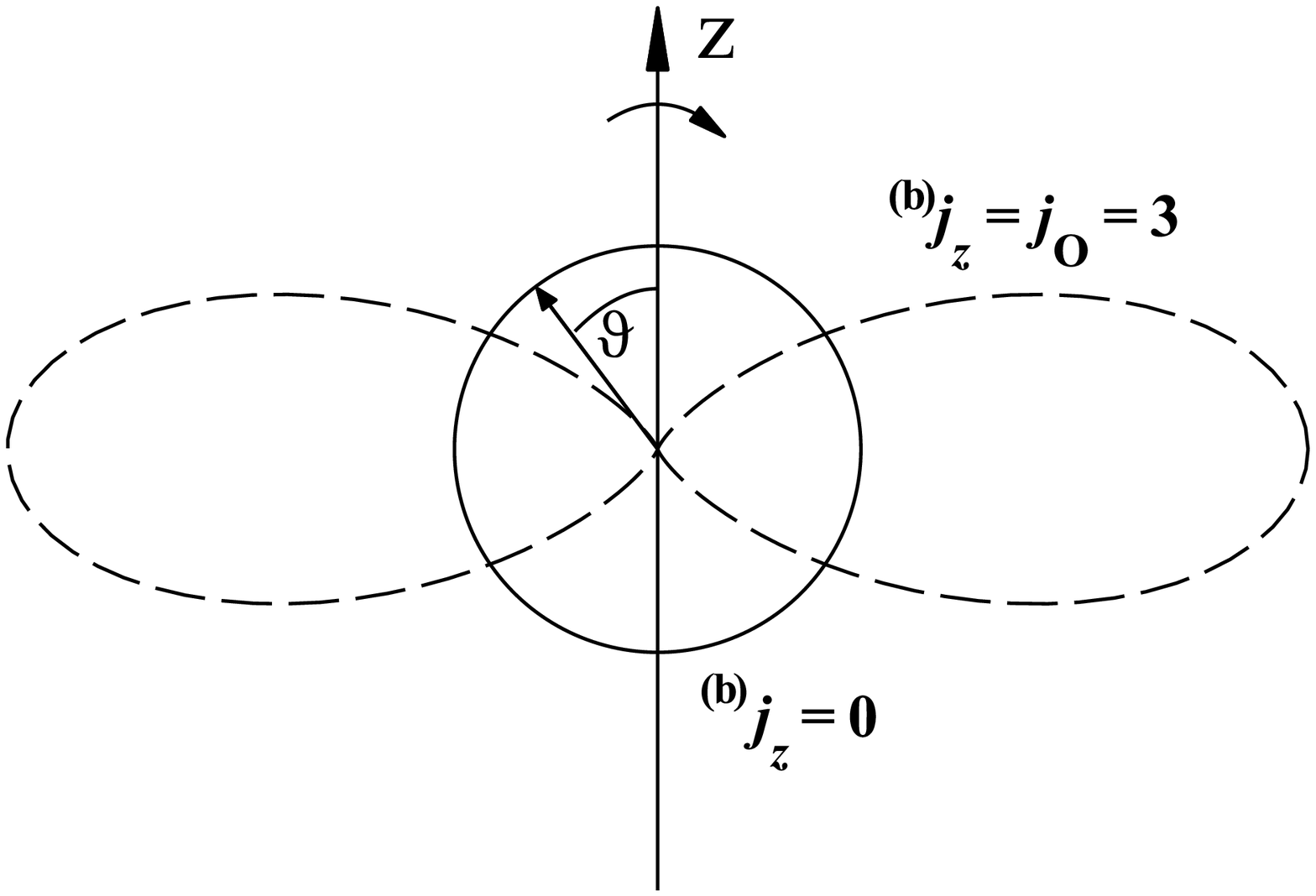,height=12cm}
\end{center}
\vspace{-1.5cm}
{\textbf{Fig.~IV.A}\hspace{5mm} \emph{\large\textbf{Rolling out of the Rotating Charge Clouds }  }\Large\boldmath{$\bko(\vartheta)$} }
\indent\\*
\label{figIV.A}

For small rotational quantum number~$\bjz$ (solid curve:~$\bjz=0$) the charge density
$\bgkn(r,\vartheta)=\bgkn(\vartheta)\cdot\bgkn(r)$ (\ref{eq:IV.20}-(\ref{eq:IV.21}) is
concentrated close to the rotation axis (i.e.\ z-axis;~$\vartheta=0$). The angular
pre-factor~$\bgkn(\vartheta)$ due to~$\bjz=0$ is the same for all values
of~$\jO\,(=0,1,2,3,\ldots)$, i.e.
\begin{equation*}
 \sin\vartheta  \cdot\bgkn(\vartheta)=\frac{1}{2\pi^2} \hspace{1cm} (\text{solid line})
\end{equation*}
see \textbf{App.A}, equation (\ref{eq:A.5}). However, for extremal value of~$\bjz$
(i.e.~$\bjz=\pm \jO$) the charge cloud becomes rolled out to a disc-like shape through the
centrifugal forces; dotted curve:~$\bjz=\jO=3$ with
\begin{equation*}
 \sin\vartheta\cdot {}^{(b)}k_0(\vartheta) = \frac{8}{5\pi^2}\cdot\sin^6\vartheta\ ,
\end{equation*}
see the table on p.~\pageref{table4c}. Such an inertial deformation of the rotating charge clouds is to
be considered the origin of the elimination of the~$j_z$-degeneracy which itself occurs
only in the spherically symmetric approximation.

\begin{center}
\label{fig4b}
\epsfig{file=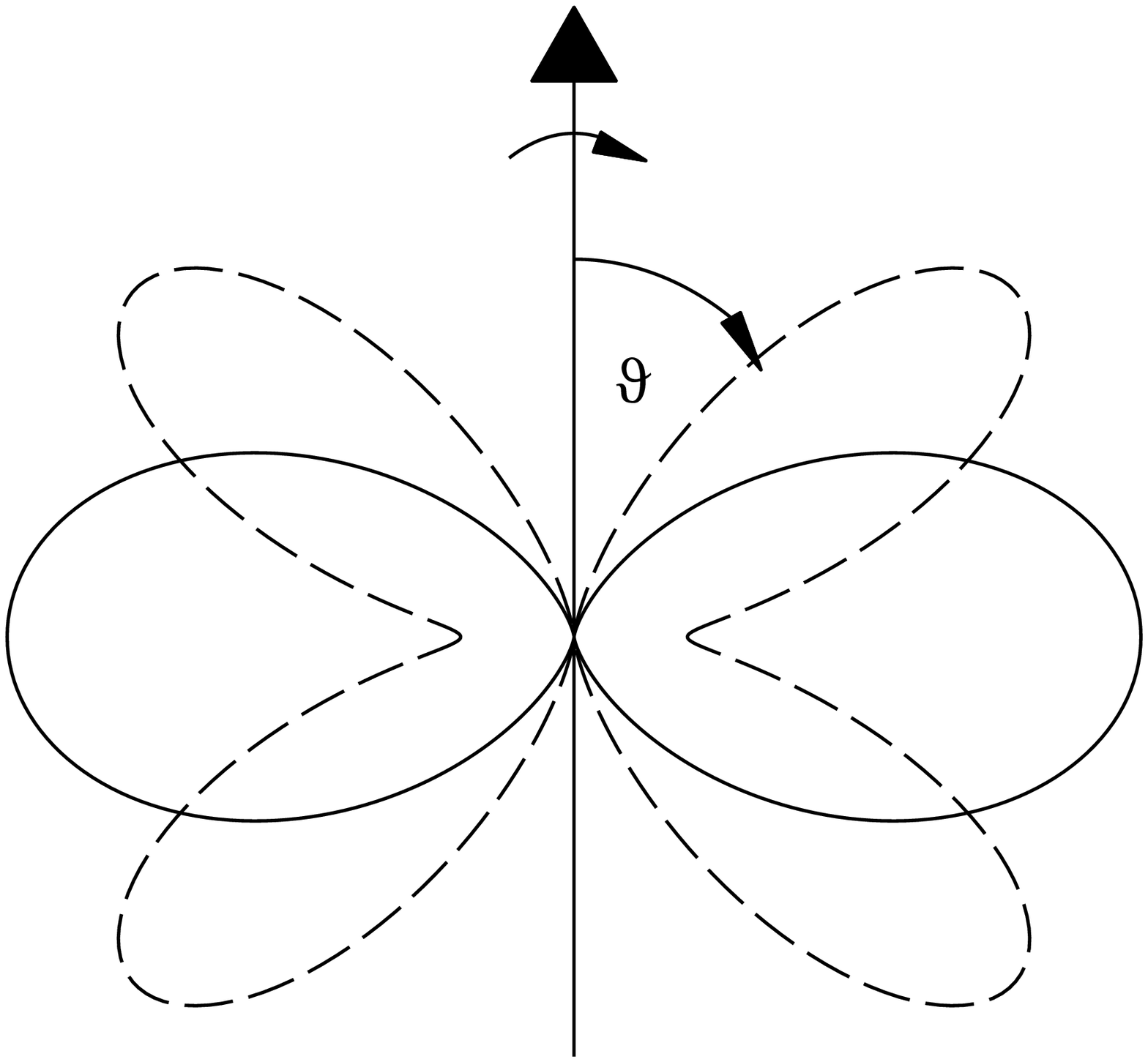,height=12cm}
\end{center}
{\textbf{Fig.~IV.B}\hspace{5mm} \emph{\large\textbf{Dimorphism of Angular Density \boldmath{$\bko(\vartheta)$}
      \\\phantom{Fig.~IV.B}   \hspace{25mm}for\large\boldmath~$\jO=3,\bjz=\pm 2$}   }
\indent\\[-5mm]
\label{figIV.B}

Even for the \emph{same} rotational quantum numbers~$\jO$ and $\bjz$ the angular
density~${}^{\{b\}}k_0(\vartheta)$ (\ref{eq:IV.20}) occurs in two different forms,
provided the quantum number~$\bjz$ is restricted by the condition~$0<|\bjz|<\jO$, see
\textbf{App.A}. For the present situation~$(\jO=3,\bjz=\pm 2)$, the two forms of angular
density are given by (see table on p.~\pageref{table4c})
\begin{equation*}
  {}^{(b)}k_0(\vartheta) =
  \begin{cases}
    \frac{4}{3\pi^2}\sin^3\vartheta\ , & \text{solid curve}\\*
    \frac{4}{\pi^2}\left(\frac{5}{3}\sin^3\vartheta-\frac{8}{5}\sin^5\vartheta\right)\ ,&
    \text{dotted curve}\ .
  \end{cases}
\end{equation*}
The different shape of the electric charge clouds entails different electrostatic energy
and therefore also different binding energies (~$\leadsto$ ``ortho-dimorphism'').

\section{Anisotropy of the Gauge Potential $\mathbf{\bAn(}\vec{\mathbf{r}}\mathbf{)}$}
\indent

Besides the matter fields, the second basic building block of RST is constituted by the
gauge fields. Since the latter fields are coupled to the matter fields via the Poisson
equations, the anisotropy of the matter fields is transferred to the gauge fields just by
these Poisson equations. For the present \emph{electrostatic approximation}, the gauge
fields enter the theory only in form of their time component $\bAn(\vr)$ due to the
corresponding four-potential $\bAmu$. The field equations for the gauge fields are the
(generally non-Abelian) Maxwell equations, which however for the present electrostatic
situation reduce to the ordinary Poisson equation (\ref{eq:IV.4}). Here, the charge
density $\bkn$ factorizes as shown by equation (\ref{eq:IV.19}); two examples of this
dimorphism of the angular factor $\bgkn(\vartheta)$ are given by the above tables at the
end of \textbf{Subsect.IV.2}. Thus it should be evident that exact solutions of the
Poisson equation (\ref{eq:IV.19}) can not easily be found. However, this should not be
considered a bad situation since there are approximation procedures at hand which are able
to mediate a rough picture of how the exact solution $\bAn(\vr)$ will look like. The point
of departure for this picture is the ``\emph{spherically symmetric approximation}'', which
has already been briefly mentioned in connection with the equations (\ref{eq:IV.18}) and
(\ref{eq:IV.25})--(\ref{eq:IV.26}).

\begin{center}
  \emph{\textbf{1.\ Spherical Symmetry as a First Approximation}}
\end{center}

The spherically symmetric approximation of the gauge potential has been based upon the
assumption that this potential is SO(3) symmetric, i.\,e. we put
\begin{align}
\label{eq:V.1}
\bAn(\vr) &\Rightarrow \bAe(r) \\
(r &= ||\vr||)\;. \nonumber
\end{align}
The corresponding Poisson equation (\ref{eq:IV.18}) for the approximative potential
$\bAe(r)$ is obtained through extremalizing the electro-static part $\etEEO$ of the full
energy functional $\tEEO$ which of course must consist of both the gauge field part
$\GtEEO$ and the (Dirac) matter part $\DtEEO$, see ref.~\cite{2}
\begin{equation}
\label{eq:V.2}
\tEEO = \DtEEO + \GtEEO \;.
\end{equation}
Here, the matter part $\DtEEO$ will be readily specified and discussed, but in the present context we are rather interested in the electric contribution $\etEEO$ to the gauge part $\GtEEO$ because the extremalization of the latter part already yields the desired electrostatic field equations.

Indeed, the wanted electrostatic part $\etEEO$ is composed of two contributions
\begin{equation}
\label{eq:V.3}
\etEEO = \ERe + \lGe \cdot \tNNGer \;.
\end{equation}
The Langrangean multiplier adopts the value $\lGe = -2$ and the first contribution ($\ERe$) is the electrostatic gauge field energy
\begin{equation}
\label{eq:V.4}
\ERe = -\frac{\hbar\mathrm{c}}{4 \pi \alpha_s} \int d^3\,\vr \left( \vec{E}_b(\vr) \sdot \vec{E}_b(\vr) \right)
\end{equation}
where the electrostatic field strength $\vec{E}_b(\vr)$ of the ortho-configurations is given as usual by
\begin{equation}
\label{eq:V.5}
\vec{E}_b(\vr) = -\vec{\nabla}\ \bAn(\vr)
\end{equation}
(this simple form of $\vec{E}_b(\vr)$ does exclusively apply to \emph{non-identical}
particles, see ref.~\cite{34}). For our presently considered spherical symmetric
approximation (\ref{eq:V.1}), the general form $\vec{E}_b(\vr)$ (\ref{eq:V.5}) does reduce
to
\begin{equation}
\label{eq:V.6}
\vec{E}_b(\vr) \Rightarrow -\vec{\nabla}\ \bAe(r) = -\vec{e}_r\ \frac{d\ \bAe(r)}{dr}\;,
\end{equation}
and consequently the electrostatic energy $\ERe$ (\ref{eq:V.4}) becomes simplified to
\begin{equation}
\label{eq:V.7}
\eER \Rightarrow \eeER = -\frac{\hbar\mathrm{c}}{\alpha_s} \int dr\ r^2\ \left( \frac{d\ \bAe(r)}{dr} \right)^2 \;.
\end{equation}

A similar conclusion does also hold for the ``\emph{Poisson constraint}'' $\tNGe$ which
generally is defined as the difference of the gauge field energy $\eER$ and its
``\emph{mass equivalent}'' $\Mee \mathrm{c^2}$~\cite{2}:
\begin{equation}
\label{eq:V.8}
\NGe = \eER - \Mee \mathrm{c^2} \;,
\end{equation}
where the mass equivalent itself is defined through [Zitat]
\begin{equation}
\label{eq:V.9}
\Mee \mathrm{c^2} \doteqdot -\hbar \mathrm{c} \int d^3\vr\ \bAn(\vr) \cdot \bkn \;.
\end{equation}
For the spherically symmetric approximation, the gauge potential is assumed to be SO(3) symmetric, cf. (\ref{eq:V.1}), and the charge density $\bkn$ is of the product form described by (\ref{eq:IV.19})--(\ref{eq:IV.21}). Therefore the mass equivalent (\ref{eq:V.9}) factorizes as follows
\begin{equation}
\label{eq:V.10}
\Mee \mathrm{c^2} \Rightarrow -\hbar\mathrm{c} \int d\Omega \ \bgkn(\vartheta) \int\limits_0^\infty dr\ r^2\ \bAe(r) \cdot \bgkn(r) \;.
\end{equation}
But here we can resort to the angular normalization condition (\ref{eq:IV.24a}) for $\bgkn(\vartheta)$ and also to the non-relativistic version of the radial charge density $\bgkn(r)$ (\ref{eq:IV.21})
\begin{equation}
\label{eq:V.11}
\bgkn(r) = \frac{\tO^2(r)}{r}
\end{equation}
so that we end up with the final form of the (``isotropic'') mass equivalent
\begin{equation}
\label{eq:V.12}
\Mee \mathrm{c^2} \Rightarrow \tMMee \mathrm{c^2} = -\hbar\mathrm{c} \int\limits_0^\infty dr\ r\ \bAe(r) \cdot \tO^2(r) \;.
\end{equation}

With all these arrangements, the ``spherically symmetric'' form of the electrostatic gauge field functional $\etEEO$ is of the shape
\begin{equation}
\label{eq:V.13}
\etEEO \Rightarrow \eetEEO = \eeER + \lGe \cdot \tNNGee
\end{equation}
with $\eeER$ being given by (\ref{eq:V.7}) and
\begin{equation}
\label{eq:V.14}
\tNNGee = \eeER - \tMMee \mathrm{c^2} \;.
\end{equation}
Now that the specific SO(3) symmetric form of all individual contributions is clarified,
one can carry through the variational process for $\eetEEO$ (\ref{eq:V.13}) and thereby
obtains the well-known Poisson equation (\ref{eq:IV.18}) for $\bAe(r)$. The standard
formal solution is here given by equation (\ref{eq:IV.26}) which we will readily exploit
for the study of the corresponding anisotropy corrections.

\begin{center}
  \emph{\textbf{2.\ Anisotropy Corrections}}
\end{center}

If one disregards for the moment the spherically symmetric approximation (\ref{eq:V.1}), the general Poisson equation is found as the extremal equation due to the general gauge field functional $\etEEO$. Furthermore, the use of the product ansatz (\ref{eq:IV.5a})--(\ref{eq:IV.5b}) recasted the general Poisson equation (\ref{eq:IV.4}) to the more specific form (\ref{eq:IV.19}) with the formal solution
\begin{equation}
\label{eq:V.15}
\bgAe\rt = \alpha_s \int d^3 \vec{r}\,'\ \frac{\bgkn(\vartheta') \cdot \bgkn(r')}{||\vr-\vr\,'||} \;.
\end{equation}
Clearly, this is the anisotropic generalization (up to a factor of $4\pi$) of the former
result (\ref{eq:IV.26}) for the spherically symmetric approximation.

It is true, the latter approximation disregards completely the anisotropic features of the
exact solution of the eigenvalue problem, but nevertheless it plays a certain role when
considering now the \emph{anisotropic} potential $\bgAe\rt$ (\ref{eq:V.15}). Namely, since
it appears rather difficult to exactly calculate the integral of that formal solution
(\ref{eq:V.15}) we are forced to think about some approximative evaluation. For this
purpose, one may resort to an expansion of the anisotropic potential $\bgAe\rt$ around the
spherically symmetric potential $\bAe(r)$ (\ref{eq:IV.26}), i.\,e. an \emph{expansion with
  respect to the magnitude of anisotropy}. In this sense, one may look upon the
spherically symmetric approximation $\bAe(r)$ as the leading term of such an expansion,
namely such that it appears as the angular average of the more general anisotropic
solution $\bgAe\rt$ (\ref{eq:V.15}), i.\,e. explicitly
\begin{align}
\label{eq:V.16}
\bAe(r) &= \int \frac{d\Omega}{4\pi}\ \bgAe(r, \vartheta) \\
(d\Omega &\doteqdot d\vartheta\,d\phi\,\sin\vartheta) \;. \nonumber
\end{align}

In order to build up now such an expansion with respect to the magnitude of anisotropy,
one starts with the following expansion of the denominator in the integral representation
(\ref{eq:V.15}) of $\bgAe\rt$~\cite{2}
\begin{align}
\label{eq:V.17}
\frac{1}{||\vr - \vr\,'||} &= \frac{1}{\left( r^2+{r'}^2 \right)^{1/2}} + \frac{r\,r'}{\left( r^2+{r'}^2 \right)^{3/2}} \; \left( \hat{\vr} \cdot \hat{\vr}\,' \right) + \frac{3}{2} \frac{r^2\,{r'}^2}{\left( r^2+{r'}^2 \right)^{5/2}} \; \left( \hat{\vr} \cdot \hat{\vr}\,' \right)^2 \\
{} &+ \frac{5}{2} \frac{r^3\,{r'}^3}{\left( r^2+{r'}^2 \right)^{7/2}} \; \left( \hat{\vr} \cdot \hat{\vr}\,' \right)^3 + \frac{35}{8} \frac{r^4\,{r'}^4}{\left( r^2+{r'}^4 \right)^{9/2}} \; \left( \hat{\vr} \cdot \hat{\vr}\,' \right)^4 + \ldots \nonumber \\
&( \hat{\vr} = \frac{\vr}{||\vr||},\ \hat{\vr}\,' = \frac{\vr\,'}{||\vr\,'||} \quad \leadsto ||\hat{\vr}|| = ||\hat{\vr}\,'|| = 1) \;. \nonumber
\end{align}
However, this expansion is not directly introduced in the integral of equation (\ref{eq:V.15}) but we first define the properly \emph{anisotropic part} $\bgAan\rt$ of $\bgAe\rt$ (\ref{eq:V.15}) through
\begin{equation}
\label{eq:V.18}
\bgAan\rt \doteqdot \bgAe\rt - \bAe(r) \;,
\end{equation}
so that the spherically symmetric approximation $\bAe(r)$ actually appears as the expected leading term of the proposed expansion:
\begin{equation}
\label{eq:V.19}
\bgAe\rt = \bAe(r) + \bgAan\rt \;.
\end{equation}
As a consistency requirement, the former averaging postulate (\ref{eq:V.16}) entails now that the average value of the anisotropic part $\bgAan(r, \vartheta)$ must vanish
\begin{equation}
\label{eq:V.20}
\int \frac{d\Omega}{4\pi}\ \bgAan(r, \vartheta) = 0 \;.
\end{equation}

But once the anisotropic part of the gauge potential is precisely defined now, its power
series expansion can be easily carried through. Here, the point of departure is the
corresponding integral representation of $\bgAan(r, \vartheta)$, which is obtained by
substituting both integral representations (\ref{eq:IV.26}) and (\ref{eq:V.15}) into
(\ref{eq:V.18}) yielding
\begin{equation}
\label{eq:V.21}
\bgAan\rt = \alpha_s \int d^3\vr\,'\ \frac{1}{||\vr - \vr\,'||} \left( \bgkn(\vartheta') - \frac{1}{4\pi} \right) \cdot \bgkn(r') \;,
\end{equation}
and if the expansion (\ref{eq:V.17}) is inserted herein one finds emerging a series expansion of the following product form:
\begin{align}
\label{eq:V.22}
\bgAan\rt &= \bgA^{\mathsf{I}}(\vartheta) \cdot \bgA^{\mathsf{I}}(r) + \bgA^{\mathsf{II}}(\vartheta) \cdot \bgA^{\mathsf{II}}(r) + \bgAiii(\vartheta) \cdot \bgAiii(r) \\
{} &+ \bgA^{\mathsf{IV}}(\vartheta) \cdot \bgA^{\mathsf{IV}}(r) + \bgAv(\vartheta) \cdot \bgAv(r) + \ldots \nonumber
\end{align}
Here, the angular factors appear first in integral form, and the lowest orders may easily be calculated exactly:
\begin{subequations}
\begin{align}
\label{eq:V.23a}
\bgA^{\mathsf{I}}(\vartheta) &\doteqdot \int d\Omega'\;\left( \bgkn(\vartheta') - \frac{1}{4\pi} \right) = 0 \\
\label{eq:V.23b}
\bgA^{\mathsf{II}}(\vartheta) &\doteqdot \int d\Omega'\;\left( \bgkn(\vartheta') - \frac{1}{4\pi} \right) \left( \hat{\vr} \cdot \hat{\vr}\,' \right) = 0 \\
&( d\Omega' \doteqdot d\vartheta'\,d\phi'\,\sin\vartheta' ) \;. \nonumber
\end{align}
\end{subequations}
The importance of these lowest-order results lies in the circumstance that they are valid for \emph{any} angular distribution $\bgkn(\vartheta)$; namely the first result (\ref{eq:V.23a}) is a consequence of the separate angular normalization condition (\ref{eq:IV.24a}) and the second result (\ref{eq:V.23b}) is an implication of the skew symmetry of the integrand with respect to the reflection of the two-sphere at the equatorial two-plane ($\vartheta = \frac{\pi}{2}$). Moreover by the latter argument, all the angular factors of odd order are trivial, i.\,e.
\begin{gather}
\label{eq:V.24}
\int d\Omega' \left( \bgkn(\vartheta') - \frac{1}{4\pi} \right) \left( \hat{\vr} \cdot \hat{\vr}\,' \right)^{2n+1} = 0 \\
( n = 0, 1, 2, 3, \ldots ) \;. \nonumber
\end{gather}

Consequently, the first non-trivial angular factor is $\bgAiii(\vartheta)$, followed by $\bgAv(\vartheta)$:
\begin{subequations}
\begin{xalignat}{2}
\label{eq:V.25a}
\bgAiii(\vartheta) &\doteqdot \frac{3}{2} \int d\Omega' \left( \bgkn(\vartheta') -
  \frac{1}{4\pi} \right) \left( \hat{\vr} \cdot \hat{\vr}\,' \right)^2\ , && \text{quadrupole} \\
\label{eq:V.25b}
\bgAv(\vartheta) &\doteqdot \frac{35}{8} \int d\Omega' \left( \bgkn(\vartheta') -
  \frac{1}{4\pi} \right) \left( \hat{\vr} \cdot \hat{\vr}\,' \right)^4 \ , && \text{octupole}~\cite{33}\ .
\end{xalignat}
\end{subequations}
However, it should be evident that the precise value of these integrals will depend on the
actual density $\bgkn(\vartheta)$; for instance, any member of a dichotomic pair of
ortho-states (due to the same quantum number $\jO$) may have its own angular factor
$\bgAiii(\vartheta)$ etc. being different from that one due to its $\jO$-partner. But a
common property of all angular factors $\bgAiii(\vartheta)$ etc. is that their average on
the two-sphere is zero
\begin{equation}
\label{eq:V.26}
\int \frac{d\Omega}{4\pi}\ \bgAiii(\vartheta) = \int \frac{d\Omega}{4\pi}\ \bgAv(\vartheta) = \ldots = 0 \;,
\end{equation}
which in turn implies that the anisotropic correction $\bgAan\rt$ (\ref{eq:V.21}) has also
zero average over the two-sphere, cf. (\ref{eq:V.20}).

For the sake of brevity, we are satisfied for the moment with a treatment of only the
lowest-order case (\ref{eq:V.25a}) for $\jO = 2,3,4$ and $\bjz = \pm 1$. But even for this
restriction to only a few values of $\jO$ we can observe the \emph{ortho-dimorphism}, see
the subsequent table which transcribes the densities $\bgkn(\vartheta)$ to the
corresponding angular factors $\bgAiii(\vartheta)$ according to the prescription
(\ref{eq:V.25a}).

\vskip 0.5cm
\begin{center}
\label{table5}
\begin{tabular}{|c|c||c|c|}
\hline
$\jO$ & $\bjz$ & $\bgkn(\vartheta)$ (\ref{eq:IV.20}) & $\bgAiii(\vartheta)$ (\ref{eq:V.25a}) \\
\hline\hline
\multirow{2}{*}{$2$} & \multirow{2}{*}{$\pm 1$} & $\frac{\sin\vartheta}{\pi^2}$ & $-\frac{3}{16}\left[ \cos^2\vartheta - \frac{1}{3} \right]$ \\
\cline{3-4}
 & & $\frac{\sin\vartheta}{3\pi^2} \cdot \left( 9 - 8\sin^2\vartheta \right)$ & $\ \ \frac{3}{16}\left[ \cos^2\vartheta - \frac{1}{3} \right]$ \\
\specialrule{1.2pt}{0pt}{0pt}
\multirow{2}{*}{$3$} & \multirow{2}{*}{$\pm 1$} & $\frac{3}{\pi^2}\sin\vartheta \left( 1 - \frac{8}{9} \sin^2\vartheta \right)$ & $\ \ \frac{3}{16}\left[ \cos^2\vartheta - \frac{1}{3} \right]$ \\
\cline{3-4}
 & & $\frac{6}{\pi^2} \sin\vartheta \left( 1 - \frac{20}{9}\sin^2\vartheta + \frac{4}{3}\sin^4\vartheta \right)$ & $\ \ \frac{9}{32}\left[ \cos^2\vartheta - \frac{1}{3} \right]$ \\
\specialrule{1.2pt}{0pt}{0pt}
\multirow{2}{*}{$4$} & \multirow{2}{*}{$\pm 1$} & $\frac{10}{3\pi^2}\sin\vartheta \left( \frac{9}{5} - 4\sin^2\vartheta + \frac{12}{5} \sin^4\vartheta \right)$ & $\ \ \frac{9}{32} \left[ \cos^2\vartheta - \frac{1}{3} \right]$ \\
\cline{3-4}
 & & $\frac{10}{\pi^2}\sin\vartheta \left( 1 - 4\sin^2\vartheta + \frac{28}{5} \sin^4\vartheta - \frac{64}{25}\sin^6\vartheta \right)$ &$\ \ \frac{51}{160} \left[ \cos^2\vartheta - \frac{1}{3} \right]$ \\
\hline
\end{tabular}
\end{center}
\vskip 0.5cm

This table demonstrates some striking features of the quadrupole correction term
$\bgAiii(\vartheta)$. First, the quadrupole corrections $\bgAiii(\vartheta)$
(\ref{eq:V.25a}) for both ortho-densities $\bgkn(\vartheta)$ due to $\jO = 2$ differ
merely in sign. This circumstance has its consequences as far as the energy (being
concentrated in the anisotropic field configuration) is concerned, see below for the
discussion of the corresponding degeneracy. But in order to see here somewhat more clearly
the origin of this strange effect, which partly (i.\,e. from the energetic point of view)
suppresses the ortho-dimorphism just elaborated, one reconsiders the original definition
(\ref{eq:V.25a}) of $\bgAiii(\vartheta)$ and observes also the separate angular
normalization (\ref{eq:IV.24a}) of the density $\bgkn(\vartheta)$, together with the
specific angular structure of the scalar product of unit vectors $\hat{\vr},
\hat{\vr}\,'$:
\begin{align}
\label{eq:V.27}
\left(\hat{\vr} \cdot \hat{\vr}\,'\right)^2 &= \sin^2\vartheta \sin^2\vartheta' \cos^2(\phi - \phi') + \cos^2\vartheta \cos^2\vartheta' \\
&\quad {} + 2\sin\vartheta \sin\vartheta' \cos\vartheta \cos\vartheta' \cos(\phi - \phi') \nonumber \;.
\end{align}
Indeed, these simple mathematical elements are sufficient in order to deduce, with a little bit algebra, the following general form of $\bgAiii(\vartheta)$:
\begin{equation}
\label{eq:V.28}
\bgAiii(\vartheta) = -\frac{3}{4} \left( 1 - 3 \cdot \Kb^{\mathsf{III}} \right) \left[ \cos^2\vartheta - \frac{1}{3} \right] \;,
\end{equation}
where the integral $\Kb^{\mathsf{III}}$ is defined by
\begin{equation}
\label{eq:V.29}
\Kb^{\mathsf{III}} \doteqdot \int d\Omega' \cos^2\vartheta'\ \bgkn(\vartheta') = 2\pi \int d\vartheta' \sin\vartheta' \cos^2\vartheta' \cdot \bgkn(\vartheta') \;.
\end{equation}
This result (\ref{eq:V.28}) explains why both ortho-configurations due to $\jO = 2$ and
$\bjz = \pm 1$ share the same angular dependence, apart from sign. The value of this
integral $\Kb^{\mathsf{III}}$ (\ref{eq:V.29}) on both densities $\bgkn(\vartheta)$ due to
$\jO =2$ (see the table of densities $\bgkn(\vartheta)$ on p.~\pageref{table5} yields now
\begin{equation}
\label{eq:V.30}
\Kb^{\mathsf{III}} = \begin{cases}
\ \,\frac{\ds 1}{\ds 4}\!\!\!&;\quad \text{first density} \quad, \quad \bgkn(\vartheta) = \frac{1}{\pi^2}\sin\vartheta \\
\ \frac{\ds 5}{\ds 12}\!\!\!&;\quad \text{second density} \quad,\ \bgkn(\vartheta) = \frac{1}{3\pi^2}\sin\vartheta \cdot \left( 9 - 8\sin^2\vartheta \right) \;.
\end{cases}
\end{equation}
Inserting now these two possible values of the integral $\Kb^{\mathsf{III}}$ into the
general relation (\ref{eq:V.28}) yields just the two quadrupole corrections
$\bgAiii(\vartheta)$ for $\jO = 2$ with their different signs, as displayed by the table
on p.~\pageref{table5}. As we shall readily see, this difference in sign of both versions
$\bgAiii(\vartheta)$ is \emph{not} sufficient in order to equip both corresponding
solutions of the eigenvalue problem with \emph{different} energies
$\EE^{\{3\}}_{\mathcal{O}}$. Thus the result will be that the lowest approximation order
(with respect to the anisotropy corrections) \emph{cannot induce an electrostatic energy
  difference for the dimorphic pair of states due to $\nO=3,\jO = 2,\bjz=\pm 1$}. (See
also some further comments on this in \textbf{App.D}).

However, as is clearly demonstrated by the lower half of that table on
p.~\pageref{table5}, this result does hold only for the non-relativistic state due to $\jO
= 2$ of ortho-positronium. For the excited ortho-states ($\jO = 3$ and $\jO = 4$), the
angular densities~$\gklo{b}{k}_0(\vartheta)$ and therefore also the quadrupole corrections
$\bgAiii(\vartheta)$ of both dimorphic partners are seen to be distinctly different from
each other (\textbf{App.D}) which then will entail also different electrostatic
interaction energies. Of course, this energetic degeneracy of the dimorphic ortho-states
($\jO = 2,\bjz=\pm 1$) does refer exclusively to the \emph{first} anisotropy correction as
expressed by $\bgAiii(\vartheta)$ (\ref{eq:V.25a})! The next higher approximation order
$\bgAv(\vartheta)$ (\ref{eq:V.25b}) will surely eliminate this first-order
degeneracy. Indeed, this expectation can easily be verified by inserting both
ortho-densities $\bgkn(\vartheta)$ due to $\jO = 2$ (see table on p.~\pageref{table5}) into the
general prescription (\ref{eq:V.25b}) for the quadrupole correction $\bgAv(\vartheta)$ in
order to yield the following results:
\begin{subequations}
\begin{align}
\label{eq:V.31a}
\bgkn(\vartheta) = \frac{\sin\vartheta}{\pi^2} &\quad \Rightarrow \quad \bgAv(\vartheta) = \frac{35}{256} \left\{ \frac{11}{10} - 3\cos^2\vartheta - \frac{1}{2}\cos^4\vartheta \right\} \\[2em]
\label{eq:V.31b}
\bgkn(\vartheta) = \frac{\sin\vartheta}{3\pi^2} &\cdot \left( \sin^2\vartheta + 9\cos^2\vartheta \right) \nonumber \\
&\quad\Rightarrow \quad \bgAv(\vartheta) = \frac{7}{256} \left\{ -7 + 30\cos^2\vartheta - 15\cos^4\vartheta \right\} \\
&\qquad\qquad\qquad\boldsymbol{(\jO = 2,\bjz=\pm 1 )} \nonumber \;.
\end{align}
\end{subequations}
Obviously, this second approximation order $\bgAv(\vartheta)$ displays no common angular
dependence for the dimorphic pair as is the case with the first approximation order
$\bgAiii(\vartheta)$ (\ref{eq:V.28})--(\ref{eq:V.30}), but nevertheless the average of all
multipole corrections $\bgAiii(\vartheta)$, $\bgAv(\vartheta)$ over the two-sphere is zero,
as required by equation (\ref{eq:V.26}). However, in contrast to their trivial mean
values, the contribution of these anisotropy corrections to the binding energy is readily
seen to be non-trivial.

\section{Energy of Matter and Gauge Fields}
\indent

It should be a matter of course that different interaction potentials $\bgAe\rt$ for both
positronium constituents will entail different binding energies
$\EEOj$. This is surely a reasonable assumption and may be expected
to hold also for the approximate solutions of the eigenvalue problem to be discussed
subsequently. The corresponding approximation method is the same as has been applied in
the precedent paper~\cite{2}, where the RST energy functional $\tekru{\mathbb{E}}{\Omega}$
is extremalized on a suitably selected set of trial configurations.

The trial amplitudes $\tO(r)$ are postulated to be of adequate form and include some
variational parameters (here $\beta, \nu$):
\begin{subequations}
\begin{align}
\label{eq:VI.1a}
\tO(r) &= \Omega_*\,r^\nu\e^{-\beta r} \\
\label{eq:VI.1b}
\Omega_*^2 &= \frac{(2\beta)^{2\nu + 2}}{\Gamma(2\nu + 2)} \;.
\end{align}
\end{subequations}
Obviously, after separating off the angular dependence (see
(\ref{eq:IV.5a})--(\ref{eq:IV.5b})), the selected class of trial amplitudes $\tO(r)$ is
here of the spherically symmetric form which itself is surely an approximation to the
exact solutions; but for the purposes of the present paper we will be satisfied with such
a spherically symmetric form of the wave amplitudes $\tO(r)$.

In contrast to this, the interaction potential $\bAn(\vr)$ is admitted to be anisotropic;
however, it is assumed to be the sum of a spherically symmetric part $\bAe(r)$ plus an
anisotropic perturbation $\bgAan\rt$, see equation (\ref{eq:V.19}). Furthermore, these two
constituents of the interaction potential $\bgAe\rt$ are assumed now to be kinematically
independent, where this independence of the anisotropic part $\bgAan\rt$ refers
exclusively to the \emph{radial} auxiliary potentials $\bgAiii(r), \bgAv(r)$ etc., cf. the
series expansion (\ref{eq:V.22}); whereas their angular counterparts $\bgAiii(\vartheta),
\bgAv(\vartheta)$ etc. are fixed through the integral relations (\ref{eq:V.25a}),
(\ref{eq:V.25b}) etc. Thus our selected trial configurations consist of a set of
spherically symmetric fields $\left\{ \tO(r); \bAe(r), \bgAiii(r), \bgAv(r), \ldots
\right\}$ where only the trial amplitudes $\tO(r)$ are of a pre-specified form
(cf. (\ref{eq:VI.1a})--(\ref{eq:VI.1b})) with undetermined variational parameters $\beta$
and $\nu$. In contrast to this, the spherically symmetric potential $\bAe(r)$ and the
radial auxiliary potentials $\bgAiii(r), \bgAv(r)$ etc. are first left unspecified and
must afterwards be determined from their extremal equations, i.\,e. from extremalization
of the energy functional $\tEEO$ to be understood as the anisotropic generalization of the
spherically symmetric version (\ref{eq:V.2}). The extremalization of $\tEEO$ with respect
to the trial amplitude $\tO(r)$ occurs then through extremalization with respect to the
variational parameters $\beta$ and $\nu$ which themselves are entering also the potentials
$\bAe(r); \bgAiii(r), \bgAv(r)$ etc., namely via the solutions of the extremal equations
for those potentials whose source is determined by just the square of the amplitude
$\tO(r)$.

Thus, summarizing the approximation procedure, we will end up with a certain \emph{energy
  function} $\EEgivbn$ of the two variational parameters $\beta$ and $\nu$; and the
stationary points of this function, being determined by the requirements
\begin{subequations}
\begin{align}
\label{eq:VI.2a}
\frac{\partial \EEgivbn}{\partial\beta} &= 0 \\
\label{eq:VI.2b}
\frac{\partial \EEgivbn}{\partial\nu} &= 0 \;,
\end{align}
\end{subequations}
yield the equilibrium values $\beta_*, \nu_*$ so that the wanted energies appear as the
values $\EEgiv(\beta_*, \nu_*)$ of that energy function $\EEgivbn$ on these special
values $\beta_*, \nu_*$.

According to the sum structure of the ortho-funtional $\tEEO$ (\ref{eq:V.2}), its value
$\EEO(\beta, \nu)$ on our selected trial configurations will appear as the sum of the
matter energy $\Ekin(\beta, \nu)$ and the anisotropic gauge field energy $\egER(\beta,
\nu)$ plus the constraint terms:
\begin{align}
\label{eq:VI.3}
\EEO(\beta, \nu) &= 2\Ekin(\beta, \nu) + \egER(\beta, \nu) + 2\lambda_s \cdot \tNNO(\beta, \nu) + \lGe \cdot \tNNGeg(\beta, \nu) \\
&\equiv \EEgivbn + 2\lambda_s \cdot \tNNO(\beta, \nu) + \lGe \cdot \tNNGeg(\beta,  \nu) \nonumber \;.
\end{align}
Whether or not the Poisson constraint term $\tNNGer$ (\ref{eq:V.8}) in its anisotropic form $\tNNGeg$, i.\,e.
\begin{equation}
\label{eq:VI.4}
\tNNGeg \doteqdot \EReg - \tMMeg \mathrm{c^2} \;,
\end{equation}
will contribute in a non-trivial way must be discussed separately (see below). In contrast to this, the other constraint term refers to the normalization condition for the non-relativistic amplitude $\tO(r)$
\begin{equation}
\label{eq:VI.5}
\tNNO \doteqdot \int\limits_0^\infty dr\,r\ \tO^2(r) - 1
\end{equation}
and can always be put to zero by simply admitting only those trial amplitudes $\tO(r)$
which are normalized to unity ($\leadsto \tNNO = 0$). However, from formal reasons it is
necessary to first include the constraint terms in the energy functional $\tEEO$ so that
\emph{both} the matter equation (\ref{eq:IV.17}) \emph{and} the gauge field equation
(\ref{eq:IV.18}) do appear as the extremal equations due to that ortho-functional $\tEEO$
(\ref{eq:V.2}) whose matter part appears now as
\begin{equation}
\label{eq:VI.6}
\DtEEO = 2\Ekin + 2 \lambda_s \cdot \tNNO
\end{equation}
and similarly its gauge field part as
\begin{equation}
\label{eq:VI.7}
\GtEEO \Rightarrow \etEEO = \EReg + \lGe \cdot \tNNGeg \;.
\end{equation}

\begin{center}
  \emph{\textbf{1.\ Energy of the Matter Fields}}
\end{center}

First, we may turn to the kinetic energy which for the two-particle system is twice the
one-particle energy $\Ekin$~\cite{34}
\begin{equation}
\label{eq:VI.8}
\Ekin = \frac{\hbar^2}{2M} \int\limits_0^\infty dr\,r\ \left\{ \left( \frac{d\tO(r)}{dr} \right)^2 + \frac{\jO^2}{r^2} \cdot \tO^2(r) \right\} \;.
\end{equation}
Here, the centrifugal term ($\sim \jO^2$) is induced by the former product ansatz
(\ref{eq:IV.5a})--(\ref{eq:IV.5b}) and evidently does account for the total angular
momentum of the considered ortho-state. Substituting herein the selected form
(\ref{eq:VI.1a})--(\ref{eq:VI.1b}) of the trial amplitude $\tO(r)$ yields for the
one-particle kinetic energy $\Ekin$ (\ref{eq:VI.8})
\begin{subequations}
\begin{align}
\label{eq:VI.9a}
\Ekin(\beta, \nu) &= \frac{\textrm{e}^2}{2a_B} (2a_B\beta)^2 \cdot \ekin(\nu) \\
\label{eq:VI.9b}
\ekin(\nu) &\doteqdot \frac{1}{2\nu + 1} \left( \frac{1}{4} + \frac{\jO^2}{2\nu} \right) \;,
\end{align}
\end{subequations}
see equation (VI.154b) of ref.~\cite{34}. This is formally the same kinetic energy function
as was used for para-positronium with merely the quantum number $\lP$ ($= j_\wp$) for
para-positronium being replaced now by $\jO$ for ortho-positronium.

\begin{center}
  \emph{\textbf{2.\ Energy of the Gauge Fields}}
\end{center}

Next, the gauge field part $\GtEEO$ is to be specified in terms of the three gauge field
constituents $\bAe(r), \bgAiii(r), \bgAv(r)$ which we treat as independent constituents of
the gauge field subsystem. Since, on account of the specific form of the Poisson
constraint $\tNNGeg$ (\ref{eq:VI.4}), the desired gauge field part $\etEEO$
(\ref{eq:VI.7}) consists essentially of the gauge field energy $\EReg$ and its mass
equivalent $\tMMeg \mathrm{c^2}$ it is sufficient to explicitly display these two objects
in terms of the newly introduced gauge fields $\bAe(r), \bgAiii(r), \bgAv(r)$.

Turning here first to the electrostatic gauge field energy $\ERe$, the spherically symmetric relation (\ref{eq:V.7}) is to be generalized for the presently considered anisotropic situation to
\begin{equation}
\label{eq:VI.10}
\vec{E}_b(\vr) = -\vec{\nabla}\ \bgAe\rt = -\vec{e}_r\;\frac{\partial\,\bgAe\rt}{\partial r} - \vec{e}_\vartheta\;\frac{1}{r}\,\frac{\partial\,\bgAe\rt}{\partial \vartheta}
\end{equation}
so that the gauge field energy (\ref{eq:V.4}) reappears as
\begin{equation}
\label{eq:VI.11}
\ERe\ \Rightarrow\ \EReg = -\frac{\hbar\mathrm{c}}{4\pi\alpha_s} \int d^3\vr\ \left\{ \left( \frac{\partial\,\bgAe\rt}{\partial r} \right)^2 + \frac{1}{r^2} \left( \frac{\partial\,\bgAe\rt}{\partial \vartheta} \right)^2 \right\} \;.
\end{equation}
But here we have to observe the splitting of the anisotropic potential $\bgAe\rt$ into the sum of the isotropic part $\bAe(r)$ and the anisotropic perturbation $\bgAan\rt$, cf. equation (\ref{eq:V.19}); and this must then induce a similar splitting of the gauge field energy $\EReg$ (\ref{eq:VI.11}) into two parts, i.\,e. we put
\begin{equation}
\label{eq:VI.12}
\EReg = \ERee + \Eegan \;.
\end{equation}
Here, the ``isotropic'' part $\ERee$ has already been specified in terms of $\bAe(r)$ by equation (\ref{eq:V.7}) and the ``anisotropic'' energy $\Eegan$ is found to be of the following form
\begin{equation}
\label{eq:VI.13}
\Eegan = -\frac{\hbar\mathrm{c}}{4\pi\alpha_s} \int d^3\vr\ \left\{ \left( \frac{\partial\,\bgAan\rt}{\partial r} \right)^2 + \frac{1}{r^2} \left( \frac{\partial\,\bgAan\rt}{\partial \vartheta} \right)^2 \right\} \;.
\end{equation}
(See the precedent paper~\cite{1} for the reason why there is no mixed term involving both
parts $\bAe(r)$ and $\bgAan\rt$).

For further specifying down this anisotropic part $\Eegan$ (\ref{eq:VI.13}) one observes
the decomposition (\ref{eq:V.22}) of the anisotropic correction $\bgAan\rt$ into the sum
of products of an angular and radial auxiliary potential; here we retain only the first
two terms of the series expansion (\ref{eq:V.22}):
\begin{equation}
\label{eq:VI.14}
\bgAe\rt = \bgAiii(r) \cdot \bgAiii(\vartheta) + \bgAv(\vartheta) \cdot \bgAv(r) \;.
\end{equation}
This arrangement lets then reappear the ``anisotropic'' gauge field energy $\Eegan$ (\ref{eq:VI.13}) in the following final form
\begin{align}
\label{eq:VI.15}
\Eegan = -\frac{\hbar c}{\as}\int\limits_0^\infty dr\,r^2\,\Bigg\{&\left[ \rklo{b}{e}_\mathsf{III}\cdot \left( \frac{d\,\bgAiii(r)}{dr} \right)^2 + \rklo{b}{f}_\mathsf{III} \cdot \left( \frac{\bgAiii(r)}{r} \right)^2 \right] \\
{} + &\left[ \rklo{b}{e}_\mathsf{V} \cdot \left( \frac{d\,\bgAv(r)}{dr} \right)^2 + \rklo{b}{f}_\mathsf{V} \cdot \left( \frac{\bgAv(r)}{r} \right)^2 \right] \nonumber \\
{} + 2 & \left[ \rklo{b}{e}_\mathsf{IV}\,\frac{d\,\bgAiii(r)}{dr} \cdot \frac{d\,\bgAv(r)}{dr} + \rklo{b}{f}_\mathsf{IV} \frac{\bgAiii(r)}{r} \cdot \frac{\bgAv(r)}{r}  \right] \Bigg\} \nonumber \;.
\end{align}
The angular coefficients $e_\mathbb{N}, f_\mathbb{N}\ (\mathbb{N}=\mathsf{III,IV,V})$ do arise here
by carrying out the corresponding angular integrations in the original energy integral
(\ref{eq:VI.13}) which yields
\begin{subequations}
\begin{align}
\label{eq:VI.16a}
\rklo{b}{e}_\mathsf{III} &= \int \frac{d\Omega}{4\pi} \bigg( \bgAiii(\vartheta) \bigg)^2\;,\quad\quad\ \rklo{b}{f}_\mathsf{III} = \int \frac{d\Omega}{4\pi} \left( \frac{d\,\bgAiii(\vartheta)}{d \vartheta} \right)^2 \\
\label{eq:VI.16b}
\rklo{b}{e}_\mathsf{IV} &= \int \frac{d\Omega}{4\pi} \bgAiii(\vartheta) \cdot \bgAv(\vartheta)\;,\  \rklo{b}{f}_\mathsf{IV} = \int\frac{d\Omega}{4\pi} \left( \frac{d\,\bgAiii(\vartheta)}{d \vartheta} \right) \cdot \left( \frac{d\,\bgAv(\vartheta)}{d \vartheta} \right) \\
\label{eq:VI.16c}
\rklo{b}{e}_\mathsf{V} &= \int\frac{d\Omega}{4\pi} \bigg( \bgAv(\vartheta) \bigg)^2\;,\ \quad\quad\, \rklo{b}{f}_\mathsf{V} = \int\frac{d\Omega}{4\pi} \left( \frac{d\,\bgAv(\vartheta)}{d \vartheta} \right)^2 \;.
\end{align}display\end{subequations}
Since the angular functions $\bgAiii(\vartheta), \bgAv(\vartheta)$ are known and depend
on both quantum numbers $\jO$ and $\bjz$ but depend also on the dimorphic type, the
angular coefficients $e_\mathbb{N},f_\mathbb{N}\ (\mathbb{N}=\mathsf{III,IV,V})$ are
functionals of the considered ambiguous quantum state $\left( \jO, \bjz \right)$ but are
otherwise to be treated as fixed numerical constants. We will readily see that through
this arrangement the \emph{principle of minimal energy} becomes reduced to a purely radial
variational problem. The numerical values of the constants are displayed by the subsequent
table for the quadrupole approximation $(\sim \bgAiii(\vartheta))$.

\vskip 0.5cm
\begin{center}
\label{table6}
\begin{tabular}{|c|c||c|c|c|c|}
%\specialrule{1.2pt}{0pt}{0pt}
\hline
 & & & \multicolumn{2}{c|}{(\ref{eq:VI.16a})} & (\ref{eq:VI.22a}) \\[-0.3em]
$\jO$ & $\bjz$ & $\bgAiii(\vartheta)$ & $\rklo{b}{e}_\mathsf{III}$ & $\rklo{b}{f}_\mathsf{III}$ & $\rklo{b}{m}_\mathsf{III}$ \\
\hline\hline
\multirow{2}{*}{$2$} & \multirow{2}{*}{$\pm 1$} & $-\frac{3}{16}\,\left[ \cos^2\vartheta - \frac{1}{3} \right]$ & $\frac{1}{320}$ & $\frac{3}{160}$ & $\frac{1}{64}$ \\
\cline{3-6}
 &  & $\frac{3}{16}\,\left[ \cos^2\vartheta - \frac{1}{3} \right]$ & $\frac{1}{320}$ & $\frac{3}{160}$ & $\frac{1}{64}$ \\
\specialrule{1.2pt}{0pt}{0pt}
\multirow{2}{*}{$3$} & \multirow{2}{*}{$\pm 1$} & $\frac{3}{16}\,\left[ \cos^2\vartheta - \frac{1}{3} \right]$ & $\frac{1}{320}$ & $\frac{3}{160}$ & $\frac{1}{64}$ \\
\cline{3-6}
 &  & $\frac{9}{32}\,\left[ \cos^2\vartheta - \frac{1}{3} \right]$ & $\frac{9}{1280}$ & $\frac{27}{640}$ & $\frac{9}{256}$ \\
\specialrule{1.2pt}{0pt}{0pt}
\multirow{2}{*}{$4$} & \multirow{2}{*}{$\pm 1$} & $\frac{9}{32}\,\left[ \cos^2\vartheta - \frac{1}{3} \right]$ & $\frac{9}{1280}$ & $\frac{27}{640}$ & $\frac{9}{256}$ \\
\cline{3-6}
 &  & $\frac{51}{160}\,\left[ \cos^2\vartheta - \frac{1}{3} \right]$ & $\frac{1}{5}\,\left( \frac{17}{80} \right)^2$ & $\frac{1}{30}\,\left( \frac{51}{40} \right)^2$ & $\left( \frac{17}{80} \right)^2$ \\
\hline
\end{tabular}
\end{center}
\vskip 0.5cm

The striking feature of these numerical values refers to the fact that the ratios $\frac{f_3}{e_3}$ and $\frac{m_3}{e_3}$ are all the same (see \textbf{App.A})
\begin{subequations}
\begin{align}
\label{eq:VI.17a}
\frac{\rklo{b}{f}_\mathsf{III}}{\rklo{b}{e}_\mathsf{III}} &= 6 \\
\label{eq:VI.17b}
\frac{\rklo{b}{m}_\mathsf{III}}{\rklo{b}{e}_\mathsf{III}} &= 5 \;.
\end{align}
\end{subequations}
The meaning of this will readily become obvious when considering below the quadrupole equation for the radial potential correction $\bgAiii(r)$.

\begin{center}
  \emph{\textbf{3.\ Mass Equivalent}}
\end{center}

Similar arguments as for the anisotropy energy $\Eean$ (\ref{eq:VI.15}) do also apply to
the mass equivalent $\Mee \mathrm{c^2}$ (\ref{eq:V.9}). Indeed, inserting therein the
series expansion (\ref{eq:V.22}) (up to the \textsf{V}-th order) in combination with the
factorized density $\bkn$ (\ref{eq:IV.20})--(\ref{eq:IV.21}) and the separate angular
normalization (\ref{eq:IV.24a}) lets emerge the (non-relativistic) mass equivalent in a
form analogous to the ``anisotropic'' gauge field energy $\EReg$ (\ref{eq:VI.12}):
\begin{equation}
\label{eq:VI.18}
\Mee \mathrm{c^2}\ \Rightarrow\ \tMMeg \mathrm{c^2} = \tMMee \mathrm{c^2} + \tMMegan \mathrm{c^2} \;.
\end{equation}
This says that the general mass equivalent splits up into the sum of the spherically
symmetric part $\tMMee \mathrm{c^2}$ (\ref{eq:V.12}) and the anisotropic correction
$\tMMegan \mathrm{c^2}$, where the latter part appears as the sum of the \textsf{III}-rd
and \textsf{V}-th order, i.\,e.
\begin{equation}
\label{eq:VI.19}
\tMMegan \mathrm{c^2} = \tMMeg_{\ \mathsf{III}} \mathrm{c^2} + \tMMeg_{\ \mathsf{V}} \mathrm{c^2} \;.
\end{equation}
This yields finally the following decomposition
\begin{equation}
\label{eq:VI.20}
\tMMeg \mathrm{c^2} = \tMMee \mathrm{c^2} + \tMMeg_{\ \mathsf{III}} \mathrm{c^2} + \tMMeg_{\ \mathsf{V}} \mathrm{c^2}
\end{equation}
where the spherically symmetric part $\tMMee \mathrm{c^2}$ has already been specified by equation (\ref{eq:V.12}) and the anisotropy corrections are given by
\begin{subequations}
\begin{align}
\label{eq:VI.21a}
\tMMeg_{\ \mathsf{III}} \mathrm{c^2} &= -\hbar\textrm{c}\,\rklo{b}{m}_\mathsf{III} \cdot \int\limits_0^\infty dr\,r\ \bgAiii(r) \cdot \tO^2(r) \\
\label{eq:VI.21b}
\tMMeg_{\ \mathsf{V}} \mathrm{c^2} &= -\hbar\textrm{c}\,\rklo{b}{m}_\mathsf{V} \cdot \int\limits_0^\infty dr\,r\ \bgAv(r) \cdot \tO^2(r) \;,
\end{align}
\end{subequations}
with the angular constants $\rklo{b}{m}_\mathsf{III}$ and $\rklo{b}{m}_\mathsf{V}$ being defined through
\begin{subequations}
\begin{align}
\label{eq:VI.22a}
\rklo{b}{m}_\mathsf{III} &\doteqdot \int d\Omega\ \bgAiii(\vartheta) \cdot \bgkn(\vartheta) \\
\label{eq:VI.22b}
\rklo{b}{m}_\mathsf{V} &\doteqdot \int d\Omega\ \bgAv(\vartheta) \cdot \bgkn(\vartheta) \;.
\end{align}
\end{subequations}
The angular coefficient $\rklo{b}{m}_\mathsf{III}$ (\ref{eq:VI.22a}) is obviously due to the formerly
considered \emph{quadrupole approximation} and some values have already been displayed in
the table on p.~\pageref{table6}.

The last point concerning the gauge field subsystem refers to the ``anisotropic'' Poisson
constraint $\tNNGeg$ (\ref{eq:VI.4}). Since both constituents $\EReg$ and $\tMMeg
\mathrm{c^2}$ have been split up into sums of the isotropic and anisotropic parts,
cf. (\ref{eq:VI.12}) and (\ref{eq:VI.18}), the constraint itself must undergo a similar
splitting:
\begin{equation}
\label{eq:VI.23}
\tNNGeg = \tNNGee + \tNNegan \;,
\end{equation}
where the isotropic part $\tNNGee$ is of course built up by the isotropic parts of the gauge field energy and its mass equivalent, cf. (\ref{eq:V.14}), and analogously for the anisotropic part $\tNNegan$
\begin{equation}
\label{eq:VI.24}
\tNNegan = \Eegan - \tMMegan \mathrm{c^2} \;.
\end{equation}

Summarizing, we ultimately arrive at a splitting of the whole energy functional $\tEEO$ (\ref{eq:V.2}) not only in two but rather three constituents which correspond in a self-evident way to the matter part and the isotropic and anisotropic gauge field parts:
\begin{align}
\label{eq:VI.25}
\tEEO &= \left[ 2 \Ekin + 2\lambda_s \cdot \tNNO \right] + \left[ \ERee + \lGe \cdot \tNNGee \right] + \left[ \Eegan + \lGe \cdot \tNNegan \right] \\
&\doteqdot \DtEEO + \eetEEO + \antEEO \nonumber \\
&\equiv \DtEEO + \GtEEO \nonumber \;.
\end{align}
These three subsystems will now be adopted as three kinematically independent (but
coupled) degrees of freedom so that we can deduce the corresponding field equations by
extremalization of this functional $\tEEO$ (\ref{eq:VI.25}) on the space of selected trial
configurations $\left\{ \tO(r); \bAe(r), \bgAiii(r), \bgAv(r) \right\}$.

\begin{center}
  \emph{\textbf{4.\ Principle of Minimal Energy}}
\end{center}

Once the energy functional for the RST configurations has been set up, cf. (\ref{eq:VI.25}), one can now look for the corresponding extremal equations. For this purpose, one may derive benefit from the fact that the various terms of $\tEEO$ do not contain all four independent quantities $\left\{ \tO; \bAe, \bgAiii(r), \bgAv(r) \right\}$. Therefore the variational process for any of these field quantities does comprehend only a certain subset of all those terms constituting $\tEEO$. For instance, the extremalization of $\tEEO$ with respect to the amplitude field $\tO(r)$ does disregard the electrostatic gauge field energy $\EReg$ (\ref{eq:VI.11}) and thus takes into account only the matter part $\DEEOe $ (\ref{eq:VI.6}) and the mass equivalent $\tMMeg \mathrm{c^2}$ (\ref{eq:VI.20}). Thus the extremalization ($\dO$, say) with respect to the amplitude $\tO(r)$ looks as follows
\begin{equation}
\label{eq:VI.26}
0 \mustbe \dO\,\tEEO = \dO\,\DtEEO - \lGe \cdot \dO \left\{ \tMMee \mathrm{c^2} + \tMMeg_\mathsf{III} \mathrm{c^2} + \tMMeg_\mathsf{V} \mathrm{c^2} \right\} \;,
\end{equation}
and this yields for $\tO(r)$ the Schr\"odinger-like equation (provided we put $\lGe = -2$):
\begin{align}
\label{eq:VI.27}
- &\frac{\hbar^2}{2M} \left\{ \frac{d^2\tO(r)}{dr^2} + \frac{1}{r} \frac{d\tO(r)}{dr} \right\} + \frac{\hbar^2}{2Mr^2} \jO^2 \cdot \tO(r) \\
{} - &\hbar \mathrm{c} \left\{ \bAe(r) + \rklo{b}{m}_\mathsf{III} \cdot \bgAiii(r) + \rklo{b}{m}_\mathsf{V} \cdot \bgAv(r) \right\} \cdot \tO(r) = E_* \cdot \tO(r) \;. \nonumber
\end{align}
Obviously, this is the anisotropic generalization of the former eigenvalue equation
(\ref{eq:IV.17}) which thus is seen to refer exclusively to the spherically symmetric
approximation (\ref{eq:V.1}). If the anisotropic corrections $\bgAiii\rt$ and $\bgAv\rt$
are taken into account, their radial parts $\bgAiii(r), \bgAv(r)$ do obviously also couple
to the spherically symmetric trial amplitude $\tO(r)$, where $\rklo{b}{m}_\mathsf{III}$
and $\rklo{b}{m}_\mathsf{V}$ (\ref{eq:VI.22a})--(\ref{eq:VI.22b}) play the role of
coupling constants.

Clearly, we are not able to solve the amplitude equation (\ref{eq:VI.27}) exactly, but
instead we will resort to the previously used trial functions $\tO(r)$
(\ref{eq:VI.1a})--(\ref{eq:VI.1b}). Thus, denoting the value of the matter functional
$\DtEEO$ (\ref{eq:VI.25}) on this normalized trial amplitude by $2\Ekin(\beta, \nu)$, one
readily is led back to the former results (\ref{eq:VI.9a})--(\ref{eq:VI.9b}).

\begin{center}
  \large{\textit{Isotropic Energy Function} $\EEeivbn$}
\end{center}

Next, concerning the extremal equation for the spherically symmetric approximation $\bAe(r)$, one observes that this isotropic potential does enter the functional $\tEEO$ (\ref{eq:VI.25}) only via its second constituent $\eetEEO$ (\ref{eq:V.13}). Therefore the ``isotropic'' extremalization process ($\dn$, say) becomes active only on this ``isotropic'' constituent $\eetEEO$
\begin{equation}
\label{eq:VI.28}
\dn\,\tEEO = \dn\,\eetEEO \mustbe 0
\end{equation}
so that of course the ``isotropic'' Poisson equation (\ref{eq:IV.18}) is recovered here. But the point with our present approximation procedure is now that we do not try some (more or less reasonable) potential $\bAe(r)$, but rather we prefer to solve \emph{exactly} that isotropic Poisson equation, albeit for the situation where the amplitude $\tO(r)$ is given by our trial ansatz (\ref{eq:VI.1a})--(\ref{eq:VI.1b}). For this purpose, one recasts the Poisson equation (\ref{eq:IV.18}) in dimensionless form
\begin{equation}
\label{eq:VI.29}
\Delta_y\;\tanu(y) = -\frac{y^{2\nu-1} \cdot \e^{-y}}{\Gamma(2\nu + 2)} \equiv -\frac{1}{\Gamma(2\nu + 2)} \frac{\tO^2_\nu(y)}{y} \;,
\end{equation}
see equation (A.1) of the precedent paper~\cite{1}; here the solution can easily be found for integer $2\nu$ as
\begin{equation}
\label{eq:VI.30}
\tanu(y) = \frac{1}{y} \left( 1 - \e^{-y} \cdot \sum^{2\nu}_{n=0} \frac{2\nu + 1 - n}{2\nu + 1} \cdot \frac{y^n}{n!} \right)\;.
\end{equation}
This can be shown to be a special case of the general situation with arbitrary real values
of the variational parameter $\nu\ (\geq -\frac{1}{2})$ where the desired solution then
adopts the following form:
\begin{equation}
\label{eq:VI.31}
\tanu(y) = \frac{1}{2\nu + 1} \left( 1 - \e^{-y} \cdot \sum^{\infty}_{n=0} \frac{n}{\Gamma(2\nu + 2 + n)} y^{2\nu + n} \right) \;.
\end{equation}

But once the isotropic potential $\bAe(r)$ (or $\tanu(y)$, resp.) is known, one can
readily calculate the associated ``isotropic'' energy $\ERee$ (\ref{eq:V.7}) and its mass
equivalent $\tMMee \mathrm{c^2}$ (\ref{eq:V.12}). In the dimensionless form, both objects
appear as~\cite{2,34}
\begin{subequations}
\begin{align}
\label{eq:VI.32a}
\ERee &= -e^2\,(2\beta) \int\limits^\infty_0 dy\,y^2 \left( \frac{d\tanu(y)}{dy} \right)^2 = -\frac{e^2}{\aB} \left( 2\aB\beta \right) \cdot \epot(\nu) \\
\label{eq:VI.32b}
\tMMee \mathrm{c^2} &= -e^2\,(2\beta)\;\frac{1}{\Gamma(2\nu + 2)} \int\limits^\infty_0 dy\,y\ \tanu(y)\,\tO^2_\nu(y) = -\frac{e^2}{\aB} \left( 2\aB\beta \right) \cdot \epot(\nu)
\end{align}
\end{subequations}
with the ``isotropic'' potential function $\epot(\nu)$ being given by~\cite{1,34}
\begin{equation}
\label{eq:VI.33}
\epot(\nu) = \frac{1}{2\nu + 1} \left( 1 - \frac{1}{2^{4\nu + 2}} \cdot \sum^\infty_{n=0} \frac{n}{2^n} \cdot \frac{\Gamma(4\nu + 2 + n)}{\Gamma(2\nu + 2) \cdot \Gamma(2\nu + 2 + n)} \right) \;.
\end{equation}
Observe here the fact that the electrostatic gauge field energy $\ERee$ (\ref{eq:VI.32a}) equals its mass equivalent $\tMMee \mathrm{c^2}$ (\ref{eq:VI.32b}) so that the ``isotropic'' Poisson constraint (\ref{eq:V.14}) is zero
\begin{equation}
\label{eq:VI.34}
\tNNGee(\beta, \nu) \equiv 0 \;,
\end{equation}
and thus this constraint term disappears from the energy function $\tEEOo(\beta, \nu)$ due to (\ref{eq:VI.25}). The reason for this is that for the isotropic potential $\bAe(r)$ (or $\tanu(y)$, resp.) we use an \emph{exact} solution of the corresponding Poisson equation, e.\,g. in the dimensionless form (\ref{eq:VI.30}) and (\ref{eq:VI.31}); and for this situation it can easily be shown through integration by parts and use of the Poisson equation (\ref{eq:VI.29}) that both integrals in (\ref{eq:VI.32a})--(\ref{eq:VI.32b}) are actually identical:
\begin{equation}
\label{eq:VI.35}
\epot(\nu) \doteqdot \int\limits^\infty_0 dy\,y^2 \left( \frac{d\tanu(y)}{dy} \right)^2 = \frac{1}{\Gamma(2\nu + 2)} \int\limits^\infty_0 dy\,y\;\tanu(y)\,\tO^2_\nu(y) \;.
\end{equation}

At this point of the discussion it may be instructive to discontinue for a moment in order to get some survey of what has been attained up to now. Evidently, with both the matter part $\DtEEO$ and the ``isotropic'' gauge field part $\eetEEO$ being at hand now, one has at one's disposal the isotropic part of the energy functional $\tEEO$ (\ref{eq:VI.25}). Indeed, with the results (\ref{eq:VI.9a})--(\ref{eq:VI.9b}) for the kinetic energy $\Ekin$ and with the ``isotropic'' gauge field energy $\ERee$ (\ref{eq:VI.32a}) one possesses knowledge of the corresponding ``isotropic'' energy function $\EEeivbn$, i.\,e.
\begin{align}
\label{eq:VI.36}
\EEeivbn &= 2 \Ekin(\beta, \nu) + \ERee(\beta, \nu) \\
&= \frac{e^2}{\aB} \left\{ \left( 2\aB\beta \right)^2 \cdot \ekin(\nu) - \left( 2\aB\beta \right) \cdot \epot(\nu) \right\} \;. \nonumber
\end{align}
This result may now be exploited in order to obtain the (non-relativistic) energy spectrum of ortho-positronium in the \emph{isotropic approximation}. To this end, one subjects this ``isotropic'' function $\EEeivbn$ (\ref{eq:VI.36}) to the first one (\ref{eq:VI.2a}) of both extremalizing conditions in order to find the corresponding equilibrium value $\beta_*$ as
\begin{equation}
\label{eq:VI.37}
2 \alpha_s \beta_* = \frac{\epot(\nu)}{2 \ekin(\nu)} \;.
\end{equation}
Substituting this back into the original energy function $\EEeivbn$ yields its reduced form as
\begin{equation}
\label{eq:VI.38}
\EEivbn \Rightarrow \EEt(\nu) = - \frac{e^2}{4\aB} \cdot S_\mathcal{O}(\nu) \;,
\end{equation}
with the ``isotropic'' spectral function $S_\mathcal{O}(\nu)$ being given by
\begin{equation}
\label{eq:VI.39}
S_\mathcal{O}(\nu) = \frac{\epot^2(\nu)}{\ekin(\nu)} \;.
\end{equation}
Finally in the last step, one looks for the extremal points of this function
\begin{equation}
\label{eq:VI.40}
\left. \frac{dS_\mathcal{O}(\nu)}{d\nu} \right|_{\nu_*} = 0
\end{equation}
in order to find the equilibrium value ($\nu_*$) of the second variational parameter ($\nu$).

In this way, one gets the energy spectrum $\left\{ \EEt^{(n)} (\nu_*) \right\} \doteqdot
\left\{ \EE_\mathcal{O}^{[n]} \right\}$ of ortho-positronium in the spherically symmetric
(i.\,e. ``isotropic'') approximation. Of course, this spectrum is numerically the same as
for para-positronium, if calculated on the same level of approximation, see the table on
p.~\pageref{tableDb}. Observe however that there is an important difference between both
energy spectra. Namely, the presently considered ortho-levels are dimorphic doublets, due
to the occurence of different charge distributions $\bgkn(r, \vartheta)$ for the same
quantum numbers $j_z, \jO \left( = n_\mathcal{O} - 1 \right)$. This dimorphism of the
ortho-states will induce a certain energy difference between the members of such a doublet
which is readily to be discussed now in greater detail (the \emph{degree of degeneracy} of
the ortho-levels will be studied in an other paper).

\begin{center}
  \large{\textit{Anisotropic Energy Function} $\EEgivbn$}
\end{center}

Naturally, the ``isotropic'' energy function $\EEeivbn$ (\ref{eq:VI.36}) must undergo certain corrections if the ``anisotropic'' constituent $\anrtEEO$ (\ref{eq:VI.25})
\begin{equation}
\label{eq:VI.41}
\anrtEEO = \Eegan + \lGe \cdot \tNNegan
\end{equation}
of the total functional $\tEEO$ is taken into account. The anisotropy corrections $\bgAiii(r)$ and $\bgAv(r)$ of the interaction potential $\bgAe\rt$ are explicitly present in this anisotropic part (\ref{eq:VI.41}) so that the extremalization (denoted by $\delta_{\mathsf{III}}$ and $\delta_{\mathsf{V}}$, resp.) of the total functional $\tEEO$ with respect to those correction potentials $\bgAiii(r)$ and $\bgAv(r)$ becomes active only for that anisotropic part, i.\,e.
\begin{subequations}
\begin{align}
\label{eq:VI.42a}
\delta_{\mathsf{III}}\,\tEEO &= \delta_{\mathsf{III}}\,\anrtEEO = 0 \\
\label{eq:VI.42b}
\delta_{\mathsf{V}}\,\tEEO &= \delta_{\mathsf{V}}\,\anrtEEO = 0 \;.
\end{align}
\end{subequations}
Carrying here through both extremalization processes yields the following coupled
Poisson-like equations for the anisotropy corrections $\bgAiii(r), \bgAv(r)$:
\begin{subequations}
\begin{align}
\label{eq:VI.43a}
\rklo{b}{e}_\mathsf{III} \cdot \Delta_r\,\bgAiii(r) + \rklo{b}{e}_\mathsf{IV}  \cdot \Delta_r\,\bgAv(r) - \rklo{b}{f}_\mathsf{III} \cdot \frac{\bgAiii(r)}{r^2} - \rklo{b}{f}_\mathsf{IV} \cdot \frac{\bgAv(r)}{r^2} \\
 = - \rklo{b}{m}_\mathsf{III} \,\alpha_s \cdot \frac{\tO^2(r)}{r} \nonumber \\[0.5cm]
\label{eq:VI.43b}
\rklo{b}{e}_\mathsf{V}  \cdot \Delta_r\,\bgAv(r) + \rklo{b}{e}_\mathsf{IV} \cdot \Delta_r\,\bgAiii(r) - \rklo{b}{f}_\mathsf{V} \cdot \frac{\bgAv(r)}{r^2} - \rklo{b}{f}_\mathsf{IV} \cdot \frac{\bgAiii(r)}{r^2} \\
 = - \rklo{b}{m}_\mathsf{V}\,\alpha_s \cdot \frac{\tO^2(r)}{r} \nonumber \;.
\end{align}
\end{subequations}

If it were possible to find \emph{exact} solutions of the present coupled system (\ref{eq:VI.27}) plus (\ref{eq:VI.43a})--(\ref{eq:VI.43b}), then the ``anisotropic'' Poisson constraint term (\ref{eq:VI.24}) would be zero
\begin{equation}
\label{eq:VI.44}
\tNNegan \equiv 0 \;.
\end{equation}
But clearly, it is very hard to get these (surely existing) exact solutions; but even if we manage to get exact solutions only of both coupled Poisson equations (\ref{eq:VI.43a})--(\ref{eq:VI.43b}) alone (i.\,e. for arbitrary source $\tO(r)$, e.\,g. (\ref{eq:VI.1a})--(\ref{eq:VI.1b})), then the ``anisotropic'' Poisson identity (\ref{eq:VI.44}) will nevertheless hold. This may easily be verified by multiplying through the coupled equations (\ref{eq:VI.43a})--(\ref{eq:VI.43b}) with $\bgAiii(r)$ and $\bgAv(r)$, resp., and integrating over whole three-space under regard of $\Eegan$ (\ref{eq:VI.15}) and its mass equivalent $\tMMegan \mathrm{c^2}$ (\ref{eq:VI.19}). For such a situation where both Poisson identities (\ref{eq:VI.34}) and (\ref{eq:VI.44}) are valid, the anisotropic part $\anrtEEO$ (\ref{eq:VI.41}) gets rid of its constraint and thus becomes reduced to its proper physical part $\Eegan$, i.\,e.
\begin{equation}
\label{eq:VI.45}
\anrtEEO\ \Rightarrow\ \Eegan \;.
\end{equation}

But the same conclusion does hold also for the isotropic part, i.\,e. the ``isotropic'' Poisson constraint term $\tNNGee$ (\ref{eq:V.14}) vanishes whenever an exact solution of the ``isotropic'' Poisson equation (\ref{eq:IV.18}) is used, see equation (\ref{eq:VI.34}). Taking this altogether we get rid of all the constraint terms in the original energy functional $\tEEO$ (\ref{eq:VI.25}) which then also becomes reduced to its proper physical part $\EEgiv$:
\begin{equation}
\label{eq:VI.46}
\tEEO \Rightarrow\ \EEgiv = 2\Ekin + \ERee + \Eegan \;.
\end{equation}
The value of this reduced functional on the set of our trial configurations, parametrized
by the variational parameters $\left\{ \beta, \nu \right\}$, yields again some energy
function $\EEgivbn$ whose stationary points according to
(\ref{eq:VI.2a})--(\ref{eq:VI.2b}) determine the energy spectrum of ortho-positronium in
the next higher orders beyond the spherically symmetric approximation. Naturally, one
expects that there will emerge some energy difference between the corresponding ortho- and
para-states and also between the dimorphic partners of ortho-positronium. With the
precedent preparations, these questions will now be discussed in some numerical detail.

\section{Energy Difference of Dimorphic Partners}
\indent

The Poisson-like equations (\ref{eq:VI.43a})--(\ref{eq:VI.43b}) provide us now with the
possibility to study the energy difference of both dimorphic partners. In order to keep
our demonstration of this effect as simple as possible it may be sufficient here to study
only the case~${}^{(b)}j_z=\pm 1$ for~$\jO=2,3,4$. Indeed the results, obtained so far,
say that the dimorphism cannot occur for~$\bjz=0$ and~$\bjz=\pm \jO,\,\forall \jO$; and
this excludes a priori its occurrence for~$\jO=0,1$. 

It should be evident that such an energy difference is in the first line due to the
corresponding anisotropy corrections $\bgAiii(r)$ and $\bgAv(r)$ which stand for the first
and second order of our anisotropy approximation, cf. the perturbation expansion
(\ref{eq:V.22}). In contrast to this, the \emph{kinetic energy} $\Ekin$ is assumed to be
still of the simple form (\ref{eq:VI.9a})--(\ref{eq:VI.9b}). Thus one may raise the
question whether perhaps the expected energy difference emerges already in the first
approximation order ($\sim \bgAiii(r)$)? If yes, is then the magnitude of the second 
order-correction ($\sim \bgAv(r)$) considerably smaller than the first-order correction ($\sim
\bgAiii(r)$)?  Obviously, these questions can now be settled by first solving one (or
both) of those Poisson-like equations (\ref{eq:VI.43a})--(\ref{eq:VI.43b}) for given
source $\tO(r)$ (e.\,g. (\ref{eq:VI.1a})--(\ref{eq:VI.1b})) and then minimalizing the
resulting energy function $\EEgivbn$ with respect to both variational parameters $\beta$
and $\nu$. This will yield the binding energy of ortho-positronium for $\bjz = \pm 1$ due
to $\jO = 2, 3, 4$, quite similarly to the discussion of the isotropic situation
(cf. (\ref{eq:VI.36})--(\ref{eq:VI.40})). The result will be that, for $\jO = 2,\bjz=\pm
1$, the energetic degeneracy of both dimorphic configurations survives the first
anisotropy correction ($\sim \bgAiii(r)$); but in the second approximation order ($\sim
\bgAv(r)$) the degeneracy becomes eliminated, albeit only by $0,04\,\text{[eV]}$
(i.e.~$0,6\,\%$ of the binding energy), see the table on p.~\pageref{table8}. \pagebreak
\begin{center}
  \emph{\textbf{1.\ Quadrupole Corrections }}$\mathbf{\left( \sim \bgAiii(r)\right);\ \jO =
    2, 3, 4;\ \bjz = \pm 1}$
\end{center}

For a first estimate of the anisotropy corrections one may accept for the moment the
hypothesis that the second-order corrections ($\sim \bgAv(r)$) will turn out as being much
smaller than their first-order counterparts ($\sim \bgAiii(r)$) so that in the
lowest-order approximation we can neglect $\bgAv(r)$ altogether and thus arrive at the
following truncated form of equation (\ref{eq:VI.43a}):
\begin{equation}
\label{eq:VII.1}
\rklo{b}{e}_\mathsf{III} \cdot \Delta_r\,\bgAiii(r) - \rklo{b}{f}_\mathsf{III} \cdot
\frac{\bgAiii(r)}{r^2} = -\alpha_s\, \rklo{b}{m}_\mathsf{III} \cdot \frac{\tO^2(r)}{r} \;.
\end{equation}
Here, the constants $\rklo{b}{e}_\mathsf{III},\rklo{b}{f}_\mathsf{III},
\rklo{b}{m}_\mathsf{III} $ are defined by the equations (\ref{eq:VI.16a}) and
(\ref{eq:VI.22a}); and this clearly demonstrates their dependency on the angular density
$\bgkn(\vartheta)$. However, since the latter object is of dimorphic nature, see the two
possibilities for $\bgkn(\vartheta)$ and the associated $\bgAiii(\vartheta)$ in the table
on p.~\pageref{table5}, one expects to get two different sets of constants
\{$\rklo{b}{e}_\mathsf{III}, \rklo{b}{f}_\mathsf{III}, \rklo{b}{m}_\mathsf{III}$\}, which
then would entail two different equations of the present form (\ref{eq:VII.1}). But such a
dimorphic equation (\ref{eq:VII.1}) would imply the occurence of two different solutions
for the anisotropy correction $\bgAiii(r)$; and this, in turn, would yield two different
anisotropy energies $\Eegan$ (\ref{eq:VI.15})
\begin{equation}
\label{eq:VII.2}
\Eegan \Rightarrow \Eegiii = -\frac{\hbar\mathrm{c}}{\alpha_s}\,\rklo{b}{e}_\mathsf{III} \int\limits_0^\infty dr\,r^2\;\left\{ \left( \frac{d\,\bgAiii(r)}{dr} \right)^2 + \frac{\rklo{b}{f}_\mathsf{III}}{\rklo{b}{e}_\mathsf{III}} \left( \frac{\bgAiii(r)}{r} \right)^2 \right\} \;.
\end{equation}
Furthermore, since the anisotropy energy is an essential part of the total energy
$\EEgiv$, cf. (\ref{eq:VI.46}), one would also end up with two different energy functions
$\EEgivbn$, namely as the values of that functional $\EEgiv$ (\ref{eq:VI.46}) on both
trial configurations $\{ \tO(r),$ $\bAe(r),$ $\bgAiii(r) \}$. But these two energy
functions would immediately include the existence of a certain energy difference between
both dimorphic partners!

Actually, in the considered quadrupole approximation \emph{there is no energy difference
  between both dimorphic states due to } $\jO = 2, \bjz = \pm 1\ $! The crucial point here
is that, despite the dimorphic nature of $\bgkn(\vartheta)$ and $\bgAiii(\vartheta)$, all
three constants~$\rklo{b}{e}_\mathsf{III},\rklo{b}{f}_\mathsf{III},
\rklo{b}{m}_\mathsf{III}$ turn out to be identical, no matter whether they are calculated
by means of the first trial configuration ($\leadsto$ first line of the table on
p.~\pageref{table5}) or the second one ($\leadsto$ second line of that table). \emph{For
  both cases one finds the same result:}
\begin{subequations}
\begin{align}
\label{eq:VII.3a}
\rklo{b}{e}_\mathsf{III} &= \int \frac{d\Omega}{4\pi} \left( \bgAiii(\vartheta) \right)^2 = \frac{1}{320} \\
\label{eq:VII.3b}
\rklo{b}{f}_\mathsf{III} &= \int \frac{d\Omega}{4\pi} \left( \frac{d\,\bgAiii(\vartheta)}{d\vartheta} \right)^2 = \frac{3}{160} \\
\label{eq:VII.3c}
\rklo{b}{m}_\mathsf{III} &= \int d\Omega\ \bgAiii(\vartheta) \cdot \bgkn(\vartheta) = \frac{1}{64} \;.
\end{align}
\end{subequations}
Of course, this identity of $\rklo{b}{e}_\mathsf{III}$ and $\rklo{b}{f}_\mathsf{III}$ for
both dimorphic configurations due to $\boldsymbol{\jO = 2}$ comes not as a surprise,
because the quadrupole correction $\bgAiii(\vartheta)$ differs only in sign for both
dimorphic partners; but on the other hand, the equality of $\rklo{b}{m}_\mathsf{III}$ (\ref{eq:VII.3c}) for
both dimorphic partners is surely somewhat amazing since the angular density
$\bgkn(\vartheta)$ is distinctly different for both cases (recall the table on
p.~\pageref{table5}). However, the situation is different for~$\jO=3$ since in this case
all three constants~$\rklo{b}{e}_\mathsf{III},\rklo{b}{f}_\mathsf{III},
\rklo{b}{m}_\mathsf{III}$ are different for both dimorphic partners; and thus it is merely
the ratio (\ref{eq:VI.17a})-(\ref{eq:VI.17b}) what remains identical. Therefore one
expects that the dimorphic partners will have different binding energy exclusively
from~$\boldsymbol{\jO=3}$ on (in the quadrupole approximation).

\begin{center}
  \large{\textit{Quadrupole Equation}}
\end{center}

But in any case, with the first-order coefficients (\ref{eq:VII.3a})--(\ref{eq:VII.3c})
being fixed in such a way, the equation (\ref{eq:VII.1}) becomes specified (for \emph{all}
cases due to $\jO =2, 3, 4$) to the same ``\emph{quadrupole equation}'' (recall the ratios
(\ref{eq:VI.17a})-\ref{eq:VI.17b})
\begin{equation}
\label{eq:VII.4}
\Delta_r\,\bgAiii(r) - 6\frac{\bgAiii(r)}{r^2} = -5\alpha_s \frac{\tO^2(r)}{r} \;,
\end{equation}
where an example for the trial amplitude $\tO(r)$ has already been specified by equations
(\ref{eq:VI.1a})--(\ref{eq:VI.1b}). This equation has been studied extensively in
connection with para-positronium in the precedent paper (see ref.~\cite{1}). It is possible
to find the \emph{exact} solution of this quadrupole equation (\ref{eq:VII.4}) for the
case where the trial amplitude $\tO(r)$ is given by equations
(\ref{eq:VI.1a})--(\ref{eq:VI.1b}). For integer values of the variational parameter $2\nu$
the appropriate solutions read in dimensionless units ($y \doteqdot 2 \beta r; \bgAcnuiii(y)
\doteqdot (2 \beta \alpha_s)^{-1}\cdot \bgAiii(r)$)
\begin{equation}
\label{eq:VII.5}
\bgAcnuiii(y) = (2\nu + 3)(\nu + 1) \left\{ \frac{1}{10} \bgAciii_1(y) - 10y^2\e^{-y} \cdot \sum_{n=6}^{2\nu + 3} \frac{1}{n!} \sum_{m=0}^{n-6} \frac{d^m}{dy^m} (y^{n-6}) \right\}
\end{equation}
where for $\nu = 1$ the second term (in the bracket) must be put to zero; and the solution
$\bgAciii_1(y)$ for $\nu = 1$ is given by
\begin{equation}
\label{eq:VII.6}
\bgAciii_1(y) = \frac{20}{y^3} \left\{ 1 - \e^{-y} \cdot \sum_{n=0}^{4} \frac{y^n}{n!} \right\} \;.
\end{equation}

Indeed, the case for $\nu = 1$ plays an important part for this set of solutions because
it fixes the boundary conditions. Namely, observe that the homogeneous (dimensionless)
form of the quadrupole equation (\ref{eq:VII.4})
\begin{equation}
\label{eq:VII.7}
\left( \Delta_y - \frac{6}{y^2} \right)\,\bgAcnuiii(y) = 0
\end{equation}
admits two simple solutions
\begin{equation}
\label{eq:VII.8}
\bgAcnuiii(y)\ \Rightarrow\ y^2\ \text{or}\ y^{-3}
\end{equation}
from which we conclude the limit behaviour
\begin{subequations}
\begin{align}
\label{eq:VII.9a}
\underset{y \rightarrow 0}{\lim}\ \bgAcnuiii(y) &= \text{const} \cdot y^2 \\
\label{eq:VII.9b}
\underset{y \rightarrow \infty}{\lim}\ \bgAcnuiii(y) &=  \frac{(2\nu + 3)(2\nu + 2)}{y^3} \;.
\end{align}
\end{subequations}
The second limit $(y \rightarrow \infty)$ is immediately obvious from the combination of
both equations (\ref{eq:VII.5}) and (\ref{eq:VII.6}); and for verifying the first limit
$(y \rightarrow 0)$ it may suffice to look at the lowest-order cases $\nu = 1,
\frac{3}{2}, 2, \frac{5}{2}$:
\begin{subequations}
\begin{align}
\label{eq:VII.10a}
\bgAciii_1(y)\ &\Rightarrow\ \frac{1}{6} \left( y^2 - \frac{5}{6} y^3 + \frac{5}{14} y^4 - \frac{5}{48} y^5 \ldots \right) \\
\label{eq:VII.10b}
\bgAciii_{3/2}(y) &= \frac{3}{2} \cdot \bgAciii_1(y) - \frac{5}{24} y^2\e^{-y}\ \Rightarrow\ \frac{1}{24} \left( y^2 - \frac{5}{14} y^4 \ldots \right) \\
\label{eq:VII.10c}
\bgAciii_2(y) &= \frac{21}{10} \cdot \bgAciii_1(y) - \frac{1}{24} y^2\e^{-y}(8 + y)\ \Rightarrow\ \frac{1}{60} \left( y^2 - \frac{5}{48} y^5 \ldots \right) \\
\label{eq:VII.10d}
\bgAciii_{5/2}(y) &= \frac{14}{5} \cdot \bgAciii_1(y) - \frac{1}{144} y^2\e^{-y}(y^2 + 10y + 66)\\
&\Rightarrow\ \frac{1}{120} \left( y^2 + o(y^6) \right) \nonumber \;.
\end{align}
\end{subequations}

Once the quadrupole solutions are \emph{exactly} known one may substitute them into the
first-order form of the anisotropy energy $\Eegiii$ (\ref{eq:VII.2}) which then must equal
its mass equivalent $\tMMeg_{\ \mathsf{III}} \mathrm{c^2}$ (\ref{eq:VI.21a}), i.\,e. the
value of the Poisson constraint $\tNNegan$ (\ref{eq:VI.24}) on our selected trial
configuration $\left\{ \tO(r), \bgAiii(r) \right\}$ is again zero
\begin{equation}
\label{eq:VII.11}
\tNNegan(\beta, \nu) \Rightarrow \tNNegiii \doteqdot \Eegiii - \tMMegiii c^2 \equiv 0 \;.
\end{equation}
Here, the anisotropy energy $\Eegiii$ (\ref{eq:VII.2}) adopts the following form
\begin{align}
\label{eq:VII.12}
\Eegiii &= -\rklo{b}{e}_\mathsf{III} \,\frac{\hbar\mathrm{c}}{\alpha_s} \int\limits_0^\infty dr\,r^2\,\left\{ \left( \frac{d\,\bgAiii(r)}{dr} \right)^2 + 6 \left( \frac{\bgAiii(r)}{r} \right)^2 \right\} \\
&\doteqdot -\rklo{b}{e}_\mathsf{III} \,\frac{\e^2}{a_B}\,\left( 2\beta a_B \right)\cdot\epot^\mathsf{III}(\nu)
\nonumber\\*
\Bigg(\epot^\mathsf{III}(\nu)&\doteqdot\int_0^\infty
dy\,y^2\left\{\left(\frac{d\,\bgAciii_\nu(y)}{dy}\right)^2 
  + 6 \left(\frac{\bgAciii_\nu(y)}{y}\right)^2   \right\} \Bigg)\nonumber
\end{align}
and equals its mass equivalent $ \tMMegiii \mathrm{c^2}$ (\ref{eq:VI.21a}) which may be rewritten as
\begin{align}
\label{eq:VII.13}
\tMMeg_{\ \mathsf{III}} \mathrm{c^2} &= -\rklo{b}{m}_\mathsf{III}\, \hbar \mathrm{c} \int\limits_0^\infty dr\,r\ \bgAiii(r) \cdot \tO^2(r) \\
&\doteqdot -\rklo{b}{m}_\mathsf{III} \,\frac{\e^2}{a_B}\,\left( 2 a_B \beta \right)\cdot\muegiii(\nu) \;, \nonumber
\end{align}
so that the claimed Poisson identity (\ref{eq:VII.11}) can here explicitly be validated by
reference to the solution (\ref{eq:VII.5}) of the quadrupole equation (\ref{eq:VII.4}) in the form
\begin{equation}
\label{eq:VII.14}
\rklo{b}{e}_\mathsf{III} \cdot \epot^\mathsf{III}(\nu) = \rklo{b}{m}_\mathsf{III} \cdot \muegiii(\nu) \;.
\end{equation}

Properly speaking, the present final form of the energy functional $\tEEO$,
i.\,e. $\EEgiv$ (\ref{eq:VI.46}), would require now the use of the anisotropy energy
$\Eegiii$ (\ref{eq:VII.12}); but since the calculation of this object is somewhat tedious
one may resort to its mass equivalent $\tMMegiii \mathrm{c^2}$ (\ref{eq:VII.13}) which is
known to be numerically identical to the desired $\Eegiii$ because of the vanishing of the
Poisson constraint $\tNNegiii$ (\ref{eq:VII.11}). A closer inspection of the mass
equivalent (\ref{eq:VII.13}) reveals the following structure for the involved
mass-equivalent function $\muegiii(\nu)$~\cite{2}:
\begin{align}
\label{eq:VII.15}
\muegiii(\nu) &\doteqdot \frac{1}{\Gamma(2\nu + 2)}\,\int_0^\infty dy\;y^{2\nu + 1}\,\e^{-y}\cdot\bgAcnuiii(y) \\
&= \frac{\Gamma(2\nu + 4) \cdot \Gamma(2\nu - 1)}{\Gamma(2\nu + 2)^2} + \frac{1}{\Gamma(2\nu + 2)^2} \cdot \sum_{n=0}^\infty p_n(\nu)\,\frac{\Gamma(4\nu + 3 + n)}{2^{4\nu + 3 + n}} \nonumber \;,
\end{align}
where the coefficients $p_n(\nu)$ are given by
\begin{equation}
\label{eq:VII.16}
p_n(\nu) = \frac{\Gamma(2\nu + 4)}{\Gamma(2\nu + 5 + n)} - \frac{\Gamma(2\nu - 1)}{\Gamma(2\nu + n)} \;.
\end{equation}

\begin{center}
  \emph{\textbf{2.\ Energy Function of the Anisotropic Configurations}}
\end{center}

Since the total gauge field energy $\egER$ (\ref{eq:VI.12}) appears as the sum of the
isotropic and the anisotropic parts, with the isotropic part $\eeER$ being given by
equations (\ref{eq:VI.32a})--(\ref{eq:VI.33}) and the anisotropic part $\Eegan$ by
equation (\ref{eq:VII.12}), one ultimately obtains the gauge field energy in the following
form:
\begin{align}
\label{eq:VII.17}
\egER &= - \frac{e^2}{a_B} \left( 2 a_B \beta \right) \left\{ \epot(\nu) + \rklo{b}{m}_\mathsf{III} \cdot \muegiii(\nu)\right\} \\
&\doteqdot -\frac{e^2}{a_B} \left( 2 a_B \beta \right) \cdot \eegtot(\nu) \;. \nonumber
\end{align}
Obviously, the dimorphism enters the total gauge field energy $\EReg$ via the
mass-equivalent parameter $\rklo{b}{m}_\mathsf{III}$ which in general has been shown to
adopt different values for the dimorphic partners, see the table on
p.~\pageref{table6}. The important point with this result is that, also after the
inclusion of the anisotropy correction, the total gauge field energy $\egER$ still
persists in form of a \emph{linear} function of the first variational parameter
$\beta$. On the other hand, the kinetic energy $\Ekin$ has already been found to be a
\emph{quadratic} term of $\beta$, cf. (\ref{eq:VI.9a})--(\ref{eq:VI.9b}); and
therefore the total energy function $\EEgivbn$ emerges as the sum of a linear and a
quadratic term of $\beta$, i.\,e.
\begin{gather}
\label{eq:VII.18}
\EEgivbn = \frac{e^2}{a_B} \left\{ \left( 2 a_B \beta \right)^2 \cdot \ekin(\nu) - \left( 2 a_B \beta \right) \cdot \eegtot(\nu)  \right\} \;.
\end{gather}

This ``anisotropic'' energy function is of the same form as its ``isotropic'' precursor
$\EEeivbn$ (\ref{eq:VI.36}) and therefore the extremalization process runs in a quite
similar way, cf. equations (\ref{eq:VI.36})--(\ref{eq:VI.40}), i.\,e. we ultimately have
to look for the minimal values of the spectral function $S_\mathcal{O}(\nu)$ being defined
through
\begin{gather}
\label{eq:VII.19}
S_\mathcal{O}^{\{j\}}(\nu) \doteqdot \frac{\left( \eegtot(\nu) \right)^2}{\ekin(\nu)} \ ,
\end{gather}
where~$\{j\}$ symbolizes the pair configuration of quantum numbers~$\{\jO,\bjz \}$.
Here $\eegtot(\nu)$ is to be taken over from equation (\ref{eq:VII.17})
\begin{gather}
\label{eq:VII.20}
\eegtot(\nu) =\epot(\nu) + \rklo{b}{m}_\mathsf{III} \cdot \muegiii(\nu) \;,
\end{gather}
and $\ekin(\nu)$ is specified by equation (\ref{eq:VI.9b}). Denoting the maximalizing
parameter for the spectral function~$S_\mathcal{O}^{\{j\}}(\nu)$ (\ref{eq:VII.19}) by
$\nu_*$
\begin{equation}
\label{eq:VII.21}
\left. \frac{dS_\mathcal{O}^{\{j\}}(\nu)}{d\nu} \right|_{\nu=\nu_*} = 0 \;,
\end{equation}
the corresponding energy spectrum $\EE_\mathcal{O}^{\{j\}}$ is then obtained again from
the local maxima of $S_\mathcal{O}^{\{j\}}(\nu)$
\begin{gather}
\label{eq:VII.22}
\EE_\mathcal{O}^{\{j\}} = - \frac{e^2}{4a_B} \cdot S_{\mathcal{O}, \text{max}}^{\{j\}} \\
\left( \frac{e^2}{4a_B} \simeq 6{,}8029\ldots\ [\text{eV}] \right) \;. \nonumber
\end{gather}
The results for~$\bjz=\pm 1,\ \jO=1,2,3,4$ are collected in the following table:
\begin{center}
 \label{table7}
  \begin{tabular}{|c||c|c|c|c|c|c|}
    \hline
    $\jO\Rightarrow$ & 0 & 1 & 2 & 3 & 4\\
    \hline
    $\bjz\Rightarrow$& 0 & $\pm 1$ & $\pm 1$ & $\pm 1$ & $\pm 1$\\
    \hline
    \multirow{2}{*}{${}^{(b)}m_\mathsf{III}\Rightarrow$} & \multirow{2}{*}{$\frac{1}{16}$}&\multirow{2}{*}{$\frac{\ds 1}{\ds 64}$} &\multirow{2}{*}{$\frac{\ds 1}{\ds 64}$} & $\frac{1}{64}$ &
    $\frac{9}{256}$  \\
    \cline{5-6}
    & & & & $\frac{9}{256}$ & $\frac{289}{6400}$\\
    \hline
    \multirow{2}{*}[1mm]{$\EE_\mathcal{O}^{\{j\}}\Rightarrow $} & \multirow{2}{*}{$-7.5378\ldots$}&\multirow{2}{*}{$-1,575\ldots$} &\multirow{2}{*}{$-0.6811\ldots$} & $-0.3808\ldots$ & $-0.2501\ldots$\\
    \cline{5-6} 
    [eV] & & & & $-0.3900\ldots$ & $-0.2533\ldots$\\
    \hline
  \end{tabular}
\end{center}

As expected, the ortho-dimorphism does not occur for quantum-numbers~$\jO=0$
and~$\jO=1$. The lowest-order possibility for its occurrence is~$\jO=2,\bjz=\pm 1$.  But
now the important point with this result is that the calculated energy
$\EE_\mathcal{O}^{\{2,1\}}$ (\ref{eq:VII.22}) is (in this first-order approximation to
anisotropy) numerically the same for both dimorphic configurations with their rather
different charge distributions $\bgkn(\vartheta)$, see the upper half ($\jO = 2$) of the
table on p.~\pageref{table5}. Clearly, such an amazing result calls for an
explanation. But here the first observation is that both \emph{angular} anisotropy
corrections $\bgAiii(\vartheta)$ differ only in sign, cf. the table on
p.~\pageref{table5}; and therefore the coefficients $\rklo{b}{m}_\mathsf{III}$
(\ref{eq:VII.3a}) and $\rklo{b}{f}_\mathsf{III}$ (\ref{eq:VII.3b}) must be numerically
identical for the dimorphic pair with $\jO = 2$. Furthermore, the coefficient
$\rklo{b}{m}_\mathsf{III}$ (\ref{eq:VII.3c}) turns also out to be the same for both
configurations. This, however, entails that the first-order equation (\ref{eq:VII.1}) must
be identical for both dimorphic configurations, see its common form equation
(\ref{eq:VII.4}). But for this situation it is clear that the corresponding solution
$\bgAiii(\vartheta)$ of that ``quadrupole equation'' is also the same, since for both
cases one takes the same trial amplitude $\tO(r)$
(\ref{eq:VI.1a})--(\ref{eq:VI.1b}). Furthermore, the latter argument does also apply to
the spherically symmetric approximation $\bAe(r)$ (\ref{eq:V.16}) so that now all three
fields $\left\{ \tO(r), \bAe(r), \bgAiii(r) \right\}$ are identical for both dimorphic
configurations ($\jO = 2$), albeit only in the lowest order of the anisotropy
approximation.

But this is sufficient in order that all three energy contributions to $\EEgiv$
(\ref{eq:VI.46}) are also identical; and thus both configurations $\{2,\pm 1\}$ carry the
same energy $\EE_\mathcal{O}^{\{2,1\}}$ (\ref{eq:VII.22}). In physical terms, this says that
the degeneracy of the dimorphic partners (occuring in the spherically symmetric
approximation) is not eliminated by the present first-order anisotropy corrections (see
\textbf{Fig.VII.A} below)! See also \textbf{App.D} for entering into particulars.

\begin{center}
  \emph{\textbf{3.\ Comparison of Ortho- and Para-Levels}}
\end{center}

The reference basis of the comparison of the various RST levels is the conventional level
system~$\Ea{E}{n}{C}$ (\ref{eq:I.18}), \textbf{solid line}. The various RST approximations
are the spherically symmetric approximation $\Ec{E}{n}{T}$~(\boxes) and the quadrupole
approximations for para-positronium~$\Ed{E}{j}{\wp}$ (\cross) and
ortho-positronium~$\Ed{E}{j}{O}$~(\points).

All three RST approximations~$\Ec{E}{n}{T},\Ed{E}{j}{O},\Ed{E}{j}{\wp}$ are based on the
partial-extremalization process which relies on the maximalization of the spectral
function~$S^{\{j\}}_\mathcal{O,\wp}$,~cf.~ (\ref{eq:VII.22})
\begin{equation}
  \label{eq:VII.23}
  S^{\{j\}}_\mathcal{O,\wp}(\nu) =
  \frac{\left[ \en{pot}(\nu) + \rklo{b,p}{m}_\mathsf{III}\cdot\muegiii(\nu) \right]^2}
  {\frac{1}{2\nu+1}\left(\frac{1}{4}+\frac{ {j_{\mathcal{O,\wp}}}^2  }{2\nu}\right)}\ .
\end{equation}
For the spherically symmetric approximation~$\Ec{E}{n}{T}$~\cite{1}, one
puts~$j_\mathcal{O,\wp}=0$; and concerning~$\rklo{b,p}{m}_\mathsf{III}$ for the quadrupole
approximation of para-positronium one reads off the mass-equivalent
parameter~$\rklo{p}{m}_\mathsf{III}$ from the table on p.\pageref{tableDb}; and finally
for ortho-positronium one takes~$\rklo{b}{m}_\mathsf{III}$ from that table on
p.~\pageref{tableDb}. The conclusions of these numerical results for the spherically symmetric
and quadrupole approximations are now the following (\textbf{Fig.VII.A}): \pagebreak
\begin{center}
\label{fig7a}
\epsfig{file=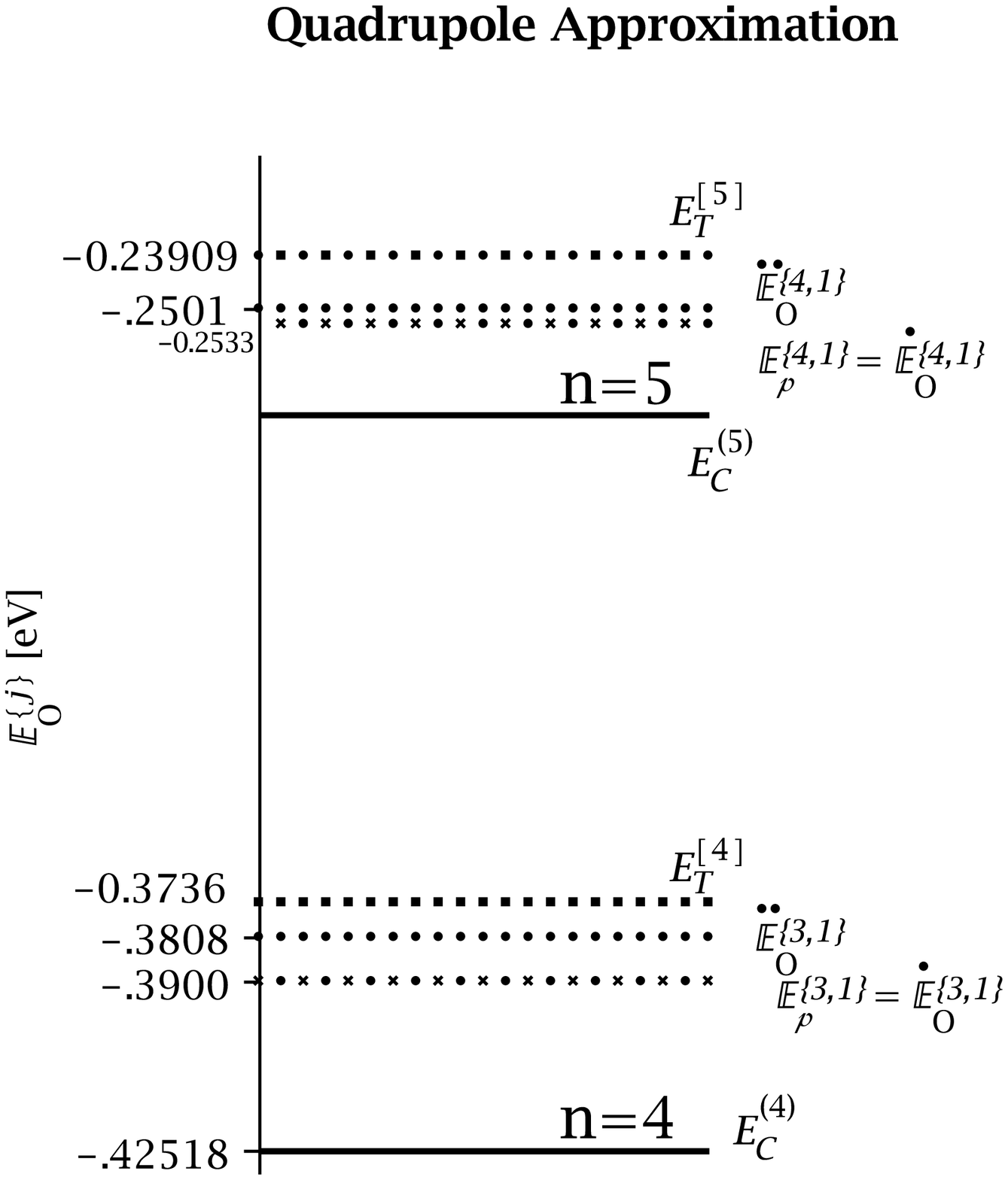,height=18cm}
\end{center}
\vspace{5mm}
{\textbf{Fig.~VII.A}\hspace{5mm} \emph{\large\textbf{Comparison of Ortho- and Para-Levels 
      for Principal \\ \centerline{\phantom{VII.A} Quantum Numbers\large\boldmath\ $n=4$ and $n=5$ }   } }}
\label{figVII.A}

\begin{itemize}
\item[\textbf{i)}] \textbf{RST and Conventional Predictions}
  
  \hspace{5mm}All three RST approximations~$\Ec{E}{n}{T} (\boxes),\Ed{E}{j}{\wp}$
  (\cross), $\Ed{E}{j}{O}$ (\points) yield numerical predictions \emph{above} the
  conventional~$\Eb{E}{n}{C}$ (\ref{eq:I.18}) (\solid)! This may be understood as a
  consequence of the fact that the RST energy functional~$\nrft{E}{\Omega}$
  (\ref{eq:VI.25}) has a \emph{lower bound} which is in the immediate neighborhood of the
  conventional levels~$\Eb{E}{n}{C}$. Therefore the value of this RST
  functional~$\nrft{E}{\Omega}$ on a (non-optimal) trial amplitude~$\tilde{\Omega}(r)$,
  such as our variational ansatz (\ref{eq:VI.1a})-(\ref{eq:VI.1b}), will yield predictions
  which always \emph{surmount} the corresponding conventional prediction~$\Ea{E}{n}{C}$
  (i.e.~$\Ec{E}{n}{T},\Ed{E}{j}{\wp},\Ee{E}{j}{O}\,\big| > \Eb{E}{n}{C}$). Thus one may
  expect that a more clever selection of the trial amplitude~$\tilde{\Omega}(r)$ will
  shift all three kinds of RST predictions further towards the
  conventional~$\Ea{E}{n}{C}$. Of course, one would like to know how close
  to~$\Eb{E}{n}{C}$ the RST predictions will turn out for the (hitherto unknown)
  \emph{exact} solution of the RST eigenvalue problem.

  However, one qualitative feature of the RST spectrum should be mentioned in this
  context: namely, a closer inspection of the larger principal quantum numbers~($n\gtrsim
  10$, say) yields the result that the RST predictions~$\Ed{E}{j}{\wp}$ become
  \emph{smaller} than their conventional counterparts~$\Ea{E}{n}{C}$
  (i.e.~$\Ed{E}{j}{\wp}\lesssim \Ea{E}{n}{C}$ for~$n\gtrsim 10$) even for the non-optimal
  trial amplitude~$\tilde{\Omega}(r)$ (VI.1). Therefore the \emph{true} RST energy
  eigenvalues should also be \emph{smaller} than their conventional
  counterparts~$\Ea{E}{n}{C}$ for sufficiently large principal number~$n$
  (or~$n_{\mathcal{O}},n_\wp$ resp.).

\item[\textbf{ii)}] \textbf{Elimination of the~$j_z$-Degeneracy}

  \hspace{5mm}The second observation concerns the elimination of the~$j_z$-degeneracy
  which has been discussed in great detail in the precedent paper~\cite{2}. This means
  that the $(2j_\mathcal{O,\wp}+1)$-fold degeneracy of the conventional and the
  spherically symmetric RST levels (due to any fixed~$j_\mathcal{O,\wp}$) becomes
  eliminated by the anisotropy of the electrostatic interaction potential. This anisotropy
  may be thought to come about by the action of the centrifugal forces on the rotating
  charge clouds~$(\rklo{b,p}{j}_z\neq 0)$. We assume that this degeneracy elimination for
  para-positronium~\cite{2} does occur in an analogous way also for ortho-positronium, but
  we do not study this phenomenon in the present paper which is rather concerned with the
  effect of \emph{dimorphism} (see also \textbf{App.D}).

\item[\textbf{iii)}] \textbf{Dimorphism of Ortho-Positronium}

  \hspace{5mm}The effect of dimorphism acts physically in a similar way as do the
  centrifugal forces in the case of the~$j_z$-degeneracy; namely by anisotropic
  deformation of the electric charge distributions, see \textbf{Fig.IV.A}. However the
  difference between both deforming mechanisms is that the ortho-dimorphism is an
  intrinsic geometric feature of the eigenvalue problem (\ref{eq:IV.6a})-(\ref{eq:IV.6b})
  itself; i.e.\ an indirect consequence of the presence of spin. If there is no angular
  momentum~($\bjz=0$,~\textbf{App.A}), then there is no dimorphism (see the tables on
  p.~\pageref{table8} and p.~\pageref{tableA}), as one may expect also intuitively. But a
  somewhat counter-intuitive effect occurs for maximal angular momentum~$\bjz=\pm \jO$,
  where there is also no dimorphism (see \textbf{App.A}). A hint at the possibility that
  the present quadrupole approximation may not display the true occurrence of the
  ortho-dimorphism comes from the quantum number~$\jO=2$. Here it is possible to satisfy
  the condition~$0<|\bjz|< \jO$ for the occurrence of dimorphism, i.e.\ by
  choosing~$\bjz=\pm 1$; but nevertheless the dimorphism does not exist (see the table on
  p.~\pageref{table8}), at least within the framework of the quadrupole
  approximation. Therefore the next step must consist in inspecting the octupole
  approximation (\textbf{Sect.VIII}), especially with respect to the dimorphism.

\item[\textbf{iv)}] \textbf{Ortho- and Para-Levels}

  \hspace{5mm} Finally, it is also interesting to notice the specific way in which the
  para-levels~$\Ed{E}{j}{\wp}$ become splitted into the ortho-levels~$\Ed{\dot{E}}{j}{O}$
  and~$\Ed{\ddot{E}}{j}{O}$ by the dimorphism. The first observation is here (see table on
  p.~\pageref{table8} and \textbf{App.D}) that the ortho-levels do agree with the para-levels
  for those states which are not subjected to the
  dimorphism:~$\Ed{E}{j}{O}=\Ed{E}{j}{\wp}$. This is exemplified here for~$\jO=0$
  and~$\jO=1$ (see the table on p.~\pageref{tableDb}), but presumably it holds for all values
  of~$\jO$. But even if the dimorphic splitting is active (here:~$\jO=3,4; \bjz=\pm 1$),
  one of the two split levels (i.e.~$\Ed{\dot{E}}{j}{O}$) does furthermore agree with the
  corresponding para-level~$\rogk{\mathbb{E}}{j}_\wp$ (\textbf{Fig.VII.A} on
  p.~\pageref{fig7a}). The other split level of the ortho-configuration
  (i.e.~$\Ed{\ddot{E}}{j}{O}$) is shifted upwards off from the residual
  ortho/para-level. This enhances the deviation from the corresponding conventional
  prediction~$\Ea{E}{n}{C}$.  It remains to be checked whether this result survives the
  consideration of the next higher (i.e.\ octupole) approximation.
\end{itemize}

\section{Multipole Solutions}
\indent

Naturally, one cannot be quite sure that all the essential features of the ortho-spectrum
are already accounted for by the quadrupole approximation. The uncertainty refers here
mainly to three questions:
\begin{itemize}
\item[\textbf{i)}] is the elimination of the $\bjz$-degeneracy a true effect or an
  artefact of the quadrupole approximation,
\item[\textbf{ii)}] if it is a true effect, is then the order of magnitude of the level splitting width
  correctly accounted for already by the quadrupole approximation, or do the \emph{higher-order}
  multipole results exceed/reduce the magnitude of the \emph{quadrupole} predictions,
\item[\textbf{iii)}] is the ``incidental'' missing of the ortho-dimorphism for $\jO = 2$; $\bjz = \pm 1$
  an artefact of the quadrupole approximation and thus disappears again when passing over
  to the multipole approximations of higher order?
\end{itemize}
Subsequently, we will try to answer these questions by a first preliminary inspection of
the multipole solutions of higher order.

\begin{center}
  \emph{\textbf{1.\ Alternative Multipole Expansion}}
\end{center}

The perturbation expansion (\ref{eq:V.22}) of the anisotropic part $\bgAan\rt$ of the
original gauge potential $\bgAe\rt$ (\ref{eq:V.19}) may be further elaborated by observing
the specific form of the two lowest-order contributions $\bgAiii(\vartheta)$ (table on
p.~\pageref{table5}) and $\bgAv(\vartheta)$ (\ref{eq:V.31a})--(\ref{eq:V.31b}). Indeed, the latter two
pertubations (\ref{eq:V.31a})--(\ref{eq:V.31b}) can be recast to the following form
\begin{subequations}
\begin{align}
\label{eq:VIII.1a}
\text{(V.31a):}\quad \bgkn(\vartheta) &= \frac{\sin\vartheta}{\pi^2} \\
&\Rightarrow \bgAv(\vartheta) = -\frac{105}{256}\,\left[ \cos^2\vartheta - \frac{1}{3} \right] - \frac{35}{512}\,\left[ \cos^4\vartheta - \frac{1}{5} \right] \nonumber \\
\label{eq:VIII.1b}
\text{(V.31b):}\quad \bgkn(\vartheta) &= \frac{\sin\vartheta}{3\pi^2}\,\left( \sin^2\vartheta + 9\cos^2\vartheta \right) \\
&\Rightarrow \bgAv(\vartheta) = \frac{105}{128}\,\left[ \cos^2\vartheta - \frac{1}{3}
\right] - \frac{105}{256}\,\left[ \cos^4\vartheta - \frac{1}{5} \right]\;. \nonumber
\end{align}
\end{subequations}
Thus, one is lead to suppose here that the original anisotropic expansion (\ref{eq:V.22})
yields an angular expansion in terms of the specific functions
$\mathbb{C}^{\{2l\}}(\vartheta)$ being defined by $(l = 1,2,3,4,\ldots)$
\begin{equation}
\label{eq:VIII.2}
\mathbb{C}^{\{2l\}}(\vartheta) \doteqdot \cos^{2l}\vartheta - \frac{1}{2l+1} \;,
\end{equation}
i.\,e. the original anisotropy expansion (\ref{eq:V.22}) would then look as follows
\begin{equation}
\label{eq:VIII.3}
\bgAan\rt = \sum_{l=1}^\infty \mathbb{C}^{\{2l\}}(\vartheta) \cdot \bgA_{2l}(r) \;.
\end{equation}

\begin{center}
  \large{\textit{Properties of the New Basis Functions}}
\end{center}

The new basis functions $\mathbb{C}^{\{2l\}}(\vartheta)$ have some interesting properties
which we will later on exploit for our purposes of attaining better approximations for the
wanted ortho-spectrum. First, observe here that the angular average of all the basis
functions is zero:
\begin{equation}
\label{eq:VIII.4}
\int\frac{d\Omega}{4\pi}\,\mathbb{C}^{\{2l\}}(\vartheta) = 0 \;,
\end{equation}
so that the anisotropic part $\bgAan\rt$ (\ref{eq:VIII.3}) of the gauge potential actually
does obey the requirement (\ref{eq:V.20}) of vanishing angular average. Next, split up
the Laplacean $\Delta_{r,\vartheta,\phi}$ into the three contributions due to the
spherical polar cooedinates $r,\vartheta,\phi$
\begin{equation}
\label{eq:VIII.5}
\Delta_{r,\vartheta,\phi} = \Delta_r + \frac{1}{r^2}\,\Delta_\vartheta + \frac{1}{r^2\sin^2\vartheta}\,\Delta_\phi
\end{equation}
with
\begin{subequations}
\begin{align}
\label{eq:VIII.6a}
\Delta_r &= \frac{\partial^2}{\partial r^2} + \frac{2}{r}\,\frac{\partial}{\partial r} \\
\label{eq:VIII.6b}
\Delta_\vartheta &= \frac{\partial^2}{\partial \vartheta^2} + \cot\vartheta\,\frac{\partial}{\partial \vartheta} \\
\label{eq:VIII.6c}
\Delta_\phi &= \frac{\partial^2}{\partial \phi^2} \;,
\end{align}
\end{subequations}
and then find that the longitudinal part $\Delta_\vartheta$ (\ref{eq:VIII.6b}) of the
Laplacean $\Delta_{r,\vartheta,\phi}$ acting on the basis functions
$\mathbb{C}^{\{2l\}}(\vartheta)$ as follows
\begin{equation}
\label{eq:VIII.7}
\Delta_\vartheta\,\mathbb{C}^{\{2l\}}(\vartheta) = 2l\,\left( 2l - 1 \right) \cdot \mathbb{C}^{\{2l-2\}}(\vartheta) - 2l\,\left( 2l + 1 \right)\,\mathbb{C}^{\{2l\}}(\vartheta) \;,
\end{equation}
i.\,e. the set of basis functions is closed under the action of the
longitudinal Laplacean $\Delta_\vartheta$. From this we conclude that
the Poisson equation for the anisotropic part $\bgAan\rt$
\begin{align}
\label{eq:VIII.8}
\Delta\,\bgAan\rt &= -4\pi\as\,\left[ \bgkn\rt - \frac{\bgkn(r)}{4\pi} \right] \\
&= -4\pi\as\,\bgkn(r)\,\left[ \bgkn(\vartheta) - \frac{1}{4\pi} \right] \nonumber
\end{align}
will transcribe to a system of coupled Poisson-like equations for the radial potential
corrections $\bgA_{2l}(r)$ (\ref{eq:VIII.3}), see below. In order to deduce that system by
means of the \emph{principle of minimal energy} it is necessary to define a scalar product
$\left\{ C_{l,m} \right\}$ for the vector space spanned by the basis functions. A nearby
choice for this is:
\begin{align}
\label{eq:VIII.9}
C_{l,m} &\doteqdot \frac{1}{2}\int_0^\pi d\vartheta \sin\vartheta\,\mathbb{C}^{\{2l\}}(\vartheta) \cdot \mathbb{C}^{\{2m\}}(\vartheta) \\
&=  \frac{1}{2m + 2n + 1} - \frac{1}{(2m + 1)(2n + 1)} \;. \nonumber
\end{align}
For instance, for a treatment including terms up to the octupole approximation
$\bgAv(\vartheta)$ (\ref{eq:VIII.1a})--(\ref{eq:VIII.1b}) one has to work with the
following (symmetric) coefficient matrix
  \begin{align}
\label{eq:VIII.10}
\left\{ C_{l,m} \right\} \Rightarrow
\begin{pmatrix}
\frac{4}{45} & \frac{8}{105} \\
\frac{8}{105} & \frac{16}{225}
\end{pmatrix} \ ,\qquad
\left\{ C^{-1}_{l,m} \right\} = \frac{45}{16}
\begin{pmatrix}
49 & -\frac{105}{2} \\
-\frac{105}{2} & \frac{245}{4}
\end{pmatrix}\ .
  \end{align}
The quadrupole approximation ($\sim \bgAiii(\vartheta)$) is a subcase hereof where only
one single matrix element $C_{1,1} = \frac{4}{45}$ is needed. For the higher-multipole
approximations one correspondingly extends the dimension of the matrix $\left\{ C_{l,m}
\right\}$ according to (\ref{eq:VIII.9}).

We will readily see the relevancy of these coefficient matrices $\left\{ C_{l,m} \right\}$
which obviously do represent a generalization of the former coefficients $\br e_\mathsf{III}$, $\br
e_\mathsf{IV}$, $\br e_\mathsf{V}$, \ldots (\ref{eq:VI.16a})--(\ref{eq:VI.16c}) occuring in connection with
the original perturbation expansion (\ref{eq:V.22}). The analogous generalization of the
former coefficients $\br f_\mathsf{III}$, $\br f_\mathsf{IV}$, $\br f_\mathsf{V}$, \ldots
(\ref{eq:VI.16a})-(\ref{eq:VI.16c}) would now obviously be a certain matrix $\left\{
  F_{l,m} \right\}$ which is to be introduced by means of the derivatives of the new basis
functions $\mathbb{C}^{\{2l\}}(\vartheta)$, i.\,e. we define:
\begin{equation}
\label{eq:VIII.11}
F_{l,m} \doteqdot \frac{1}{2} \int_0^\pi d\vartheta\;\sin\vartheta\,\left( \frac{d\,\mathbb{C}^{\{2l\}}(\vartheta)}{d\vartheta} \right) \cdot \left( \frac{d\,\mathbb{C}^{\{2m\}}(\vartheta)}{d\vartheta} \right) = \frac{8ml}{(2l + 2m)^2 - 1} \;.
\end{equation}
The two-dimensional case ($l \leq 2, m \leq 2$) is again needed for the octupole
approximation and is found to look as follows:
\begin{equation}
\label{eq:VIII.12}
\left\{ F_{l,m} \right\} \Rightarrow
\begin{pmatrix}
\frac{8}{15} & \frac{16}{35} \\
\frac{16}{35} & \frac{32}{63}
\end{pmatrix} \;.
\end{equation}
The case of the higher-multipole approximations should be self-evident now.

Finally, one has to inquire after the higher-multipole generalization of the
mass-equivalent parameters $\br m_\mathsf{III}$, $\br m_\mathsf{V}$, \ldots
(\ref{eq:VI.22a})--(\ref{eq:VI.22b}). For this purpose, one restarts from the general
definition of the mass equivalent $\Mee\crm^2$ (\ref{eq:V.9}) and inserts therein the new
anisotropy expansion (\ref{eq:VIII.3}), i.e.\ we then obtain the ``anisotropic'' mass
equivalent $\tMMegan\crm^2$ in the following form
\begin{equation}
\label{eq:VIII.13}
\tMMegan \,c^2 = -\hbar\crm\sum_{l=1}^\infty {}^{\{b\}}\!m_l \int_0^\infty dr\;r\,\tO^2(r) \,\bgA_{2l}(r)
\doteqdot \sum_{l=1}^\infty \tilde{\mathbb{M}}^{\{e\}}_{2l} c^2
\end{equation}
where the new mass-equivalent parameters ${}^{\{b\}}\!m_l$ are given by
\begin{equation}
\label{eq:VIII.14}
{}^{\{b\}}\!m_l \doteqdot \int d\Omega\;\bgkn(\vartheta)\,\mathbb{C}^{\{2l\}}(\vartheta) \;.
\end{equation}
In contrast to the precedent matrices $C_{l,m}$ and $F_{l,m}$ the new parameters
${}^{\{b\}}\!m_l$ depend now on the angular distribution $\bgkn(\vartheta)$ and are thus
not merely numerical objects of general character.

The new definition (\ref{eq:VIII.14}) of the mass-equivalent parameters looks very similar
to the old one (\ref{eq:VI.22a})--(\ref{eq:VI.22b}); and indeed, both definitions are
almost (but not really) identical. Indeed, there is a certain (albeit technical)
difference between the original perturbation formalism (\ref{eq:V.22}) and the new one
(\ref{eq:VIII.3}) if one stops at a \emph{finite} perturbation order; although both
formalisms would yield the same result if the \emph{infinite} number of perturbation
contributions could be taken into account. In order to see this technical difference
somewhat more clearly, consider (e.\,g.) the peculiar situation with $\boldsymbol{\jO =
  2,}\,\boldsymbol{\bjz = \pm 1}$ (see the table on p.~\pageref{table5}). Here, the new prescription
(\ref{eq:VIII.14}) yields for both dimorphic densities $\bgkn(\vartheta)$ the following
results for the mass-equivalent parameter $\bg m_l$ up to the octupole approximation
\begin{subequations}
\begin{align}
\label{eq:VIII.15a}
\bgkn(\vartheta) = \frac{\sin\vartheta}{\pi^2} &\Rightarrow \begin{cases}
\bg m_1 = -\frac{1}{12}\ ,\qquad\text{ quadrupole} \\
\bg m_2 = -\frac{3}{40}\ ,\qquad\text{ octupole}    
\end{cases} \\
\label{eq:VIII.15b}
\bgkn(\vartheta) = 3\,\frac{\sin\vartheta}{\pi^2}\,\left( 1 - \frac{8}{9}\,\sin^2\vartheta \right)
&\Rightarrow \begin{cases}
  \ \bg m_1 = \frac{1}{12}\ ,\ \qquad\text{ quadrupole} \\
  \ \bg m_2 = \frac{1}{20}\ ,\ \qquad\text{ octupole}
\end{cases} \\
&\left( \boldsymbol{\jO = 2, \bjz = \pm 1} \right) \;. \nonumber
\end{align}
\end{subequations}
Thus the new quadrupole parameters are found to differ in sign ($\bg m_1 = \pm
\frac{1}{12}$) and are some three times larger than their counterpart $\br m_\mathsf{III}$
of the original formalism
\begin{equation}
\label{eq:VIII.16}
\bg m_1 = \mp \frac{16}{3} \cdot \br m_\mathsf{III} = \mp \frac{16}{3} \cdot \frac{1}{64} = \mp \frac{1}{12} \;,
\end{equation}
see the table on p.~\pageref{tableDb}. Thus, on the level of the \emph{quadrupole} approximation there
is actually no essential difference of both formalisms because the objects are of the same
order of perturbation and differ merely by some unsubstantial numerical factor. It remains
to be shown (see below) that such a difference of the intermediate objects of both
approaches does not generate any difference of the final predictions of physical relevance.

\begin{center}
  \large{\textit{Comparison of the Expansion Coefficients}}
\end{center}

However, the analogous comparison of the \emph{octupole} contribution of both formalisms
reveals now some (albeit purely formal) distinctions. For the sake of simplicity, we
restrict ourselves again to the specific state in question ($\jO = 2, \bjz = \pm 1$) and
observe that the octupole potential term $\bgAv(\vartheta)$
(\ref{eq:VIII.1a})--(\ref{eq:VIII.1b}) emerges as a decomposition with respect to the new
basis $\mathbb{C}^{\{2l\}}(\vartheta)$ (\ref{eq:VIII.2}):
\begin{equation}
\label{eq:VIII.17}
\bgAv(\vartheta) = \sum_{l=1}^2\,a^\mathsf{(V)}_l\,\mathbb{C}^{\{2l\}}(\vartheta)\;.
\end{equation}
Such a finite series expansion of the original angular modes $\bgAiii(\vartheta)$,
$\bgAv(\vartheta)$, $\bgA^\mathsf{VII}(\vartheta)$, \ldots in terms of the new modes
$\mathbb{C}^{\{2l\}}(\vartheta)$ seems to occur for any perturbation order. The expansion
coefficients $a^\mathsf{(III)}_l$, $a^\mathsf{(V)}_l$, $a^\mathsf{(VII)}_l$, \ldots are
functionals of the angular density $\bgkn(\vartheta)$, for instance for the present
octupole cases (\ref{eq:VIII.1a})--(\ref{eq:VIII.1b})
\begin{subequations}
\begin{align}
\label{eq:VIII.18a}
a^\mathsf{(V)}_1 &= \begin{cases}
-\frac{105}{256}\ ,\qquad \text{first density (\ref{eq:VIII.1a})} \\
\frac{105}{128}\ ,\qquad\ \ \text{second density (\ref{eq:VIII.1b})}
\end{cases} \\
\label{eq:VIII.18b}
a^\mathsf{(V)}_2 &= \begin{cases}
-\frac{35}{512}\ ,\qquad \text{first density (\ref{eq:VIII.1a})} \\
-\frac{105}{256}\ ,\qquad \text{second density (\ref{eq:VIII.1b})} \;.
\end{cases}
\end{align}
\end{subequations}

Naturally, in view of the definitions of both mass-equivalent parameters $\br
m_\mathsf{V}$ (\ref{eq:VI.22b}) and $\bg m_l$ (\ref{eq:VIII.14}) the expansion
(\ref{eq:VIII.17}) can now be transcribed also to these parameters:
\begin{equation}
\label{eq:VIII.19}
\rklo{b}{m}_\mathsf{V} = \sum_{l=1}^2\,a^\mathsf{(V)}_l\,\bg m_l\;,
\end{equation}
i.\,e. concretely for the present demonstration (\ref{eq:VIII.15a})-(\ref{eq:VIII.15b})\\
\centerline{($\boldsymbol{\jO = 2$, $\bjz = \pm 1}$)}
\begin{equation}
\label{eq:VIII.21}
\rklo{b}{m}_\mathsf{V} = \begin{cases}
\ -\frac{105}{256} \cdot \left( - \frac{1}{12} \right) \\
\ \frac{105}{128} \cdot \frac{1}{12}
\end{cases} + \begin{cases}
\ -\frac{35}{512} \cdot \left( -\frac{3}{40} \right) \\
\ -\frac{105}{256} \cdot \frac{1}{20}
\end{cases} = \begin{cases}
\frac{161}{4096}\ ,\ \ \text{first density} \\
\frac{49}{1024}\ ,\ \ \text{second density} \;.
\end{cases}
\end{equation}
It should become now obvious from the numerical comparison of both approaches ($\br
m_\mathsf{III} \Leftrightarrow \bg m_1$, $\br m_\mathsf{V} \Leftrightarrow \bg m_2$) that
their corresponding perturbation orders are of equivalent numerical magnitude ($\sim
10^{-2}$), at least concerning the present lowest orders ($l = 1,2$). This suggests that
none of both competitors is numerically more advantageous, as far as the rapid convergence
of their series expansion is concerned. However, the basis elements
$\mathbb{C}^{\{2l\}}(\vartheta)$ (\ref{eq:VIII.2}) of the second expansion (\ref{eq:VIII.3})
are more favourable from the purely technical viewpoint because, in contrast to their
counterparts $\bgAiii(\vartheta)$ (\ref{eq:V.25a}) etc. due to the original expansion
(\ref{eq:V.22}), they are independent of the angular density $\bgkn(\vartheta)$ and
therefore the second approach appears to be more manageable for dealing with the higher
orders of the multipole expansion (see below).

\begin{center}
  \emph{\textbf{2.\ Multipole Equations}}
\end{center}

Once a new basis system of some promise is at hand now, one would like to see how the
former quadrupole and octupole equations (\ref{eq:VI.43a})--(\ref{eq:VI.43b}) transcribe
to the corrsponding new radial potentials $\bgA_{2l}(r)$ (\ref{eq:VIII.3}). The desired
Poisson-like equations for $\bgA_{2l}(r)$ may be deduced by inserting the new expansion
(\ref{eq:VIII.3}) of the anisotropic part $\bgAan\rt$ of the gauge potential $\bgAe\rt$
(\ref{eq:V.19}) in the ``anisotropic'' gauge-field energy $\Eegan$ (\ref{eq:VI.13}) and
also in the ``anisotropic'' Poisson constraint $\tNNegan$ (\ref{eq:VI.24}). Since the
latter two contributes to the energy functional $\tEEO$ (\ref{eq:VI.25}) are the sole ones
which contain the anisotropic part $\bgAan\rt$ it is sufficient to consider the
anisotropic part $\antEEO$ of the energy functional alone, i.\,e.
\begin{equation}
\label{eq:VIII.22}
\antEEO = \rogk{E}{e}_\mathrm{an} + \lGe \cdot \tNNegan \;,
\end{equation}
namely in order to deduce hereof the Poisson-like equations for the new radial potentials
$\bgA_{2l}(r)$ by means of \emph{the principle of minimal energy}
\begin{equation}
\label{eq:VIII.23}
\delta_l \antEEO = 0 \;.
\end{equation}
Here, $\delta_l$ is the functional derivative with respect to any potential $\bgA_{2l}(r)$
$(l = 1,2,3,4,\ldots)$ where the derivative of the mass-equivalent $\tMMegan$
(\ref{eq:VIII.13}) is immediately evident:
\begin{equation}
\label{eq:VIII.24}
\delta_e\,\tMMegan = \int_0^\infty dr\;r^2\,\Bigg(\delta\,\bgA_{2l}(r)\Bigg) \cdot \left\{ -\hbar \crm \, \bg m_l \, \frac{\tO^2(r)}{r} \right\} \;.
\end{equation}

\begin{center}
  \large{\textit{Anisotropy Energy}}
\end{center}

The case of the anisotropy energy $\Eegan$ (\ref{eq:VI.13}) is a little bit more complicated. It is
the sum of a radial derivative term $E^{\{r\}}_\text{an}$ and a longitudinal term
$E^{\{\vartheta\}}_\text{an}$, i.\,e.
\begin{equation}
\label{eq:VIII.25}
\Eegan = E^{\{r\}}_\text{an} + E^{\{\vartheta\}}_\text{an} \;,
\end{equation}
with the radial contribution being given by
\begin{equation}
\label{eq:VIII.26}
E^{\{r\}}_\text{an} = -\frac{\hbar\crm}{4\pi\as} \int d^3\vr \cdot \left( \frac{\partial\,\bgAan\rt}{\partial r} \right)^2
\end{equation}
and its longitudinal counterpart by
\begin{equation}
\label{eq:VIII.27}
E^{\{\vartheta\}}_\text{an} = -\frac{\hbar\crm}{4\pi\as} \int d^3\vr \cdot \left( \frac{1}{r}\,\frac{\partial\,\bgAan\rt}{\partial \vartheta} \right)^2 \;.
\end{equation}
Inserting herein the new series expansion (\ref{eq:VIII.3}) of the anisotropic part
$\bgAan\rt$ yields for the radial energy $\rogk{E}{r}_\mathrm{an}$ (\ref{eq:VIII.26})
\begin{equation}
\label{eq:VIII.28}
E^{\{r\}}_\text{an} = -\frac{\hbar\crm}{\as} \sum\limits_{l,m|=1}^\infty C_{l,m} \int_0^\infty dr\;r^2\,\frac{d\,\bgA_{2l}(r)}{dr} \cdot \frac{d\,\bgA_{2m}(r)}{dr} \;,
\end{equation}
and analogously for the longitudinal part $E^{\{\vartheta\}}_\text{an}$ (\ref{eq:VIII.27})
\begin{equation}
\label{eq:VIII.29}
E^{\{\vartheta\}}_\text{an} = -\frac{\hbar\crm}{\as} \sum\limits_{l,m|=1}^\infty F_{l,m} \int dr\;r^2\,\frac{\bgA_{2l}(r) \cdot \bgA_{2m}(r)}{r^2} \;.
\end{equation}
The functional derivatives of both energy contributions are easily seen to look as follows
\begin{subequations}
\begin{align}
\label{eq:VIII.30a}
\delta_l\,E_\text{an}(r) &= \int dr\;r^2\,\Bigg(\delta\,\bgA_{2l}(r)\Bigg) \cdot \left\{ 2\,\frac{\hbar\crm}{\as} \sum_k C_{l,k}\, \Delta_r \bgA_{2m}(r) \right\} \\
\label{eq:VIII.30b}
\delta_l\,E_\text{an}(\vartheta) &= \int dr\;r^2\,\Bigg(\delta\,\bgA_{2l}(r)\Bigg) \cdot \left\{ -2\,\frac{\hbar\crm}{\as} \sum_k F_{l,k}\, \frac{\bgA_{2k}(r)}{r^2} \right\} \;.
\end{align}
\end{subequations}
But with all three functional derivatives being at hand now, one can substitute them in
the anisotropic part of the \emph{principle of minimal energy} (\ref{eq:VIII.23}) which
then reads explicitly
\begin{equation}
\label{eq:VIII.31}
\delta_l\,\anEEO \equiv \left( 1 + \lGe \right) \cdot \delta_l\Eegan - \lGe\cdot\delta_l\,\left( \tMMegan \crm^2 \right)
\end{equation}
with the Lagrangean multiplier $\lGe = -2$.

\begin{center}
  \large{\textit{General Multipole Equation}}
\end{center}

This then yields the following Poisson-like equations for the radial correction potentials $\bgA_{2l}(r)$ (``\emph{multipole equations}'')
\begin{equation}
\label{eq:VIII.32}
\Delta_r\,\bgA_{2l}(r) - \sum_{k=1}^\infty F'_{l,k} \cdot \frac{\bgA_{2k}(r)}{r^2} = -\as \bg m'_l \cdot \frac{\tO^2(r)}{r} \;,
\end{equation}
where the transformed mass-equivalent parameter $\bg m'_l$ is given by
\begin{equation}
\label{eq:VIII.33}
\bg m'_l = \sum_{k=1}^\infty C^{-1}_{l,k} \cdot \bg m_k
\end{equation}
and the transformed matrix $\left\{ F'_{l,k} \right\}$ by
\begin{equation}
\label{eq:VIII.34}
F'_{l,k} = \sum_{j=1}^\infty C^{-1}_{l,j} \cdot F_{j,k} \;.
\end{equation}
For instance, for the octupole approximation where the matrix $\left\{ C^{-1}_{l,m}
\right\}$ is given by equation (\ref{eq:VIII.10}) and $\left\{ F_{l,m} \right\}$ by
(\ref{eq:VIII.12}) one finds for $\left\{ F'_{l,k} \right\}$ (\ref{eq:VIII.34})
\begin{equation}
\label{eq:VIII.35}
\left\{ F'_{l,k} \right\} = \begin{pmatrix}
6 & -12 \\
0 & 20
\end{pmatrix} \;.
\end{equation}

Observe also that all the higher-order multipole potentials~$\logk{b}{A}_{2l}(r)$ become
coupled by (\ref{eq:VIII.32}) to the \emph{spherically-symmetric}
amplitude~$\tilde{\Omega}(r)$!

\begin{center}
  \large{\textit{Quadrupole Approximation}}
\end{center}

As a brief demonstration of the alternative perturbation expansion (\ref{eq:VIII.3}) one
may resort again to the quadrupole approximation. Here, some results have already been
worked out within the framework of the original perturbation expansion (\ref{eq:V.22}),
see \textbf{Sect.VII}; and it may be instructive now to see how the same results can be
obtained also by use of the alternative expansion (\ref{eq:VIII.3}). Once we thus have
gained some confidence in this alternative approach, one may then tackle the octupole
approximation (or even the higher-multipole approximations) which will then appear in a
more pleasant shape than in the original perturbation formalism.

The quadrupole approximation is the lowest order of perturbation for which the system of
multipole equations (\ref{eq:VIII.32}) becomes reduced to one single equation ($l = m =
1$). Thus the multipole matrices $\left\{ C_{l,m} \right\}$ (\ref{eq:VIII.10}) and
$\left\{ F_{l,m} \right\}$ (\ref{eq:VIII.12}) become reduced to their components $C_{1,1}
= \frac{4}{45}$ and $F_{1,1} = \frac{8}{15}$; hence the transformed matrix simplifies to
\begin{equation}
\label{eq:VIII.36}
F'_{1,1} \Rightarrow C^{-1}_{1,1} \cdot F_{1,1} = \frac{45}{4} \cdot \frac{8}{15} = 6 \;.
\end{equation}
Furthermore, the mass-equivalent parameter $\bg m'_1$ (\ref{eq:VIII.33}) becomes
\begin{equation}
\label{eq:VIII.37}
\bg m'_1 = C_{1,1}^{-1} \cdot \bg m_1 = \frac{45}{4} \cdot \left( \mp \frac{1}{12} \right) = \mp \frac{15}{16} \;,
\end{equation}
cf. (\ref{eq:VIII.15a})--(\ref{eq:VIII.15b}). As a result of all these quadrupole
specifications, the general multipole equation (\ref{eq:VIII.32}) becomes cut down to the
\emph{quadrupole equation}:
\begin{equation}
\label{eq:VIII.38}
\Delta_r\,\bgA_2(r) - 6\,\frac{\bgA_2(r)}{r^2} = \pm \frac{15}{16} \,\as\,\frac{\tO^2(r)}{r} \;,
\end{equation}
where the upper/lower sign refers to the first/second density $\bgkn(\vartheta)$
(\ref{eq:VIII.15a})/(\ref{eq:VIII.15b}). Already from this form of the quadrupole equation
it is possible to conclude that both quadrupole correction potentials $\bgA_2(r)$ (for
$\boldsymbol{\jO = 2}, \boldsymbol{\bjz = \pm 1}$) due to both densities
(\ref{eq:VIII.15a}) and (\ref{eq:VIII.15b}) will \emph{differ only in sign}, just as was
the case for the corresponding quadrupole potentials $\bgAiii(r)$, see the table on
p.~\pageref{table5}. This difference in sign has already been shown to represent the
origin of the missing of the dimorphism for the states with $\jO = 2, \bjz \pm 1$, so that
this effect must occur also in the quadrupole approximation of the present alternative
approach. Furthermore, the quadrupole equations of both approaches, cf. (\ref{eq:VII.4})
vs. (\ref{eq:VIII.38}), are almost of the same shape and differ merely in the coupling
parameters, cf. (\ref{eq:VI.17b}) and (\ref{eq:VIII.37}):
\begin{align}
\text{(\ref{eq:VI.17b}):}\ \frac{\br m_\mathsf{III}}{\br e_\mathsf{III}} &=5 \Leftrightarrow\text{(\ref{eq:VIII.37}):}\ \bg m'_1 &=
   \begin{cases}
\ -\frac{15}{16}\quad\text{, first density (\ref{eq:VIII.15a})} \\
\ +\frac{15}{16}\quad\text{, second density (\ref{eq:VIII.15b})}
\end{cases}\ .
\end{align}
This similarity of both quadrupole equations then entails that their solutions must be proportional, i.\,e.
\begin{equation}
\label{eq:VIII.40}
\bgA_2(r) = \frac{\bg m'_1}{\left( \frac{\br m_\mathsf{III}}{\br e_\mathsf{III}} \right)} \cdot \bgAiii(r) = \mp \frac{3}{16}\,\bgAiii(r) \;.
\end{equation}

Concerning now the energy being stored in this quadrupole mode, one specialises the general
mass equivalent $\tMMegan\crm^2$ (\ref{eq:VIII.13}) of the alternative approach down to
the quadrupole appoximation ($l=1$), i.\,e.
\begin{equation}
\label{eq:VIII.41}
\tMMegan\crm^2 \Rightarrow \tMMeg_2\crm^2 = - \hbar\crm\,\bg m_1 \int\limits_0^\infty dr\;r^2\,\bgA_2(r)\,\frac{\tO^2(r)}{r} \;,
\end{equation}
or by use of the link (\ref{eq:VIII.40}) to the original quadrupole potential $\bgAiii(r)$
\begin{equation}
\label{eq:VIII.42}
\tMMeg_2\crm^2 = - \hbar\crm\,\frac{\bg m_1 \cdot \bg m'_1}{\left( \frac{\br m_\mathsf{III}}{\br e_\mathsf{III}} \right)} \int\limits_0^\infty dr\;r^2\,\bgAiii(r)\,\frac{\tO^2(r)}{r} \;.
\end{equation}
Here, the pre-factor is found by means of (\ref{eq:VIII.37}) and (\ref{eq:VI.17b})
\begin{equation}
\label{eq:VIII.43}
\frac{\bg m_1 \cdot \bg m'_1}{\left( \frac{\br m_\mathsf{III}}{\br e_\mathsf{III}} \right)} = \frac{\frac{45}{4} \cdot \left( \pm \frac{1}{12} \right)^2}{5} = \frac{1}{64} \;,
\end{equation}
so that the quadrupole mass equivalent in the alternative approach $\tMMeg_2\crm^2$
(\ref{eq:VIII.42}) actuall equals the original form $\tMMegiii\crm^2$,
cf. (\ref{eq:VI.22a}):
\begin{equation}
\label{eq:VIII.44}
\tMMegiii\crm^2 \equiv \tMMeg_2\crm^2 \;.
\end{equation}
For the specific form of $\MMeg_\mathsf{III}\crm^2$ see equation (\ref{eq:VII.13}).

Of course, this identity of the mass equivalents must have its counterpart for the quadrupole energies, i.\,e. one concludes
\begin{equation}
\label{eq:VIII.45}
\Eegiii \equiv E^\textrm{\{e\}}_2 \;,
\end{equation}
where the original form of $\Eegiii$ is given by equation (\ref{eq:VII.12}), with $\br
e_\mathsf{III} = \frac{1}{320}$ (table on p.~\pageref{table6}), and its alternative form~$\rogk{E}{e}_2$
is the specialization of $\Eegan$ (\ref{eq:VIII.25}) to the quadrupole case ($l=1$), i.e.
\begin{align}
\label{eq:VIII.46}
\Eegan \Rightarrow E^\textrm{\{e\}}_{2l}\big|_{l=1} \doteqdot E^\textrm{\{e\}}_2 = -\frac{\hbar\crm}{\as} \int\limits_0^\infty dr\;r^2\,&\bigg\{ C_{1,1}\,\left( \frac{d\,\bgA_2(r)}{dr} \right)^2 \\
&+ F_{1,1}\,\left( \frac{\bgA_2(r)}{r} \right)^2 \bigg\} \;. \nonumber
\end{align}
Inserting here again the required matrix elements from the matrices $C_{l,m}$
(\ref{eq:VIII.10}) and $F_{l,m}$ (\ref{eq:VIII.12}) in combination with the tranformation
of the quadrupole potentials (\ref{eq:VIII.40}) does actually validate the claimed
equality (\ref{eq:VIII.45}). For the precise form of $\Eegiii$ see equation
(\ref{eq:VII.12}).

This complete equivalence of the original and the alternative perturbation approach
(within the framework of the \emph{quadrupole} approximation) should provide now sufficient
confidence for tackling the next higher order of the perturbation expansion, i.\,e. the
\emph{octupole} approximation, just by means of the \emph{alternative} technique.
\pagebreak

\begin{center}
  \emph{\textbf{3.\ Octupole Approximation}}
\end{center}

Surely, it is reasonable to suppose that a certain improvement of the quadrupole results
(\textbf{Fig.VII.A}) can be attained by taking into account also the octupole character of
the charge distribution~$\logk{b}{k}_0(r,\vartheta)$, i.e.\ we consider now the multipole
equations (\ref{eq:VIII.32}) with restriction of the multipole order to $l,m| \leq 2$. The
required octupole matrices $\left\{ C_{l,m} \right\}$, $\left\{ C^{-1}_{l,m} \right\}$,
$\left\{ F_{l,m} \right\}$ and $\left\{ F'_{l,m} \right\}$ have already been specified by
equations (\ref{eq:VIII.10}), (\ref{eq:VIII.12}) and (\ref{eq:VIII.35}). Thus, we obtain
from the general multipole system (\ref{eq:VIII.32}) for $l=1$ the \emph{modified
  quadrupole equation}
\begin{equation}
\label{eq:VIII.47}
\Delta_r\,\bgA_2(r) - 6\,\frac{\bgA_2(r)}{r^2} + 12\,\frac{\bgA_4(r)}{r^2} = -\as \bg m'_1 \cdot \frac{\tO^2(r)}{r}
\end{equation}
and for $l=2$ the \emph{pure octupole equation}
\begin{equation}
\label{eq:VIII.48}
\Delta_r\,\bgA_4(r) - 20\,\frac{\bgA_4(r)}{r^2} = -\as \bg m'_2 \cdot \frac{\tO^2(r)}{r} \;.
\end{equation}
Any individual multipole equation of order $l$ owns its specific coupling constant $\bg
m'_l$ (\ref{eq:VIII.33}); especially for the present octupole approximation ($l \leq 2$)
we have for the \emph{quadrupole coupling constant} in equation (\ref{eq:VIII.47})
\begin{equation}
\label{eq:VIII.49}
\bg m'_1 = C^{-1}_{1,1} \cdot \bg m_1 + C^{-1}_{1,2} \cdot \bg m_2 = \begin{cases}
\ -\frac{105}{256}\ ,\quad&\text{first density (\ref{eq:VIII.15a})} \\
\ \frac{525}{128}\ ,\quad&\text{second density (\ref{eq:VIII.15b})}
\end{cases}
\end{equation}
and similarly for the \emph{octupole coupling constant} in equation (\ref{eq:VIII.48})
\begin{equation}
\label{eq:VIII.50}
\bg m'_2 = C^{-1}_{2,1} \cdot \bg m_1 + C^{-1}_{2,2} \cdot \bg m_2 = \begin{cases}
\ -\frac{315}{512}\ ,\quad&\text{first density (\ref{eq:VIII.15a})} \\
\ -\frac{945}{256}\ ,\quad&\text{second density (\ref{eq:VIII.15b})}\ .
\end{cases}
\end{equation}

Observe here that, on account of the particular form of the matrix $\left\{ F'_{l,m}
\right\}$ (\ref{eq:VIII.35}), the octupole equation (\ref{eq:VIII.48}) expresses the
decoupling from the quadrupole influence whereas the quadrupole equation
(\ref{eq:VIII.47}) still contains the octupole potential $\bgA_4(r)$. Surely, it would be
more aesthetic if one could deduce also a \emph{pure} quadrupole equation which formally does
contain only the proper quadrupole terms (see below).

\begin{center}
  \large{\textit{Octupole Energy}}
\end{center}

Concerning the energy being stored in the quadrupole and octupole modes, one may first
turn to the mass equivalent $\tMMegan\crm^2$ (\ref{eq:VIII.13}) which is the sum of the
individual multipole contributions, without any interference term:
\begin{equation}
\label{eq:VIII.51}
\tMMegan\crm^2 \Rightarrow \tMMeg_{2+4}\crm^2 = \tMMeg_2\crm^2 + \tMMeg_4\crm^2 \;,
\end{equation}
with the quadrupole contribution being given by
\begin{equation}
\label{eq:VIII.52}
\tMMeg_2\crm^2 = - \hbar\crm\,\bg m_1 \cdot \int\limits_0^\infty dr\;r^2\,\bgA_2(r)\,\frac{\tO^2(r)}{r}
\end{equation}
and analogously for the octupole contribution
\begin{equation}
\label{eq:VIII.53}
\tMMeg_4\crm^2 = -\hbar\crm\,\bg m_2 \cdot \int\limits_0^\infty dr\;r^2\,\bgA_4(r)\,\frac{\tO^2(r)}{r} \;.
\end{equation}
(For the mass-equivalent parameters $\bg m_1$ and $\bg m_2$ see equations (\ref{eq:VIII.15a})--(\ref{eq:VIII.15b})).

Unfortunately, such a pleasant separation of the various multipole contributions does not
occur for the anisotropy energy $\Eegan$ (\ref{eq:VI.13}). For its radial part $\Ergan$
(\ref{eq:VIII.28}) one finds
\begin{align}
\label{eq:VIII.54}
\Ergan \Rightarrow \rogk{E}{r}_{2+4} = &-\left( \frac{4}{15} \right)^2
\frac{\hbar\crm}{\as} \int\limits_0^\infty dr\;r^2\,\bigg\{ \frac{5}{4}
\left( \frac{d\,\bgA_2(r)}{dr} \right)^2 \\
&+ \frac{15}{7} \cdot \frac{d\,\bgA_2(r)}{dr} \, \frac{d\,\bgA_4(r)}{dr} + \left( \frac{d\,\bgA_4(r)}{dr} \right)^2 \bigg\} \;, \nonumber
\end{align}
and similarly for the longitudinal part $\Evgan$ (\ref{eq:VIII.29})
\begin{align}
\label{eq:VIII.55}
\Evgan \Rightarrow \rogk{E}{\vartheta}_{2+4} = -8\,\frac{\hbar\crm}{\as} \int\limits_0^\infty dr\;\bigg\{ \frac{1}{15} \, \left( \bgA_2(r) \right)^2 &+ \frac{4}{35}\,\bgA_2(r) \cdot \bgA_4(r) \\
&+ \frac{4}{63}\,\left( \bgA_4(r) \right)^2 \bigg\} \;. \nonumber
\end{align}
But now that all contributions to the ``anisotropic'' energy are explicitly known one can
of course carry through the variational \emph{principle of minimal energy}
(\ref{eq:VIII.23}) for $l=1,2$ and would then regain the quadrupole-octupole system
(\ref{eq:VIII.47})--(\ref{eq:VIII.48}). Though this result provides sufficient confidence
in the consistency of the general multipole approach (\ref{eq:VIII.32}) it is nevertheless
very instructive to consider also a slight modification thereof.

\begin{center}
  \large{\textit{Conjugate Potentials}}
\end{center}

Reconsidering for a moment the radial anisotropy energy $\Ergan$ (\ref{eq:VIII.28}), one
is tempted to introduce the \emph{conjugate potentials} $\bgA'_{2l}(r)$ through
\begin{equation}
\label{eq:VIII.56}
\bgA'_{2l}(r) = \sum_{m=1}^\infty C_{l,m}\,\bgA_{2m}(r) \;,
\end{equation}
or conversely
\begin{equation}
\label{eq:VIII.57}
\bgA_{2l}(r) = \sum_{m=1}^\infty C^{-1}_{l,m}\,\bgA'_{2m}(r) \;.
\end{equation}
In terms of these conjugate potentials, the former multipole equation (\ref{eq:VIII.32}) reappears now in the following form
\begin{equation}
\label{eq:VIII.58}
\Delta_r\,\bgA'_{2l}(r) - \sum_{k=1}^\infty F''_{l,k}\,\frac{\bgA'_{2k}(r)}{r^2} = -\as\,\bg m_l \,\frac{\tO^2(r)}{r} \;.
\end{equation}

The matrix $\left\{ F''_{l,k} \right\}$ emerging here is in general asymmetric and is defined through
\begin{equation}
\label{eq:VIII.59}
F''_{l,k} = \sum_{m=1}^\infty F_{l,m} \cdot C^{-1}_{m,k} \;.
\end{equation}
For instance, for the octupole approximation (i.\,e. $m,l,k| = 1,2$) this matrix looks as follows
\begin{equation}
\label{eq:VIII.60}
\left\{ F''_{l,k} \right\} = \begin{pmatrix}
6 & 0 \\
-12 & 20
\end{pmatrix}
\end{equation}
and thus is the transpose of the former matrix $\left\{ F'_{l,k} \right\}$
(\ref{eq:VIII.35}). Furthermore, the octupole approximation $(l,k| \leq 2)$ of the general
multipole system (\ref{eq:VIII.58}) consists now of a \emph{pure} quadrupole equation
($l=1$)
\begin{equation}
\label{eq:VIII.61}
\Delta_r\,\bgA'_2(r) - 6\,\frac{\bgA'_2(r)}{r^2} = -\as\,\bg m_1 \cdot \frac{\tO^2(r)}{r}
\end{equation}
and a \emph{modified} octupole equation ($l=2$)
\begin{equation}
\label{eq:VIII.62}
\Delta\,\bgA'_4(r) - 20\,\frac{\bgA'_4(r)}{r^2} + 12\,\frac{A'_2(r)}{r^2} = -\as\,\bg m_2 \cdot \frac{\tO^2(r)}{r} \;,
\end{equation}
which is just the other way round if compared to the original case (\ref{eq:VIII.47})--(\ref{eq:VIII.48}).

Of course, the conjugate multipole system (\ref{eq:VIII.58}) can also be deduced from the
principle of minimal energy
\begin{equation}
\label{eq:VIII.63}
\delta'_l\,\Big(\antEEO'\Big) = 0
\end{equation}
where $\delta'_l$ denotes the functional derivative of $\antEEO'$ with respect to the
conjugate potentials $\bgA'_{2l}(r)$ (\ref{eq:VIII.56}); and $\antEEO'$ is the former
functional $\antEEO$ expressed in terms of the conjugate potentials $\bgA'_{2l}(r)$,
cf. (\ref{eq:VI.25}). In order to validate this claim, one first transcribes the mass
equivalent $\tMMegan\crm^2$ (\ref{eq:VIII.13}) to its conjugate form:
\begin{equation}
\label{eq:VIII.64}
\tMMegan\crm^2 \Rightarrow {}'\tMMegan\crm^2 = -\hbar\crm \sum_{l=1}^\infty \bg m'_l \cdot \int\limits_0^\infty dr\;r^2\,\bgA'_{2l}(r)\,\frac{\tO^2(r)}{r} \;.
\end{equation}
Next, the radial energy $\Ergan$ (\ref{eq:VIII.28}) reappears as
\begin{equation}
\label{eq:VIII.65}
\Ergan \Rightarrow {}'\!\Ergan = -\frac{\hbar\crm}{\as} \sum_{l,k|=1}^{\infty} C^{-1}_{l,k} \int\limits_0^\infty dr\;r^2\,\frac{d\,\bgA'_{2l}(r)}{dr} \cdot \frac{d\,\bgA'_{2k}(r)}{dr} \;,
\end{equation}
and finally the longitudinal energy $\Evgan$ (\ref{eq:VIII.29}) adopts the following shape
\begin{equation}
\label{eq:VIII.66}
\Evgan \Rightarrow {}'\!\Evgan = -\frac{\hbar\crm}{\as} \sum_{l,k|=1}^\infty F'''_{l,k}
\int\limits_0^\infty dr\;r^2\,\frac{\bgA'_{2l}(r) \cdot \bgA'_{2k}(r)}{r^2}
\end{equation}
with the symmetric matrix $\left\{ F'''_{l,k} \right\}$ being given by
\begin{equation}
\label{eq:VIII.67}
F'''_{l,k} = \sum_{m,n|=1}^\infty C^{-1}_{m,k}\;F_{m,n}\;C^{-1}_{n,l} \;.
\end{equation}
Now one can carry through again the variational principle (\ref{eq:VIII.63}) (see
(\ref{eq:VIII.31}) for a more explicit form of this) in order to find the corresponding
extremal equations just coinciding with the result (\ref{eq:VIII.58}), as expected.

For a concrete demonstration of the conjugate formalism, one resorts again to the octupole
approximation ($j,k,l,m| \leq 2$). Here, the mass equivalent ${}'\tMMegan\crm^2$
(\ref{eq:VIII.64}) appears again as the sum of the quadrupole and the octupole contributions
\begin{equation}
\label{eq:VIII.68}
{}'\tMMegan\crm^2 \Rightarrow {}'\tMMeg_{2+4}\crm^2 = {}'\tMMeg_2\crm^2 + {}'\tMMeg_4\crm^2 \;.
\end{equation}
The conjugate quadrupole part is given here by
\begin{equation}
\label{eq:VIII.69}
'\tMMeg_2\crm^2 = -\hbar\crm\,\bg m'_1 \int\limits_0^\infty dr\;r^2\,\bgA'_2(r) \cdot \frac{\tO^2(r)}{r}
\end{equation}
and the corresponding octupole part by
\begin{equation}
\label{eq:VIII.70}
'\tMMeg_4\crm^2 = -\hbar\crm\,\bg m'_2 \int\limits_0^\infty dr\;r^2\,\bgA'_4(r) \cdot \frac{\tO^2(r)}{r} \;.
\end{equation}
Next, the radial energy ${}'\!\Ergan$ (\ref{eq:VIII.65}) becomes specified down to
\begin{align}
\label{eq:VIII.71}
{}'\!\Ergan \Rightarrow {}'\!\rogk{E}{r}_{2+4} = &- \frac{45}{16}\,\frac{\hbar\crm}{\as} \int\limits_0^\infty dr\;r^2\,\bigg\{ 49 \left( \frac{d\,\bgA'_2(r)}{dr} \right)^2 \\
&- 105\,\frac{d\,\bgA'_2(r)}{dr} \cdot \frac{d\,\bgA'_4(r)}{dr} + \frac{245}{4}\,\left( \frac{d\,\bgA'_4(r)}{dr} \right)^2 \bigg\} \;. \nonumber
\end{align}
And finally, the longitudinal energy ${}'\!\Evgan$ (\ref{eq:VIII.66}) is found as
\begin{equation}
\label{eq:VIII.72}
\begin{split}
  &{}'\!\Evgan \Rightarrow {}'\!\Evg_{2+4} =  \\*
  &-\frac{45}{16}\,\frac{\hbar\crm}{\as} \int_0^\infty dr\;r^2\,\frac{924\,\left( \bgA'_2(r)
    \right)^2 - 2100\, \bgA'_2(r) \cdot \bgA'_4(r)+1225 \left(\logk{b}{A}'_4(r)\right)^2  }  {r^2}
\end{split}
\end{equation}

For the desired deduction of the octupole system (\ref{eq:VIII.61})--(\ref{eq:VIII.62}) by
the extremalization of the ``anisotropic'' energy functional $\antEEO$ (\ref{eq:VI.25}) we
have to write down now that functional in the octupole approximation again in terms of the
conjugate potentials $\bgA'_{2l}(r)$, i.e.
\begin{align}
\label{eq:VIII.73}
\antEEO \Rightarrow \gkloit{2+4}{\mathbb{E}}{[\Omega]}' &\doteqdot {}'\rogk{E}{e}_{2+4} + \lGe \cdot {}'\tNN^\mathrm{\{e\}}_{2+4} \\
&= {}'\!\rogk{E}{r}_{2+4} + {}'\!\rogk{E}{\vartheta}_{2+4} + \lGe \cdot \bigg\{
{}'\rogk{E}{r}_{2+4} + {}'\rogk{E}{\vartheta}_{2+4} -
{}'\tMMeg_2 - {}'\tMMeg_4 \bigg\} \nonumber \\
&= \left( 1 + \lGe \right) \cdot \bigg\{ {}'\!\rogk{E}{r}_{2+4} +
{}'\!\rogk{E}{\vartheta}_{2+4} \bigg\} - \lGe \cdot \bigg\{ {}'\tMMeg_2 + {}'\tMMeg_4
\bigg\} \;. \nonumber
\end{align}
By means of the standard variational technique, it is then easy to see that both extremal equations
\begin{subequations}
\begin{align}
\label{eq:VIII.74a}
\delta'_2\,\gkloit{2+4}{\mathbb{E}' }{[\Omega]} &= 0 \\
\label{eq:VIII.74b}
\delta'_4\,\gkloit{2+4}{\mathbb{E}' }{[\Omega]} &= 0
\end{align}
\end{subequations}
actually do lead us back to just the claimed octupole system (\ref{eq:VIII.61})--(\ref{eq:VIII.62})!

\begin{center}
  \large{\textit{Hybrid Method}}
\end{center}

There exists also a kind of hybrid formulation of the octupole system which then contains
both kinds of potentials $\bgA_{2l}(r)$ and $\bgA'_{2l}(r)$. Indeed, it is easy to see
that the radial energy $\Ergan$ (\ref{eq:VIII.28}) can also be written in the following
form
\begin{equation}
\label{eq:VIII.75}
\Ergan \Rightarrow \Erg_{hy} \doteqdot -\frac{\hbar\crm}{\as} \sum_{l=1}^\infty \int\limits_0^\infty dr\;r^2\,\frac{d\,\bgA_{2l}(r)}{dr} \cdot \frac{d\,\bgA'_{2l}(r)}{dr} \;,
\end{equation}
and similarly for the longitudinal contribution $\Evgan$ (\ref{eq:VIII.29})
\begin{equation}
\label{eq:VIII.76}
\Evgan \Rightarrow \Evg_{hy} \doteqdot -\frac{\hbar c}{\as}\sum_{k,l|=1}^\infty F^{\prime\prime}_{l,k} \int\limits_0^\infty dr\;r^2\,\frac{\bgA_{2l}(r) \cdot \bgA'_{2k}(r)}{r^2} \;,
\end{equation}
where the matrix $\left\{ F^{\prime\prime}_{l,k} \right\}$ has already been specified by
equation (\ref{eq:VIII.59}). Concerning the mass equivalent, one takes half the sum of its
two forms given by (\ref{eq:VIII.13}) and (\ref{eq:VIII.64}), i.\,e.
\begin{equation}
\label{eq:VIII.77}
\tMMegan\crm^2 \Rightarrow \tMMeg_{hy}\crm^2 \doteqdot - \frac{\hbar\crm}{2} \sum_{l=1}^\infty \int\limits_0^\infty dr\;r^2\,\frac{\tO^2(r)}{r}\,\left\{ \bg m_l \cdot \bgA_{2l}(r) + \bg m'_l \cdot \bgA'_{2l}(r)  \right\} \;.
\end{equation}
The corresponding hybrid form of the energy functional $\antEEO$ (\ref{eq:VI.25}) is
\begin{equation}
\label{eq:VIII.78}
\antEEO \Rightarrow {}^{\{hy\}}\!\tEEO = - \left( \Erg_{hy} + \Evg_{hy} \right) + 2\,\tMMeg_{hy}\crm^2
\end{equation}
and its extremalization, cf.\ (\ref{eq:VIII.31}) and
(\ref{eq:VIII.74a})-(\ref{eq:VIII.74b}), with respect to both potentials $\bgA_{2l}(r)$
and $\bgA'_{2l}(r)$ leads us again back to both former multipole systems
(\ref{eq:VIII.47})-(\ref{eq:VIII.48}) and (\ref{eq:VIII.61})-(\ref{eq:VIII.62}).
\vspace{2cm}
\begin{center}
*\quad*\quad*
\end{center}
\vspace{1cm}

At this point, it seems worth while to pause for a moment in order to briefly survey the
results of the octupole approximation obtained so far. Perhaps the most interesting
feature is that one can work with two alternative formalisms, namely the one being based
upon the alternative multipole potentials $\bgA_{2l}(r)$, which were introduced in
connection with the new basis system, cf. (\ref{eq:VIII.3}), or one may work with the
conjugate potentials $\bgA'_{2l}(r)$ (\ref{eq:VIII.56}). Clearly, both formalisms are
completely equivalent and one may prefer the first or the second one. Indeed, both
approaches lead us to an \emph{octupole system} of equations, i.\,e. in the first case
(\ref{eq:VIII.47})-(\ref{eq:VIII.48}) and in the second case
(\ref{eq:VIII.61})-(\ref{eq:VIII.62}). Each one of both systems consists of a
\emph{quadrupole equation}, i\,e. (\ref{eq:VIII.47}) and (\ref{eq:VIII.61}), and of an
\emph{octupole equation}, i.\,e. (\ref{eq:VIII.48}) and (\ref{eq:VIII.62}). Furthermore,
both approaches admit the existence of a variational principle and thus are equipped with
an energy functional, $\gkloit{2+4}{\mathbb{E}}{[\Omega]}$ and
$\gkloit{2+4}{ \mathbb{E} }{[\Omega]}'$ (\ref{eq:VIII.73}), whose extremalization yields just
the corresponding octupole systems.

But a difference (albeit only in a technical respect) exists concerning the specific shape
of the multipole equations: whereas in the first case
(\ref{eq:VIII.47})-(\ref{eq:VIII.48}) the second equation (\ref{eq:VIII.48}) for the
octupole potential $\bgA_4(r)$ appears decoupled from the quadrupole potential
$\bgA_2(r)$, the first equation (\ref{eq:VIII.47}) for the quadrupole potential
$\bgA_2(r)$ contains a coupling to the octupole potential $\bgA_4(r)$, i.\,e. the octupole
potential $\bgA_4(r)$ appears to be independent of the quadrupole potential, but the
converse is not true. This quadrupole-octupole interrelationship is just the opposite of
the second case (\ref{eq:VIII.61})-(\ref{eq:VIII.62}); here the quadrupole potential
$\bgA'_2(r)$ seems to be independent of its octupole counterpart $\bgA'_4(r)$, whereas the
latter is coupled back to the quadrupole potential $\bgA'_2(r)$, cf. (\ref{eq:VIII.62}).

\begin{center}
  \large{\textit{Separative Method}}
\end{center}

This coupling between the quadrupole $( \bgA_2(r)$, $\bgA'_2(r))$ and octupole
($\bgA_4(r)$, $\bgA'_4(r)$) potentials cannot be removed by resorting to the hybrid method
since the latter does embrace both coupled octupole systems,
i\,e. (\ref{eq:VIII.47})-(\ref{eq:VIII.48}) and
(\ref{eq:VIII.61})-(\ref{eq:VIII.62}). However, it would be very desirable to have a
completely decoupled system because in this case it would become possible to associate
well-defined anisotropy energies to both the quadrupole mode $(l=1)$ and to the octupole
mode~$(l=2)$. Indeed, only for such a situation it would be legitimate to talk about a
``quadrupole energy'' and an ``octupole energy''. But such a \emph{mode separation} of the
anisotropy energy is not possible up to now because the anisotropy energy $\Eegan =
\Erg_{2+4} + \Evg_{2+4}$ of all three methods do contain a mixed term consisting of both
the quadrupole potential $\bgA_2(r) / \bgA'_2(r)$ and the octupole potential $\bgA_4(r) /
\bgA'_4(r)$, cf. equations (\ref{eq:VIII.54})-(\ref{eq:VIII.55}),
(\ref{eq:VIII.71})-(\ref{eq:VIII.72}), and (\ref{eq:VIII.75})-(\ref{eq:VIII.76}).

Fortunately, there exists a possibility for the \emph{complete} separation of the
quadrupole and octupole modes, namely by introducing suitable new potentials
$\bgA^{\prime\prime}_2(r)$ and $\bgA^{\prime\prime}_4(r)$:
\begin{subequations}
\begin{align}
\label{eq:VIII.79a}
\bgA^{\prime\prime}_2(r) &\equiv \bgA'_2(r) \\
\label{eq:VIII.79b}
\bgA^{\prime\prime}_4(r) &\equiv \bgA_4(r) \;.
\end{align}
\end{subequations}
This says that the new (separative) quadrupole potential $\bgA^{\prime\prime}_2(r)$ is
identified with the conjugate potential $\bgA'_2(r)$ (\ref{eq:VIII.56}) and the new
(separative) octupole potential $\bgA^{\prime\prime}_4(r)$ is identified with the
alternative octupole potential $\bgA_4(r)$ being formerly introduced by the transition
(\ref{eq:VIII.3}) to the new basis system $\left\{ \mathbb{C}^{\{2l\}}_{(\vartheta)}
\right\}$. Since the new separative potentials are composed of \emph{both} the alternative
and the conjugate potentials, the present separative method is obviously the true hybrid
method. Clearly, the separative potentials may be expressed also solely in terms of the
alternative potentials $\bgA_{2l}(r)$
\begin{subequations}
\begin{align}
\label{eq:VIII.80a}
\bgA^{\prime\prime}_2(r) &= \frac{4}{45} \cdot \bgA_2(r) + \frac{8}{105} \cdot \bgA_4(r) \\
\label{eq:VIII.80b}
\bgA^{\prime\prime}_4(r) &= \bgA_4(r)
\end{align}
\end{subequations}
or soleley in terms of the conjugate potentials $\bgA'_{2l}(r)$:
\begin{subequations}
\begin{align}
\label{eq:VIII.81a}
\bgA^{\prime\prime}_2(r) &= \bgA'_2(r) \\
\label{eq:VIII.81b}
\bgA^{\prime\prime}_4(r) &= -\frac{4725}{32} \cdot \bgA'_2(r) + \frac{11025}{64} \cdot \bgA'_4(r) \;.
\end{align}
\end{subequations}
Conversely, the alternative potentials $\bgA_{2l}(r)$ read in terms of the separative
potentials $\bgA^{\prime\prime}_{2l}(r)$ as
\begin{subequations}
\begin{align}
\label{eq:VIII.82a}
\bgA_2(r) &= \frac{45}{4} \cdot \bgA^{\prime\prime}_2(r) - \frac{6}{7}\,\bgA^{\prime\prime}_4(r) \\
\label{eq:VIII.82b}
\bgA_4(r) &= \bgA^{\prime\prime}_4(r) \;,
\end{align}
\end{subequations}
or similarly, the conjugate potentials $\bgA'_{2l}(r)$ do reappear in terms of the
separative potentials $\bgA^{\prime\prime}_{2l}(r)$ as
\begin{subequations}
\begin{align}
\label{eq:VIII.83a}
\bgA'_2(r) &= \bgA^{\prime\prime}_2(r) \\
\label{eq:VIII.83b}
\bgA'_4(r) &= \frac{6}{7} \cdot \bgA^{\prime\prime}_2(r) + \frac{64}{11025} \cdot \bgA^{\prime\prime}_4(r) \;.
\end{align}
\end{subequations}

By means of these results, one can now eliminate the alternative (or conjugate) potentials
in favour of the separative potentials so that the corresponding systems, i.\,e. the
``alternative'' system (\ref{eq:VIII.47})-(\ref{eq:VIII.48}) or the ``conjugate'' system
(\ref{eq:VIII.61})-(\ref{eq:VIII.62}), are both recast, in terms of the separative
potentials $\bgA^{\prime\prime}_2(r)$ and $\bgA^{\prime\prime}_4(r)$ to the following
form:
\begin{subequations}
\begin{align}
\label{eq:VIII.84a}
\Delta_r\,\bgA^{\prime\prime}_2(r) - 6\,\frac{\bgA^{\prime\prime}_2(r)}{r^2} &= - \bg m_1 \cdot \as\,\frac{\tO^2(r)}{r} \\
\label{eq:VIII.84b}
\Delta_r\,\bgA^{\prime\prime}_4(r) - 20\,\frac{\bgA^{\prime\prime}_4(r)}{r^2} &= - \bg m'_2 \cdot \as\,\frac{\tO^2(r)}{r} \;.
\end{align}
\end{subequations}
For the values of the mass-equivalent parameters $\bg m_1$ and $\bg m'_2$ see equations
(\ref{eq:VIII.16}) and (\ref{eq:VIII.50}), and observe also their relationship
\begin{equation}
\label{eq:VIII.85}
\bg m'_2 = \frac{11025}{64}\,\left( \bg m_2 - \frac{6}{7}\,\bg m_1 \right) \;.
\end{equation}
Thus the separative method takes over both the pure quadrupole equation (\ref{eq:VIII.61})
and the pure octupole equation (\ref{eq:VIII.48}), inclusive the corresponding
mass-equivalent parameters. Evidently, our attempt
(\ref{eq:VIII.79a})--(\ref{eq:VIII.79b}) of decoupling the octupole systems has thus been
successful; and consequently we can now look for the \emph{separative form} of the energy
functional so that the present octupole system (\ref{eq:VIII.84a})--(\ref{eq:VIII.84b})
represents just the corresponding extremal equations thereof.

\begin{center}
  \large{\textit{Separative Energy Functional}}
\end{center}

From the purely formal viewpoint, the desired ``separative'' energy functional
(${}^{\{2,4\}}\!\tEEO^{\prime\prime}$, say) must of course be of the same shape as its
``conjugate'' and ``hybrid'' predecessors ${}^{\{2+4\}}\!\tEEO'$ (\ref{eq:VIII.73}) and
${}^{\{hy\}}\!\tEEO$ (\ref{eq:VIII.78}); i.\,e. one expects again the splitting of the
anisotropy energy in a radial and a longitudinal part:
\begin{equation}
\label{eq:VIII.86}
{}^{\prime\prime}\!\Eegan \Rightarrow {}^{\prime\prime}\!\rogk{E}{e}_{2,4} \doteqdot {}^{\prime\prime}\!\Erg_{2,4} + {}^{\prime\prime}\!\Evg_{2,4} \;,
\end{equation}
just as was the case with all the predecessors, e.\,g. (\ref{eq:VIII.25}) for the
alternative choice of basis system. Thus the ``separative'' functional
${}^{\{2,4\}}\!\tEEO^{\prime\prime}$ will appear again as
(cf. ${}^{\{2+4\}}\!\tEEO'$ (\ref{eq:VIII.73}) and ${}^{\{hy\}}\!\tEEO$
(\ref{eq:VIII.78}))
\begin{equation}
\label{eq:VIII.87}
{}^{\{2,4\}}\!\EEOe^{\prime\prime} = - {}^{\prime\prime}\!\Erg_{2,4} - {}^{\prime\prime}\!\Evg_{2,4} + 2\,{}^{\prime\prime}\tMMeg_{2,4} \;.
\end{equation}
But the pleasant property of this functional is now that it decays in a sum of
subfunctionals, i.\,e. the quadrupole and octupole contributions:
\begin{equation}
\label{eq:VIII.88}
{}^{\{2,4\}}\!\tEEO^{\prime\prime} = {}^{\{2\}}\!\tEEO^{\prime\prime} + {}^{\{4\}}\!\tEEO^{\prime\prime}
\end{equation}
where the quadrupole contribution is given by
\begin{equation}
\label{eq:VIII.89}
{}^{\{2\}}\!\tEEO^{\prime\prime} = - {}^{\prime\prime}\!\Erg_2 - {}^{\prime\prime}\!\Evg_2 + 2\,{}^{\prime\prime}\tMMeg_2
\end{equation}
and the octupole contribution by
\begin{equation}
\label{eq:VIII.90}
{}^{\{4\}}\!\tEEO^{\prime\prime} = - {}^{\prime\prime}\!\Erg_4 - {}^{\prime\prime}\!\Evg_4 + 2\,{}^{\prime\prime}\tMMeg_4 \;.
\end{equation}
Here, the individual energy contributions are obtained by simply inserting those inverse
transformation relations (\ref{eq:VIII.82a})--(\ref{eq:VIII.82b}) or
(\ref{eq:VIII.83a})--(\ref{eq:VIII.83b}) in the corresponding ``conjugated'' or
``separative'' results; this yields for the quadrupole contributions (\ref{eq:VIII.89})
\begin{subequations}
\begin{align}
\label{eq:VIII.91a}
{}^{\prime\prime}\!\Erg_2 &= -\frac{45}{4} \cdot \frac{\hbar\crm}{\as} \int\limits_0^\infty dr\;r^2\,\left( \frac{d\,\bgA^{\prime\prime}_2(r)}{dr} \right)^2 \\
\label{eq:VIII.91b}
{}^{\prime\prime}\!\Evg_2 &= -\frac{45}{4} \cdot \frac{\hbar\crm}{\as} \int\limits_0^\infty dr\;6\,\left( \bgA^{\prime\prime}_2(r) \right)^2 \\
\label{eq:VIII.91c}
{}^{\prime\prime}\tMMeg_2 c^2 &= -\frac{45}{4} \bg m_1 \cdot \hbar\crm \int\limits_0^\infty dr\;r^2\,\bgA^{\prime\prime}_2(r)\,\frac{\tO^2(r)}{r} \;,
\end{align}
\end{subequations}
and similarly for the octupole contributions (\ref{eq:VIII.90})
\begin{subequations}
\begin{align}
\label{eq:VIII.92a}
{}^{\prime\prime}\!\Erg_4 &= -\frac{64}{11025} \cdot \frac{\hbar\crm}{\as} \int\limits_0^\infty dr\;r^2\,\left( \frac{d\,\bgA^{\prime\prime}_4(r)}{dr} \right)^2 \\
\label{eq:VIII.92b}
{}^{\prime\prime}\!\Evg_4 &= -\frac{64}{11025} \cdot \frac{\hbar\crm}{\as} \int\limits_0^\infty dr\;20\,\left( \bgA^{\prime\prime}_4(r) \right)^2 \\
\label{eq:VIII.92c}
{}^{\prime\prime}\tMMeg_4 c^2 &= -\frac{64}{11025}\cdot \bg m'_2 \; \hbar c \int\limits_0^\infty dr\;r^2\,\bgA^{\prime\prime}_4(r)\,\frac{\tO^2(r)}{r} \;.
\end{align}
\end{subequations}

Such a pleasant decay of the energy functional (as being given by equation
(\ref{eq:VIII.88})) into pure quadrupole and octupole contributions
${}^{\{2\}}\!\tEEO^{\prime\prime}$ and ${}^{\{4\}}\!\tEEO^{\prime\prime}$,
cf. (\ref{eq:VIII.89})--(\ref{eq:VIII.90}), does not occur for the other approaches,
i.\,e. ``alternative'', ``conjugate'', and ``hybrid''. And the consequence is that any
subfunctional is sufficient in order to deduce the corresponding multipole
equation. Indeed, it is easy to see that the extremalization of the quadrupole
subfunctional ${}^{\{2\}}\!\tEEO^{\prime\prime}$ (\ref{eq:VIII.89}) yields
just the quadrupole equation (\ref{eq:VIII.84a}), and analogously the extremalization of
the octupole subfunctional ${}^{\{4\}}\!\tEEO^{\prime\prime}$ (\ref{eq:VIII.90}) yields
the octupole equation (\ref{eq:VIII.84b}). This neat separation of the quadrupole and
octupole effects can now be used in order to clarify the question of octupole dimorphism
for those states with quantum numbers $\nO = \jO + 1 = 3$, $\bjz = \pm 1$ which in the
quadrupole approximation misses this effect of dimorphism (see the discussion of this
effect on p.~\pageref{table5}).

\begin{center}
  \emph{\textbf{4.\ Magnitude of Octupole Splitting}}
\end{center}

For a closer inspection of the critical energy level $\left\{ \nO = 3, \jO = 2, \bjz = \pm
  1 \right\}$ it is very instructive to first consider the \emph{quadrupole part}
(\ref{eq:VIII.84a}) of the \emph{octupole approximation}
(\ref{eq:VIII.84a})-(\ref{eq:VIII.84b})! 
\pagebreak

\begin{center}
  \large{\textit{Quadrupole Part}}
\end{center}

Observe here, that this is not identical to the former quadrupole approximation
(\ref{eq:VIII.38}), being due to the ``alternative'' approach, since the coupling constant
is now $\bg m_1$ (\ref{eq:VIII.15a})--(\ref{eq:VIII.15b}) in place of $\bg m'_1$
(\ref{eq:VIII.37}). But a common property is the fact that the quadrupole ambiguity shows
here up again in form of the sign ambiguity of the mass-equivalent parameter $\bg m_1$
(\ref{eq:VIII.15a})--(\ref{eq:VIII.15b}), just as is the case with $\bg m'_1$
(\ref{eq:VIII.37}); and this ambiguity transcribes then also quite correspondingly to both
solutions $\bg A_2(r)$ and $\bg A^{\prime\prime}_2(r)$. The consequence of this is that,
also within the framework of the separative method, there can occur \emph{no dimorphic
  energy splitting} of those states specified by the quantum numbers $\left\{ \nO = 3, \jO
  = 2, \bjz = \pm 1 \right\}$. The reason is again that the corresponding anisotropy
energy $\ppE^\mathrm{\{e\}}_2\,\left( = \ppE^{\{r\}}_2 + \ppE^{\{\vartheta\}}_2 \right)$
(\ref{eq:VIII.91a})-(\ref{eq:VIII.91b}) is quadratic with respect to the quadrupole
potential $\bgA^{\prime\prime}_2(r)$, and is therefore insensitive with respect to a
change of sign of $\bgA^{\prime\prime}_2(r)$!

A further common property of the former \emph{quadrupole approximation} (\ref{eq:VIII.38})
and the present \emph{quadrupole part} (\ref{eq:VIII.84a}) is the fact that both
approaches do yield the same quadrupole energy, i.\,e. we have not only the energy
identity (\ref{eq:VIII.45}) but also
\begin{equation}
\label{eq:VIII.93}
\Eegiii = \ppE^\mathrm{\{e\}}_2
\end{equation}
where $\Eegiii$ has been defined in terms of the original potential correction
$\bgAiii(r)$ by equation (\ref{eq:VII.12}) and the present energy
$\ppE^\mathrm{\{e\}}_2\,\left( = \ppE^{\{r\}}_2 + \ppE^{\{\vartheta\}}_2 \right)$ by the
equations (\ref{eq:VIII.91a})-(\ref{eq:VIII.91b}). In order to validate this claim,
simply observe that when $\bgAiii(r)$ is the solution of the former original quadrupole
equation (\ref{eq:VII.4}) then the solution of the present quadrupole part
(\ref{eq:VIII.84a}) is
\begin{equation}
\label{eq:VIII.94}
\bgA^{\prime\prime}_2(r) = \frac{1}{5}\,\bg m_1 \cdot \bgAiii(r) \;.
\end{equation}
On the other hand, the quadrupole part (\ref{eq:VIII.84a}) of the octupole approximation
is uniquely linked to the quadrupole energy $\ppE^\mathrm{\{e\}}_2\,\left( =
  \ppE^{\{r\}}_2 + \ppE^{\{\vartheta\}} \right)$ (\ref{eq:VIII.91a})-(\ref{eq:VIII.91b})
so that we merely need to substitute therein the solution $\bgA^{\prime\prime}_2(r)$
(\ref{eq:VIII.94}) in order to validate the claimed identification (\ref{eq:VIII.93}).

\begin{center}
  \large{\textit{Octupole Energy}}
\end{center}

Finally, let us mention also the fact that the decoupling of the quadrupole and octupole
potentials $\bgA^{\prime\prime}_2(r)$ and $\bgA^{\prime\prime}_4(r)$ by the separative
method, as expressed by (\ref{eq:VIII.84a})-(\ref{eq:VIII.84b}) entails the separate
validity of the octupole identity
\begin{equation}
\label{eq:VIII.95}
\ppE^\mathrm{\{e\}}_4 \equiv \pptMMeg_4 \crm^2 \;,
\end{equation}
where the octupole energy $\ppE^\mathrm{\{e\}}_4\,\left(= \ppE^{\{r\}}_4 +
  \ppE^{\{\vartheta\}}_4 \right)$ is given by equations
(\ref{eq:VIII.92a})-\ref{eq:VIII.92b}) and the corresponding mass equivalent $\pptMMeg_4
\crm^2$ by (\ref{eq:VIII.92c}). If we are satisfied here again with the former trial
amplitude $\tO(r)$ (\ref{eq:VI.1a})--(\ref{eq:VI.1b}), both objects (\ref{eq:VIII.95})
become functions of the trial parameters $\beta$ and $\nu$, i.\,e.
\begin{subequations}
\begin{align}
\label{eq:VIII.96a}
\ppE^\mathrm{\{e\}}_4 &= -\frac{\e^2}{a_B}\,\left( 2\beta a_B \right) \left(\gklo{b}{m}'_2 \right)^2\cdot \eeg_4(\nu) \\
\label{eq:VIII.96b}
\pptMMeg_4 \crm^2 &= -\frac{\e^2}{a_B}\,\left( 2\beta a_B \right) \left(\gklo{b}{m}'_2 \right)^2\cdot \mueg_4(\nu)
\end{align}
\end{subequations}
where both functions $\eeg_4(\nu)$ and $\mueg_4(\nu)$ are explicitly determined in
\textbf{App.E}. Thus, the octupole identity (\ref{eq:VIII.95}) reads in coefficient form
\begin{equation}
\label{eq:VIII.97}
\eeg_4(\nu) \equiv \mueg_4(\nu) \;.
\end{equation}

The latter results (\ref{eq:VIII.96a})-(\ref{eq:VIII.96b}) can now be used in order to
convince oneself that the absence of the energy splitting becomes eliminated by the
octupole approximation: namely, the mass-equivalent $\pptMMeg_4\crm^2$ (\ref{eq:VIII.96b})
shows that this object is \textbf{quadratic} with respect to the mass-equivalent parameter
$\bg m'_2$, i.\,e.
\begin{equation}
\label{eq:VIII.101}
\pptMMeg_4\crm^2 \sim \left( \bg m'_2 \right)^2 \;,
\end{equation}
because $\mueg_4(\nu)$ is independent of $\bg m'_2$, see equation (\ref{eq:E.27}) of
\textbf{App.E}. On the other hand, it has already been shown through the equations
(\ref{eq:VIII.50}) that $\bg m'_2$ must adopt two different values which differ by a
factor of six due to the existence of two different angular densities
$\bgkn(\vartheta)$, see equations (\ref{eq:VIII.15a})--(\ref{eq:VIII.15b}). Therefore the
octupole correction energy (\ref{eq:VIII.96a})-(\ref{eq:VIII.96b}) must lower the gauge
field energy of the considered ortho-state (with $\nO = \jO + 1 = 3$, $\bjz = \pm 1$) in
\emph{two} ways and thus creates the dimorphic energy splitting for this state! The
corresponding two (energy lowering) octupole-corrections will differ by a factor of six,
see below.

\begin{center}
  \large{\textit{Schr\"odinger Equation with Octupole Interaction}}
\end{center}

Properly speaking, the Schr\"odinger-like equation for the amplitude field $\tO(r)$ has
already been deduced from the energy functional $\tEEO$ (\ref{eq:VI.25}) and yielded the
equation (\ref{eq:VI.27}). However, that deduction had been based on the original
decomposition (\ref{eq:V.22}) of the anisotropic gauge field $\bgAan\rt$ so that the
corresponding Schr\"odinger equation (\ref{eq:VI.27}) refers to the potential corrections
$\bgAiii(r)$ and $\bgAv(r)$, etc. However, in the meantime it has turned out to be more
advantageous to decompose the anisotropic gauge field $\bgAan\rt$ as shown by equation
(\ref{eq:VIII.3}) where the alternative potentials $\bgA_{2l}(r)$ have been further
transcribed to the conjugated gauge fields $\bgA^\prime_{2l}(r)$ and ultimately to the
separative potentials $\bgA^{\prime\prime}_2(r)$ and $\bgA^{\prime\prime}_4(r)$
(\ref{eq:VIII.79a})--(\ref{eq:VIII.79b}). But since the separative method provided us with
a clear decoupling of the quadrupole and octupole interactions, it will surely be helpful
to write down the Schr\"odinger equation for the amplitude field also in terms of the
separative potentials $\bg A^{\prime\prime}_2(r)$ and $\bgA^{\prime\prime}_4(r)$.

For this purpose, we have to go back to the original energy functional $\tEEO$
(\ref{eq:VI.25}) and have to recast the anisotropic gauge field part $\antEEO$ in terms of
the separative potentials $\bgA^{\prime\prime}_2(r)$ and $\bgA^{\prime\prime}_4(r)$,
whereas the matter part $\DtEEO$ and the ``isotropic'' gauge field part $\eetEEO$ remains
unchanged. The latter fact then implies also that the Poisson equation (\ref{eq:IV.18})
for the \emph{isotropic} potential $\beA_0(r)$ remains unchanged. Thus, it is only the
Schr\"odinger equation (\ref{eq:VI.27}) which must get a new shape.

The latter is to be deduced from the energy functional (\ref{eq:VI.25}) by extremalization
with respect to the amplitude field $\tO(r)$, which is contained only in the matter part
$\DtEEO\,( \doteqdot 2\,\Ekin + 2\,\lambda_s \cdot \tNNO )$ and in all the mass
equivalents $\tMMee\crm^2$ (\ref{eq:V.12}), $\pptMMeg_2\crm^2$ (\ref{eq:VIII.91c}), and
$\pptMMeg_4$ (\ref{eq:VIII.92c}). Consequently, denoting the functional derivative with
respect to the amplitude field $\tO(r)$ by $\delta_\Omega$, the corresponding extremal
equation reads in abstract form
\begin{align}
\label{eq:VIII.102}
0 = \delta_\Omega\,\pptEEO \equiv \;&\delta_\Omega\,\left( \DtEEO \right) - \lGe\,\delta_\Omega\,\left( \tMMee\crm^2 \right) \\
- &\lGe\,\delta_\Omega\,\left( \pptMMeg_2\crm^2 \right) - \lGe\,\delta_\Omega\,\left( \pptMMeg_4\crm^2 \right) \;, \nonumber
\end{align}
or concretely
\begin{align}
\label{eq:VIII.103}
-&\frac{\hbar^2}{2M}\,\left( \frac{d^2}{dr^2} + \frac{1}{r}\,\frac{d}{dr} \right)\,\tO(r) + \frac{\hbar^2}{2Mr^2}\cdot\jO^2\,\tO(r) \\
- &\hbar\crm\,\left\{ \beA_0(r) + \frac{45}{4}\,\bg m_1 \cdot \bgA^{\prime\prime}_2(r) + \frac{64}{11025}\,\bg m'_2 \cdot \bgA^{\prime\prime}_4(r) \right\}\,\tO(r) \nonumber \\
&\qquad\qquad\qquad\qquad\qquad\qquad\qquad\qquad\qquad = -\lambda_s\cdot\tO(r)\ , \nonumber
\end{align}
where the Lagrangean parameter $\lambda_s$ plays the part of the Schr\"odinger energy
eigenvalue $E_*\,(\doteqdot - \lambda_s)$. Neglecting the octupole interactions
$(\bgA^{\prime\prime}_4(r) \Rightarrow 0)$ and observing
\begin{equation}
\label{eq:VIII.104}
\frac{45}{4}\,\bg m_1 \cdot \bgA^{\prime\prime}_2(r) \Rightarrow \rklo{b}{m}_\mathsf{III} \cdot \bgAiii(r)
\end{equation}
lets the present octupole Schr\"odinger equation (\ref{eq:VIII.103}) becoming identical to the former quadrupole Schr\"odinger equation (\ref{eq:VI.27}).

\begin{center}
  \large{\textit{Variational Procedure}}
\end{center}

Unfortunately, we are not smart enough in order to find the exact solution of the
eigenvalue problem (\ref{eq:VIII.103}) which must be complemented by the monopole equation
(\ref{eq:IV.18}) for the spherically symmetric approximation $\beA_0(r)$, by the
quadrupole equation (\ref{eq:VIII.84a}) for $\bgA^{\prime\prime}_2(r)$ and finally by the
octupole equation (\ref{eq:VIII.84b}) for $\bgA^{\prime\prime}_4(r)$. Therefore we must be
satisfied here with the extremalization of the energy functional $\tEEO$ (\ref{eq:VI.25})
on the subspace of the trial amplitudes $\tO(r)$ (\ref{eq:VI.1a})--(\ref{eq:VI.1b}) and its
descendants $\beA_0(r)$ ($\Rightarrow \tanu(y)$ (\ref{eq:VI.30})),
$\bgA^{\prime\prime}_2(r)$ ($\Rightarrow \bgAciii_\nu(y)$ (\ref{eq:VII.5})), and
$\bgA^{\prime\prime}_4(r)$ ($\Rightarrow \bgP_4(y)$ (\ref{eq:E.26})). Each of these
spherically-symmetric multipole potentials carries its own energy content, i.e.
\begin{subequations}
\begin{align}
\label{eq:VIII.105a}
\text{(\ref{eq:VI.32a}):}\qquad \beA_0(r) &\Rightarrow \ERee = -\frac{\e^2}{a_B}\,(2\beta a_B)\cdot \epot(\nu) \\
\label{eq:VIII.105b}
\text{(\ref{eq:VII.12}):}\qquad \bgA^{\prime\prime}_2(r) &\Rightarrow \Eegiii = \ppE^\mathrm{\{e\}}_2 = -\frac{\e^2}{a_B}\,(2\beta a_B) \cdot \bg m_\mathsf{III} \cdot \muegiii(\nu) \\
\label{eq:VIII.105c}
\text{(\ref{eq:VIII.90}):}\qquad \bgA^{\prime\prime}_4(r) &\Rightarrow \ppE^\mathrm{\{e\}}_4 = -\frac{\e^2}{a_B}\,(2\beta a_B)\,\left( \bg m'_2 \right)^2 \cdot \mueg_4(\nu) \;.
\end{align}
\end{subequations}
(Observe here that the quadrupole energy $\Eegiii = \ppE^\mathrm{\{e\}}_2$
(\ref{eq:VIII.105b}) appears to be \emph{proportional} to the mass-equivalent parameter
$\br m_\mathsf{III}$ whereas the octupole energy $\ppE^\mathrm{\{e\}}_4$
(\ref{eq:VIII.105c}) is \emph{quadratic} with respect to the corresponding parameter $\bg
m'_2$. The reason is that the universality of the ratio $\frac{\br m_\mathsf{III}}{\br
  e_\mathsf{III}} = 5$ (\ref{eq:VI.17b}) admits to absorb this into the quadrupole
ptential $\bgAiii(r)$, cf. the quadrupole equation (\ref{eq:VII.4}); see also the
discussion in \textbf{Appendix A}.

The total gauge field energy $\EReg$ (\ref{eq:VI.12}) in the octupole approximation is now
due to the sum of all three multipole components
(\ref{eq:VIII.105a})--(\ref{eq:VIII.105c}), i.\,e.
\begin{equation}
\label{eq:VIII.106}
\EReg = - \frac{\e^2}{a_B}\,(2\beta a_B) \cdot \eegtot(\nu)
\end{equation}
with the total potential function $\eegtot(\nu)$ being defined through
\begin{equation}
\label{eq:VIII.107}
\eegtot(\nu) = \epot(\nu) + \br m_\mathsf{III} \cdot \muegiii(\nu) + \left( \bg m'_2 \right)^2 \cdot \mueg_4(\nu) \;.
\end{equation}
Recall here also that the functions $\epot(\nu)$ (\ref{eq:VI.33}), $\muegiii(\nu)$
(\ref{eq:VII.15}), and $\mueg_4(\nu)$ (\ref{eq:E.27}) are independent of the quantum
numbers $\nO, \jO, \bjz$! Furthermore, the kinetic energy $\Ekin$ (\ref{eq:VI.8}) is still
given by (\ref{eq:VI.9a})--(\ref{eq:VI.9b}); and the three poisson identities
(\ref{eq:VI.34}), (\ref{eq:VII.11}), and (\ref{eq:VIII.95}) are satisfied since we always
use the \emph{exact} solutions of the corresponding multipole equations. Therefore these
constraints can be omitted in the energy functional $\tEEO$ (\ref{eq:VI.25}) together with
the normalization condition (\ref{eq:VI.5}): $\tNNO = 0$, provided that we restrict
ourselves to the use of normalized trial amplitudes $\tO(r)$. This cuts down the energy
functional $\tEEO$ (\ref{eq:VI.25}) to its physical part $\tEEO^\mathsf{(IV)}$:
\begin{equation}
\label{eq:VIII.108}
\tEEO \Rightarrow \tEEO^\mathsf{\{IV\}} = 2\,\Ekin + \ERee + \Eegiii + \ppE^\mathrm{\{e\}}_4 \;,
\end{equation}
and the value of this reduced functional on the subspace of RST fields \{$\tO(r)$;
$\beA_0(r)$, $\bgA^{\prime\prime}_2(r)$, $\bgA^{\prime\prime}_4(r)$\} due to our trial
ansatz (\ref{eq:VI.1a})-(\ref{eq:VI.1b}) yields then an energy function $\tEEgiv(\beta,
\nu)$ which depends on the trial parameters $\beta$ and $\nu$:
\begin{equation}
\label{eq:VIII.109}
\tEEgiv(\beta, \nu) = \frac{\e^2}{a_B}\,(2\beta a_B)^2 \cdot \ekin(\nu) - \frac{\e^2}{a_B}\,(2\beta a_B) \cdot \eegtot(\nu) \;,
\end{equation}
where $\ekin(\nu)$ is given by (\ref{eq:VI.9b}) and $\eegtot(\nu)$ by (\ref{eq:VIII.107}).

The quantum number $\jO \,(=\nO - 1)$ is contained explicitly in the kinetic function
$\ekin(\nu)$ (\ref{eq:VI.9b}) and implicitly in $\eegtot(\nu)$ (\ref{eq:VIII.107}) via
$\br m_\mathsf{III}$ and $\bg m'_2$, whereas the azimuthal quantum number $\bjz$ is
(implicitly) contained only in the potential function $\eegtot(\nu)$
(\ref{eq:VIII.107}). But after having fixed both quantum numbers $\jO\,(= \nO - 1)$ and
$\bjz$ with $-\jO \leq \bjz \leq \jO$, we can obtain the non-relativistic energy spectrum
of ortho-positronium by simply looking for the minimal values of the energy function
$\tEEgiv(\beta, \nu)$ (\ref{eq:VIII.109}) with respect to both trial parameters $\beta$
and $\nu$. By means of partial extremalization (i.\,e. with respect to $\beta$), there
remains to minimalize the reduced energy function $\EEOj(\nu)$ with respect to the
residual parameter $\nu$
\begin{equation}
\label{eq:VIII.110}
\EEOj(\nu) = -\frac{\e^2}{4 a_B} \cdot \SOjg(\nu) \;,
\end{equation}
i.\,e. the desired energy value $\EEOj$, associated to the quantum numbers $\{j\}
\doteqdot \{ \jO, \bjz \}$, is given by the maximal value of the spectral function
$\SOjg(\nu)$:
\begin{equation}
\label{eq:VIII.111}
\EEOj = - \frac{\e^2}{4 a_B}\, \Sjg_{\mathcal{O},\text{max}} \;.
\end{equation}
Here, the spectral function $\SOjg(\nu)$ contains the relevant quantum numbers $\{ \jO, \bjz \}$ in the following way
\begin{equation}
\label{eq:VIII.112}
\SOjg(\nu) = \frac{\left[ \eegtot(\nu) \right]^2}{\ekin(\nu)} = \frac{\left[ \epot(\nu) + \left( \bg m_1 \right)^2 \cdot \mueg_2(\nu) + \left( \bg m'_2 \right)^2 \cdot \mueg_4(\nu) \right]^2}{\frac{1}{2\nu + 1}\,\left( \frac{1}{4} + \frac{\jO^2}{2\nu} \right)} \;.
\end{equation}
The simple potential function $\epot(\nu)$ due to the spherically symmetric approximation
is given by equation (\ref{eq:VI.33}); its octupole analogue $\eeg_4(\nu)\,\left(\equiv
  \mueg_4(\nu)\right)$ is also explicitly prepared in \textbf{App.E}, cf. (\ref{eq:E.27});
and thus it is only the quadrupole mass-equivalent function $\mueg_2(\nu)$ which
necessitates a brief comment. However, the quadrupole approximation has already been
studied thoroughly in the preceding sections; and if we identify the corresponding mass
equivalent $\tMMegiii\crm^2$ (\ref{eq:VII.13}) with its separative counterpart
$\pptMMeg_2\crm^2$ (\ref{eq:VIII.91c}) one is immediately led to the identification of the
mass-equivalent functions:
\begin{equation}
\label{eq:eq:VIII.113}
\br m_\mathsf{III} \cdot \muegiii(\nu) = \left( \bg m_1 \right)^2 \cdot \mueg_2(\nu) \;.
\end{equation}
But from here it is a simple exercise to find the direct link of both mass-equivalent
functions: inserting the general form (\ref{eq:A.1}) of the quadrupole potential
$\bgAiii(\nu)$ in the definition (\ref{eq:VI.22a}) of the mass equivalent $\bg
m_\mathsf{III}$ yields
\begin{equation}
\label{eq:VIII.114}
\br m_\mathsf{III} = a_\mathsf{III} \cdot \bg m_1 \;.
\end{equation}
Next, observe equation (\ref{eq:A.4}) and conclude from this
\begin{equation}
\label{eq:VIII.115}
\bg m_1 = \frac{4}{9}\,a_\mathsf{III}
\end{equation}
which then finally yields
\begin{equation}
\label{eq:VIII.116}
\br m_\mathsf{III} = \frac{9}{4}\,\left( \bg m_1 \right)^2
\end{equation}
and thus
\begin{equation}
\label{eq:VIII.117}
\mueg_2(\nu) = \frac{9}{4}\,\muegiii(\nu) \;.
\end{equation}
So we can ultimately express the spectral function $\SOjg(\nu)$ in terms of the three
known-functions $\epot(\nu)$ (\ref{eq:VI.33}), $\muegiii(\nu)$ (\ref{eq:VII.15}), and
$\mueg_4(\nu)$ (\ref{eq:E.27}):
\begin{equation}
\label{eq:VIII.118}
\SOjg(\nu) = \frac{\left[ \epot(\nu) + \left( \frac{3}{2}\,\bg m_1 \right)^2 \cdot
    \muegiii(\nu) + \left( \bg m'_2 \right)^2 \cdot \mueg_4(\nu) \right]^2}{\frac{1}{2\nu
    + 1}\,\left( \frac{1}{4} + \frac{\jO^2}{2\nu} \right)}\ .
\end{equation}
If desired, the three functions~$\epot(\nu), \muegiii(\nu), \mueg_4(\nu)$ may be
alternatively used here in their closed analytic form, cf.~(\ref{eq:D.3a})-(\ref{eq:D.3b})
and (\ref{eq:E.50}). The subsequent table (see~p.~\pageref{table8}) represents a collection of the
\textbf{numerical results}:
\begin{itemize}
\item[\textbf{i)}] The \emph{groundstate}~($\jO=0$, first line) is not doubled
  (\textbf{App.s A} and \textbf{B}) and receives a relatively bad numerical
  prediction. The conventional value, cf.~(\ref{eq:I.18}), is
  \mbox{$\rork{E}{1}_\mathrm{C}\simeq -6,8029\,\text{[eV]}$} whereas the present RST
  octupole approximation yields $-7,59507\,\text{[eV]}$ which is a deviation of $-11,6\%$.
  Why the groundstate has such a large \emph{negative} deviation in contrast to the
  \emph{positive} deviations of the excited states~$(\jO=1,2,3,\ldots)$ must be clarified
  elsewhere. For instance, the first excited state $\jO=1,\bjz=\pm 1$ (second line) has a
  deviation of $+7,4\%$.
\item[\textbf{ii)}] The exceptional case $\{\boldsymbol{\nO=3,\jO=2,\bjz=\pm 1}\}$, which
  misses the level doubling in the \emph{quadrupole approximation} (see the discussion of
  the table below equation (\ref{eq:VII.22}) and \textbf{App.D}), is found to undergo now
  also the level doubling in the \emph{octupole approximation} (third line). The energy
  difference of the emerging dimorphic pair is found as $0,68566 - 0,68124\simeq 0,00442\,
  \text{[eV]}$ which amounts to (roughly) 0,65\% of the binding energy of the dimorphic
  partners. The corresponding energy difference for the other two dimorphic pairs, i.e.\
  $\jO=3$ and $\jO=4$, show up in the same order of magnitude (1,65\%).
\item[\textbf{iii)}] It is also interesting to see how the predictions change when one
  passes beyond the quadrupole approximation (fourth column) to the octupole approximation
  (sixth column). As a convenient measure of this change one may introduce the relative
  quadrupole-octupole deviation through
  \begin{equation}
    \label{eq:VIII.119}
    \rogk{\Delta}{j} =
    \frac{\rogk{\mathbb{E}}{j}_\mathcal{O}|_\mathrm{qu}-\rogk{\mathbb{E}}{j}_\mathcal{O}|_\mathrm{oc}}{\rogk{\mathbb{E}}{j}_\mathcal{O}|_\mathrm{oc}}\ ,
  \end{equation}
see the seventh column of the subsequent table. Here it is interesting to note that the
quadrupole predictions (fourth column) become corrected at most by (roughly) 0,5\% or
less. This may be understood as a signal that our separative perturbation method can be
expected to converge very rapidly. In any case, the level splitting due to the ortho-
dimorphism seems to be at least twice the present octupole corrections; and therefore the
doubling of the ortho-levels can hardly be interpreted as an artefact occurring only in the
low perturbation orders (i.e.\ quadrupole and octupole). Of course, this question should be
studied in greater detail by an extra treatment.
\end{itemize}

\begin{landscape}
\begin{center}
\label{table8}
\begin{tabular}{|c||c|c|c||c|c||c|c|}
\hline
\multirow{1}{*}{$\jO = $} & \multirow{2}{*}{$\bjz$} & $\rklo{b}{m}_\mathsf{III}=(\frac{3}{2}\gklo{b}{m}_1)^2$ & $\rogk{\mathbb{E}}{j}_\mathcal{O}\;[\textrm{eV}]$ & $(\gklo{b}{m}'_2)^2$ & $\rogk{\mathbb{E}}{j}_\mathcal{O}\;[\textrm{eV}]$ & $\rogk{\Delta}{j}\;\%$ & $\rork{E}{n}_C$ \\
$(\nO-1)$ &   &  (\ref{eq:VI.22a}), (\ref{eq:VIII.116}) & Quadrupole, (\ref{eq:VII.22}) &  (\ref{eq:VIII.85}) & Octupole, (\ref{eq:VIII.111}) &  (\ref{eq:VIII.119}) &  (\ref{eq:I.18}) \\
\hline\hline
\multirow{2}{*}{0} & \multirow{2}{*}{0} & \multirow{2}{*}{$\frac{1}{16}=0,0625$} &
\multirow{2}{*}{$-7,53786$} & \multirow{2}{*}{$(\frac{19845}{3584})^2\simeq 30,6595$} 
& \multirow{2}{*}{$-7.59507$} & \multirow{2}{*}{$-0,75\,\%$} & \multirow{2}{*}{$-6,8029\ldots$}\\
 & & & & & & & \\
\specialrule{1.5pt}{0pt}{0pt}
\multirow{2}{*}{1} & \multirow{2}{*}{$\pm 1$} & \multirow{2}{*}{$\frac{1}{64}=0,015625$} &
\multirow{2}{*}{$-1,57527$} & \multirow{2}{*}{$(\frac{315}{512})^2\simeq 0,3785$} 
& \multirow{2}{*}{$-1.57551$} & \multirow{2}{*}{$-0,015\,\%$} & \multirow{2}{*}{$-1,7007\ldots$}\\
 & & & & & & & \\
\specialrule{1.5pt}{0pt}{0pt}
\multirow{3}{*}{2} & \multirow{3}{*}{$\pm 1$} & \multirow{3}{*}{$\frac{1}{64}=0,015625$} &
\multirow{3}{*}{$-0.68112$} & \multirow{2}{*}{$(\frac{315}{512})^2\simeq 0,3785$} 
& \multirow{2}{*}{$-0.68124$} & \multirow{2}{*}{$-0,017\,\%$} & \multirow{3}{*}{$-0,7558\ldots$}\\[5mm]
\cline{5-7}
 & & & &$(\frac{945}{256})^2\simeq 13,6264$ & $-0.68566$ &$-0,66\,\%$ & \\[2mm]
\specialrule{1.5pt}{0pt}{0pt}
\multirow{3}{*}{3} & \multirow{3}{*}{$\pm 1$} & \multirow{2}{*}{$\frac{1}{64}=0,015625$} &
\multirow{2}{*}{$-0.38089$} & \multirow{2}{*}{$(\frac{945}{256})^2\simeq 13,6264$} 
& \multirow{2}{*}{$-0.38374$} & \multirow{2}{*}{$-0,74\,\%$} & \multirow{3}{*}{$-0,4251\ldots$}\\[5mm]
\cline{3-7}
 & &$\frac{9}{256}\simeq 0,03515$ & $-0.39001 $ &$(\frac{945}{1024})^2\simeq 0,8516$ &$-0.39019$ &$-0,046\,\%$ & \\[2mm]
\specialrule{1.5pt}{0pt}{0pt}
\multirow{4}{*}{4} & \multirow{3}{*}{$\pm 1$} & \multirow{2}{*}{$\frac{9}{256}\simeq0,03515$} &
\multirow{2}{*}{$-0.25010$} & \multirow{2}{*}{$(\frac{945}{1024})^2\simeq 0,8516$} 
& \multirow{2}{*}{$-0.25023$} & \multirow{2}{*}{$-0,051\,\%$} & \multirow{3}{*}{$-0,2721\ldots$}\\[5mm]
\cline{3-7}
 & &$(\frac{17}{80})^2\simeq 0,04515$ & $-0.25330$ & $(\frac{2835}{1024})^2\simeq 7,664$ &$-0.25443$ &$-0,44\,\%$ & \\[2mm]
\specialrule{1.5pt}{0pt}{0pt}
\end{tabular}
\end{center}
\end{landscape}

\renewcommand{\theequation}{\Alph{section}.\arabic{equation}}
\setcounter{section}{1}
\setcounter{equation}{0}
  \begin{center}
  {\textbf{\Large Appendix A}}\\[1.5em]
  \emph{\textbf{\Large General Properties of the Quadrupole Approximation}}
  \end{center}
  \vspace{2ex}

  For getting some survey of all properties of the quadrupole approximation it is most
  instructive to consider also a further example, i.\,e. the case $\jO = 4$. The
  subsequent table (at the end of this appendix) presents a collection of the
  corresponding relevant data for $\jO = 4$ and thus complements the precedent tables as
  our basis of inductive reasoning.

\begin{center}
  \large{\textit{General Validity of the Ratios (\ref{eq:VI.17a})--(\ref{eq:VI.17b})}}
\end{center}

The first and most striking observation refers to the fact that the ratios
(\ref{eq:VI.17a})--(\ref{eq:VI.17b}) are valid for all considered cases ranging from $\jO
= 1$ up to $\jO = 4$. Thus the self-suggesting conclusion from this limited number of
cases is that these ratios could perhaps be valid for all $\jO$ (i.\,e. $\jO =$ 1, 2, 3,
4, \ldots). If this conclusion could be shown to be true, the \emph{quadrupole equation}
(\ref{eq:VII.4}) for the potential correction $\bgAiii(r)$ would be generally valid, as
well as the general form of the \emph{quadrupole energy} $\Eegiii$ (\ref{eq:VII.12}).

Indeed, it is a rather simple matter to validate the first one of these ratios,
i.\,e. (\ref{eq:VI.17a}). For this purpose, observe first that the general form of the
angular potential factor $\bgAiii(\vartheta)$ (\ref{eq:V.25a}) is
\begin{equation}
\label{eq:A.1}
\bgAiii(\vartheta) = a_\mathsf{III} \cdot \left[ \cos^2\vartheta - \frac{1}{3} \right]
\end{equation}
with the constant $a_\mathsf{III}$ being given by
\begin{equation}
\label{eq:A.2}
a_\mathsf{III} = -\frac{3}{4}\,\left( 1 - 3\,\Kb^\mathsf{III} \right) \;,
\end{equation}
see equation (\ref{eq:V.28}). Next, calculate both constants $\rklo{b}{e}_\mathsf{III}$ and $\rklo{b}{f}_\mathsf{III}$
(\ref{eq:VI.16a}) by use of the present form of $\bgAiii(\vartheta)$ (\ref{eq:A.1}) and
find
\begin{subequations}
\begin{align}
\label{eq:A.3a}
\rklo{b}{e}_\mathsf{III} &= \frac{4}{45} \cdot a_\mathsf{III}^2 \\
\label{eq:A.3b}
\rklo{b}{f}_\mathsf{III} &= \frac{8}{15} \cdot a_\mathsf{III}^2 \;.
\end{align}
\end{subequations}
Consequently, this verifies the first ratio (\ref{eq:VI.17a}). Observe here that this
ratio is a purely geometric relation, independent of the specific angular density
$\bgkn(\vartheta)$.

A quite similar (albeit somewhat more complicated) reasoning yields for the constant
$\rklo{b}{m}_\mathsf{III}$ (\ref{eq:VI.22a})
\begin{equation}
\label{eq:A.4}
\rklo{b}{m}_\mathsf{III} = \left( \frac{2}{3} a_\mathsf{III} \right)^2
\end{equation}
which then validates the second ratio (\ref{eq:VI.17b}). Thus both ratios
(\ref{eq:VI.17a})--(\ref{eq:VI.17b}) are found to represent quite general features of the
quadrupole approximation and thus turn out to be independent of the chosen trial amplitude
$\tO(r)$ (\ref{eq:VI.1a}). It is also a rather easy exercise to verify that all states
with $\jO = 4$ (see end of this appendix) actually do satisfy the claimed relations
(\ref{eq:VI.17a})--(\ref{eq:VI.17b}). Especially, these relations must hold also for the
exact (unknown) solution of the eigenvalue problem \emph{in the quadrupole approximation}.

\begin{center}
  \large{\textit{No Dimorphism for $\boldsymbol{\bjz = 0}$}}
\end{center}

The subsequent table for $\jO = 4$, p.~\pageref{tableA}, suggests in combination with the
precedent table on p.~\pageref{table4b} that for $\boldsymbol{\bjz = 0}$ and all values of
$\jO$ (i.\,e. $\jO = $ 0, 1, 2, 3, 4, \ldots) there does occur \emph{no} dimorphism; and
furthermore the angular density $\bgkn(\vartheta)$ (\ref{eq:IV.20}) is always given by
\begin{equation}
\label{eq:A.5}
\bgkn(\vartheta) = \frac{1}{2\pi^2} \cdot \frac{1}{\sin\vartheta} \;,
\end{equation}
so that the angular potential factor $\bgAiii(\vartheta)$ (\ref{eq:V.25a}) must always appear for $\bjz = 0$ as
\begin{equation}
\label{eq:A.6}
\bgAiii(\vartheta) = \frac{3}{8}\,\left[ \cos^2\vartheta - \frac{1}{3} \right] \;.
\end{equation}
Clearly, once the claim (\ref{eq:A.5}) for the angular density $\bgkn(\vartheta)$ is
verified, then the result (\ref{eq:A.6}) for the angular factor $\bgAiii(\vartheta)$ is an
inevitable consequence thereof, see equations (\ref{eq:V.28})--(\ref{eq:V.29}). Therefore,
we merely have to show that the claim (\ref{eq:A.5}) for $\boldsymbol{\bjz = 0}$ is
correct.

However, the desired proof is very simple: First, specify the original angular system (\ref{eq:IV.6a})--(\ref{eq:IV.6b}) down to $\boldsymbol{\bjz = 0}$ in order to find
\begin{subequations}
\begin{align}
\label{eq:A.7a}
\frac{d\,g_R(\vartheta)}{d\vartheta} - \cot\vartheta \cdot g_R(\vartheta) &= \mp \jO\,\sin^2\vartheta \cdot g_S(\vartheta) \\
\label{eq:A.7b}
\frac{d\,g_S(\vartheta)}{d\vartheta} + \cot\vartheta \cdot g_S(\vartheta) &= \pm \jO \cdot \frac{g_R(\vartheta)}{\sin^2\vartheta} \;.
\end{align}
\end{subequations}
This system is to be understood as actually representing two different coupled systems
according to whether one takes the upper (u) or lower (l) signs. Let the solution of the
upper case be denoted by $\ugR(\vartheta)$, $\ugS(\vartheta)$ and the solutions of the
lower case by $\lgR(\vartheta)$, $\lgS(\vartheta)$. Then it is found that the only
difference between both sets of solutions refers to a change of the signs, i.\,e.
\begin{subequations}
\begin{align}
\label{eq:A.8a}
\lgR(\vartheta) &= - \ugR(\vartheta) \\
\label{eq:A.8b}
\lgS(\vartheta) &= \ugS(\vartheta) \;.
\end{align}
\end{subequations}
However, as the very definition of angular density $\bgkn(\vartheta)$ (\ref{eq:IV.20})
says, such a difference in sign does not imply a difference in the angular density itself;
and this then entails also that the angular potential $\bgAiii(\vartheta)$ is invariant
under that change of sign (\ref{eq:A.8a})--(\ref{eq:A.8b}). Such a change may be brought
about also in the first-order system (\ref{eq:A.7a})--(\ref{eq:A.7b}) by simply putting
$\jO \Rightarrow -\jO$. Finally, the Schr\"odinger-like equation (\ref{eq:IV.17}) remains
unchanged by the latter replacement ($\jO \Rightarrow -\jO$); and thus the solution
$\tO(r)$ cannot react to it either, nor the solution $\bgAe\rt$ of the Poisson equation
(\ref{eq:IV.18}). The result is that the physically relevant objects are insensitive to
the change of sign (\ref{eq:A.8a})--(\ref{eq:A.8b}) and thus all the quantum states due to
$\bjz = 0$ are physically unique ($\leadsto$ \emph{no dimorphism}).

Finally, it remains to explain why all these states (with $\bjz = 0$) share the same
angular density $\bgkn(\vartheta)$ (\ref{eq:A.5}), independently of the quantum number
$\jO$. To this end, reconsider the definition of angular density $\bgkn(\vartheta)$
(\ref{eq:IV.20}) and find by use of the present first-order system
(\ref{eq:A.7a})--(\ref{eq:A.7b}) the differential equation
\begin{equation}
\label{eq:A.9}
\frac{d\,\bgkn(\vartheta)}{d\vartheta} + \cot\vartheta \cdot \bgkn(\vartheta) = 0 \;.
\end{equation}
Of course, the solution hereof (being normalized to unity according to the prescription (\ref{eq:IV.24a})) is just identical to our claim (\ref{eq:A.5}).

\begin{center}
  \large{\textit{No Dimorphism for $\boldsymbol{\bjz =} \pm \boldsymbol{\jO}$}}
\end{center}

As suggested by the tabulated cases for $\jO = 2, 3, 4$~p.~\pageref{table4b} there is also
sufficient motivation to suppose that no dimorphism can occur quite generally for $\bjz =
\pm \jO$. This supposition can easily be validated by writing down the eigenvalue
equations for angular momentum (\ref{eq:IV.6a})--(\ref{eq:IV.6b}) in terms of the
eigenvalues $\bjz$ and $\jO$:
\begin{subequations}
\begin{align}
\label{eq:A.10a}
\frac{d\,g_R(\vartheta)}{d\vartheta} - \left( \bjz + 1 \right)\cot\vartheta \cdot g_R(\vartheta) &= \left( \bjz \mp \jO \right) \sin^2\vartheta \cdot g_S(\vartheta) \\
\label{eq:A.10b}
\frac{d\,g_S(\vartheta)}{d\vartheta} + \left( \bjz + 1 \right)\cot\vartheta\cdot g_S(\vartheta) &= \left( \bjz \pm \jO \right) \cdot \frac{g_R(\vartheta)}{\sin^2\vartheta} \;.
\end{align}
\end{subequations}
Now for $\boldsymbol{\bjz = \jO}$ this reduces to either
\begin{subequations}
\begin{align}
\label{eq:A.11a}
\frac{d\,g_R(\vartheta)}{d\vartheta} - \left( \jO + 1 \right) \cot\vartheta \cdot g_R(\vartheta) &= 0 \\
\label{eq:A.11b}
\frac{d\,g_S(\vartheta)}{d\vartheta} + \left( \jO + 1 \right) \cot\vartheta \cdot g_S(\vartheta) &= 2\jO \cdot \frac{g_R(\vartheta)}{\sin^2\vartheta}
\end{align}
\end{subequations}
or to
\begin{subequations}
\begin{align}
\label{eq:A.12a}
\frac{d\,g_R(\vartheta)}{d\vartheta} - \left( \jO + 1 \right) \cot\vartheta \cdot g_R(\vartheta) &= 2\jO\, \sin^2\vartheta \cdot g_S(\vartheta) \\
\label{eq:A.12b}
\frac{d\,g_S(\vartheta)}{d\vartheta} + \left( \jO + 1 \right) \cot\vartheta \cdot g_S(\vartheta) &= 0 \;,
\end{align}
\end{subequations}
according to whether we take the upper or the lower signs in (\ref{eq:A.10a})--(\ref{eq:A.10b}). The dimorphism under consideration can occur if \emph{both} systems (\ref{eq:A.11a})--(\ref{eq:A.11b}) and (\ref{eq:A.12a})--(\ref{eq:A.12b}) do admit normalizable solutions. This, however, can not be true.

Obviously, the critical equations are here (\ref{eq:A.11a}) and (\ref{eq:A.11b}) because these equations cannot admit simultaneously the required non-singular and normalizable solution. Indeed, the general solution of (\ref{eq:A.11a}) is found as
\begin{gather}
\label{eq:A.13}
g_R(\vartheta) = g_*\, \left( \sin\vartheta \right)^{\jO + 1} \;. \\
(g_* = \text{const.}) \nonumber
\end{gather}
It is true, this is an absolutely reasonable solution, but if this is inserted on the right-hand side of the second equation (\ref{eq:A.11b}) we get
\begin{align}
\label{eq:A.14}
\frac{d}{d\vartheta}\,\left[ (\sin\vartheta)^{\jO + 1} \cdot g_S(\vartheta) \right] &= 2 \jO\,(\sin\vartheta)^{\jO - 1} \cdot g_R(\vartheta) \\
&= 2 g_* \, \jO \, (\sin\vartheta)^{2\jO} \nonumber \;,
\end{align}
with the obvious solution being given by
\begin{equation}
\label{eq:A.15}
g_S(\vartheta) = 2\jO \, g_ *\, \frac{1}{(\sin\vartheta)^{\jO + 1}} \int\limits_0^\vartheta d\vartheta'\; (\sin\vartheta')^{2\jO} \;.
\end{equation}
But this solution for the second angular function $g_S(\vartheta)$ is obviously singular for $\vartheta \Rightarrow \pi$ which thus spoils the required normalization condition for the angular density $\bgkn(\vartheta)$ (\ref{eq:IV.24a})! Consequently, we have to reject the first system (\ref{eq:A.11a})--(\ref{eq:A.11b}); and a regular solution can be expected exclusively from the second set of equations (\ref{eq:A.12a})--(\ref{eq:A.12b}).

However, the rejected system (\ref{eq:A.11a})--(\ref{eq:A.11b}) admits also an exceptional case, i.\,e. $\jO = \bjz = 0$. For this case, the second equation (\ref{eq:A.11b}) becomes homogeneous
\begin{equation}
\label{eq:A.16}
\frac{d\,g_S(\vartheta)}{d\vartheta} + \cot\vartheta \cdot g_S(\vartheta) = 0 \;,
\end{equation}
and thus admits the trivial solution $g_S(\vartheta) \equiv 0$. Furthermore, the solution $g_R(\vartheta)$ (\ref{eq:A.13}) becomes specified to
\begin{equation}
\label{eq:A.17}
g_R(\vartheta) = g_* \cdot \sin\vartheta \;,
\end{equation}
and we end up with the angular density
\begin{equation}
\label{eq:A.18}
\bgkn(\vartheta) = \frac{1}{4\pi} \cdot \frac{g_R^2(\vartheta)}{\sin^3\vartheta} = \frac{g_*^2}{4\pi} \cdot \frac{1}{\sin\vartheta} \;,
\end{equation}
or in normalized form, resp.
\begin{equation}
\label{eq:A.19}
\bgkn(\vartheta) = \frac{1}{2\pi^2} \cdot \frac{1}{\sin\vartheta} \;.
\end{equation}
This is just the one single solution listed at the end of the subsequent table for $\jO = 4$, whose general emergence for all values of $\jO$ has been proven by the above arguments below equation (\ref{eq:A.5}).

Thus, there remains to be considered the second system (\ref{eq:A.12a})--(\ref{eq:A.12b}). Here, the second equation (\ref{eq:A.12b}) admits only the trivial solution $g_S(\vartheta) \equiv 0$, and this leaves (\ref{eq:A.12a}) for $g_R(\vartheta)$ also as a homogeneous equation, namely just the precedent case (\ref{eq:A.11a}) whose solution has already been specified by equation (\ref{eq:A.13}). Combining now both solutions $g_S(\vartheta) \equiv 0$ and (\ref{eq:A.13}) yields the angular density $\bgkn(\vartheta)$ as
\begin{equation}
\label{eq:A.20}
\bgkn(\vartheta) = \frac{1}{4\pi}\,\frac{g_R^2(\vartheta)}{\sin^3\vartheta} = \frac{1}{4\pi}\,g_*^2\,\left( \sin\vartheta \right)^{2\jO - 1} \;,
\end{equation}
i.\,e. in normalized form
\begin{equation}
\label{eq:A.21}
\bgkn(\vartheta) = \frac{1}{2\pi^2} \, \frac{\left( 2\jO \right)!!}{\left( 2\jO - 1 \right)!!} \, \left( \sin\vartheta \right)^{2\jO - 1} \;.
\end{equation}
Special cases ($\jO = 2,3$) of this result are found in the table on p.~\pageref{table4c};
for $\jO = 4$ see the table below.

\begin{landscape}
\begin{center}
\label{tableA}
\begin{tabular}{|c|c||c|c|c|c|c|}
\hline
$\jO$ & $\bjz$ & $\bgkn(\vartheta)$ & $\bgAiii(\vartheta)$ & $\rklo{b}{e}_\mathsf{III}$ & $\rklo{b}{f}_\mathsf{III}$ & $\rklo{b}{m}_\mathsf{III}$ \\
\hline\hline
$4$ & $\pm 4$ & $\frac{64}{35\pi^2}\sin^7\vartheta$ & $-\frac{21}{40}\,\left[ \cos^2\vartheta - \frac{1}{3} \right]$ & $\frac{49}{2000}$ & $\frac{147}{1000}$ & $\frac{49}{400}$ \\
\specialrule{1.2pt}{0pt}{0pt}
\multirow{2}{*}{$4$} & \multirow{2}{*}{$\pm 3$} & $\frac{8}{5\pi^2}\sin^5\vartheta$ & $-\frac{15}{32}\,\left[ \cos^2\vartheta - \frac{1}{3} \right]$ & $\frac{5}{256}$ & $\frac{15}{128}$ & $\frac{25}{256}$ \\
\cline{3-7}
 & & $\frac{56}{5\pi^2}\sin^5\vartheta\left( 1 - \frac{48}{49}\sin^2\vartheta \right)$ & $-\frac{21}{160}\,\left[ \cos^2\vartheta - \frac{1}{3} \right]$ & $\frac{49}{32000}$ & $\frac{147}{16000}$ & $\frac{49}{6400}$ \\
\specialrule{1.2pt}{0pt}{0pt}
\multirow{2}{*}{$4$} & \multirow{2}{*}{$\pm 2$} & $\frac{20}{3\pi^2}\sin^3\vartheta\left( 1 - \frac{24}{25}\sin^2\vartheta \right)$ & $0$ & $0$ & $0$ & $0$ \\
\cline{3-7}
 & & $\frac{20}{\pi^2}\,\left(\sin^3\vartheta - \frac{56}{25}\sin^5\vartheta + \frac{32}{25}\sin^7\vartheta \right)$ & $\frac{3}{20}\,\left[ \cos^2\vartheta - 1 \right]$ & $\frac{1}{500}$ & $\frac{3}{250}$ & $\frac{1}{100}$ \\
\specialrule{1.2pt}{0pt}{0pt}
\multirow{2}{*}{$4$} & \multirow{2}{*}{$\pm 1$} & $\frac{6}{\pi^2}\sin\vartheta\left( 1 - \frac{20}{9}\sin^2\vartheta + \frac{4}{3} \sin^4\vartheta \right)$ & $\frac{9}{32}\,\left[ \cos^2\vartheta - \frac{1}{3} \right]$ & $\frac{9}{1280}$ & $\frac{27}{640}$ & $\frac{9}{256}$ \\
\cline{3-7}
 & & $\frac{10}{\pi^2}\sin\vartheta\left( 1 - 4\sin^2\vartheta + \frac{28}{5}\sin^4\vartheta - \frac{64}{25}\sin^6\vartheta \right)$ & $\frac{51}{160}\,\left[ \cos^2\vartheta - \frac{1}{3} \right]$ & $\frac{289}{32000}$ & $\frac{867}{16000}$ & $\frac{289}{6400}$ \\
\specialrule{1.2pt}{0pt}{0pt}
$4$ & $0$ & $\frac{1}{2\pi^2} \cdot \frac{1}{\sin\vartheta}$ & $\frac{3}{8}\,\left[ \cos^2\vartheta - \frac{1}{3} \right]$ & $\frac{1}{80}$ & $\frac{3}{40}$ & $\frac{1}{16}$ \\
\hline
\end{tabular}
\end{center}
\vskip 1cm
\begin{center}
  {\large{\textit{\textbf{Angular Density $\boldsymbol{\bgkn(}\vartheta)$ (\ref{eq:IV.20})\\
        and Angular Potential $\boldsymbol{\bgAiii(\nu)}$  (\ref{eq:V.25a}) for $\boldsymbol{\jO=4}$ }}}}
\end{center}
\end{landscape}

\setcounter{section}{2}
\setcounter{equation}{0}
  \newpage
  \setcounter{section}{2}
  \setcounter{equation}{0}

  \begin{center}
  {\textbf{\Large Appendix \Alph{section}}}\\[1.5em]
  \emph{\textbf{\Large Groundstate ($\boldsymbol{n_\mathcal{O} = 1 \Rightarrow \jO = 0;\ \bjz = 0}$)}}
  \end{center}
  \vspace{2ex}

  The absence of the ortho-dimorphism for both $\bjz = 0$ and $\bjz = \pm \jO$ implies now
  that this effect of dimorphism can not occur especially for $\jO = 0$ and $\jO = 1$. For
  the first case $(\jO = 0)$, the quantum number $\bjz$ can assume exclusively the value
  $\bjz = 0$ (because of $-\jO \leq \bjz \leq \jO$), and for $\bjz = 0$ we already have
  proven that the dimorphism cannot occur (see \textbf{App.A}). Nevertheless, this case
  $\jO = 0$ owns some peculiarities so that it may appear worthwhile to reconsider it in
  some detail.

  First, observe here that the angular eigenvalue system
  (\ref{eq:IV.6a})--(\ref{eq:IV.6b}) becomes decoupled because both constants $\dlO$ and
  $\ddlO$ (\ref{eq:IV.15a})--(\ref{eq:IV.15b}) must vanish together with $\bjz$ and
  $\jO$. Thus, we are left with the following decoupled system
\begin{subequations}
\begin{align}
\label{eq:B.1a}
\frac{d\,g_R(\vartheta)}{d\vartheta} - \cot\vartheta \cdot g_R(\vartheta) &= 0 \\
\label{eq:B.1b}
\frac{d\,g_S(\vartheta)}{d\vartheta} + \cot\vartheta \cdot g_S(\vartheta) &= 0 \;.
\end{align}
\end{subequations}
The solution of the first equation (\ref{eq:B.1a}) is
\begin{equation}
\label{eq:B.2}
g_R(\vartheta) = g_{R,*} \cdot \sin\vartheta \;,
\end{equation}
and similarly, the solution of the second equation (\ref{eq:B.1b}) is obtained as
\begin{equation}
\label{eq:B.3}
g_S(\vartheta) = g_{S,*} \cdot \frac{1}{\sin\vartheta}
\end{equation}
where $g_{R,*}$ and $g_{S,*}$ are the integration constants. But actually, the coupling of
these eigenvalue equations (\ref{eq:IV.6a})--(\ref{eq:IV.6b}) does enforce the presence of
only one integration constant. However, for the present exceptional situation
(i.e.~$\jO=\bjz=0$) we have to admit the existence of two independent integration
constants~$g_{R,*}$ and~$g_{S,*}$ which become restricted by the normalization condition
(\ref{eq:IV.24a}) to the constraint
\begin{equation}
  \label{eq:B.4}
  g_{R,*}^2 + g_{S,*}^2 = \frac{2}{\pi}\ .
\end{equation}
This restriction admits the parametrization by only one constant(~$\gamma_*$, say)
\begin{subequations}
  \begin{align}
    \label{eq:B.5a}
    g_{R,*} &= \sqrt{\frac{2}{\pi}}\cdot\cos\gamma_*\\*
    \label{eq:B.5b}
    g_{S,*} &= \sqrt{\frac{2}{\pi}}\cdot\sin\gamma_*\ ,
  \end{align}
\end{subequations}
so that the solution (\ref{eq:B.2})-(\ref{eq:B.3}) becomes
\begin{subequations}
  \begin{align}
    \label{eq:B.6a}
    g_R(\vartheta) &= \sqrt{\frac{2}{\pi}}\cos\gamma_*\cdot\sin\vartheta\\*
    \label{eq:B.6b}
    g_S(\vartheta) &= \sqrt{\frac{2}{\pi}}\cdot\frac{\sin\gamma_*}{\sin\vartheta}
  \end{align}
\end{subequations}
with the angular density~$\bgkn(\vartheta)$ being found in agreement with the former claim
(\ref{eq:A.19}). As a special case, one recovers here the former solution (\ref{eq:A.17})
for~$\gamma_*=0$.

Concerning now the energy of the \emph{groundstate} of ortho-positronium ($n_\mathcal{O} =
1, \jO = 0, \bjz = 0$), one first observes that the angular density $\bgkn(\vartheta)$
(\ref{eq:A.19}) is the same as for the groundstate of para-positronium~\cite{2}; and
therefore both the first angular potential correction $\bgAiii(\vartheta)$
(\ref{eq:V.25a}) and the constants $\rklo{b}{e}_\mathsf{III}$, $\rklo{b}{f}_\mathsf{III}$
(\ref{eq:VI.16a}), $\rklo{b}{m}_\mathsf{III}$ (\ref{eq:VI.22a}) must then also be the
same, i.\,e.
\begin{subequations}
\begin{align}
\label{eq:B.7a}
\bgAiii(\vartheta) &= \frac{3}{8}\,\left[ \cos^2\vartheta - \frac{1}{3} \right] \\
\label{eq:B.7b}
\rklo{b,p}{e}_\mathsf{III} &= \frac{1}{80} \\
\label{eq:B.7c}
\rklo{b,p}{f}_\mathsf{III} &= \frac{3}{40} \\
\label{eq:B.7d}
\rklo{b,p}{m}_\mathsf{III} &= \frac{1}{16} \;.
\end{align}
\end{subequations}
Furthermore, these numerical results ensure again the validity of the quadrupole equation
(\ref{eq:VII.4}) which then leads us to the gauge field energy $\EReg$
(\ref{eq:VII.17}). This is to be added to the kinetic energy $\Ekin$
(\ref{eq:VI.9a})--(\ref{eq:VI.9b}) for the groundstate ($\jO = 0$)
\begin{gather}
\label{eq:B.8}
\Ekin \Rightarrow \frac{e^2}{2a_B}\,\left( 2a_B\beta \right)^2 \cdot \frac{1}{4(2\nu + 1)} \\
\left( \ekin(\nu) = \frac{1}{4(2\nu + 1)} \right) \nonumber
\end{gather}
in order to obtain for the reduced form of the groundstate energy function, cf. (\ref{eq:VII.22})
\begin{equation}
\label{eq:B.9}
\EEOon(\nu) = -\frac{e^2}{a_B}\,\left( 2\nu + 1 \right)\,\left[ \epot(\nu) + \frac{1}{16}\, \muegiii(\nu) \right]^2
\end{equation}
According to the \emph{principle of minimal energy}, the (approximate) groundstate energy
$\EEOon$ of ortho-positronium is now given by the minimal value of the energy function
$\EEOon(\nu)$ (\ref{eq:B.9}) and is found as
\begin{equation}
\label{eq:B.10}
\EEOon \simeq -7,5378\,\text{[eV]} \;.
\end{equation}
This groundstate energy is unique because the corresponding angular density
$\bgkn(\vartheta)$ turned out to be \emph{unique} despite of the continuous set of
solutions (\ref{eq:B.6a})-(\ref{eq:B.6b}). This confirms the former assertion that the
ortho-dimorphism cannot occur for angular states due to $\bjz = 0$, i.\,e. especially for
$\jO = 0$.

Furthermore, it happens that the ortho-groundstate energy $\EE_\mathcal{O}^{\{0,0\}}$
(\ref{eq:B.10}) is (in the quadrupole approximation) the same as the groundstate energy
$\EE_\mathcal{P}^{\{0,0\}}$ of para-positronium \cite{2}. The reason for this is the
simple fact that all ingredients of the groundstate energy function
$\EE^{\{\mathsf{IV}\}}(\beta, \nu)$ are the same for ortho- and para-positronium,
i.\,e. especially the angular-dependent functions $\bgkn(\vartheta) \equiv
\pgko(\vartheta)$, $\peA^\mathsf{III}(\vartheta) \equiv
\beA^\mathsf{III}(\vartheta),\bAe(r)\equiv\peAo(r)$. The equality of these angular functions
ensures then also the equality of the coefficients
\begin{subequations}
\begin{align}
\label{eq:B.12a}
\pr e_\mathsf{III} &= \br e_\mathsf{III} = \frac{1}{80} \\
\label{eq:B.12b}
\pr f_\mathsf{III} &= \br f_\mathsf{III} = \frac{3}{40} \\
\label{eq:B.12c}
\pr m_\mathsf{III} &= \br m_\mathsf{III} = \frac{1}{16} \;.
\end{align}
\end{subequations}
And finally, the equality of the angular functions plus that of the associated
coefficients admits then also the equality of the corresponding groundstate energies,
i.\,e.
\begin{equation}
\label{eq:B.13}
\EEOon \equiv \EE_\mathcal{P}^{\{0,0\}} = -7,5378\, \text{[eV]} \;,
\end{equation}
cf. the process of partial extremalization (\ref{eq:VII.18})--(\ref{eq:VII.22}). Indeed by
means of the above mentioned equalities, both spectral functions $S^{\{0,0\}}(\nu)$
become identified for the groundstate ($\nO = n_\mathcal{P} \equiv \elp + 1=1$, $\bjz = \pjz
\equiv \elz = 0$)
\begin{equation}
\label{eq:B.14}
S_\mathcal{O}^{\{0,0\}}(\nu) \equiv S_\mathcal{P}^{\{0,0\}}(\nu) = 4(2\nu + 1)\,\left[
  \epot(\nu) + \frac{1}{16}\cdot\muegiii(\nu) \right]^2 \;,
\end{equation}
see the precedent paper~\cite{2} for the corresponding treatment of para-positronium.

\setcounter{section}{3}
\setcounter{equation}{0}
  \begin{center}
  {\textbf{\Large Appendix \Alph{section}}}\\[1.5em]
  \emph{\textbf{\Large No Dimorphism for the first Excited State ($\boldsymbol{\nO = 2}$)}}
  \end{center}
  \vspace{2ex}

  Since the ortho-dimorphism can occur only if the quantum number $\bjz$ is in the range
  $0 < \left| \bjz \right| < \jO$, that phenomenon of dimorphism can surely not emerge for
  the first excited state $\nO = 2$ ($\Rightarrow \jO = 0, 1$; $\bjz = 0, \pm
  1$). However, the mathematical mechanism for the exclusion is quite
  different. Whereas for $\boldsymbol{\bjz = 0}$ both configurations $\{1, -1\}$ and $\{-1, 1\}$ of
  $\{\dlO, \ddlO\}$ are allowed but yield one and the same physical state, in the case
  $\boldsymbol{\bjz = 1}$ one has to reject the configuration $\{2, 0\}$ and to resort to $\{0, 2\}$;
  and analogously for $\boldsymbol{\bjz = -1}$ one can conversely admit only the configuration $\{2,
  0\}$ and has to reject~$\{0, 2\}$.

\begin{center}
$\boldsymbol{\jO = 1,\ \bjz = 0}$
\end{center}

For $\{\jO, \bjz\} = \{1,0\}$ one first deduces from equation (\ref{eq:IV.36}) that $G_S(x)$ must be constant
\begin{equation}
\label{eq:C.1}
G_S(x) = \sigma_0 = \text{const.}
\end{equation}
Indeed, the second-order equation (\ref{eq:IV.29b}) is satisfied for this conclusion
(\ref{eq:C.1}). The corresponding angular function $g_S(\vartheta)$ (\ref{eq:IV.28c}) must
then also be a constant
\begin{equation}
\label{eq:C.2}
g_S(\vartheta) = \sigma_0 \;,
\end{equation}
and for the associated angular function $g_R(\vartheta)$ one deduces from the second
first-order equation (\ref{eq:IV.6b})
\begin{equation}
\label{eq:C.3}
g_R(\vartheta) = \frac{\sigma_0}{\dlO}\,\sin\vartheta\cos\vartheta \;.
\end{equation}
But both possibilities $\dlO = \pm 1$ (\ref{eq:IV.15a}) yield here merely a change of sign
which may be compensated for by the integration constant $\sigma_0$. Thus, one obtains for
the (normalized) angular density $\bgkn(\vartheta)$ (\ref{eq:IV.20}) just the
\emph{unambigous} result (\ref{eq:A.19}), as claimed above.

\begin{center}
$\boldsymbol{\jO = 1,\ \bjz = 1}$
\end{center}

The question of unambiguity is somewhat different for $\bjz = \pm 1$ (and still $\jO =
1$). First, one concludes again from equation (\ref{eq:IV.35}) that the angular
eigenfunction $g_R(\vartheta)$ must look as follows
\begin{equation}
\label{eq:C.4}
g_R(\vartheta) = \rho_2 \cdot \sin^2\vartheta \;.
\end{equation}
If this solution is substituted in the first eigenvalue equation (\ref{eq:IV.6a}), one finds
\begin{equation}
\label{eq:C.5}
\ddlO \cdot g_S(\vartheta) \equiv 0
\end{equation}
so that either $\ddlO$ or $g_S(\vartheta)$ must vanish. Furthermore, the solution (\ref{eq:C.4}) for $g_R(\vartheta)$ recasts the second eigenvalue equation (\ref{eq:IV.6b}) to the following form
\begin{equation}
\label{eq:C.6}
\frac{d\,g_S(\vartheta)}{d\vartheta} + 2\cot\vartheta \cdot g_S(\vartheta) = \dlO \cdot \rho_2
\end{equation}
whose solution is
\begin{equation}
\label{eq:C.7}
g_S(\vartheta) = \rho_2\,\left\{ \frac{\vartheta}{\sin^2\vartheta} - \frac{1}{2}\,\frac{\sin(2\vartheta)}{\sin^2\vartheta} \right\} \;.
\end{equation}
However, this solution must be rejected because it is singular at $\vartheta = \pi$. As a consequence, we have to demand $\dlO = 0$, so that the equation (\ref{eq:C.6}) for $g_S(\vartheta)$ admits the trivial solution in place of the former $g_S(\vartheta)$ (\ref{eq:C.7}):
\begin{equation}
\label{eq:C.8}
g_S(\vartheta) \equiv 0 \;.
\end{equation}
This result says that the configuration $\dlO = 2$, $\ddlO = 0$ must be rejected which
leaves us with $\dlO = 0$, $\ddlO = 2$. Thus, both angular eigenfunctions do generate the
\emph{unambiguous} (normalized) density $\bgkn(\vartheta)$ in the form
\begin{equation}
\label{eq:C.9}
\bgkn(\vartheta) = \frac{\sin\vartheta}{\pi^2}
\end{equation}
which validates again the above mentioned claim of \emph{unambiguity}.

\newpage
\begin{center}
$\boldsymbol{\jO = 1,\ \bjz = -1}$
\end{center}

The remaining case $\bjz = -1$ (and $\jO = 1$) can be settled now by a very brief
argument; namely by simply evoking the symmetry replacements
(\ref{eq:IV.47a})-(\ref{eq:IV.47b}). This transcribes the present solution for $\bjz = 1$
to the corresponding \emph{unambiguous} solution for $\bjz = -1$:
\begin{subequations}
\begin{align}
\label{eq:C.10a}
\dlO &= 2 \\
\label{eq:C.10b}
\ddlO &= 0 \\
\label{eq:C.10c}
g_R(\vartheta) &\equiv 0 \\
\label{eq:C.10d}
g_S(\vartheta) &= \sqrt{\frac{4}{\pi}} = \text{const.}
\end{align}
\end{subequations}
Indeed it is easy to see that this is a solution of both eigenvalue equations
(\ref{eq:IV.6a})--(\ref{eq:IV.6b}) with the angular density $\bgkn(\vartheta)$
(\ref{eq:IV.20}) coinciding just with the former result (\ref{eq:C.9}) for $\bjz =
1$. Thus, both cases $\bjz = 0$ and $\bjz = \pm 1$ for $\jO = 1$ are actually not
subjected to the ortho-dimorphism!

Concerning now the energy $\EE_\mathcal{O}^{\{1, \pm 1\}}$ of the excited states due to
$\nO = 2$, $\jO = 1$, $\bjz = \pm 1$, one can apply again the method of partial
extremalization which is based upon the spectral function $S_\mathcal{O}^{\{j\}}(\nu)$
(\ref{eq:VII.19}). For the present case ($\jO = 1$, $\bjz = \pm 1$, $\nO = 2$) the
spectral function adopts the following form
\begin{equation}
\label{eq:C.11}
S_\mathcal{O}^{\{1, \pm 1\}}(\nu) = \frac{\left[ \epot(\nu) + \frac{1}{64}\,\muegiii(\nu) \right]^2}{\frac{1}{2\nu + 1}\,\left( \frac{1}{4} + \frac{1}{2\nu} \right)} \;.
\end{equation}
This is the same spectral function as was found for the corresponding first excited state
of para-positronium ($\elz = \elp = 1$), see ref.~\cite{2}, equation (IV.161). Therefore
the present \emph{unambiguous (!)} ortho-state with quantum numbers $\{\nO = 2$, $\jO =
1$, $\bjz = \pm 1\}$ has the same RST binding energy as the para-state $\{n_\mathcal{P} =
2$, $j_\mathcal{P} \equiv \elp = 1$, $\pjz \equiv \elz = \pm 1\}$, namely~\cite{2}
\begin{equation}
\label{eq:C.12}
\EE_\mathcal{O}^{\{1, \pm 1\}} = -1,57527\ldots \text{[eV]} \;.
\end{equation}

\setcounter{section}{4}
\setcounter{equation}{0}
  \begin{center}
  {\textbf{\Large Appendix \Alph{section}}}\\[1.5em]
  \emph{\textbf{\Large Dimorphism vs. Elimination of $\boldsymbol{j_z}$-Degeneracy}}
  \end{center}
  \vspace{2ex}

  It is important to observe that the occurence of the ortho-dimorphism and the effect of
  degeneracy elimination are two separate things which are independent of each other and
  have different origins. Nevertheless, they may occur in a combined way. For such a
  combined situation, there naturally arises the question of magnitude of those energy
  differences being induced by both kinds of level splitting (i.\,e. degeneracy
  elimination vs. dimorphism).

  In order to elucidate this question a little bit more, it may be instructive to consider
  a somewhat larger value of the quantum number $\jO$ ($\jO = 4$, say) where both effects
  can occur simultaneously. Here, the quantum number $\bjz$ can adopt nine values
  (i.\,e. $-4 \leq \bjz \leq 4$) which, at first glance, would have to be linked to five
  different angular densities $\bgkn(\vartheta)$, namely any one density being associated
  with any one value of $|\bjz|$ (= 0, 1, 2, 3, 4); and therefore one expects the
  occurence of five different energy levels due to $\jO = 4$ (yielding the electrostatic
  fine structure). But the effect of dimorphism consists now in the additional splitting
  of three of these five energy levels (i.\,e. $|\bjz| = 1,2,3$) so that two of them
  (i.\,e. those due to $|\bjz| = 0,4$) remain unsplitted! Thus, the splitting effect says
  that there should ultimately be left over \emph{eight} different energy levels for the
  quantum number $\jO = 4$ (i.\,e. \emph{one} level for any $|\bjz| = 0,4$ and \emph{two}
  levels for any $|\bjz| = 1,2,3$). The subsequent table presents a collection of the
  actual results.

\begin{landscape}
\begin{center}
\label{tableDa}
$\boldsymbol{\jO = 4}$
\begin{tabular}{|c||c|c|c|}
%\specialrule{1.2pt}{0pt}{0pt}
\hline
$\bjz$ & $g_R(\vartheta)$ & $g_S(\vartheta)$ & $\bgkn(\vartheta)$ (\ref{eq:IV.20}) \\
\hline\hline
\multirow{2}{*}{$4$} & \multirow{2}{*}{$\frac{16}{\sqrt{35\pi}} \cdot \sin^5\vartheta$} & \multirow{2}{*}{$0$} & \multirow{2}{*}{$\frac{64}{35\pi^2} \cdot \sin^7\vartheta$} \\
 & & & \\
\specialrule{2.0pt}{0pt}{0pt}
\multirow{2}{*}{$3$} & $\sqrt{\frac{32}{5\pi}} \cdot \sin^4\vartheta\cos\vartheta$ & $\sqrt{\frac{32}{5\pi}} \cdot \sin^3\vartheta$ & $\frac{8}{5\pi^2} \cdot \sin^5\vartheta$ \\
\cline{2-4}
 & $-7\,\sqrt{\frac{32}{35\pi}} \cdot \sin^4\vartheta\cos\vartheta$ & $\sqrt{\frac{32}{35\pi}} \cdot \sin^3\vartheta$ & $\frac{56}{5\pi^2} \cdot \sin^5\vartheta\,\left( 1 - \frac{48}{49}\,\sin^2\vartheta \right)$ \\
\specialrule{2.0pt}{0pt}{0pt}
\multirow{2}{*}{$2$} & $\sqrt{\frac{80}{3\pi}} \cdot \sin^3\vartheta\,\left( 1 - \frac{6}{5}\,\sin^2\vartheta \right)$ & $\frac{6}{5}\sqrt{\frac{80}{3\pi}} \cdot \sin^2\vartheta \cos\vartheta$ & $\frac{4}{\pi^2} \cdot \sin^3\vartheta\,\left( \frac{5}{3} - \frac{8}{5}\,\sin^2\vartheta \right)$ \\
\cline{2-4}
 & $\sqrt{\frac{80}{\pi}} \cdot \sin^3\vartheta\,\left( 1 - \frac{6}{5}\,\sin^2\vartheta \right)$ & $\frac{-8}{\sqrt{5\pi}} \cdot \sin^2\vartheta\cos\vartheta$ & $\frac{20}{\pi^2}\,\sin^3\vartheta\,\left( 1 - \frac{56}{25}\,\sin^2\vartheta + \frac{32}{25}\,\sin^4\vartheta \right)$ \\
\specialrule{2.0pt}{0pt}{0pt}
\multirow{2}{*}{$1$} & $6\,\sqrt{\frac{2}{3\pi}} \cdot \sin^2\vartheta\cos\vartheta\,\left( 1 - 2\,\sin^2\vartheta \right)$ & $10\,\sqrt{\frac{2}{3\pi}}\,\sin\vartheta\,\left( 1 - \frac{6}{5}\,\sin^2\vartheta \right)$ & $\frac{6}{\pi^2} \cdot \sin\vartheta\,\left( 1 - \frac{20}{9}\,\sin^2\vartheta + \frac{4}{3}\,\sin^4\vartheta \right)$ \\
\cline{2-4}
 & $-2\,\sqrt{\frac{10}{\pi}} \cdot \sin^2\vartheta\cos\vartheta\,\left( 1 - 2\,\sin^2\vartheta \right)$ & $2\,\sqrt{\frac{10}{\pi}}\,\sin\vartheta\,\left( 1 - \frac{6}{5}\,\sin^2\vartheta \right)$ & $\frac{10}{\pi^2} \cdot \sin\vartheta\,\left( 1 - 4\,\sin^2\vartheta + \frac{28}{5}\,\sin^4\vartheta - \frac{64}{25}\,\sin^6\vartheta \right)$ \\
\specialrule{2.0pt}{0pt}{0pt}
\multirow{2}{*}{$0$} & \multirow{2}{*}{$\sqrt{\frac{2}{\pi}} \cdot \sin\vartheta\,\left( 1 - 8\,\sin^2\vartheta + 8\,\sin^4\vartheta \right)$} & \multirow{2}{*}{$\sqrt{\frac{32}{\pi}} \cdot \cos\vartheta\,\left( 1 - 2\,\sin^2\vartheta \right)$} & \multirow{2}{*}{$\frac{1}{2\pi^2} \cdot \frac{1}{\sin\vartheta}$} \\
 & & & \\
\specialrule{2.0pt}{0pt}{0pt}
\multirow{2}{*}{$-1$} & $-10\,\sqrt{\frac{2}{3\pi}} \cdot \sin^3\vartheta \,\left( 1 - \frac{6}{5}\,\sin^2\vartheta \right)$ & $6\,\sqrt{\frac{2}{3\pi}}\,\cos\vartheta\,\left( 1 - 2\,\sin^2\vartheta \right)$ & $\frac{6}{\pi^2} \cdot \sin\vartheta\,\left( 1 - \frac{20}{9}\,\sin^2\vartheta + \frac{4}{3}\,\sin^4\vartheta \right)$ \\
\cline{2-4}
 & $-2\,\sqrt{\frac{10}{\pi}} \cdot \sin^3\vartheta \,\left( 1 - \frac{6}{5}\,\sin^2\vartheta \right)$ & $-2\,\sqrt{\frac{10}{\pi}} \cdot \cos\vartheta\,\left( 1 - 2\,\sin^2\vartheta \right)$ & $\frac{10}{\pi^2} \cdot \sin\vartheta\,\left( 1 - 4\,\sin^2\vartheta + \frac{28}{5}\,\sin^4\vartheta - \frac{64}{25}\,\sin^6\vartheta \right)$ \\
\specialrule{2.0pt}{0pt}{0pt}
\multirow{2}{*}{$-2$} & $-\frac{6}{5}\,\sqrt{\frac{80}{3\pi}} \cdot \sin^4\vartheta\cos\vartheta$ & $\sqrt{\frac{80}{3\pi}} \cdot \sin\vartheta\,\left( 1 - \frac{6}{5}\,\sin^2\vartheta \right)$ & $\frac{4}{\pi^2} \cdot \sin^3\vartheta\,\left( \frac{5}{3} - \frac{8}{5}\,\sin^2\vartheta \right)$ \\
\cline{2-4}
 & $\frac{2}{5}\,\sqrt{\frac{80}{\pi}} \cdot \sin^4\vartheta\cos\vartheta$ & $\sqrt{\frac{80}{\pi}} \cdot \sin\vartheta\,\left( 1 - \frac{6}{5}\,\sin^2\vartheta \right)$ & $\frac{20}{\pi^2} \cdot \sin^3\vartheta\,\left( 1 - \frac{56}{25}\,\sin^2\vartheta + \frac{32}{25}\,\sin^4\vartheta \right)$ \\
\specialrule{2.0pt}{0pt}{0pt}
\multirow{2}{*}{$-3$} & $\sqrt{\frac{32}{5\pi}} \cdot \sin^5\vartheta$ & $-\sqrt{\frac{32}{5\pi}} \cdot \sin^2\vartheta\cos\vartheta$ & $\frac{8}{5\pi^2} \cdot \sin^5\vartheta$ \\
\cline{2-4}
 & $\sqrt{\frac{32}{35\pi}} \cdot \sin^5\vartheta$ & $7\,\sqrt{\frac{32}{35\pi}} \cdot \sin^2\vartheta\cos\vartheta$ & $\frac{56}{5\pi^2} \cdot \sin^5\vartheta\,\left( 1 - \frac{48}{49}\,\sin^2\vartheta \right)$ \\
\specialrule{2.0pt}{0pt}{0pt}
\multirow{2}{*}{$-4$} & \multirow{2}{*}{$0$} & \multirow{2}{*}{$\frac{16}{\sqrt{35\pi}} \cdot \sin^3\vartheta$} & \multirow{2}{*}{$\frac{64}{35\pi^2} \cdot \sin^7\vartheta$} \\
 & & & \\
\specialrule{2.0pt}{0pt}{0pt}
\end{tabular}
\end{center}
\end{landscape}

This table demonstrates clearly the symmetries of the ortho-spectrum:
\begin{enumerate}
\item[\textbf{(i)}] the transition from $\bjz$ to $-\bjz$ leaves invariant the angular density
  $\bgkn(\vartheta)$ (\ref{eq:IV.20}), but not the angular functions $g_R(\vartheta)$ and
  $g_S(\vartheta)$
\item[\textbf{(ii)}] concerning the angular functions $g_R(\vartheta)$ and
  $g_S(\vartheta)$ themselves, the transition from $\bjz$ to $-\bjz$ rearranges them
  according to the recipe (\ref{eq:IV.47a})-(\ref{eq:IV.47b}), apart eventually from an
  overall change of sign (i.\,e. $g_R(\vartheta) \Rightarrow -g_R(\vartheta)$;
  $g_S(\vartheta) \Rightarrow -g_S(\vartheta)$). Observe that the latter is admitted by
  the eigenvalue system (\ref{eq:IV.6a})--(\ref{eq:IV.6b})
\item[\textbf{(iii)}] observe also that the symmetry rearrangement
  (\ref{eq:IV.47a})-(\ref{eq:IV.47b}) generates a second solution for $\boldsymbol{\bjz =
    0}$, namely
\begin{subequations}
\begin{align}
\label{eq:D.1a}
g_R(\vartheta) &= \sqrt{\frac{32}{\pi}} \cdot \sin^2\vartheta\cos\vartheta\,\left( 1 - 2\,\sin^2\vartheta \right) \\
\label{eq:D.1b}
g_S(\vartheta) &= -\sqrt{\frac{2}{\pi}} \cdot \frac{1}{\sin\vartheta}\,\left( 1 - 8\,\sin^2\vartheta + 8\,\sin^4\vartheta \right) \;,
\end{align}
\end{subequations}
but this leads us to the same angular density $\bgkn(\vartheta)$ and therefore yields no further level splitting.
\end{enumerate}

For a preliminary synopsis of those cooperative effects of degeneracy elimination plus
dimorphism it is instructive to collect the results for the lowest energy levels
(i.\,e. for principal quantum number $n \leq 5$) in a further table, see below. Here it is
especially interesting to compare the RST results of the spherically symmetric
approximation, the results of the para/ortho dichotomy, and those of the dimorphism to the
conventional results $E_\text{C}^{(n)}$ (\ref{eq:I.18}). Concerning the effect of
degeneracy, the RST spherically symmetric approximation $\EEt^{[n]}$ (\emph{fourth
  column}) respects the conventional degree of degeneracy (\ref{eq:I.18}) but yields
numerically only a moderate approximation to those conventional results $E_\text{C}^{(n)}$
(\emph{third column}).

\begin{landscape}
\begin{center}
\label{tableDb}
\begin{tabular}{|c||c|c|c|c|c|c|c|}
\hline
\multirow{2}{*}{$n_\mathcal{O,P} = j_\mathcal{O,P}+1$} & \multirow{2}{*}{$\bjz = \pjz$} & $\Ea{E}{n}{C}\,\text{[eV]}$ & $\EE_T^{[n]}\,\text{[eV]}$ & $\pr m_\mathsf{III}$ & $\EE_\mathcal{P}^{\{j\}}\,\text{[eV]}$ & $\br m_\mathsf{III}$ & $\EE_\mathcal{O}^{\{j\}}\,\text{[eV]}$ \\
& &  (\ref{eq:I.18}) & \cite{2} & \cite{2} & \cite{2} &   (\ref{eq:VI.22a}) & (\ref{eq:VII.22}) \\
\hline\hline
\multirow{2}{*}{1} & \multirow{2}{*}{0} & \multirow{2}{*}{$-6{,}80290\ldots$} & \multirow{2}{*}{${-7,2311}\ldots$} & \multirow{2}{*}{$\frac{1}{16}$} & \multirow{2}{*}{$-7,5378\ldots$} & \multirow{2}{*}{$\frac{1}{16}$} & \multirow{2}{*}{${-7,5378}\ldots$} \\
 & & & & & & & \\
\hline
\multirow{2}{*}{2} & \multirow{2}{*}{$\pm 1$} & \multirow{2}{*}{$-1{,}70072\ldots$} & \multirow{2}{*}{${-1,5510}\ldots$} & \multirow{2}{*}{$\frac{1}{64}$} & \multirow{2}{*}{$-1{,}5752\ldots$} & \multirow{2}{*}{$\frac{1}{64}$} & \multirow{2}{*}{$-1{,}5752\ldots$} \\
 & & & & & & & \\
\hline
\multirow{2}{*}{3} & \multirow{2}{*}{$\pm 1$} & \multirow{2}{*}{$-0{,}75588\ldots$} & \multirow{2}{*}{${ -0,6692}\ldots$} & \multirow{2}{*}{$\frac{1}{64}$} & \multirow{2}{*}{$-0{,}68113\ldots$} & \multirow{2}{*}{$\frac{1}{64}$} & \multirow{2}{*}{$-0{,}6811\ldots$} \\
 & & & & & & & \\
\hline
\multirow{2}{*}{4} & \multirow{2}{*}{$\pm 1$} & \multirow{2}{*}{$-0{,}42518\ldots$} & \multirow{2}{*}{$-0,37369\ldots$} & \multirow{2}{*}{$\frac{9}{256}$} & \multirow{2}{*}{$-0,3900\ldots$} & $\frac{1}{64}$ & ${-0,3808}\ldots$ \\
\cline{7-8}
 & & & & & & $\frac{9}{256}$ & ${-0,3900}\ldots$ \\
\hline
\multirow{2}{*}{5} & \multirow{2}{*}{$\pm 1$} & \multirow{2}{*}{$-0{,}27211\ldots$} & \multirow{2}{*}{${-0,23909}\ldots$} & \multirow{2}{*}{$\frac{289}{6400}$} & \multirow{2}{*}{$-0,2533\ldots$} & $\frac{9}{256}$ & ${-0,2501}\ldots$ \\
\cline{7-8}
 & & & & & & $\frac{289}{6400}$ & $-0{,}2533\ldots$ \\
\hline
\end{tabular}
\end{center}
All RST predictions \textbf{in the quadrupole approximation} are based upon the method of
partial extremalizing, i.\,e. maximalization of the spectral function
$S_\mathcal{O,P}^{\{j\}}(\nu)$
\begin{equation}
\label{eq:D.2}
S_\mathcal{O,P}^{\{j\}}(\nu) = \frac{\left[ \epot(\nu) + \bpr m_\mathsf{III} \cdot \muegiii(\nu) \right]^2}{\frac{1}{2\nu + 1}\,\left( \frac{1}{4} + \frac{ j_{\sigma,p}^2}{2\nu} \right)} \;.
\end{equation}
Whenever both quantum numbers of para- and ortho-positronium do agree (i.e.\
$\jO=j_\wp;\bjz= \pjz\Leftrightarrow \gklo{b}{m}_\mathsf{III}=\gklo{p}{m}_\mathsf{III}$),
the corresponding energies are identical
($\rogk{\mathbb{E}}{j}_\mathcal{O}=\rogk{\mathbb{E}}{j}_\wp$). The ortho-dimorphism arises
because for the same quantum numbers $\jO=j_\wp$ and $\bjz=\pjz$ there can sometimes exist
two values for $\rklo{b}{m}_\mathsf{III}$ (but not for $\rklo{p}{m}_\mathsf{III}$), see
the last two lines of the table,~$n=4,5$.
\end{landscape}

The functions~$\epot(\nu)$ (\ref{eq:VI.33}) and~$\rogk{\mu}{e}_\mathsf{III}$
(\ref{eq:VII.15}), which enter the numerator of the spectral function (\ref{eq:D.2}), may
also be written down in a closed analytic form:

\begin{subequations}
  \begin{align}
  \label{eq:D.3a}
\epot(\nu) &=  \left( 1-{\frac {\Gamma  \left( 4\,\nu+3 \right) }{{2}^{4\,\nu+2}
 \left[ \Gamma  \left( 2\,\nu+2 \right)  \right] ^{2}}} \right)\cdot 
 \left( 2\,\nu+1 \right)^{-1} \\*[5mm]
\label{eq:D.3b}
\rogk{\mu}{e}_\mathsf{III}(\nu) &= \frac{2\nu^2+5\nu+3}{\nu(4\nu^2-1)} -
\frac{10\nu^2+15\nu+6}{\nu(4\nu^2-1)}\cdot
\frac{\Gamma(2\nu+3/2)}{\sqrt{\pi}\cdot\Gamma(2\nu+2)}\ .\\ \nonumber
  \end{align}
\end{subequations}

\begin{center}
  \large{\textit{Lowering of Groundstate Energy}}
\end{center}

There is some curious effect with the groundstate ($n = 1$): whereas all the RST
predictions $\EEt^{[n]}$, $\EEP^{\{j\}}$, $\EE_\mathcal{O}^{\{j\}}$ for the groundstate
energy ($n = n_\mathcal{P} = \nO = 1$) are \emph{lower} than their conventional
counterpart $E_\text{conv}^{(1)}$ (third column), the excited states ($n > 1$) are
equipped by RST with \emph{higher} energy. For instance, for $\jO = j_\mathcal{P} = 2$,
$\bjz = \pjz = \pm 1$ the conventional prediction is $E_\text{conv}^{(3)}$ =
-0{,}75588\ldots\,\text{[eV]} whereas the RST prediction is $\EE_\mathcal{O}^{\{2,1\}} =
\EE_\mathcal{P}^{\{2,1\}} = -0{,}68113\,\text{[eV]}$. Thus one may expect that better
trial configurations will further lower the RST energy of the excited states and thus
shift it towards the conventional predictions. Of course, it is highly desirable to know
the \emph{exact} RST solutions in order to see more clearly how close the RST predictions
do approach the conventional results. But those ``better'' trial configurations will then
rather \emph{lower the groundstate} prediction then raise it, quite in agreement with the
true spirit of the principle of \emph{minimal} energy!

\begin{center}
  \large{\textit{No Dimorphism for $\jO = 2$}}
\end{center}

The next effect to be discussed concerns the unexpected missing of the dimorphism for the
level being specified by the quantum numbers $\nO = 3$, $\jO = 2$, $\pjz = \bjz = \pm
1$. Indeed for this value (n = 3) of the principal quantum number $\nO$, the ortho-number
$\jO$ can adopt the values $\jO = 0, 1, 2$ where it is clear that for the possibilities
$\jO = 0$ ($\Rightarrow \bjz = 0$), $\jO = 1$, and $\jO = |\bjz| = 2$ the ortho-dimorphism
cannot occur (see \textbf{App.A}+\textbf{B}). But for $\jO = 2$ and $|\bjz| = 1$ the
dimorphism could principally occur; nevertheless it ``incidentally'' does not occur in the
present quadrupole approximation! What is the reason for this? It seems near at hand that
this is the very logical structure of the quadrupole approximation itself.

In order to see this more clearly, reconsider the spectral function $S_\mathcal{O,
  P}^{\{j\}}(\nu)$ (\ref{eq:D.2}) whose general form does apply to both
ortho-($\mathcal{O}$) and para-($\mathcal{P}$) positronium. Since the RST energy
predictions are obtained by extremalizing that spectral function
$S_\mathcal{O,P}^{\{j\}}(\nu)$ with respect to the continuous variable $\nu$, the quantum
numbers $\jO$ and $\bjz$ $(\Rightarrow\!\rklo{b,p}{m}_\mathsf{III})$ being held fixed, the
corresponding energy predictions can be different not only if at least one of the two
quantum numbers $\jO$ and $\bjz$ is different, but also if for the same pair $\jO, \bjz$
the parameter $\br m_\mathsf{III}$ (\ref{eq:VI.22a}) can adopt more than one definite
value. Recalling here the route back from $\br m_\mathsf{III}$ to the angular density
$\bgkn(\vartheta)$ via the relations (\ref{eq:A.4}), (\ref{eq:A.2}) and (\ref{eq:V.29}):
\begin{equation}
\label{eq:D.4}
\br m_\mathsf{III} = \left[ \frac{1 - 3\int d\Omega\;\cos^2\vartheta\,\bgkn(\vartheta)}{2} \right]^2 \;,
\end{equation}
the two possible values for $\br m_\mathsf{III}$ are seen to become induced by the two possible
angular densities $\bgkn(\vartheta)$ due to the same pair $\jO, \bjz$. But the latter two
possibilities for $\bgkn(\vartheta)$ are of course induced by the two possibilities for
$g_R(\vartheta)$ and $g_S(\vartheta)$ (table on p.~\pageref{table4b}) according to the prescription
(\ref{eq:IV.20}). Especially for the present case $\jO = 2$, $\bjz = \pm 1$ we find the
following two angular densities
\begin{subnumcases}{\bgkn(\vartheta) = }
\label{eq:D.5a}
\ \frac{1}{\pi^2} \cdot \sin\vartheta \\
\label{eq:D.5b}
\ \frac{3}{\pi^2} \cdot \sin\vartheta\,\left( 1 - \frac{8}{9}\,\sin^2\vartheta \right) \;, 
\end{subnumcases}
see the tables on p.~\pageref{table4c} and \pageref{table5}. And the crucial point with
these two densities (\ref{eq:D.5a})-(\ref{eq:D.5b}) is now that their substitution in the
prescription (\ref{eq:D.4}) for $\rklo{b,p}{m}_\mathsf{III}$ yields ``incidentally''
\emph{one and the same} mass-equivalent parameter $\br m_\mathsf{III}$:
\begin{equation}
\label{eq:D.6}
\br m_\mathsf{III} = \frac{1}{64} \;.
\end{equation}
But clearly, when there is only one parameter $\br m_\mathsf{III}$ at hand for given values of $\jO$
and $\bjz$ the associated spectral function $S_\mathcal{O}^{\{j\}}(\nu)$ is unique and
provides us with one unique maximum, i.\,e. we end up with one single energy level. This
is the reason why there is only one energy level associated with the quantum numbers $\jO
= 2$, $\bjz = \pm 1$.

\begin{center}
  \large{\textit{Equality of Ortho- and Para-Levels}}
\end{center}

Even when the ortho-dimorphism can occur (e.\,g. for the combinations $\jO = 3, 4$ and
$\bjz = \pm 1$, see the table on p.~\pageref{tableDb}), then the ortho- and para-levels
are not necessarily different in the quadrupole approximation. For instance, take $\jO =
j_\mathcal{P} = 3$ together with $\bjz = \pm 1$ and read from the table on
p.~\pageref{tableDb} that one of the two ortho-parameters ($\br m_\mathsf{III}$) of a
dimorphic pair does agree with the corresponding para-parameter $\pr m_\mathsf{III}$,
e.\,g. for $\jO = j_\mathcal{P} = 3$:
\begin{subequations}
\begin{align}
\label{eq:D.7a}
\jO = 3,\;\bjz = \pm 1 &\Longrightarrow \begin{cases} \ \br m_\mathsf{III} = \frac{1}{64} \\
\ \br m_\mathsf{III} = \frac{9}{256} \end{cases} \\
\label{eq:D.7b}
j_\mathcal{P} = 3,\;\pjz = \pm 1 &\Longrightarrow \pr m_\mathsf{III} = \frac{9}{256} \;,
\end{align}
\end{subequations}
or similarly for $\jO = j_\mathcal{P} = 4$:
\begin{subequations}
\begin{align}
\label{eq:D.8a}
\jO = 4,\;\bjz = \pm 1 &\Longrightarrow \begin{cases} \ \br m_\mathsf{III} = \frac{9}{256} \\
\ \br m_\mathsf{III} = \frac{289}{6400} \end{cases} \\
\label{eq:D.8b}
j_\mathcal{P} = 4,\;\pjz = \pm 1 &\Longrightarrow \pr m_\mathsf{III} = \frac{289}{6400} \;.
\end{align}
\end{subequations}
Consequently, one of the two spectral ortho-functions $S_\mathcal{O}^{\{j\}}(\nu)$ must
agree with the corresponding para-function $S_\mathcal{P}^{\{j\}}(\nu)$ and therefore the
corresponding ortho- and para-levels must also agree, i.\,e. for $\jO = j_\mathcal{P} =
3$, $(\nO = n_\mathcal{P} = 4)$:
\begin{equation}
\label{eq:D.9}
\EE_\mathcal{O}^{\{3,1\}} \Longrightarrow
\begin{cases}
\ \ddot{\EE}_\mathcal{O}^{\{3,1\}} = -0{,}3808\ldots\,\text{[eV]}  \\
\ \dot{\EE}_\mathcal{O}^{\{3,1\}} = -0{,}3900\ldots\,\text{[eV]} = \EE_\mathcal{P}^{\{3,1\}}\;,
\end{cases}
\end{equation}
and analogously for $\jO = j_\mathcal{P} = 4$, $(\nO = n_\mathcal{P} = 5)$:
\begin{equation}
\label{eq:D.10}
\EE_\mathcal{O}^{\{4,1\}} \Longrightarrow
\begin{cases}
\ \ddot{\EE}_\mathcal{O}^{\{4,1\}} = -0{,}2501\ldots\,\text{[eV]}  \\
\ \dot{\EE}_\mathcal{O}^{\{4,1\}} = -0{,}2533\ldots\,\text{[eV]} = \EE_\mathcal{P}^{\{4,1\}} \;,
\end{cases}
\end{equation}
see \textbf{Fig.VII.A}. Thus, the result of this ortho-splitting is that the lower one of
the two ortho-levels does agree with the corresponding para-level whereas the solitary
ortho-level being left over is higher and therefore farer away from to the conventional
prediction $E_\text{c}^{(n)}$. However, the width of this ortho-splitting is clearly
smaller than the deviation of the ortho- and para-levels from their conventional
counterpart: in (\ref{eq:D.9}) the dimorphic splitting width is $(0{,}3900 - 0{,}3808) =
0{,}0092\,\text{[eV]}$ whereas the average deviation of the ortho/para-levels from their
corresponding conventional level is $(0{,}425\ldots$ -- $0{,}385\ldots)\,\text{[eV]}$
$\simeq 0,04\,\text{[eV]}$, which is (roughly) four times larger than the dimorphic
splitting width! See also \textbf{Fig.VII.A}. This result says that the dimorphic splitting
effect is of the same order of magnitude as the degeneracy-elimination effect which for
para-positronium has been estimated by $0{,}42003\,\text{[eV]}$ --
$0{,}37369\,\text{[eV]}$ $\simeq 0,046\ldots\,\text{[eV]}$, see the table on p.~96 of
ref.~\cite{2}.

\emph{Thus, the difference of ortho- and para-energies comes about (in the quadrupole
  approximation) solely through the dimorphic effect!}

  \newpage
  \setcounter{section}{5}
  \setcounter{equation}{0}

  \begin{center}
  {\textbf{\Large Appendix \Alph{section}}}\\[1.5em]
  \emph{\textbf{\Large Octupole Solution}}
  \end{center}
  \vspace{2ex}

  For the calculation of the octupole energy $\ppE^\mathrm{\{e\}}_4$
  (\ref{eq:VIII.92a})--(\ref{eq:VIII.92b}) and its mass equivalent $\pptMMeg_4 \crm^2$
  (\ref{eq:VIII.92c}) we obviously have first to determine the octupole potential
  $\bgA^{\prime\prime}_4(r)$ as the solution of the octupole equation (\ref{eq:VIII.84b}),
  whereas for the amplitude field $\tO(r)$ we will resort to our variational ansatz
  (\ref{eq:VI.1a})--(\ref{eq:VI.1b}). For these purposes, it is very helpful to pass over
  to dimensionless variables; i.\,e. we put again
\begin{subequations}
\begin{align}
\label{eq:E.1a}
y &\doteqdot 2\beta r \\
\label{eq:E.1b}
\bgP_4(y) &\doteqdot \frac{\Gamma(2\nu + 2)}{\gklo{b}{m}'_2} \cdot \frac{\bgA^{\prime\prime}_4(r)}{2\beta \as} \;,
\end{align}
\end{subequations}
so that the octupole equation (\ref{eq:VIII.84b}) adopts its dimensionless form as follows:
\begin{equation}
\label{eq:E.2}
\Delta_y\,\bgP_4(y) - 20\,\frac{\bgP_4(y)}{y^2} = - y^{2\nu - 1}\,\e^{-y} \;.
\end{equation}
Furthermore, the octupole energy $\ppE^\mathrm{\{e\}}_4$ reappears in terms of the dimensionless variables as
\begin{align}
\label{eq:E.3}
\ppE^\mathrm{\{e\}}_4 = -\frac{\e^2}{a_B}\,\left( 2 \beta a_B \right) \cdot
\frac{64}{11025}\, \left(\frac{\gklo{b}{m}'_2}{\Gamma(2\nu + 2)}\right)^2
\int\limits_0^\infty dy\;y^2\,\bigg\{ &\left( \frac{d\,\bgP_4(y)}{dy} \right)^2 \\
+ &20\,\left( \frac{\bgP_4(y)}{y} \right)^2 \bigg\} \;. \nonumber
\end{align}
Comparing now this result to the more concise one (\ref{eq:VIII.96a}) one deduces thereof
the ``anisotropic'' potential function $\eeg_4(\nu)$ as
\begin{align}
\label{eq:E.4}
\eeg_4(\nu) = \frac{64}{11025}\,\frac{1}{\Gamma(2\nu + 2)^2} \int\limits_0^\infty dy\;y^2\,\bigg\{ \left( \frac{d\,\bgP_4(y)}{dy} \right)^2 
+ 20\,\left( \frac{\bgP_4(y)}{y} \right)^2 \bigg\} \ .
\end{align}
By a similar calculation one finds the mass-equivalent function $\mueg_4(\nu)$ of equation (\ref{eq:VIII.96b}) as
\begin{equation}
\label{eq:E.5}
\mueg_4(\nu) = \frac{64}{11025} \cdot \frac{1}{\Gamma(2\nu + 2)^2} \int\limits_0^\infty dy\;y^{2\nu + 1}\,\e^{-y}\cdot\bgP_4(y) \;.
\end{equation}
The claimed octupole identity in coefficient form (\ref{eq:VIII.97}) is now easily
verified by partial integration in (\ref{eq:E.4})-(\ref{eq:E.5}) and use of the octupole equation (\ref{eq:E.2}).
\vskip 0.4cm

\begin{center}
  \large{\textit{Boundary Conditions}}
\end{center}

Before turning to the elaboration of the desired exact solution $\bgP_4(y)$ of the
octupole equation (\ref{eq:VIII.84b}), or (\ref{eq:E.2}), resp., it is necessary to regard
the right boundary conditions. The right-hand side of this octupole equation is the
``octupole source'' and does vanish very fast at infinity ($r \Rightarrow \infty$, or $y
\Rightarrow \infty$, resp.), because we adopt the exponential trial function $\tO(r)$
(\ref{eq:VI.1a})--(\ref{eq:VI.1b}). In contrast to this exponential decay at infinity, our
octupole solution may be assumed to decay much slower (via some power law $y^{-w},
w>0$). But also for sufficiently large value of the trial parameter $\nu$, the trial
amplitude $\tO(r)$ may be assumed to be much smaller near the origin~$(y=0)$ than the
octupole solution $\bgP_4(y)$. Thus we have two regions of three-space where the solution
$\bgP_4(y)$ is to be expected to obey the \emph{homogeneous} octupole equation:
\begin{equation}
\label{eq:E.6}
\Delta_y\,\bgP_4(y) - 20\,\frac{\bgP_4(y)}{y^2} = 0\;.
\end{equation}
From here it is easy to deduce the following limiting behaviour in these asymptotic regions
\begin{subequations}
\begin{align}
\label{eq:E.7a}
\lim_{y \rightarrow \infty} \bgP_4(y) &= \frac{B^{\{4\}}_\infty}{y^5} \\
\label{eq:E.7b}
\lim_{y \rightarrow 0} \bgP_4(y) &= B^{\{4\}}_0 \cdot y^4 \;,
\end{align}
\end{subequations}
where both $B^{\{4\}}_\infty$ and $B^{\{4\}}_0$ are constants.

The constant $B^{\{4\}}_\infty$ characterizes the behaviour at infinity ($y \rightarrow
\infty$) and may be determined by a first integration of the octupole equation
(\ref{eq:E.2}). This can be realized as follows: First, rewrite the octupole equation
(\ref{eq:E.2}) in the following form
\begin{equation}
\label{eq:E.8}
\frac{1}{y^6}\,\frac{d}{dy}\,\left\{ y^{10} \cdot \frac{d}{dy}\,\left( \frac{\bgP_4(y)}{y^4} \right) \right\} = -y^{2\nu - 1}\,\e^{-y} \;.
\end{equation}
Next, integrate here from $y$ to $\infty$ with observation of the boundary conditions
(\ref{eq:E.7a})-(\ref{eq:E.7b}) and find
\begin{equation}
\label{eq:E.9}
y^{10} \cdot \frac{d}{dy}\,\left[ \frac{\bgP_4(y)}{y^4} \right] =  \int\limits_y^\infty dy' \; y'^{2\nu + 5} \cdot \e^{-y'} - \int\limits_0^\infty dy' \;y'^{2\nu + 5}\,\e^{-y'} \;.
\end{equation}
Now take this equation at infinity ($y \rightarrow \infty$) and observe also the boundary
condition (\ref{eq:E.7a}) which then yields
\begin{equation}
\label{eq:E.10}
\lim_{y \rightarrow \infty} \left\{ y^{10} \cdot \frac{d}{dy}\,\left[ \frac{\bgP_4(y)}{y^4} \right] \right\} = -9\,B^{\{4\}}_\infty =- \Gamma(2\nu + 6) \;.
\end{equation}
Thus, the asymptotic constant $B^{\{4\}}_\infty$ (\ref{eq:E.7a}) becomes hereby fixed to
\begin{equation}
\label{eq:E.11}
B^{\{4\}}_\infty = \frac{1}{9} \cdot \Gamma(2\nu + 6) \;.
\end{equation}

In order to try out whether perhaps the first integration step (\ref{eq:E.9}) contains also some
information about the second boundary condition (\ref{eq:E.7b}) one further carries out
the integral on the right-hand side of that equation, i.\,e.
\begin{equation}
\label{eq:E.12}
\int\limits_y^\infty dy'\;y'^{2\nu + 5}\,\e^{-y'} = \left( \int\limits_0^\infty dy'\;y'^{2\nu + 5}\,\e^{-y'} \right) \cdot \left\{ 1 - \e^{-y} \sum_{n=0}^\infty \frac{y^{2\nu + 6 + n}}{\Gamma(2\nu + 7 + n)} \right\} \;.
\end{equation}
By use of this result, the first step of integration (\ref{eq:E.9}) adopts the following shape
\begin{equation}
\label{eq:E.13}
\frac{d}{dy}\,\left[ \frac{\bgP_4(y)}{y^4} \right] = -\Gamma(2\nu + 6)\,\e^{-y} \sum_{n=0}^\infty \frac{y^{2\nu - 4 + n}}{\Gamma(2\nu + 7 + n)} \;.
\end{equation}
The conclusion from this result is that we must demand $\boldsymbol{\nu > 2}$ in order to
have the boundary condition (\ref{eq:E.7b}) satisfied! But more detailed conclusions,
concerning that boundary condition, cannot be drawn from the present result
(\ref{eq:E.13}).
\newpage
\begin{center}
  \large{\textit{Exact Octupole Solution}}
\end{center}

In order to finally get the solution $\bgP_4(y)$ of the first-order equation (\ref{eq:E.13}) it
suggests itself to try a power series of the following form:
\begin{equation}
\label{eq:E.14}
\frac{\bgP_4(y)}{y^4} = \Gamma(2\nu + 6)\e^{-y} \sum_{n=0}^\infty p_n \cdot y^{2\nu - 3 + n} \;.
\end{equation}
This ansatz satisfies the equation (\ref{eq:E.13}) provided the coefficients $p_n$ do obey the following recurrence formula
\begin{equation}
\label{eq:E.15}
p_{n+1} = \frac{p_n}{2\nu - 2 + n} - \frac{1}{(2\nu - 2 + n) \cdot \Gamma(2\nu + 8 + n)} \;,
\end{equation}
with the lowest-order coefficient being given by
\begin{equation}
\label{eq:E.16}
p_0 = - \frac{1}{(2\nu - 3) \cdot \Gamma(2\nu + 7)} \ .
\end{equation}
The solution of (\ref{eq:E.15}) is ($n = 1,2,3,4,\ldots$)
\begin{equation}
\label{eq:E.17}
p_n = \frac{1}{9}\,\left\{ \frac{1}{\Gamma(2\nu + 7 + n)} - \frac{\Gamma(2\nu - 3)}{\Gamma(2\nu + 6)} \cdot \frac{1}{\Gamma(2\nu - 2 + n)} \right\} \;.
\end{equation}
Thus, the (preliminary) solution appears ultimately as
\begin{equation}
\label{eq:E.18}
\bgP_4(y) = \Gamma(2\nu + 6)\,\e^{-y} \sum_{n=0}^\infty p_n \cdot y^{2\nu + 1 + n}
\end{equation}
with the coefficients $p_n$ (\ref{eq:E.16})--(\ref{eq:E.17}). However, observe now that
this solution behaves in the vicinity of the origin ($y=0$) like
\begin{equation}
\label{eq:E.19}
\lim_{y \rightarrow 0} \bgP_4(y) = \Gamma(2\nu + 6)\,p_0\,y^{2\nu + 1 \;,}
\end{equation}
in contrast to the second boundary condition (\ref{eq:E.7b}) (recall here the former
demand $\boldsymbol{\nu > 2}$). But it is possible to add some fourth-order term ($\sim y^4$) to the
solution (\ref{eq:E.18}) because the latter is a solution of the homogeneous version
(\ref{eq:E.6}) of the octupole equation (\ref{eq:E.2})!

For the determination of the correct fourth-order term, to be added to the preliminary
solution (\ref{eq:E.18}), we simplify for a moment the problem by resorting to integer
values of the variational parameter $2\nu$, i.\,e. $2\nu = 5,6,7,8,\ldots$. For this
simplified situation, the preliminary solution (\ref{eq:E.18}) can be transcribed to the
following form:
\begin{align}
\label{eq:E.20}
\bgP_4(y) \Rightarrow \frac{1}{9}\,\Bigg[ \frac{\Gamma(2\nu + 6)}{y^5}\,\left\{ 1 - \e^{-y} \sum_{n=0}^{2\nu + 5} \frac{y^n}{n!} \right\} 
- \Gamma(2\nu - 3) \cdot y^4\,\left\{ 1 - \e^{-y} \sum_{n=0}^{2\nu - 4} \frac{y^n}{n!}
\right\} \Bigg] \ .
\end{align}
But in this form the deficiency of the preliminary solution (\ref{eq:E.18}) becomes now
evident: for approaching infinity ($y \Rightarrow \infty$) we have for integer $2\nu$ in
(\ref{eq:E.20})
\begin{equation}
\label{eq:E.21}
\lim_{y \rightarrow \infty}\bgP_4(y) = - \frac{1}{9}\,\Gamma(2\nu - 3) \cdot y^4 \;,
\end{equation}
which says that the preliminary solution (\ref{eq:E.20}) does not obey the required boundary condition
(\ref{eq:E.7a}) at infinity! We can now easily remedy this deficiency of the preliminary
solution by simply adding to it the negative of the false limit term (\ref{eq:E.21}) which
itself is a solution of the homogeneous form (\ref{eq:E.6}) of the octupole equation
(\ref{eq:E.2}). Thus, the correct octupole solution for integer $2\nu$ can be read off
from equation (\ref{eq:E.20}) as
\begin{align}
\label{eq:E.22}
\bgP_4(y) = \frac{1}{9}\,\Bigg[ \frac{(2\nu + 5)!}{y^5}\,\left\{ 1 - \e^{-y} \sum_{n=0}^{2\nu + 5} \frac{y^n}{n!} \right\} 
+ (2\nu - 4)!\,y^4\,\e^{-y} \sum_{n=0}^{2\nu - 4} \frac{y^n}{n!} \Bigg] \ .
\end{align}

Obviously, this ``integer-valued'' solution $\bgP_4(y)$ consists of two parts:
\begin{equation}
\label{eq:E.23}
\bgP_4(y) = \bgP_4^{(\infty)}(y) + \bgP_4^{(0)}(y) \;,
\end{equation}
with
\begin{subequations}
\begin{align}
\label{eq:E.24a}
\bgP_4^{(\infty)}(y) &= \frac{1}{9}\,\frac{(2\nu + 5)!}{y^5}\,\left\{ 1 - \e^{-y}
  \sum_{n=0}^{2\nu + 5} \frac{y^n}{n!} \right\} =
\frac{(2\nu+5)!}{9\,y^5}\,e^{-y}\cdot\sum_{n=2\nu+6}^\infty \frac{y^n}{n!} \\[5mm]
\label{eq:E.24b}
\bgP_4^{(0)}(y) &= \frac{1}{9}\,(2\nu - 4)!\,y^4\,\e^{-y} \sum_{n=0}^{2\nu - 4}
\frac{y^n}{n!} =
\frac{(2\nu-4)!}{9}\,y^4\left(1-e^{-y}\sum_{n=2\nu-3}^\infty\frac{y^n}{n!} \right)\ ,
\end{align}
\end{subequations}
such that the first part $\bgP_4^{(\infty)}(y)$ (\ref{eq:E.24a}) is responsible for
satisfying the boundary condition (\ref{eq:E.7a}) at infinity; and the second part
$\bgP_4^{(0)}(y)$ (\ref{eq:E.24b}) is responsible for satisfying the other boundary
condition (\ref{eq:E.7b}) near the origin. Both parts $\bgP_4^{(\infty)}(y)$ and
$\bgP_4^{(0)}(y)$ do cooperate in order that their sum $\bgP_4(y)$ (\ref{eq:E.23}) can
actually satisfy the octupole equation (\ref{eq:E.2}), namely via the following
differential relations:
\begin{subequations}
\begin{align}
\label{eq:E.25a}
\left( \Delta_y - \frac{20}{y^2} \right)\,\bgP_4^{(\infty)}(y) &= \frac{1}{9}\,\left\{ (2\nu - 3)\,\e^{-y} \cdot y^{2\nu - 1} - \e^{-y} \cdot y^{2\nu} \right\} \\
\label{eq:E.25b}
\left( \Delta_y - \frac{20}{y^2} \right)\,\bgP_4^{(0)}(y) &= -\frac{1}{9}\,\left\{ (2\nu + 6)\,\e^{-y} \cdot y^{2\nu - 1} - \e^{-y} \cdot y^{2\nu} \right\} \;.
\end{align}
\end{subequations}

Finally, it remains to recast that ``integer-valued'' solution (\ref{eq:E.22}) to the
general case of arbitrary (but real) values of the variational parameter $\nu$. This may
simply be done by adding the negative of the false limit term (\ref{eq:E.21}) to the more
general (but preliminary) solution (\ref{eq:E.18}) which then ultimately yields the
correct solution as
\begin{equation}
\label{eq:E.26}
\bgP_4(y) =  \frac{1}{9}\,\Gamma(2\nu - 3) \cdot y^4 + \Gamma(2\nu + 6)\,\e^{-y}
\sum_{n=0}^\infty p_n\,y^{2\nu + 1 + n} \ .
\end{equation}
\vskip 0.5cm

\begin{center}
  \large{\textit{Mass-Equivalent Function $\mueg_4(\nu)$}}
\end{center}

Once the exact octupole solution $\bg P_4(y)$ for arbitrary value of the variational
parameter $\nu$ is now at hand, cf. (\ref{eq:E.26}), one can go the last step and insert
this solution in the former result (\ref{eq:E.5}) for the mass-equivalent function
$\mueg_4(\nu)$. This then fixes the latter object definitely as follows:
\begin{equation}
\label{eq:E.27}
\mueg_4(\nu) = \frac{64}{11025}\cdot\frac{\Gamma(2\nu+6)}{\Gamma(2\nu+2)^2}
\left\{ \frac{\Gamma(2\nu-3)}{9}+\frac{1}{2^{4\nu+3}}\cdot\sum_{n=0}^{\infty} 
\left(\frac{p_n(\nu)}{2^n}\Gamma(4\nu+3+n) \right) \right\}
\end{equation}
where the coefficients $p_n(\nu)$ have already been specified by equations
(\ref{eq:E.16})-(\ref{eq:E.17}). For integer values of $2\nu$, one substitutes the
solution $\bgP_4(\nu)$ (\ref{eq:E.22}) into the right-hand side of equation (\ref{eq:E.5})
and thus finds the ``integer-valued'' version of (\ref{eq:E.27}) as
\begin{equation}
\label{eq:E.28} 
\rogk{\mu}{e}_4(\nu) = {}^{(\infty)}\rogk{\mu}{e}_4(\nu) + {}^{(0)}\rogk{\mu}{e}_4(\nu)
\end{equation}
with the first part~${}^{(\infty)}\rogk{\mu}{e}_4(\nu)$, being due
to~$\bgP_4^{(\infty)}(y)$ (\ref{eq:E.24a}), being given by
\begin{equation}
\begin{split}
  \label{eq:E.29}
   {}^{(\infty)}\rogk{\mu}{e}_4(\nu) &=
   \frac{64}{99225}\frac{(2\nu+5)!}{[(2\nu+1)!]^2}\left((2\nu-4)!-\sum_{n=0}^{2\nu+5}\frac{(2\nu-4+n)!}{n!\,2^{2\nu-3+n}}\right)\\
   &\equiv \frac{64}{99225}\frac{(2\nu+5)!}{[(2\nu+1)!]^2}\sum_{n=2\nu+6}^\infty\frac{(2\nu-4+n)!}{n!\,2^{2\nu-3+n}}
\end{split}
\end{equation}
and, analogously, the second part~${}^{(0)}\rogk{\mu}{e}_4(\nu)$, being due
to~$\bgP_4^{(0)}(y)$ (\ref{eq:E.24b}), by
\begin{equation}
  \begin{split}
  \label{eq:E.30}
  {}^{(0)}\rogk{\mu}{e}_4(\nu) &=
  \frac{1}{99225}\frac{(2\nu-4)!}{[(2\nu+1)!]^2}\sum_{n=0}^{2\nu-4}
  \frac{(2\nu+5+n)!}{n!\,2^{2\nu+n}}\\*
  &\equiv \frac{64}{99225}\frac{(2\nu-4)!}{[(2\nu+1)!]^2} \left((2\nu+5)! -
    \sum_{n=2\nu-3}^\infty  \frac{(2\nu+5+n)!}{n!\,2^{2\nu+6+n}}\right)\ .
  \end{split}
\end{equation}
The identities emerging in both equations (\ref{eq:E.29}) and (\ref{eq:E.30}) may easily
be validated by tracing them back to the following identity, needed here for
$z=2\nu-4$ and~$z=2\nu+5$:
\begin{equation}
  \label{eq:E.31}
  2^{z+1}\cdot \Gamma(z+1) = \sum_{n=0}^\infty\,\frac{\Gamma(z+n+1)}{2^n\cdot n!}\ .
\end{equation}

\begin{center}
  \large{\textit{Relative Magnitude of Quadrupole and Octupole Corrections}}
\end{center}

In order to better understand the numerical differences between the quadrupole and
octupole corrections (see the table on p.~\pageref{tableDb}) it is helpful to oppose both
perturbation orders $l = 1$ and $l = 2$ to each other on the same notational footing. In
this sense, we recast the quadrupole potential $\gklo{b}{A}''_2(r)$ to its dimensionless
form $\bgP_2(y)$ by putting:
\begin{equation}
\label{eq:E.32}
\gklo{b}{A}''_2(r) \doteqdot \frac{\bg m_1}{\Gamma(2\nu + 2)}\,(2\beta\as) \cdot \bgP_2(y) \;,
\end{equation}
in close analogy to the octupole case (\ref{eq:E.1a})--(\ref{eq:E.1b}). By this rescaling,
the quadrupole equation (\ref{eq:VII.4}) or (\ref{eq:VIII.84a}), resp., becomes simplified
to
\begin{equation}
\label{eq:E.33}
\left( \Delta_y - \frac{6}{y^2} \right)\,\bgP_2(y) = -y^{2\nu - 1} \cdot \e^{-y}
\end{equation}
which says that the dimensionless quadrupole potential $\bgP_2(y)$ feels the same source
($\sim$ right-hande side) as does its octupole counter-part $\bgP_4(y)$,
cf.~(\ref{eq:E.2}).

Indeed, there are many similarities between the former octupole and present quadrupole
modes (presumably these similarities can be generalized to arbitrary $l =
1,2,3,4,\ldots$). For instance, the former octupole boundary conditions
(\ref{eq:E.7a})--(\ref{eq:E.7b}) transcribe to the present quadrupole case as
\begin{subequations}
\begin{align}
\label{eq:E.34a}
\lim_{y \rightarrow \infty} \bgP_2(y) &= \frac{B^{\{2\}}_\infty}{y^3} \\
\label{eq:E.34b}
\lim_{y \rightarrow 0} \bgP_2(y) &= B^{\{2\}}_0 \cdot y^2 \;.
\end{align}
\end{subequations}
The constant $B^{\{2\}}_\infty$ may again be determined by a first integration step for
the quadrupole equation (\ref{eq:E.32}), quite analogously to the octupole situation
(\ref{eq:E.9}):
\begin{equation}
\label{eq:E.35}
y^6\,\frac{d}{dy}\,\left[ \frac{\bgP_2(y)}{y^2} \right] = \int\limits_y^\infty dy'\;y^{\prime\, 2\nu + 3} \cdot \e^{-y'} - \int\limits_0^\infty dy\;y^{2\nu + 3} \cdot \e^{-y} \;.
\end{equation}
From this result, one concludes for the quadrupole limit at infinity (\ref{eq:E.34a})
\begin{equation}
\label{eq:E.36}
B^{\{2\}}_\infty = \frac{1}{5} \cdot \Gamma(2\nu + 4) \;,
\end{equation}
quite analogously to the octupole limit (\ref{eq:E.11}).

The exact solution of the quadrupole equation (\ref{eq:E.32}) can also be determined in
form of a power-series expansion which is found to look as follows~\cite{2}
\begin{equation}
\label{eq:E.37}
\bgP_2(y) = \frac{1}{5}\, \Gamma(2\nu - 1) \cdot y^2 + \frac{1}{5}\,\e^{-y} \sum_{n=0}^\infty b_n(\nu) \cdot y^{2\nu + 1 + n}
\end{equation}
with the coefficients $b_n(\nu)$ being given by
\begin{equation}
\label{eq:E.38}
b_n(\nu) = \frac{\Gamma(2\nu + 4)}{\Gamma(2\nu + 5 + n)} - \frac{\Gamma(2\nu - 1)}{\Gamma(2\nu + n)} \;.
\end{equation}
Thus, the constant $B^{\{2\}}_0$ (\ref{eq:E.34b}) is deduced hereof as
\begin{equation}
\label{eq:E.39}
B^{\{2\}}_0 = \frac{1}{5}\,\Gamma(2\nu - 1) \;,
\end{equation}
which thus turns out as the quadrupole analogue of the octupole limit (\ref{eq:E.7b}). And
finally, the ``integer-valued'' ($2\nu = 2,3,4,\ldots$) version of the general result
(\ref{eq:E.36}) reads
\begin{equation}
\label{eq:E.40}
\bgP_2(y) = \frac{(2\nu + 3)!}{5}\cdot\frac{1 - \e^{-y} \sum\limits_{n=0}^{2\nu + 3}
  \frac{\ds y^n}{\ds n!}}{ y^3} + \frac{(2\nu - 2)!}{5} \cdot y^2\e^{-y} \sum_{n=0}^{2\nu - 2} \frac{y^n}{n!} \;.
\end{equation}
This is obviously the quadrupole analogue of the former octupole result (\ref{eq:E.22}) and
consists again of two parts
\begin{equation}
\label{eq:E.41}
\bgP_2(y) = \bgP^{(\infty)}_2(y) + \bgP^{(0)}_2(y)
\end{equation}
with the first part $\bgP^{(\infty)}_2(y)$ being again responsible for the behaviour (\ref{eq:E.34b}) at infinity
\begin{equation}
\label{eq:E.42}
\bgP^{(\infty)}_2(y) = \frac{(2\nu + 3)!}{5} \cdot \frac{1 - \e^{-y} \sum\limits_{n=0}^{2\nu + 3} \frac{y^n}{n!}}{y^3} \;,
\end{equation}
and the second part $\bgP^{(0)}_2(y)$ near the origin ($y=0$):
\begin{equation}
\label{eq:E.43}
\bgP^{(0)}_2(y) = \frac{(2\nu - 2)!}{5} \,y^2\e^{-y} \cdot \sum_{n=0}^{2\nu - 2} \frac{y^n}{n!} \;.
\end{equation}
Clearly, the sum $\bgP_2(y)$ (\ref{eq:E.40}) must again obey the quadrupole equation
(\ref{eq:E.32}) which, however, is ensured just by the separate differential relations for
the long-range part $\bgP^{(\infty)}_2(y)$ (\ref{eq:E.42})
\begin{equation}
\label{eq:E.44}
\left( \Delta_y - \frac{6}{y^2} \right)\,\bgP^{(\infty)}_2(y) = \frac{2\nu - 1}{5}\, \e^{-y} \cdot y^{2\nu - 1} - \frac{1}{5}\,\e^{-y} \cdot y^{2\nu}
\end{equation}
and the short-range part $\bgP^{(0)}_2(y)$ (\ref{eq:E.42})
\begin{equation}
\label{eq:E.45}
\left( \Delta_y - \frac{6}{y^2} \right)\,\bgP^{(0)}_2(y) = -\frac{2\nu + 4}{5}\,\e^{-y} \cdot y^{2\nu - 1} + \frac{1}{5}\,\e^{-y} \cdot y^{2\nu} \;,
\end{equation}
cf. the octupole case (\ref{eq:E.25a})--(\ref{eq:E.25b}).

The quadrupole energy $\ppEE^\mathrm{\{e\}}_2 = \ppEE^{\{r\}}_2 + \ppEE^{\{\vartheta\}}_2$
(\ref{eq:VIII.91a})-(\ref{eq:VIII.91b}) and also its mass-equivalent $\pptMMeg_2\crm^2$
(\ref{eq:VIII.91c}) may now be re-expressed in terms of the dimensionless quadrupole
potential $\bgP_2(y)$ and thus read as follows
\begin{subequations}
\begin{align}
\label{eq:E.46a}
\ppEE^\mathrm{\{e\}}_2 &= -\frac{\e^2}{a_B}\,(2\beta a_B)\,\left( \bg m_1 \right)^2 \cdot \eeg_2(\nu) \\
\label{eq:E.47b}
\pptMMeg_2\crm^2 &= -\frac{\e^2}{a_B}\,(2\beta a_B)\,\left( \bg m_1 \right)^2 \cdot \mueg_2(\nu) \;,
\end{align}
\end{subequations}
with the electrostatic function $\eeg_2(\nu)$ being given by
\begin{equation}
\label{eq:E.48}
\eeg_2(\nu) \doteqdot \frac{45}{4}\,\frac{1}{\Gamma(2\nu + 2)^2} \int\limits_0^\infty dy\;y^2\,\left\{ \left( \frac{d\,\bgP_2(y)}{dy} \right)^2 + 6\,\left( \frac{\bgP_2(y)}{y} \right)^2 \right\}
\end{equation}
and the corresponding mass-equivalent function $\mueg_2(\nu)$ by
\begin{equation}
\label{eq:E.49}
\mueg_2(\nu) = \frac{45}{4}\,\frac{1}{\Gamma(2\nu + 2)^2} \int\limits_0^\infty dy\;y^{2\nu + 1}\,\e^{-y} \cdot \bgP_2(y) \;.
\end{equation}
Just as in the octupole case (\ref{eq:VIII.97}), there is here also a corresponding quadrupole identity
\begin{equation}
\label{eq:E.50}
\eeg_2 \equiv \mueg_2
\end{equation}
which may easily be verified by integrating by parts in (\ref{eq:E.48}) and using the
quadrupole equation (\ref{eq:E.33}). Observe here that these \emph{separate} identities
(\ref{eq:VIII.97}) and (\ref{eq:E.48}) can occur only because the \emph{separative} method
admitted us to \emph{decouple completely} the quadrupole and octupole modes! 

Now that both modes are elaborated in great detail, one can proceed to the intended study
of their numerical relationships.  

The qualitative features of these multipole solutions~$\logk{b}{{\cal P}}_2(y)$
and~$\logk{b}{{\cal P}}_4(y)$ may perhaps be best seen from their sketch,
cf.~\textbf{Fig.E.I} below. For~$y\gtrsim 20$ one actually observes the asymptotic
behaviour (\ref{eq:E.34a}) and (\ref{eq:E.7a}) for~$\logk{b}{{\cal P}}_{2/4}(y)$. In a
similar way, the behavior (\ref{eq:E.7b}) and (\ref{eq:E.34b}) near the origin~$(y=0)$ is
also neatly displayed by \textbf{Fig.E.I}. Furthermore, one expects that any multipole
solution~$\logk{b}{{\cal P}}_{2l'}(y)$ of higher order~$l'$ is smaller (for fixed value
of~$\nu$) than its counterpart~$\logk{b}{{\cal P}}_{2l}(y)$ due to a lower multipole
order~$l\,(<l')$. For~$l'=2,l=1$ this expectation is also clearly expressed by
\textbf{Fig.E.I} below. Observe, however, that the magnitudes of the multipole
solutions~$\logk{b}{{\cal P}}_{2,4}(y)$ are not directly responsible for the magnitudes of
the corresponding multipole energy corrections.  \

\begin{center}
\epsfig{file=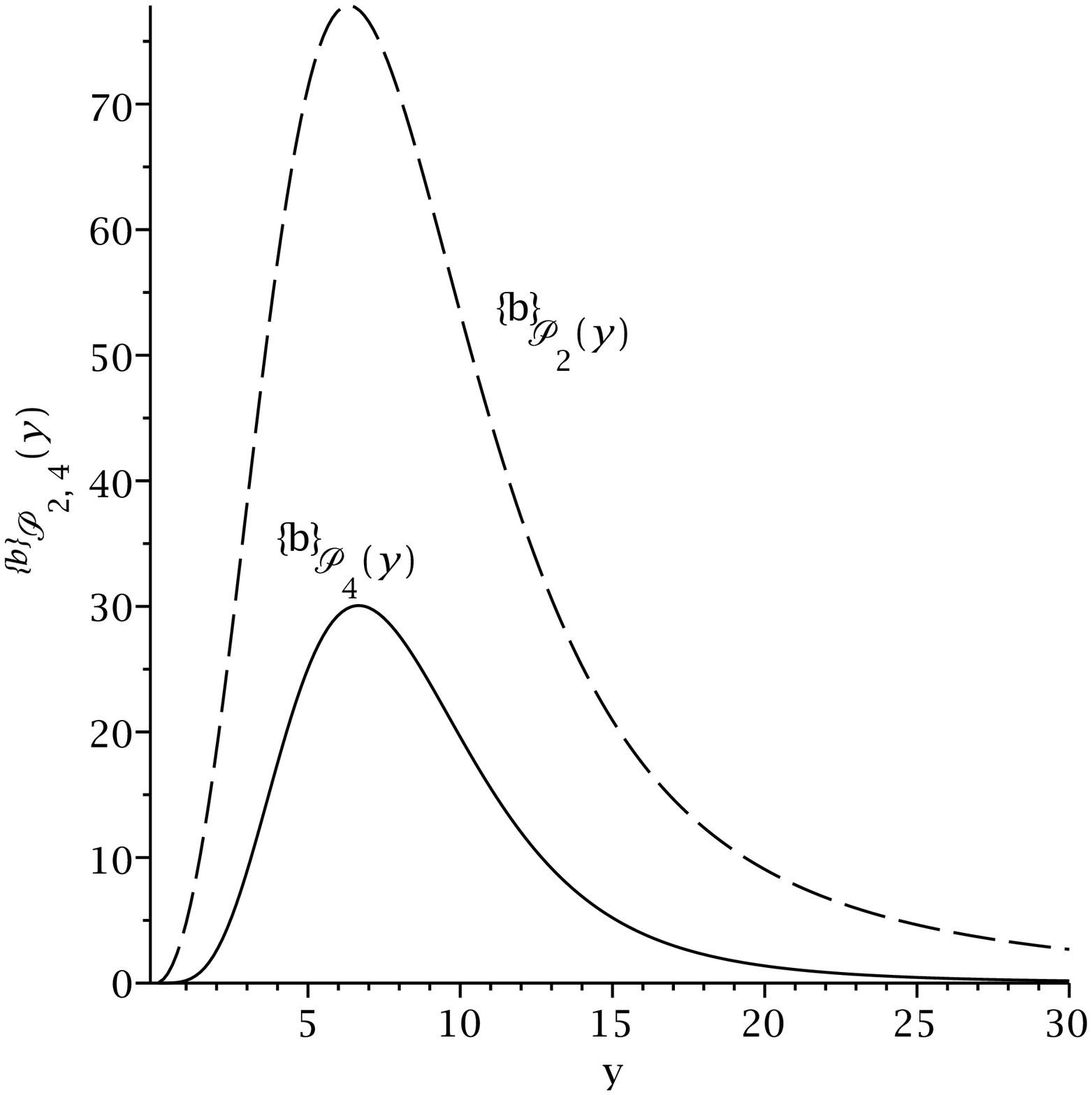,height=18cm}
\end{center}
%\vspace{-1.5cm}
{\textbf{Fig.E.I}\hspace{5mm} \emph{\large\textbf{\boldmath Quadrupole Solution 
      $\logk{b}{\cal P}_2(y)$ (\ref{eq:E.36}) and Octupole \\
      \phantom{Fig.E.1  }Solution
       $\logk{b}{\cal P}_4(y)$  (\ref{eq:E.26}) for~$\nu=3$  } } }

\label{figE.1}

%%% Local Variables: 
%%% mode: latex
%%% TeX-master: "main"
%%% End: 
 
\newpage 

Both multipole energy corrections~$\pptMMeg_2 c^2$ (\ref{eq:E.47b}) and~$\pptMMeg_4 c^2$
(\ref{eq:VIII.96b}) do essentially appear as the product of some reference
energy~$\rogk{\mu}{e}_2(\nu)$ and~$\rogk{\mu}{e}_4(\nu)$ times the mass-equivalent
parameter~$(\logk{b}{m}_1)^2$ or~$(\logk{b}{m}'_2)^2$, resp. Here, the reference
energies~$\rogk{\mu}{e}_{2/4}(\nu)$ (\ref{eq:E.5}) and (\ref{eq:E.48}) do not depend on
the quantum number~$\bjz$ of angular momentum but are determined solely by the spherically
symmetric trial density~$\tilde{\Phi}^2(r)$ (\ref{eq:VI.1a})-(\ref{eq:VI.1b}). Thus the
``anisotropic'' energy corrections are composed of two different effects, one of
``radial'' and the other of ``angular'' type. The latter one is measured by the
mass-equivalent parameters~$\logk{b}{m}_1$ and~$\logk{b}{m}'_2$ which then play the part
of quadrupole and octupole ``strengths'' and are displayed by the table
on~p.~\pageref{table8}. From the entries of that table it becomes clear that the octupole
strength~$\logk{b}{m}'_2$ is larger than its quadrupole counterpart~$\logk{b}{m}_1$
(i.e.~$\logk{b}{m}'_2>\logk{b}{m}_1$) which is of course due to the specific anisotropy of
the charge distributions, see \textbf{Fig.IV.A}. But this does not yet admit the
conclusion that the octupole corrections are larger than their quadrupole counterparts.

Indeed, it is necessary to take into account also the second influence on the
``anisotropic'' energy corrections, i.e.\ the reference
energies~$\rogk{\mu}{e}_{2/4}(\nu)$. The subsequent \textbf{Fig.E.II} presents a plot of
both functions $\rogk{\mu}{e}_2(\nu)=\frac{9}{4}\,\rogk{\mu}{e}_{\sf III}(\nu)$
(\ref{eq:VII.15}) and~$\rogk{\mu}{e}_4(\nu)$ (\ref{eq:E.27}). Observe here that, just as
for $\rogk{\mu}{e}_{\sf III}(\nu)$ (\ref{eq:D.3b}), there does exist also a closed
analytic form for~$\rogk{\mu}{e}_4(\nu)$~(\ref{eq:E.27}):
\begin{equation}
  \label{eq:E.51}
  \begin{split}
    \rogk{\mu}{e}_4(\nu) &=
    \frac{64}{99225}\frac{4\nu^4+28\nu^3+71\nu^2+77\nu+30}
    {\nu\,(2\,\nu-3)\,(\nu-1)\left(4\,\nu^2-1\right)} \\ &-
\frac{64}{33075}\frac{\left(12\,\nu^4+44\,\nu^3+81\,\nu^2 
  +67\,\nu+20\right)\Gamma(2\,\nu+3/2)}
{ \sqrt{\pi}\,\nu\,(2\,\nu-3)\,(\nu-1)\left(4\,\nu^2-1\right)\Gamma(2\,\nu+2) }\ .
  \end{split}
\end{equation}
\begin{center}
\epsfig{file=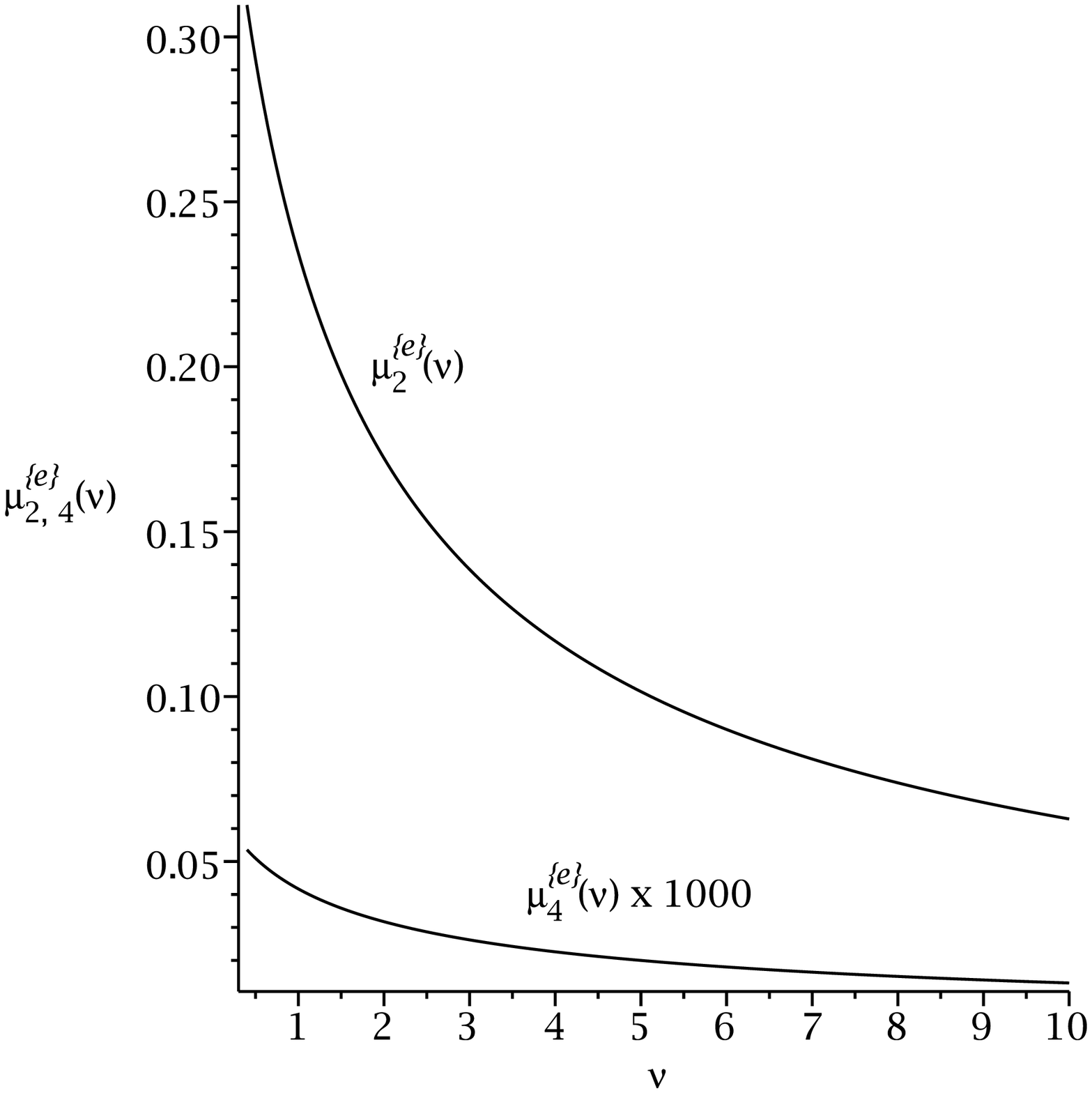,height=18cm}
\end{center}
%\vspace{-1.5cm}
{\textbf{Fig.E.II}\hspace{5mm} \emph{\large\textbf{\boldmath Quadrupole
      Function~$\rogk{\mu}{e}_2(\nu)=\frac{9}{4}\rogk{\mu}{e}_{\sf III}(\nu)$
      (\ref{eq:VII.15}) and \phantom{Fog.E.2} Octupole Function $\rogk{\mu}{e}_4(\nu)$ (\ref{eq:E.27}) } } }

\label{figE.2}

%%% Local Variables: 
%%% mode: latex
%%% TeX-master: "main"
%%% End: 

\renewcommand{\refname}{{\bfseries \Large References}}

\end{document}